\documentclass[12pt]{article}

\usepackage{lineno,hyperref}
\modulolinenumbers[5]

\usepackage[T1]{fontenc} 
\usepackage[utf8]{inputenc}
\usepackage{newtxtext}
\usepackage{newtxmath}
\usepackage{amsmath}
\usepackage{graphicx}
\usepackage[dvipsnames]{xcolor}
\numberwithin{equation}{section}
\usepackage{array}
\usepackage{mathtools}
\usepackage{dsfont}
\usepackage{mathrsfs}
\usepackage{tikz}
\usepackage{multirow}
\usetikzlibrary{matrix,fit,decorations.pathmorphing,decorations.markings,calc, angles,positioning,intersections}
\definecolor{e1}{RGB}{255,0,0}
\definecolor{e6}{RGB}{255,165,0}
\definecolor{e4}{RGB}{255,255,0}
\definecolor{e3}{RGB}{0,255,0}
\definecolor{e2}{RGB}{0,0,255}
\definecolor{e7}{RGB}{75,0,130}
\definecolor{e5}{RGB}{238,130,238}
\definecolor{e8}{RGB}{0,191,255}
\newcommand{\cir}[1]{\tikz\draw[black, fill = #1, radius=3pt] (0,0) circle ;}%
\definecolor{mygreen1}{RGB}{0, 204, 102}

\usepackage{varwidth}
\usepackage{enumerate}
\usepackage{empheq}
\usepackage{youngtab}
\usepackage{setspace}
\usepackage{upgreek}
\usepackage{color}
\usepackage{slashed}
\usepackage{stmaryrd}

\newcolumntype{L}[1]{>{\raggedright\let\newline\\\arraybackslash\hspace{0pt}}m{#1}}
\newcolumntype{C}[1]{>{\centering\let\newline\\\arraybackslash\hspace{0pt}}m{#1}}
\newcolumntype{R}[1]{>{\raggedleft\let\newline\\\arraybackslash\hspace{0pt}}m{#1}}

\newcommand{\boldlangle}{\left\langle}
\newcommand{\boldrangle}{\right\rangle}

\newdimen\mytextwidth
\newcommand\rem[2][cyan!40!green]{\noindent\nobreak\hfil\penalty1000\hfilneg
\mytextwidth=\linewidth\advance\mytextwidth by 2mm%
\begin{tikzpicture}[baseline=-\the\dimexpr\fontdimen22\textfont2\relax]\node[outer sep=0pt,draw=black,fill=#1,fill opacity=1,text opacity=1,rectangle,rounded corners]{\begin{varwidth}{\mytextwidth}\textcolor{white}{#2}\end{varwidth}};
\end{tikzpicture}\allowbreak%
}

\newcommand{\dd}{\partial}
\newcommand{\bd}{\overline{\partial}}
\newcommand{\CP}{\mathds{CP}}
\newcommand{\CC}{\mathds{C}}

\renewcommand{\bar}{\overline}
\renewcommand{\tilde}{\widetilde}

\newcommand{\bea}{\begin{equation}}
\newcommand{\eea}{\end{equation}}
\newcommand{\bear}{\begin{eqnarray}}
\newcommand{\eear}{\end{eqnarray}}
\newcommand{\bearr}{\begin{eqnarray*}}
\newcommand{\eearr}{\end{eqnarray*}}

\newcommand{\fl}{\mathcal{F}}
\newcommand{\Zz}{\mathbb{Z}}

\newcommand{\tr}{\mathrm{Tr}}

\newcommand{\fO}{\mathcal{O}}

\newcommand{\im}{\mathrm{i}\,}
\newcommand{\dx}{\partial_x}

\newcommand{\as}{\mathrm{a}}
\newcommand{\bs}{\mathrm{b}}
\newcommand{\cs}{\mathrm{c}}
\newcommand{\xs}{\mathrm{x}}
\newcommand{\ys}{\mathrm{y}}
\newcommand{\zs}{\mathrm{z}}
\newcommand{\ol}{\frac{1}{L}}
\newcommand{\olsq}{\frac{1}{L^2}}
\newcommand{\fL}{\mathcal{L}}
\newcommand{\bDelta}{\boldsymbol{\Delta}}

\newcommand{\SU}{\text{SU}}

\newcommand{\tPSU}{\text{PSU}}
\newcommand{\tSU}{\text{SU}}    
\newcommand{\tGL}{\text{GL}}
\newcommand{\tU}{\text{U}}

\renewcommand{\bar}{\overline}
\renewcommand{\tilde}{\widetilde}
\newcommand{\Rc}{\mathcal{R}}

\DeclareFontFamily{U}{solomos}{}
\DeclareErrorFont{U}{solomos}{m}{n}{10}
\DeclareFontShape{U}{solomos}{m}{n}{
  <-> s*[1.1]  gsolomos8r
}{}
\newcommand{\vkappa}{\text{\usefont{U}{solomos}{m}{n}\symbol{'153}}}

\newcommand{\diff}{\mathrm{d}}

\newcommand{\Diff}{{\mathcal{D}}}

\makeatletter

\newlength{\fboxhsep}
\newlength{\fboxvsep}

\newlength{\fboxtoprule}
\newlength{\fboxbottomrule}
\newlength{\fboxleftrule}
\newlength{\fboxrightrule}

\def\@frameb@xother#1{%
  \@tempdima\fboxtoprule
  \advance\@tempdima\fboxvsep
  \advance\@tempdima\dp\@tempboxa
  \hbox{%
    \lower\@tempdima\hbox{%
      \vbox{%
        \hrule\@height\fboxtoprule
        \hbox{%
          \vrule\@width\fboxleftrule
          #1%
          \vbox{%
            \vskip\fboxvsep
            \box\@tempboxa
            \vskip\fboxvsep}%
          #1%
          \vrule\@width\fboxrightrule}%
        \hrule\@height\fboxbottomrule}%
    }%
  }%
}

\long\def\fboxother#1{%
  \leavevmode
  \setbox\@tempboxa\hbox{%
    \color@begingroup
    \kern\fboxhsep{#1}\kern\fboxhsep
    \color@endgroup}%
  \@frameb@xother\relax}

\makeatother

\let\mysub=\subset

\usepackage{tocloft}

\addtocontents{toc}{%
    \protect}

\let \savenumberline \numberline
\def \numberline#1{\savenumberline{#1.}}

\usepackage{etoolbox}
\patchcmd{\tableofcontents}{\@starttoc}{\vspace{-0.3cm}\@starttoc}{}{}

\usepackage{hyperref}
\hypersetup{
    colorlinks=false,
    linkbordercolor	=BlueViolet,
citebordercolor	=OliveGreen,
urlbordercolor=RoyalBlue
}

\usepackage{onimage} 

\newcommand\scalemath[2]{\scalebox{#1}{\mbox{\ensuremath{\displaystyle #2}}}}

\usepackage[
    backend=bibtex,
maxbibnames=99,
giveninits=true,
minalphanames=1,
maxalphanames=3,
url=false,
isbn=false
  ]{biblatex}
  
  \bibliography{reviewrefs}

\newbibmacro{string+doi}[1]{%
  \iffieldundef{doi}{#1}{\href{http://dx.doi.org/\thefield{doi}}{#1}}}
\DeclareFieldFormat{title}{\usebibmacro{string+doi}{\mkbibemph{#1}}}
\DeclareFieldFormat[article]{title}{\usebibmacro{string+doi}{\mkbibquote{#1}}}

\title{\Huge  Flag manifold sigma models\\
{\Large \emph{Spin chains and integrable theories}}}

\usepackage{titling}
\usepackage{authblk}

  \makeatletter
\def\@maketitle{%
\newpage%
\null%
\begin{center}%
    \let\footnote\thanks %
    {\LARGE \@title %
      \par 
    }
   \vspace{1cm} 
    {\large
     \lineskip .5em
     \begin{tabular}[t]{c}
        \baselineskip=12pt
        \@author
     \end{tabular}
     \par
    }
    {\large \@date}
\end{center}
\par 
\vskip 1.5em} 
\makeatother

\renewbibmacro{in:}{%
  \ifentrytype{article}
    {}
    {\printtext{\bibstring{in}\intitlepunct}}}

\begin{document}
\begin{titlingpage}

\author{Ian Affleck}
\author{Kyle Wamer}
\affil{\small Department of Physics and Astronomy and Stewart Blusson Quantum Matter Institute, \\University of British Columbia, Vancouver, B.C., Canada, V6T1Z1}
\affil[]{\emph{iaffleck@phas.ubc.ca, kylewamer@phas.ubc.ca}}
\author{Dmitri Bykov}
\affil{\small Steklov Mathematical Institute of Russian Academy of Sciences, Moscow, Russia}
\affil[]{\emph{bykov@mi-ras.ru, dmitri.v.bykov@gmail.com}}
\date{}
   \maketitle
    \begin{abstract}
This review is dedicated to two-dimensional sigma models with flag manifold target spaces, which are generalizations of the familiar $\CP^{n-1}$ and Grassmannian models. They naturally arise in the description of continuum limits of spin chains, and their phase structure is sensitive to the values of the topological angles, which are determined by the representations of spins in the chain. Gapless phases can in certain cases be explained by the presence of discrete 't Hooft anomalies in the continuum theory. We also discuss integrable flag manifold sigma models, which provide a generalization of the theory of integrable models with symmetric target spaces. These models, as well as their deformations, have an alternative equivalent formulation as bosonic Gross-Neveu models, which proves useful for demonstrating that the deformed geometries are solutions of the renormalization group (Ricci flow) equations, as well as for the analysis of anomalies and for describing potential couplings to fermions.
\end{abstract}

\vspace{2cm}
\centering
Prepared for \emph{Physics Reports}
\end{titlingpage}

\tableofcontents

\pagebreak

\vspace{0.5cm}
\noindent
\rule{\textwidth}{1pt}
    \vspace{1ex}
\begin{center}
\vspace{-0.3cm}
{\Large    Introduction}
\end{center}

\noindent
\vspace{-0.5ex}%
\rule{\textwidth}{1pt}

\addcontentsline{toc}{section}{\bfseries Introduction}

\vspace{2cm}
Haldane's conjecture is the prediction that antiferomagnetic spin chains with integer spin have a gap above the ground state, while those with half-odd integer spin are gapless~\cite{Haldane1983}. The distinction between these two cases can be seen by taking a large spin limit, in which case the quantum fluctuations of the antiferromagnet are governed by the O(3) nonlinear sigma model, with topological angle $\theta=2\pi S$.
It was surprising to condensed matter physicists that spin chains were gapped for 
integer spin and surprising to high energy theorists that the O(3) non-linear sigma 
model was massless for $\theta =\pi$.  
Recently, this paradigm of mapping spin chains to relativistic quantum field theories has been generalized to SU($n$) chains in various representations~\cite{BykovHaldane1,BykovHaldane2,HaldaneSU3,Wamer2019,HaldaneSUN,Wamer2020}. For chains that have a rank-$p$ symmetric representation at each site, the corresponding field theory is a sigma model with target space $\SU(n)/[\tU(1)]^{n-1}$. This space is an example of a flag manifold, which generalizes the familiar notions of complex projective space and Grassmannian manifolds, and in this case may be parametrized in terms of $n$ mutually orthonormal fields $z_A \in \mathbb{C}^n$. To each of these fields there is an associated topological angle $\theta_A = 2\pi p A/n$, which extends Haldane's original result, since $p=2S$ in SU(2). Based on this sigma model formulation, a generalization of Haldane's conjecture was discovered for these SU($n$) chains: When $p$ is coprime with $n$, gapless excitations will be present above the ground state; for all other values of $p$, a finite energy gap will occur, with a ground state degeneracy equal to $n/\gcd(n,p)$.

The arguments leading to this SU($n$) version of Haldane's conjecture draw from many areas of mathematical physics. This reflects the fact that the underlying flag manifold $\SU(n)/[\tU(1)]^{n-1}$ has a rich geometric structure. Indeed, flag manifolds in their own right are a fascinating subject, and for this reason we commence this review in Chapter 1 by discussing generic flag manifolds at great length. In particular, we will explain their symplectic, K\"ahler, and Riemannian geometries, as well as their cohomology, the latter being the key object for the description of topological terms. In addition to providing the reader with an overview of the general theory of flag manifolds, this chapter will allow us to introduce the necessarily technology to properly explain the mathematical underpinnings of deriving a flag manifold sigma model from an SU($n$) spin chain. Along these lines, we also review various quantization schemes of flag manifolds, and how a coherent state path integral is constructed in this context.

In Chapter 2, we turn to SU($n$) spin chains. In the interest of being self-contained, we begin by introducing the SU($n$) Heisenberg Hamiltonian, and listing various exact results that are known for these models. Then, armed with the mathematical formalism of Chapter~1, we review in great detail how the $\SU(n)/[\tU(1)]^{n-1}$ flag manifold arises as a low-energy sigma model description of the SU($n$) chain. In particular, we show how the topological angle $\theta_A$ arises as the coefficient of a Fubini-Study two-form, pulled back from $\CP^{n-1}$ to the flag manifold. We then proceed to discuss a technical issue that is related to the absence of Lorentz invariance when starting with a generic SU($n$) chain Hamiltonian. Having done this, we may then finally review the constituent arguments that make up the SU($n$) Haldane conjecture. In particular, we discuss the notion of`t Hooft anomaly matching, which is related to the inability of gauging the physical PSU($n$) symmetry of the chain while maintaining a discrete $\mathbb{Z}_n$ translation symmetry~\cite{Tanizaki:2018xto,ohmori2019sigma}. We also discuss topological excitations in the sigma model, which have fractional charge and give rise to a mass generating mechanism except for the special values of $p$ with $\gcd(n,p)=1$~\cite{WamerMerons}. Finally, we conclude Chapter 2 by listing other representations of SU($n$) that may also be mapped to the same flag manifold, $\SU(n)/[\tU(1)]^{n-1}$.

One might expect that this would be a natural point to conclude this review: We have covered the general properties of flag manifolds, and explained in great detail the relationship between said manifolds and SU($n$) spin chains, allowing for a generalization of Haldane's famed conjecture. However, this work on SU($n$) chains has very recently initiated an entirely new research program, related to integrable flag manifold sigma models. This is the subject of Chapter 3.

The history of integrable models with an `infinite number of degrees of freedom' is rather long. It has spanned most of the second part of the 20th century, starting with the study of the Korteweg-de-Vries equation~\cite{GGKM}, and continues to evolve up to the present day. Already by the end of the 1970s the classical theory saw remarkable developments based on  algebro-geometric methods and the tools of finite-gap integration, as summarized in the book~\cite{NMPZ}. On the other hand, the study of integrable structures of relativistic sigma models only started around the same time~\cite{Pohlmeyer, Zakharov1}, and the mathematical results on the classification of classical solutions were obtained substantially later~\cite{Uhlenbeck, HitchinTori}. See~\cite{HarmonicMapsBook, GuestBook} for a review of these findings.

Whereas the classical integrability theory  quickly came to be part of mathematics, the quantum theory was developed by rather different methods by physicists, starting with the famous conjecture for the $S$-matrix in  the $S^{n-1}$ sigma model~\cite{Zamolodchikov}. The development of this theory then went in two directions: towards the calculation of the spectrum in finite volume, using the so-called thermodynamic Bethe ansatz~\cite{YangYang, LiebLiniger, ZamolodchikovTBA, DoreyTateo, BazhanovExcited}, and towards investigating the full range of theories, to which such methods would be applicable. Within the latter research program remarkable results were achieved for models with $\tSU(n)$ symmetry, most importantly for the $\CP^{n-1}$-model~\cite{Cremmer, DAdda1, AddaSUSY, WittenCPN, BergWeisz}. First of all, it was found that quantum-mechanically integrability in this model is destroyed by  anomalies of a very peculiar kind. Technically these are anomalies in a certain non-local charge first constructed by L\"uscher~\cite{LuscherNonlocal}, which, when unobstructed, may be shown to generate the Yangian that underpins the integrability of these models~\cite{Bernard1, Bernard2} (see~\cite{Loebbert} for a review). This would as well lead to anomalies in the `higher' local charges, as anticipated earlier in~\cite{PolyakovCharges, Goldschmidt} based on simple dimensional analysis. It was also found that, by adding fermions to the pure bosonic models in various ways, one can cancel the anomalies, although at the conceptual level the mechanism behind these cancellations remained unclear.

Another major stumbling block was that the theory of integrable sigma models  -- both classical and quantum -- seemed to require that the target space is a \emph{symmetric} space, which substantially narrows the space of admissible models, even within the class of homogeneous spaces. In recent years the latter issue has been resolved, at least in the classical theory, since it was shown~\cite{BykovFlag1, BykovFlag2, BykovFlag3, BykovGLSM1, BykovGLSM2} that there exist canonical models with flag manifold target spaces (which in general are not symmetric)  that admit a Lax representation and share the virtues of the models with symmetric target spaces. This also allows one to make a connection to the models that emerge from the spin chains discussed in Chapter 2. Although the integrable models are not exactly identical to the ones that arise from the spin chains, they nevertheless share many common features with the latter.  Even more recently the paper~\cite{CYa} appeared, which provides a broad and unified framework for constructing classical integrable models starting from a rather exotic `four-dimensional semi-holomorphic Chern-Simons theory'. In particular, the flag manifold models may as well be obtained from that construction.

Quite unexpectedly, it turned out that the approach of~\cite{CYa}, combined with the gauged linear sigma model approach developed earlier in~\cite{BykovGLSM1, BykovGLSM2}, allows one to prove the equivalence of a wide class of sigma models with complex homogeneous target spaces (as well as their deformations) to bosonic and mixed bosonic/fermionic Gross-Neveu models~\cite{BykovGN}. This novel formulation provides insights into many facets of sigma model theory. For example, one can obtain a new way of constructing supersymmetric sigma models~\cite{BykovSUSYCPN}, and the obscure integrability anomalies are now conjectured to be related to the familiar chiral anomalies, which are otherwise not visible in the old approach. The Gross-Neveu formulation provides another window into the quantum domain, related to the analysis of the $\beta$-function of the theory. This is especially vivid in the deformed case. Since the deformation preserves only a small fraction of the original symmetries of the model, the explicit calculations in the geometric framework would be extremely cumbersome, if at all doable. In contrast, the Gross-Neveu formulation results in spectacular simplifications, which ultimately allow one to solve the generalized Ricci flow equations for the deformed geometries in a very wide class of sigma models. This is particularly important, since in the study of models with target spaces $S^2$ and $S^3$, the one-loop renormalizability of the deformed models was linked to their integrability~\cite{FOZ, Fateev} (see also the more recent discussion in~\cite{ValentKlimcik, HoareTseytlin} and references therein). We mention in passing that the subject of integrable deformations is in itself very vast, and for more on this we refer the reader to the well-known papers~\cite{Klimcik, Klimcik2, VicedoSymm, Sfetsos, Schmidtt}.

It is unlikely that all of these exciting inter-relations are purely a coincidence.  Instead, one can be optimistic that from this point the construction of the proper quantum theory of such models is within reach. Additionally, the inclusion of the non-trivial $\theta$-angles would allow one to study the phase diagram and draw parallels to the massless/massive phases of spin chains, which would then close the logical circle that we are aiming to reflect in this review article.

\newpage

\section*{Notation}

Before we begin, we comment on the various notational choices that we have made in this review. 
\begin{itemize}
    \item[$\circ$] A generic flag manifold can be embedded into a copy of $m$ Grassmanians (this will be explained in detail in Chapter 1). We use upper case Roman letter to index these copies. In the case $m=n$, each Grassmanian is $\CP^{n-1}$, which we parametrize with $z_A\in\mathbb{C}^n$. When $m<n$, and multiple $n$-component fields are required to parametrize the Grassmanians, we use lower-case roman letters, i.e. $z_A^{(k)}$.
    \item[$\circ$]  The $n$ components of $z_A$ are indexed using lower case greek letters: $z_A^\alpha$
    \item[$\circ$] Often, we will normalize the $n$-component fields to satisfy $|z_A|=1$. In this case, we write $u_A$ instead of $z_A$. 
    \item[$\circ$] We use the labels $a,b,c...$ for discrete time coordinates, and the labels $i,j,k,...$ for discrete spatial coordinates. For a field $z$ that is a function of $a$ and $j$, we write $z = z(a,j)$. When the continuum limit is taken, we write $z = z(\tau,x)$.
    \item[$\circ$] The vector complex conjugate to $z$ is written $\overline z$. We write inner products in $\mathbb{C}^n$ according to
    \begin{equation}
        \overline{w} \circ z = \sum_{\alpha=1}^n \overline{w}^\alpha z^\alpha.
    \end{equation}
    The norm of a vector is denoted by $|z|$, so that $|z|^2=\bar{z}\circ z$.
    \item[$\circ$] From time to time we will be using the notation $\mathrm{Hom}(\CC^p, \CC^q)$ (linear maps from $\CC^p$ to~$\CC^q$) for the space of $p\times q$-matrices. This notation makes it clear that $p$ is the number of columns, and $q$ the number of rows in a matrix. Accordingly $\mathrm{End}(\CC^p)$ is the space of square matrices of size $p$.
\end{itemize}

\pagebreak

\vspace{0.5cm}
\noindent
\rule{\textwidth}{1pt}
    \vspace{1ex}
\begin{center}
\vspace{-0.3cm}
{\Large    Chapter 1. Flag manifolds: geometry and first applications}
\end{center}

\noindent
\vspace{-0.5ex}%
\rule{\textwidth}{1pt}

\addcontentsline{toc}{section}{\bfseries Chapter 1. Flag manifolds: geometry and first applications}
\label{geomsec}

\vspace{2cm}
In the first chapter of this review we recall the main facts about the rather rich geometric structures on flag manifolds (mostly symplectic structures and metrics), and we explain how flag manifolds naturally arise in representation theory. Due to this tight relation, flag manifolds inevitably appear in the theory of spin chains, to which the next chapter is dedicated. As a bridge between abstract mathematical structures and applications to representations of spin operators, we describe the example of a spin carried by a mechanical particle charged w.r.t. a non-Abelian gauge group: in this case the motion of the spin is again described by a flag manifold.

\section{The geometry of SU($n$) flag manifolds}

Flag manifolds are natural generalizations of both projective spaces and Grassmanians, so we start by recalling these more familiar entities first.

The complex projective space $\CP^{n-1}$ is defined as the space of $n$-tuples of complex numbers, which are not all zero, defined up to multiplication by an overall factor, i.e. $(z^1, \cdots, z^n)\sim \uplambda  (z^1, \cdots, z^n)$. Another interpretation, which allows for generalizations more easily, is that $\CP^{n-1}$ is the space of lines in $\CC^n$, passing through the origin. Clearly, the line is defined by a non-zero vector $(z^1, \cdots, z^n)$, and two vectors that differ by an overall constant multiple define the same line.

This construction can be generalized by considering $k$-dimensional planes in $\CC^n$, passing through the origin. This leads to the notion of a Grassmannian $Gr_{k, n}$, which may be defined as the space of $k\times n$-matrices $Z$ of rank $k$, taken up to multiplication by matrices from $\tGL(k, \CC)$. The meaning of $Z$ is that it comprises $k$ vectors spanning a given $k$-dimensional plane in $\CC^n$, and multiplication by GL$(k, \CC)$ amounts to a change of basis and does not affect the plane itself. Setting $k=1$ one gets back to the projective space $\CP^{n-1}$.

We should point out that the equivalence relations just mentioned -- the quotients w.r.t. $\CC^\ast$ or $\tGL(k, \CC)$ -- are of course well-known in physics as `gauge redundancies'. In fact, more than once in our narrative we will encounter the formulation of the corresponding field theories as systems with gauge fields (the so-called `gauged linear sigma models'). From the mathematical perspective, choosing a gauge amounts to picking coordinates on the respective manifolds. The unrestricted coordinates $(z^1, \cdots, z^n)$ mentioned above, which are subject to the equivalence relation, are known as the \emph{homogeneous coordinates} on the projective space. If, say, $z^1\neq 0$, then by a $\CC^\ast$ scaling -- a gauge transformation -- we may set $z^1=1$. This fixes the gauge freedom completely at the expense of effectively excluding from consideration the part of the space where $z^1=0$. The corresponding coordinates $(1, z^2, \cdots, z^n)$ are then known as the \emph{inhomogeneous coordinates}. For example, on a sphere $S^2\simeq \CP^1$ there is a single complex inhomogeneous coordinate, which is the complex coordinate on a plane of stereographic projection (the excluded region in this case being the point from which the projection is performed). Finally, another option is to fix the gauge redundancy partially by normalizing the coordinates $\sum\limits_{\alpha=1}^n \,|z^\alpha|^2=1$, so that the coordinates are restricted to a sphere $S^{2n-1}$. This leaves the freedom of multiplying all $z^\alpha$'s by the same phase, so the remaining gauge group is $\tU(1)$. This formulation is nothing but the celebrated Hopf fibration $S^{2n-1}\to \CP^{n-1}$, with fiber $\tU(1)$. Its advantage is that the global symmetry group $\SU(n)$ is explicitly maintained. Similar choices of homogeneous, inhomogeneous and other types of coordinates may be performed for Grassmannians and flag manifolds as well.

\subsection{The flag manifold as a homogeneous space}\label{flaghomspacesec}

Both of the above examples may be concisely formulated as `spaces of subspaces' $0\mysub L\mysub \CC^n$, where $L$ is a linear subspace of a given dimension. This naturally leads to the notion of a flag. A flag in $\CC^n$ is the sequence of nested subspaces
\bea\label{complexflag}
0\subset L_1 \subset \ldots \subset L_{m-1} \subset L_m=\CC^n \quad\quad\quad\quad\quad \textrm{(Flag)}
\eea
of given dimensions $\mathrm{dim}\,L_A=d_A$. Accordingly, the flag manifold in $\CC^n$ may be defined as the manifold of such nested linear complex subspaces\footnote{Reviews of the mathematical properties of flag manifolds include \cite{Alekseevsky, Arvanito}.}:
\bea\label{embeddedspaces}
\mathcal{F}(d_1, \ldots, d_m)=\{0\subset L_1 \subset \ldots \subset L_{m-1} \subset L_m=\CC^n\}\,.
\eea

\begin{figure}[h]
    \centering 
    \includegraphics[width=0.3\textwidth]{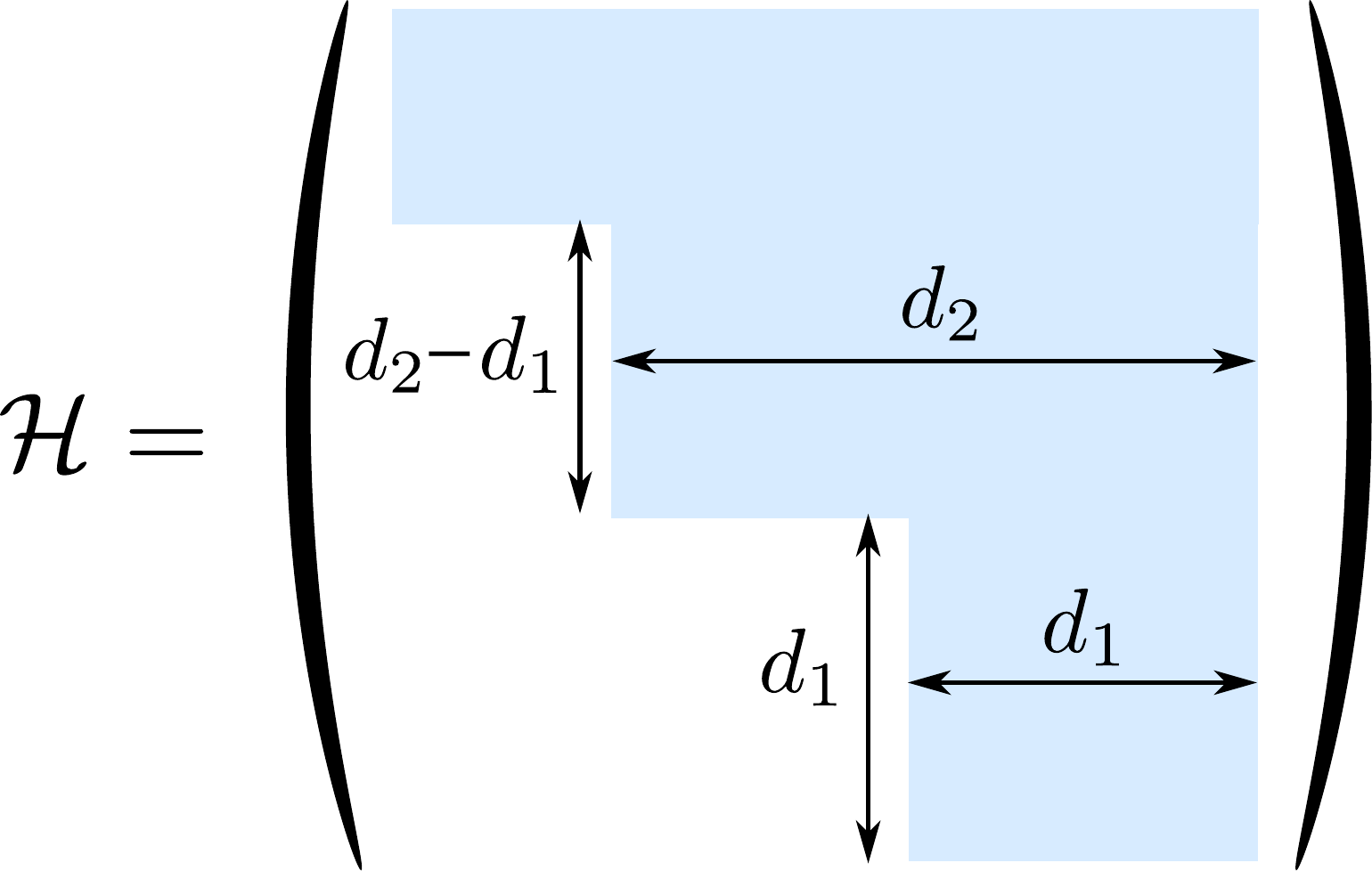}
  \caption{The parabolic subgroup, stabilizing a flag.}
  \label{staircase}
\end{figure}

The next important fact is that the projective space, Grassmannians and flag manifolds are all examples of homogeneous spaces. Moreover, there are two ways to express these manifolds as homogeneous spaces: either w.r.t. the complex symmetry group GL$(n, \CC)$, or w.r.t. its unitary subgroup U($n)\mysub \text{GL}(n, \CC)$. Let us first start with the complex parametrization. 
The group GL($n, \mathbb{C})$ acts transitively on the space of flags of a given type: given two flags, one can first rotate the subspaces of largest dimension $d_{m-1}$ into each other, then the next-to-largest subspaces, etc. The stabilizer of any given flag is a  so-called parabolic subgroup (`a staircase'), consisting of matrices, depicted in Fig.~\ref{staircase}. The reason for the non-diagonal structure is that, given a basis of $L_A$ and a (larger) basis of $L_{A+B}$, adding vectors of the first basis to vectors of the second one produces a new basis of the same sequence of spaces $L_A\mysub L_{A+B}$. Therefore we can view the flag manifold as a homogeneous space
\bea\label{flagdefhol}
\mathcal{F}(d_1, \ldots, d_m)=\text{GL}(n, \mathbb{C})/\mathcal{H}
\eea
of complex dimension
$
\mathrm{dim}_\CC\,\mathcal{F}(d_1, \ldots, d_m)=n^2-\sum\limits_{A=1}^m\,d_A\,(d_A-d_{A-1})$. 

As mentioned earlier, there is a second -- unitary -- parametrization of the flag manifold. To obtain it, one picks a metric in the ambient space $\CC^n$ and orthogonalizes the subspaces of the flag. For example, we split $L_2=L_1\oplus (L_1)^{\perp}$, $L_3=L_2\oplus (L_2)^{\perp}$ and so on. Altogether this splits $\CC^n$ into $m$ mutually orthogonal subspaces of dimensions $n_A=d_{A}-d_{A-1}$, $A=1, \ldots, m$ (where we set $d_0=0$). This allows presenting the flag manifold as a quotient space of the unitary group:
\begin{empheq}[box=\fbox]{align}
\hspace{1em}\vspace{1em}\label{flagunitary}
\mathcal{F}_{n_1, \ldots, n_m}=\frac{\tU(n)}{\tU(n_1)\times \ldots \times \tU(n_m)},\quad \sum\limits_{A=1}^m\,n_A=n\,.
\end{empheq}
Using the dimensions of the groups in the numerator and denominator, one easily computes the real dimension of this space, which, as expected, turns out to be twice the complex dimension of~(\ref{flagdefhol}), computed earlier. Note that sometimes we will denote by $\fl_n$ the complete flag manifold, i.e. the manifold~(\ref{flagunitary}), where all $n_A=1$.

We have just seen that, starting from the complex definition~(\ref{flagdefhol}), one can unambiguously proceed to the unitary one~(\ref{flagunitary}). It should be pointed out that the reverse procedure is in general not unique and involves a certain choice, namely a choice of a complex structure $\mathscr{J}$ on the flag manifold. Since it does play a role for the integrable models introduced in Chapter 3, this is explained in detail in section~\ref{compstructsec0}. For most of the exposition in the first two chapters, in order not to dwell on this subtle issue, we will simply assume that we have both definitions at hand, and we may use any of them at our will. 

Throughout this paper we will mostly be interested in relativistic sigma models with flag manifold target spaces. Such models feature two main ingredients: the metric $\mathbb{G}$ and the skew-symmetric two-form $\Omega$ (which is also called the $B$-field, Kalb-Ramond form, etc.) on the target space. Particularly important are the so-called \emph{topological terms}, which correspond to closed two-forms, i.e. to the case $d\Omega=0$. These do not affect the classical equations of motion, but might substantially alter the quantum theory. As we shall see in Chapter 2, it is precisely such topological terms that are responsible for the presence or absence of a mass gap in the spectrum of the models, and of the related spin chains as well. It is therefore very important to understand in detail, how such terms may be written in the case of flag manifolds. To this end, note that the condition $d\Omega=0$, together with an additional non-degeneracy assumption $\Omega^{\mathrm{dim}_\CC\,\mathcal{F}}\neq 0$ (i.e. $\mathrm{det} \,\Omega\neq 0$), defines what is called a \emph{symplectic} form. If, in addition, $\Omega$ is positive in a certain sense, then $\Omega$ is called a \emph{K\"ahler} form. `Positivity' means that the corresponding symmetric tensor $\mathbb{G}=-\Omega\circ\mathscr{J}$, obtained by contracting $\Omega$ with a complex structure $\mathscr{J}$, is positive-definite and therefore a Riemannian metric on the flag manifold\footnote{Technically for the tensor $\mathbb{G}$ so defined to be symmetric one also needs that $\Omega$ is of type $(1, 1)$, i.e. a Hermitian form. This always holds in our applications, and we will not elaborate on this aspect further.}. To summarize we have the following embeddings:
\bea\label{2formsdiag}
\textrm{K\"ahler forms}\quad \mysub \quad \textrm{Symplectic forms} \quad \mysub \quad \textrm{Closed two-forms (topological terms)}
\eea
Although these three sets do not coincide, they may all be described in a uniform manner. In particular, for a flag manifold of type~(\ref{flagunitary}) they all have real dimension $m-1$, and restricting to non-degenerate forms, or positive forms, amounts to simple relations among the $m-1$ parameters. For this reason we proceed to describe all of these structures at once: keeping this unified picture in mind will be useful for the foregoing exposition.

\vspace{-0.2cm}
\subsection{Symplectic structures}\label{symplman}

We start by describing symplectic forms on  $\mathcal{F}_{n_1, \ldots, n_m}$, i.e.  non-degenerate, closed 2-forms $\omega$, with $d\omega=0$. Since we will mainly be interested in SU($n$)-invariant models in what follows, we will accordingly restrict ourselves to SU($n$)-invariant symplectic forms. The main tool that we will use is the theorem of Kirillov-Kostant that coadjoint orbits of a Lie group $G$ admit natural symplectic forms (for a review see~\cite{KirillovReview}). In our applications the Lie algebra of $G$=SU($n)$ admits a Killing metric, which may be used to relate coadjoint orbits with adjoint orbits, and so we will always be talking of the latter. As the name suggests, these adjoint orbits are defined as follows: one picks a diagonal element $p=\mathrm{Diag}(p_1\,\mathds{1}_{n_1}, \ldots, p_m \mathds{1}_{n_m})\in \mathfrak{su}(n)$, where the $p_A$'s are distinct and $\mathrm{Tr}(p)=0$. In this case the flag manifold is the orbit
\bea\label{orbit}
\mathcal{F}_{n_1, \ldots, n_m}=\{g\,p\,g^{-1},\quad g\in \SU(n)\}\,,
\eea
since there is an obvious gauge invariance $g\to g\cdot h$, where $h\in H= \text{S}(\tU(n_1)\times \cdots\times \tU(n_m))$, so that the orbit is really the quotient $G\over H$. Introducing the Maurer-Cartan current
\bea
j=-g^{-1}\,dg,\quad\quad g\in \tU(n),
\eea
one may write the Kirillov-Kostant symplectic form on the orbit~(\ref{orbit}) as 
\begin{empheq}[box=\fbox]{align}
\label{symplformflag}
\hspace{1em}\vspace{5em}
\Omega =\mathrm{Tr}(p\,j\wedge j)\;\;.\quad
\end{empheq}
One can check that its non-degeneracy is equivalent to the condition that all $p_A$'s are distinct. Due to the condition $\mathrm{Tr}(z)=0$ there are exactly $m-1$ parameters entering the symplectic form~(\ref{symplformflag}). 

Another important observation about the formula~(\ref{orbit}) is that it  gives an embedding of the flag manifold into the Lie algebra $\mathfrak{su}(n)$. Moreover, this embedding may be identified with the image of the moment map
\bea\label{mommap1}
\mu=g\,p\,g^{-1}\,.
\eea
Let us recall what a moment map is, since it will be ubiquitous in the foregoing exposition. Whenever one has a symplectic manifold $\mathbf{\Phi}$ with an action of a Lie group $G$ on it that preserves the symplectic form, one can construct Hamiltonian functions for the action of this group. The action of the group on $\mathbf{\Phi}$  is generated by the vector fields $v_a$, $a=1\cdots \mathrm{dim}\,G$, whose commutators satisfy the Lie algebra relations of $\mathfrak{g}$: $[v_a, v_b]=f_{ab}^c\,v_c$. To each vector field $v_a$ one can put in correspondence a Hamiltonian function $h_a$. It turns out that all of these Hamiltonian functions may be collected in a single matrix-valued object $\mu\in \mathfrak{g}$, called the \emph{moment map}, in such a way that $h_a=\mathrm{Tr}(\mu T_a)$. Here $T_a$ is the $a$th generator of $\mathfrak{g}$. One can check from the definitions that the moment map defined in~(\ref{mommap1}) leads to the vector fields generating the action of $\SU(n)$ and preserving the symplectic form~(\ref{symplformflag}), cf.~\cite{BykovHaldane2}.

As a simple exercise, let us write out explicitly the moment map for the Grassmannian $Gr_{s, n}$. To this end we set $p=\mathrm{Diag}(\underbracket[0.6pt][0.6ex]{1, \cdots, 1}_{s}, \underbracket[0.6pt][0.6ex]{0, \cdots, 0}_{n-s})-{s\over n}\,\mathds{1}_n$, which gives
\bea\label{mugrass}
\mu_{s}=\sum\limits_{k=1}^s\,u^{(k)}\otimes \bar{u}^{(k)}-{s\over n}\,\mathds{1}_n\,,
\eea
where by $u^{(k)}$ we have denoted the (orthonormal) column vectors of the group element $g$.

The moment map~(\ref{mommap1}) is the classical analogue of the $\SU(n)$-spin and therefore will play an important role in our treatment of spin chains in Chapter 2.

\subsection{K\"ahler structures}\label{kahlstructsec}

Following the diagram in~(\ref{2formsdiag}), we now turn to the discussion of K\"ahler forms. As explained earlier, these involve the complex structure $\mathscr{J}$ in their definition, so we will shift to the complex definition of flag manifolds~(\ref{flagdefhol}).  The K\"ahler structures  can be  characterized geometrically in at least two equivalent ways:
\begin{itemize}
	\item[$\circ$] As parameters of a linear combination of the so-called quasipotentials~\cite{AzadKobayashiQureshi1997,AzadBiswas2003} that appear in the physics literature in~\cite{BandoKugoYamawaki1988}.
	\item[$\circ$] As Fayet-Iliopoulos parameters related to the gauged linear sigma model representations for flag manifolds~\cite{Nitta:2003dv,Donagi:2007hi}.
\end{itemize}
These approaches are discussed below in sections \ref{explicitkah}, \ref{kahquiv} respectively.

\subsubsection{Explicit K\"ahler metrics on flag manifolds}\label{explicitkah}

Recall that K\"ahler metrics and K\"ahler forms are in one-to-one correspondence, and are related by contraction with a complex structure $\mathscr{J}$. It is easiest to define a K\"ahler metric through the so-called K\"ahler potential $\mathscr{K}$, which in plain terms is a function of the complex coordinates $\{w^a, \bar{w^a}\}$, such that the line element takes the form $ds^2=\sum\,\frac{\dd^2 \mathscr{K}}{\dd w^a \dd \bar{w}^b}\,dw^a\,d\bar{w}^b$. A very direct way of constructing a K\"ahler potential of the most general $\SU(n)$-invariant K\"ahler metric on the flag manifold~(\ref{flagunitary}) is as follows: consider the matrix
\begin{equation}\label{Wmat}
W=(w_1,..,w_n) \in \tGL(n;\mathbb{C}),
\end{equation}
where each $w_i$ is a column vector. We also define an $n \times d_A$-matrix $W_A$  
of rank $d_A$ by truncating the matrix $W$ to the first $d_A$ columns:
\begin{equation}\label{Zmat}
W_A = (w_1,...,w_{d_A}), \ \ \ \text{where } \ d_A = \sum_{l=1}^{A} n_l\,.
\end{equation}
The columns of $W_A$ span the vector space $L_A$ in the flag~(\ref{complexflag}). Next we 
 introduce the function
\begin{equation}
t_A =   \det \left( W_A^\dagger W_A  \right)\,.
\end{equation}
One can check that $\log(t_A)$, called the quasipotential, is the K\"ahler potential for the $\pi$-normalized canonical metric\footnote[1]{This is the same normalization as that of the Fubini-Study metric on $\mathbb{CP}^{n-1}$, i.e. the volume of a holomorphic 2-sphere generating $H_2(Gr_{d_A, n},\mathbb{Z})$ is $\pi$.} on the Grassmannian $Gr_{d_A, n}$. The potential of an arbitrary $\SU(n)$ invariant K\"ahler metric on the flag manifold~\cite{AzadKobayashiQureshi1997,AzadBiswas2003} may then be written as~
\bea \label{KahlerPotFlagMfds}
\mathscr{K}_{\mathscr{F}}=\sum\limits_{A=1}^{m-1}\,\gamma_A\,\log(t_A)\,,\quad\quad \gamma_A>0\,.
\eea
For a detailed discussion of the geometric properties of these metrics (including the special case of Kähler-Einstein metrics)  cf.~\cite{AlekPerelomov, AZB}. 

As a simplest application of formula~(\ref{KahlerPotFlagMfds}) let us consider the case when the flag manifold is the complex projective space $\CP^{n-1}$. In this case $W_1$ is a column vector, and we label its components $z^1, \ldots, z^n$. The K\"ahler potential is therefore (we set $\gamma_1=2$)
\bea
\label{KahlerPotCPN}
\mathscr{K}_{\CP^{n-1}}=2\,\log{\left(\bar{z} \circ z \right)}\,.
\eea
The resulting K\"ahler form is
the familiar Fubini-Study form: 
\begin{empheq}[box=\fbox]{align}
\label{FSform}
\hspace{1em}\vspace{5em}
\Omega_{FS}=\frac{\im}{\bar{z} \circ z }\,\left(dz^\alpha\wedge d\bar{z}^\alpha-\frac{\bar{z}^\alpha d z^\alpha \wedge z^\beta d\bar{z}^\beta }{\bar{z} \circ z }\right)\,,\quad\quad\quad \int\limits_{\CP^1}\,\Omega_{FS}=2\pi\,.
\end{empheq}
In the second formula the integral is taken over a $\CP^1\subset \CP^{n-1}$ defined by the equations $z^\alpha=0,\; \alpha>2$. We will frequently use the $2\pi$-normalized Fubini-Study form later on in our narrative.

\subsubsection{The K\"ahler quotient quiver}\label{kahquiv}

An attentive reader might have noticed that at the beginning of this chapter we introduced the projective space as the quotient by the group of non-zero complex numbers $\CC^\ast$, and the Grassmannians as a quotient by $\tGL(k, \CC)$, but no similar presentation was provided for the case of flag manifolds. Indeed, the quotient by a subgroup $\mathcal{H}$ of the form found in Figure~\ref{staircase} is not the same thing, as can be readily seen in the example of $\CP^{n-1}$, where the corresponding group is certainly different from $\CC^\ast$. A suitable  formulation for flag manifolds, however, does exist, and may be formulated in terms of a so-called `quiver'.  The quiver in question has the following form: 

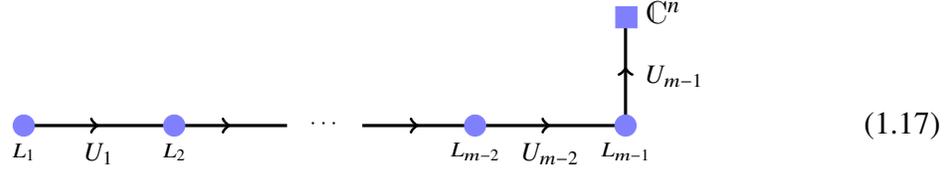
\begin{figure}
    \centering

\bea\label{flagquiv1}
\begin{tikzpicture}[
baseline=-\the\dimexpr\fontdimen22\textfont2\relax,scale=1]
\begin{scope}[very thick,decoration={
    markings,
    mark=at position 0.5 with {\arrow{>}}}
    ] 
\draw[postaction={decorate}] ([yshift=0pt,xshift=0pt]0,0) --  node [below, yshift=-2pt] {\footnotesize $U_1$} ([yshift=0pt,xshift=0pt]2,0);
\draw[postaction={decorate}] ([yshift=0pt,xshift=0pt]2,0) --    ([yshift=0pt,xshift=0pt]3.5,0);
\draw[postaction={decorate}] ([yshift=0pt,xshift=0pt]4.5,0) --    ([yshift=0pt,xshift=0pt]6,0);
\draw[postaction={decorate}] ([yshift=0pt,xshift=0pt]6,0) --  node[below, yshift=-2pt] {{\footnotesize$ U_{m-2}$}} ([yshift=0pt,xshift=0pt]8,0);
\end{scope}

\begin{scope}[very thick,decoration={
    markings,
    mark=at position 0.6 with {\arrow{>}}}
    ] 
\draw[postaction={decorate}] ([yshift=0pt,xshift=0pt]8,0) --  node[right, xshift=3pt, yshift=0pt] {{\footnotesize$ U_{m-1}$}} ([yshift=0pt,xshift=0pt]8,1.3);
\end{scope}

\filldraw[blue!50] (7.87,1.3) rectangle ++(8pt,8pt);
\filldraw[blue!50] (0,0) circle (4pt); \filldraw[blue!50] (2,0) circle (4pt); \filldraw[blue!50] (6,0) circle (4pt); \filldraw[blue!50] (8,0) circle (4pt);
\node at (8.5,1.5) {$\CC^n$};
\node at (0,-0.35) {\scriptsize $L_1$};
\node at (2,-0.35) {\scriptsize $L_2$};
\node at (6,-0.35) {\scriptsize $L_{m-2}$};
\node at (8,-0.35) {\scriptsize $L_{m-1}$};
\node at (4,0) {\scriptsize $\cdots$};

\end{tikzpicture}
\eea
\caption{The quiver describing the flag manifold as a K\"ahler quotient.}
    \label{flagquivpic}
\end{figure}
Here $L_A$ are the vector spaces defining the flag~(\ref{embeddedspaces}), so that $\mathrm{dim}\,L_A=d_A$, and each arrow corresponds to the space of matrices  $\mathrm{Hom}(L_A, L_{A+1})$, with $U_A$ being the (linear) complex coordinates in this space. At each circular node there is an action of the gauge group $\tGL(d_A, \CC)$.  The main idea is that the flag manifold may be identified with the quotient of the space of such matrices (with the requirement that each is of maximal rank) by the gauge group acting at the node. The projective space and the Grassmannians correspond in this language to a quiver with just two nodes, corresponding to the flag $L_1\mysub \CC^n$. To understand why this can be true, consider the case of complete flags in $\CC^3$, i.e. the manifold $\tU(3)\over \tU(1)^3$. One way to parametrize this manifold is as follows. Let $l, p \in \CC^3$ be two linearly independent vectors. 

These vectors define a plane
\bea\label{L2}
L_2=\mathrm{Span}(l, p)\simeq \CC^2 \subset \CC^3.
\eea
A line $L_1\subset L_2$ may be defined as
\bea\label{L1}
L_1=\mathrm{Span}(u_1\,l+u_2\,p)\mysub L_2\mysub\CC^3
\eea
with $(u^1, u^2)$ a fixed non-zero two-vector.

Clearly, $(u^1, u^2) \in \CC^2$, $l\in \CC^3$, $p \in \CC^3$ uniquely define a given flag $L_1\subset L_2\subset \CC^3$, however the map is not one-to-one. Indeed, the rotated set 
\bear\nonumber
\left( \begin{array}{c}
\widetilde{u}^1   \\
\widetilde{u}^2    \end{array} \right) = \uplambda\,g^{-1}\circ \left( \begin{array}{c}
u^1   \\
u^2    \end{array} \right),\quad\quad \left( \begin{array}{c c}
\widetilde{l} &   
\widetilde{p}    \end{array} \right) =  \left( \begin{array}{c c}
l &  
p    \end{array} \right)\circ g
\eear
with $g \in \tGL(2,\CC)$ and $\uplambda \in \CC^\ast$ defines the same flag. 
Therefore one has the gauge group
\begin{figure}
    \centering
    \includegraphics[width=0.6\textwidth]{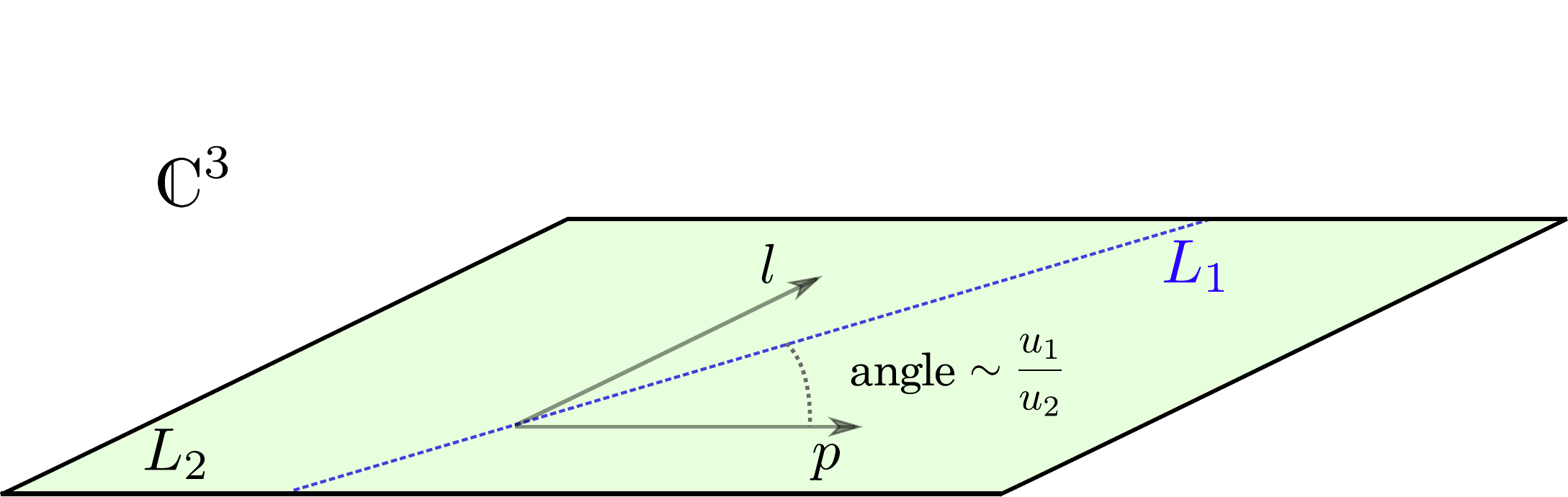}
    \caption{Parametrization of the flag manifold introduced in (\ref{L2})-(\ref{L1}).}
    \label{flagparam}
\end{figure}
$
\mathcal{G}=\CC^\ast \times \tGL(2, \CC)
$
acting on the `matter fields' constituting the linear space
$
V=(\CC^2)_u\oplus (\CC^3\otimes \CC^2)_{l, p}\,.
$
To make a connection to the quiver~(\ref{flagquiv1}), we identify $U_1=\left( \begin{array}{c}
u^1   \\
u^2    \end{array} \right)$ and $U_2=\left( \begin{array}{c c}
l &   
p    \end{array} \right)$. This is the desired generalization of the well-known presentation for the projective space and Grassmannians that we used as our starting point at the beginning of the chapter.

The quiver formulation may as well be used to describe K\"ahler metrics on the flag manifold by performing a symplectic reduction. This entails associating to each gauge node of the quiver a real constant (in the supersymmetric setup~\cite{Donagi:2007hi} these constants are called Fayet-Iliopoulos parameters), so the resulting metric depends on $m-1$ parameters. These are of course in one-to-one correspondence with the parameters $\gamma_A$ used in~(\ref{KahlerPotFlagMfds}). The reader will find the details in Appendix~\ref{kahpotapp}.

\subsection{Cohomology}\label{cohsec}

It was already emphasized in the diagram~(\ref{2formsdiag}) that K\"ahler and symplectic structures provide examples of closed two-forms. Such forms are elements of the second cohomology group $H^2(\mathcal{F}_{n_1, \cdots, n_m}, \mathbb{R})$, which is the cohomology group most relevant for   sigma model applications, since its elements are the topological terms in the action. In this section we describe another way of expressing the elements of this cohomology group, which is a very convenient model to be used in the applications discussed in subsequent chapters. Let us start by writing out the answer for the second cohomology group with integer coefficients:
\bea
H^2(\mathcal{F}_{n_1, \cdots, n_m}, \mathbb{Z})=\mathbb{Z}^{m-1}\,.
\eea
One can obtain a convenient model for this cohomology group if one notes the existence of an embedding
\bea\label{flaggenembed}
 \mathcal{F}_{n_1,\ldots, n_m}\hookrightarrow Gr_{n_1, n}\times \cdots \times Gr_{n_m, n}
\eea
of the flag manifold into a product of Grassmannians. Indeed, a point in a flag manifold is a collection of pairwise orthogonal planes of dimensions $n_1, \cdots, n_m$ (see section~\ref{flaghomspacesec}), each of which is a point in the corresponding Grassmannian.

To proceed, we will need the definition of a Lagrangian submanifold $\mathcal{M}\subset \mathcal{N}$ in a symplectic manifold $(\mathcal{N}, \omega)$, which we now recall.  $\mathcal{M}$ is Lagrangian if $\omega\big|_{\mathcal{M}}=0$ and $\mathrm{dim}\,\mathcal{M}={1\over 2}\,\mathrm{dim}\,\mathcal{N}$. Let us now consider $\mathcal{N}=Gr_{n_1, n}\times \cdots \times Gr_{n_m, n}$ as a symplectic manifold with a product symplectic form $
\omega=\sum\limits_{A=1}^m\;\omega_A
$, where all $\omega_A$ are normalized in the same way. In this case, as we shall now prove, $\mathcal{M}=\mathcal{F}_{n_1,\ldots, n_m}$ in (\ref{flaggenembed}) is a Lagrangian submanifold, i.e.
\bea\label{lagr1}
\omega\big|_{\mathcal{F}_{n_1,\ldots, n_m}}=0 .
\eea
Identifying $\Omega_A=\omega_A \big|_{\mathcal{F}_{n_1,\ldots, n_m}}$ and taking into account (\ref{lagr1}), we obtain the relation
\bea\label{formsumzero}
\sum\limits_{A=1}^m\,\Omega_A=0\,.
\eea
The cohomology group $H^2(\mathcal{F}_{n_1, \cdots, n_m}, \mathbb{Z})$ is then described as the quotient
\begin{empheq}[box=\fbox]{align}
\hspace{1em}\vspace{1em}\label{H2quot}
H^2(\mathcal{F}_{n_1, \cdots, n_m}, \mathbb{Z})=\left.\mathbb{Z}[\Omega_1, \cdots, \Omega_m]\;\middle/\; \left(\sum\limits_{A=1}^m\,\Omega_A\right)\right.\quad
\end{empheq}

To prove that the flag manifold is a Lagrangian submanifold in the product of Grassmannians, first let us perform a dimensionality check. Using $
\mathrm{dim}_{\mathbb{R}}(\mathcal{F}_{n_1,\,\cdots\, , n_m}) = N^2-\sum\limits_{A=1}^m\,n_A^2 $ \;and\; $\mathrm{dim}_{\mathbb{R}}(Gr_{s,n})=2(s\cdot n-s^2)$, we obtain
\bea
 \mathrm{dim}_{\mathbb{R}}\left( \prod\limits_{A=1}^m\;Gr_{n_A,n}\right)=2\;\sum\limits_{A=1}^m\,(n_A\cdot N-n_A^2)=2\;\left(n^2-\sum\limits_{A=1}^m n_A^2 \right)=2\cdot \mathrm{dim}_{\mathbb{R}}\,\mathcal{F}_{n_1,\,\cdots\, , n_m}
\eea
We see that the dimensions match correctly. For the rest we use the following fact (which is easy to prove starting from the definition): if $\mu$ is the moment map for the action of a group $G$, the restriction of a symplectic form to a $G$-orbit in $\mu^{-1}(0)$ vanishes. We will now construct a moment map for the diagonal action of $\SU(n)$ on the product of Grassmannians and prove that $\mu^{-1}(0)$ is the flag manifold under consideration. The moment map for a single Grassmannian was written out in~(\ref{mugrass}), so now we sum over all Grassmannians to obtain
\bea
\mu=\sum\limits_{A=1}^m\,\mu_{n_A}=\sum\limits_{A=1}^m\;\left(\sum\limits_{k=1}^{n_A}\;u_A^{(k)}\otimes \bar{u}_A^{(k)}\right)-\mathds{1}_n\,.
\eea
In this formula the vectors $u_A^{(k)}$ inside the same group $A$ are  orthonormal: $\bar{u}_A^{(k)}\circ u_A^{(k')}=\delta_{kk'}$. On the other hand, it is easy to convince oneself that the set $\mu^{-1}(0)$ is composed of $n$-tuples of orthogonal $u$-vectors. It follows that the $u$-vectors representing different $n_A$-dimensional planes in $\CC^n$ ($A=1\cdots m$) are mutually orthogonal as well. The set of such orthogonal subspaces is precisely the flag manifold $\mathcal{F}_{n_1,\,\cdots\, , n_m}$\,.

Before concluding this section, let us specialize these results to the case that we will encounter most frequently below, namely the case of the complete flag manifold, when all $n_A=1$. The second cohomology group of the complete flag manifold is
\bea
H^2(\fl_n, \Zz)=\Zz^{n-1},
\eea
hence there exist  $n-1$ linearly independent 2-forms, which are the generators of $H^2(\fl_n, \Zz)$. In order not to repeat ourselves, let us consider here a slightly different model for $H^2(\fl_n, \Zz)$. On $\fl_n$ there are $n$ standard line bundles $\mathcal{L}_1, \cdots , \mathcal{L}_n$, and their sum is a trivial bundle:
\bea\label{triv}
\overset{n}{\underset{A=1}{\oplus}}\;\mathcal{L}_A = \fl_n\, \times \CC^n\; .
\eea
The first Chern classes of these bundles are represented by $n$ closed 2-forms: $[\Omega_A] = c_1(\mathcal{L}_A), \; A=1\cdots n$. Due to the condition (\ref{triv}) and the additivity of the first Chern classes $c_1(E\oplus F)=c_1(E)+c_1(F)$ it is clear that the forms $\Omega_A$ are not independent but rather satisfy the relation
\bea\label{triv2}
\sum\limits_{A=1}^n\, [\Omega_A] = 0
\eea
This is clearly in correspondence with~(\ref{formsumzero}), and the two-forms $\Omega_A\;(A=1 \,\cdots\, n)$ satisfying the relation (\ref{triv2}), generate $H^2(\fl_n, \Zz)$. Higher cohomology groups of general flag manifolds could as well be obtained from the relations that follow from the triviality of a sum of certain vector bundles, i.e. from a generalization of~(\ref{triv}).

In the present review we will only make use of cohomology, with almost no reference to the homotopy of flag manifolds. One reason for this is that flag manifolds are simply connected, $\pi_1(\mathcal{F}_{n_1, \cdots, n_m})=0$, which  implies $\pi_2(\mathcal{F}_{n_1, \cdots, n_m})\simeq H^2(\mathcal{F}_{n_1, \cdots, n_m}, \mathbb{Z})\simeq \mathbb{Z}^{m-1}$ by Hurewicz theorem, so that the two notions coincide in dimension two. In higher dimensions this is no longer the case. For example, $H^3(\mathcal{F}_{n_1, \cdots, n_m}, \mathbb{Z})\simeq 0$, whereas for complete flag manifolds $\pi_3(\mathcal{F}_n)\simeq \mathbb{Z}$. The latter is a higher-dimensional generalization of the Hopf invariant $\pi_3(S^2)\simeq \mathbb{Z}$ and leads to the existence of topologically non-trivial Hopfion solutions~\cite{Amari, AmariNew} relevant for the Faddeev-Niemi model~\cite{FaddeevNiemiSUN}~(see also~\cite{Cho}).

\subsection{General (non-K\"ahler) metrics and $B$-fields on the flag manifold}\label{genmetr}

So far we have discussed $\SU(n)$-invariant \emph{closed} forms on flag manifolds, as well as the related question of invariant K\"ahler metrics. This is not the end of the story, however, as on a general flag manifold~(\ref{flagunitary}) there will be large families of invariant metrics, and typically only a small subfamily corresponds to K\"ahler metrics. Moreover, the metrics that will actually enter the sigma models  that we discuss in Chapters 2 and 3, are in general \emph{not} K\"ahler. In a similar way, the $B$-fields also come in large families and are not required to be topological in general, as on a general flag manifold there exist invariant two-forms that are not closed.

To construct the general metric and $B$-field,  we denote the flag manifold ${\SU(n)\over \text{S}(\tU(n_1)\times \cdots \times \tU(n_m))}$ as ${G\over H}$ and introduce the corresponding Lie algebra decomposition $\mathfrak{g}=\mathfrak{h}\oplus \mathfrak{m}$. Since $[\mathfrak{h}, \mathfrak{m}]\subset \mathfrak{m}$, the subgroup $H$ is represented in the space $\mathfrak{m}$, and this representation may be decomposed into irreducibles:
\bea\label{mirreps}
\mathfrak{m}_\CC= \oplus_{A\neq B} \,V_{AB},\quad\quad\textrm{where}\quad\quad V_{AB}=\CC^{n_A n_B}\,.
\eea
The space $V_{AB}$ of $n_A\times n_B$-matrices is the vector space of the bi-fundamental representation of the group $\tU(n_A)\times \tU(n_B)$, and moreover  $V_{BA}=\bar{V_{AB}}$. We decompose  the Maurer-Cartan current $
j=-g^{-1}\,dg,\, g\in \tU(n),
$ entering~(\ref{symplformflag}) accordingly:
\bea\label{jdecomp}
j=[j]_\mathfrak{h}+[j]_\mathfrak{m}=[j]_\mathfrak{h}+\sum\limits_{A\neq B}\,j_{AB},\quad\quad j_{AB}\in V_{
AB}\,.
\eea
The most general invariant two-form may then be written as
\bea
\Omega=\sum\limits_{A<B}\,b_{AB}\,\mathrm{Tr}(j_{AB}\wedge j_{BA})\,.
\eea
Using the zero-curvature equation for $j$, one can check that $\Omega$ is closed if and only if $
b_{AB}=p_A-p_B$, in which case it is exactly the symplectic form~(\ref{symplformflag}) (see~Appendix~\ref{adjorbsec}). Quite analogously, the line element of the most general metric is
\bea
ds^2=-\sum\limits_{A<B}\,a_{AB}\,\mathrm{Tr}(j_{AB}\cdot j_{BA})\,,
\eea
where for positivity we have to require $a_{AB}>0$. We conclude that there are $m(m-1)\over 2$ real parameters defining the most general metric, as well as $m(m-1)\over 2$ additional parameters defining the most general $B$-field.

As discussed earlier, the space of K\"ahler metrics is an $(m-1)$-dimensional subspace in the full space of metrics. In order to formulate the corresponding condition on the coefficients $a_{ij}$ more explicitly, one would have to specify the complex structure $\mathscr{J}$ (these are discussed in Chapter 3, section~\ref{compstructsec0}). In any case, the metric that will be most important for us in Chapter~3 (and features in some of the most prominent examples in chapter 2) is in general not K\"ahler. It is the so-called normal, or reductive, metric (cf.~\cite{FlagEinsteinMetrics}), with line element $ds^2=-{1\over 2}\mathrm{Tr}([j]_\mathfrak{m}^2)$, which corresponds to $a_{AB}=1$ for all $A, B$. This metric is not a K\"ahler metric, unless the flag manifold is a Grassmannian (i.e. unless $m=2$), see section~\ref{symmspace}. In contrast, K\"ahler metrics are encountered in other applications of flag manifold sigma models, for example in the description of worldsheet theories of non-Abelian vortices in certain four-dimensional supersymmetric theories~\cite{IresonShifman, Ireson} -- in this case K\"ahler metrics are required by supersymmetry.

\section{Flag manifolds and elements of representation theory}\label{appBWB}

Now that we are done with some formal aspects, we wish to present the first example of a well-known physical situation where  flag manifolds naturally arise. Incidentally this makes a neat connection to the applications of flag manifolds in representation theory, discussed below in Section~\ref{appQuant}. We will need the latter for our discussion of spin chains in Chapter 2.

\subsection{Mechanical particle in a non-Abelian gauge field}

It is well-known how one can describe the motion of a classical particle on a Riemannian manifold with metric~$\mathbb{G}$, interacting with an external electromagnetic field $A_\mu$. The action has the form
\bea
\mathcal{S}=\int\,dt\,\frac{\mathbb{G}_{\mu\nu} \dot{x}^\mu \dot{x}^\nu}{2}-\int\,A=\int\,dt\,\left(\frac{\mathbb{G}_{\mu\nu} \dot{x}^\mu \dot{x}^\nu}{2}-A_\mu \dot{x}^\mu\right)\,.
\eea
The question is, how do we write an analogous action for the case when the gauge field is non-Abelian, or, simply speaking, when it has additional gauge indices $A_\mu^{\alpha\beta}$. The answer is that the particle should possess additional degrees of freedom. For example, in the case of $\SU(2)$, the additional variables correspond to a unit vector $\vec{n}\in S^2 = \CP^1$ that couples to the $\SU(2)$ gauge field $\vec{A}$. More generally, the  degrees of freedom associated to the `internal spin' take values in a certain flag manifold, corresponding to the representation in which the particle transforms. In other words, one should enlarge the phase space of the mechanical system~\cite{Sternberg}:
\bea\label{Phasespace}
T^\ast\mathcal{M}\to T^\ast\mathcal{M}\times \mathcal{F}\,.
\eea
Here $\mathcal{M}$ is the configuration space, $T^\ast \mathcal{M}$ is the cotangent bundle (i.e. phase space), and $\mathcal{F}$ is the flag manifold. In SU(2), $\mathcal{F}=\CP^1$, but for larger non-Abelian groups, it is not clear \emph{a priori} what the appropriate choice of $\mathcal{F}$ should be, since there is now choice in the parameters $n_i$ appearing in (\ref{flagunitary}). We will see below that this choice is related to the different families of representations under which the particle transforms.

We start by rewriting the standard action of a particle in first-order form:
\bea
\mathcal{S}=\int\,dt\,\left(p_\mu\dot{x}^\mu-\frac{\mathbb{G}^{\mu\nu}p_\mu p_\nu}{2}-A_\mu \dot{x}^\mu\right).
\eea
Upon enlarging the phase space we can analogously write down the non-Abelian action as follows ($A$ is assumed Hermitian, and $\mathcal{H}$ is the Hamiltonian):
\bea\label{flagaction1}
\mathcal{S}=\int\,p_\mu\,dx^\mu-\int\,dt\,\mathcal{H}(x, p)+\int\,\left(\theta-\tr(A\,\mu) \right)\,.
\eea
Here $\theta$ is the canonical (Poincar\'e-Liouville) one-form, defined by the condition
\bea
d\theta=\Omega\quad\quad (=\textrm{the symplectic form on}\;\;\mathcal{F})\,,
\eea
and $\mu$ is the moment map for the action of the group $G$ on $\mathcal{F}$. The integral $\int \theta$ is sometimes called the \emph{Berry phase} and will be an essential ingredient of the spin chain path integrals in the next chapter. We note that the form $\theta$ is defined up to the addition of a total derivative, $\theta\to\theta+dh$, but the difference only affects the boundary terms in the action. In the case of periodic boundary conditions one may even write
\bea
\int\limits_{\Gamma}\,\theta=\int\limits_D\,\Omega,
\eea
where $D$ is a disc, whose boundary is the curve $\Gamma$: $\dd D=\Gamma$. In fact this term is nothing but the one-dimensional version of the Wess-Zumino-Novikov-Witten term~\cite{WZW, NovCurr, WittenCurr}.

One needs to show that the expression in~(\ref{flagaction1}) is gauge-invariant. For simplicity let us take as $\mathcal{F}$ the projective space, $\CP^{n-1}$. Later, we will see that this corresponds to the particle transforming in the defining representation of $\SU(n)$. Let us normalize\footnote{Throughout the review we will be mostly using the variable $z$ to denote unconstrained complex coordinates, such as the homogeneous or inhomogeneous coordinates on $\CP^{n-1}$, and the variable $u$ to denote unit-normalized vectors.} the homogeneous coordinates on $\CP^{n-1}$:
\bea\label{zknorm}
\sum\limits_{\alpha=1}^n\,|u^\alpha|^2=1\,.
\eea
One still has the remaining gauge group $\tU(1)$, which acts by multiplication of all coordinates $u^\alpha$ by a common phase. The Fubini-Study form~(\ref{FSform}) on $\CP^{n-1}$ may be simplified if one uses the above normalization:
\bea\label{omegaFSunitary}
\Omega_{FS}=i\,du^\alpha\wedge d\bar{u}^\alpha\,.
\eea
Then we have the following expressions for $\theta$ and $\mu$:
$
\theta=i\,u^\alpha\,d\bar{u}^\alpha$,\; $\mu=u\otimes \bar{u}-\frac{1}{n}\,\mathds{1}_n\,.
$ This expression for the moment map is a special case of~(\ref{mugrass}). The part of the action corresponding to the motion in the `internal' space (in this case the projective space) has the form
\begin{empheq}[box=\fbox]{align}
\label{flagpartaction}
\hspace{1em}\vspace{5em}
\tilde{\mathcal{S}}=-\int\,dt\,\bar{u}^\alpha (i\,\dot{u}^\alpha+(A_{\mu})^{\alpha\beta}\,\dot{x}^\alpha\,u^\beta)\,,
\end{empheq}
and one should take into account that the normalization condition (\ref{zknorm}) is also implied. It is evident that it is gauge-invariant w.r.t. the transformations
\bea
u\to g(x(t))\circ u\,\quad\quad A_\mu\to g A_\mu g^{-1}-i\,\dd_\mu g\,g^{-1}\,.
\eea
To make it even more obvious, we note that the exterior derivative of the one-form $\theta-\tr(A\,\mu)$ (viewed as a form on the enlarged phase space~(\ref{Phasespace})) produces a two-form, which is explicitly gauge-invariant:
\bear\label{gaugeinv2form}
&&d(\theta-\tr(A\,\mu))=i\,\mathscr{D}u^{\alpha}\wedge \mathscr{D}\bar{u}^{\alpha}-\tr(F\,\mu)\,,\\ \nonumber
&&\mathscr{D}u=du-i\,A\,u,\quad\quad \mathscr{D}\bar{u}=d\bar{u}+i\,\bar{u}\,A,\quad\quad 
F=dA-i A\wedge A\,.
\eear
Each of the two terms in (\ref{gaugeinv2form}) is separately gauge-invariant, however~(\ref{gaugeinv2form}) is the only linear combination of them, which is closed (and therefore locally is an exterior derivative of a one-form).

\subsubsection{Equations of motion for the spin}
Now that we've written down a gauge-invariant action for a particle coupled to a non-Abelian gauge field, let us next write out the equations of motion on the flag manifold, $\mathcal{F}$.

To simplify the discussion, let us begin by carrying out these steps for the case of SU(2), which corresponds to $\mathcal{F} = \CP^1 = S^2$. Instead of using the spinor $(u^1, u^2) \in \mathbb{C}^2$, we can parametrize $\mathcal{F}$ in a more standard way with the help of a unit vector $\vec{n}\in \mathbb{R}^3$. The equations of motion then take the form
\bea
\dot{\vec{n}}=\vec{A}\times \vec{n},\quad\quad \textrm{where}\quad\quad \vec{A}=\{A_\mu^a\,\dot{x}^\mu\}_{a=1, 2, 3}
\eea
is a vector of components of the gauge field in the basis of Pauli matrices. We see that the equations are \emph{linear} in $\vec{n}$, and the condition
\bea
\vec{n}^2=\mathrm{const.}
\eea
is a consequence of the equations, i.e. the motion takes place on a sphere in $\mathbb{R}^3$. This is a general fact. Indeed, in the case of a general compact simple Lie algebra $\mathfrak{g}$ with basis $\{T_a\}$ we can introduce a variable $n=\sum \,n^a\,T_a\in \mathfrak{g}$, and the equations will then take the form
\bea\label{adjmotion}
\dot{n}=[A_\mu \dot{x}^\mu, n]\,,
\eea
or, in terms of the variables $n^a$,
\bea
\dot{n}^a=f_{bc}^a\,(A_\mu \dot{x}^\mu)^b\,n^c\,,
\eea
where $f_{bc}^a$ are the structure constants of $\mathfrak{g}$. It is in this form that this system of equations was discovered in~\cite{Wong}. The motion defined by these equations in reality takes place on flag manifolds embedded in $\mathfrak{g}$, since the  `Casimirs'
\bea
C_J=\tr(n^J),\quad\quad J=1, 2, \ldots
\eea
are integrals of motion of the system (\ref{adjmotion}), and specifying the Casimirs is effectively the same as specifying the parameter~$p$ of the orbit~(\ref{orbit}). We have thus established a connection with the formulation through flag manifolds used earlier.

\subsection{`Quantization' of the symplectic form on flag manifolds}\label{appQuant}

The next question that we address is how to quantize an action of the type~(\ref{flagaction1}). Quantization of the particle phase space coordinates  $(p, x)$ is standard, so the non-trivial question is how to quantize the spin phase space $\mathcal{F}$ -- the flag manifold. In the case of SU(2), this will lead to the notion of spin quantization, i.e., that the particle transforms under some definite representation of SU(2), labeled by a single integer.

One of the approaches to quantization is related to considering path integrals of the  form\footnote{Another approach to the quantization of coadjoint orbits, which is also based on the path integral, was developed in~\cite{AFS}.}
\bea\label{pathint}
\int\,\prod\limits_{j,\,  t}\,d\varphi_j(t)\,e^{i\,\mathcal{S}}
\eea
where  the exponent contains the action (\ref{flagaction1}), and $\varphi_i$ parametrize $\mathcal{F}$. The subtlety comes from the fact that the connection $\theta$ is not a globally-defined one-form on the flag manifold. Indeed, let us consider the simplest case of $\mathcal{F}=\CP^1=S^2$. The most general invariant symplectic form is as follows\footnote{It can be also written in the form $\Omega=-\frac{p}{2}\, dz\wedge d\phi$, where $z=\cos{\vartheta}$ is the  $z$-coordinate of a given point on the sphere. Since the latter form is nothing but the area element of a cylinder, it implies that the projection of a sphere to the cylinder preserves the area.}:
\bea
\Omega=\frac{p}{2}\,\sin{\vartheta}\, d\vartheta\wedge d\phi.
\eea
Here $p$ is an arbitrary constant, and $\vartheta, \phi$ are the standard angles on the sphere. 

Since the action $\mathcal{S}$ entering the exponent in (\ref{pathint}) involves a term $\int \,\theta$, where $\theta$ is a connection satisfying $d\theta=\Omega$, a standard argument familiar from Wess-Zumino-Novikov-Witten theory~\cite{WZW, NovCurr, WittenCurr} leads to the requirement that the coefficient $p$ is quantized according to $\int \Omega\in 2\pi  \mathbb{Z}$. Let us recall the argument. To start with, we write a one-form $\theta$, well-defined on the northern hemisphere, such that $d\theta=\Omega$:
\bea
\theta_{NH}=p\,\sin{\left(\vartheta\over 2\right)}^2\,d\phi\,.
\eea
\begin{figure}[h]
    \centering 
    \includegraphics[width=0.25\textwidth]{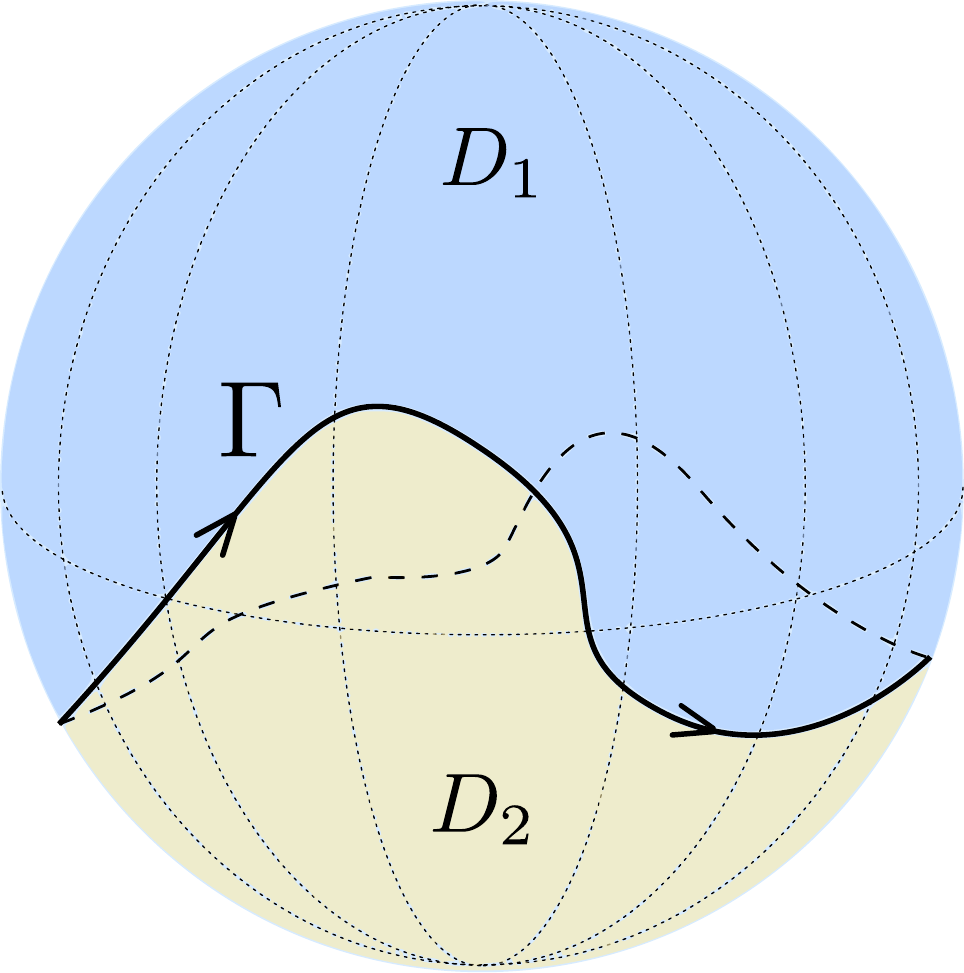}
    \caption{To apply Stokes' theorem to the integral $\int\limits_\Gamma \,\theta$, we choose a disc with boundary~$\Gamma$. There are two ways to do this, which lead to the domains $D_1, D_2\subset S^2$. One has $\int\limits_\Gamma \,\theta_{NH}=\int\limits_{D_1}\,\Omega$, $\int\limits_\Gamma \,\theta_{SH}=-\int\limits_{D_2}\,\Omega=-\int\limits_{S^2}\,\Omega+\int\limits_\Gamma \,\theta_{NH}$ (since in order to use Stokes' theorem, one has to pick a one-form that is well-defined in the interior of the domain). Since $\int\limits_{S^2}\,\Omega=2\pi$, and the choice of north/south poles was arbitrary, the integral $\int\limits_\Gamma \,\theta$ is only well-defined modulo $2\pi$.}
\label{contour}
\end{figure}
\noindent
It is well-defined at the north pole, $\vartheta=0$, since at that point the prefactor of $d\phi$ vanishes. On the other hand, at $\vartheta=\pi$ it remains constant. Another way to see this is to introduce the usual round metric on the sphere $ds^2=d\vartheta^2+\sin{\vartheta}^2\,d\phi^2$ and to calculate the norm of the differential $\theta$: $\|\theta\|^2=p^2\,\sin{\left(\vartheta\over 2\right)}^4\,\frac{1}{\sin{\vartheta}^2}$. One sees that it is bounded at $\theta=0$ but blows up at $\theta=\pi$. If one views $\theta$ as a connection on a line bundle, it is nevertheless well-defined, as on the southern hemisphere we may define a gauge-transformed $\theta_{SH}=\theta_{NH}-p \,d\phi=-p\,\cos{\left(\vartheta\over 2\right)}^2\,d\phi$, which is well-behaved at $\theta=\pi$. Therefore the integral $\int\limits_\Gamma\,\theta$ depends on which formula for the connection we take, $\theta_{SH}$ or $\theta_{NH}$, the difference being equal to $2\pi p$: $\int\limits_\Gamma\,\theta_{SH}=\int\limits_\Gamma\,\theta_{NH}-2\pi p$ (see Fig.~\ref{contour}). If $p\in \mathbb{Z}$, however, the quantity $e^{i\,\int\,\theta}$ is defined unambiguously. We say that $|p|$ labels the representation of SU(2) under which the particle transforms, and $s = {|p|\over 2}$ is called the `spin' of the particle.

Let us turn to the flag manifolds of SU($n$), with $n>2$. Now there are multiple contours $\Gamma_i$ that must be considered, corresponding to the hemispheres of homologically distinct spheres $C_i\in H_2(\mathcal{F},\mathbb{Z})$ in $\mathcal{F}$. 
We require that each of the terms $e^{i\int_{\Gamma_i} \theta}$ is well defined, i.e. 
\begin{empheq}[box=\fbox]{align}
\label{symplintquant}
\hspace{1em}\vspace{5em}
\int\limits_{C_i}\,\Omega  \in 2\pi \mathbb{Z}\quad\quad\textrm{for every 2-cycle}\quad C_i\in H_2(\mathcal{F}, \mathbb{Z})\,.\quad
\end{empheq}
These quantization conditions correspond to particular representations of SU($n$). Let us construct these 2-cycles explicitly for the case when $\mathcal{F}$ is a complete flag manifold $\mathcal{F}_n$. Later on, we can analyze the remaining (smaller) flag manifolds by use of a forgetful projection.

The manifold $\mathcal{F}_n$ can be parametrized using $n$ orthonormal vectors $u_A, A=1\ldots n$, $u_A\circ \bar{u}_B=\delta_{AB}$, defined modulo phase transformations: $u_A\sim e^{i\alpha_A}\,u_A$. As we showed in sections~\ref{symplman} and~\ref{genmetr}, the most general symplectic form on  $\mathcal{F}_n$ may be written as follows:
\bea\label{flagsymplform2}
\Omega=i\,\sum\limits_{A<B}\,(p_A-p_B)\,j_{AB}\wedge j_{BA},\quad\quad\textrm{where}\quad\quad j_{AB}=u_A \circ d\bar{u}_B
\eea
To construct the cycles $C_A$, note that if one fixes $n-2$ out of $n$ lines defined by the vectors $u_1, \ldots, u_n$, the remaining free parameters define the configuration space of ordered pairs of mutually orthogonal lines, passing through the origin and laying in a plane, orthogonal to the $n-2$ fixed lines. This configuration space is nothing but the sphere $\CP^1$:
\bear\nonumber
&&\{u_{A_1}, \ldots , u_{A_{n-2}}\quad\textrm{are fixed},\quad\quad u_{A_{n-1}}, u_{A_n}\in (u_{A_1}, \ldots , u_{A_{n-2}})^\perp \quad \\ \nonumber&&\textrm{are mutually orthogonal and otherwise generic}\}\simeq (\CP^1)_{A_{n-1}, A_{n}}\,.
\eear

\vspace{0.3cm} 
Let us now fix a permutation $(A_1, \ldots, A_n)$ in such a way that the $p_{A_i}$  form a non-increasing sequence, i.e. $p_{A_i}\geq p_{A_j}$ for $i<j$. The fact that $\Omega$ is non-degenerate requires that this sequence is actually strictly decreasing. In this case the rearrangement of $p$'s amounts to choosing a complex structure on $\mathcal{F}_n$, but we will not dwell on this fact here (see Section~\ref{compstructsec0} for details).   
After such a permutation we may choose $(\CP^1)_{{A_i},A_{i+1}}$ as a basis in the homology group $H_2(\mathcal{F}_n, \mathbb{Z})$. Then the integrals of the symplectic form over these cycles will be positive:\begin{footnote}{The orientation of the spheres is induced by the complex structure on $\mathcal{F}_n$.}\end{footnote}
\bea\label{Omegaint}
\int\limits_{(\CP^1)_{{A_i},A_{i+1}}}\,\Omega=p_{A_i}-p_{A_{i+1}}\in 2\pi \mathbb{Z}^+\,,\quad\quad i=1\ldots n-1\,.
\eea

In order for the value of the integral to be an integer, one should choose $p_A$ in the form
\bea\label{ptildep}
(p_1, \ldots, p_n)=\lambda (1, \ldots, 1)+(\tilde{p}_1, \ldots, \tilde{p}_n)\,, \quad \lambda \in \mathbb{R},\quad \tilde{p}_i\in \mathbb{Z}\,.
\eea
This freedom in adding a vector $\lambda(1,\dots,1)$ allows us to work with values $\{\tilde p_A\}$ that sum to zero. According to the general theory of adjoint orbits (see Section~\ref{symplman}), the flag manifold under consideration is then the orbit of the element
\bea
p=\left( \begin{array}{ccc}
\tilde{p}_1 & 0 & 0 \\
\vdots & \ddots  & \vdots \\
0 & 0 & \tilde{p}_n \end{array} \right)\in \mathfrak{su}(n)\,.
\eea
Let us observe what happens when some of these variables $\tilde p_A$ coincide. On the one hand, the 2-form $\Omega$ now becomes degenerate. On the other, we see that the corresponding adjoint orbit is no longer the complete flag manifold, $\mathcal{F}_n$. For example, if there are only two distinct values of $\tilde p_i$, i.e. we have $p_s=\mathrm{Diag}(\underbracket[0.6pt][0.6ex]{1, \ldots , 1}_{s}, \underbracket[0.6pt][0.6ex]{0, \ldots , 0}_{n-s})-{s\over n}\,\mathds{1}_n
$, so that the corresponding adjoint orbit is the Grassmannian $Gr_{s,n}$. This demonstrates the point that we alluded to earlier, namely that the flag manifold $\mathcal{F}$ encoding the degrees of freedom of a particle coupled to an SU($n$) gauge field is not uniquely determined by $n$ for $n>2$. Indeed, choosing different values of $p_A$ leads to different flag manifolds. In such cases when $\mathcal{F}$ is strictly smaller than $\mathcal{F}_n$, one may view the (degenerate) 2-form (\ref{flagsymplform2}) on the complete flag manifold as a non-degenerate form on the smaller $\mathcal{F}$. This amounts to a forgetful projection. The general theory that we have described is nothing but `geometric quantization' for the case of flag manifolds.

The canonical quantization of the system given by the action $S=\int\,\theta$ will be  treated in detail in the next section and, as we shall see, the non-negative integers $p_A$ are equal to the lengths of the rows of the Young diagram characterizing a given representation of $\mathfrak{su}(n)$. For this to make sense, one should choose $\lambda$ in~(\ref{ptildep}) in such a way that $p_n=0$.

\subsection{Schwinger-Wigner quantization}\label{SWbosons}

Having discussed the quantization of the symplectic form $\Omega$ on $\mathcal{F}$, we are now ready to canonically quantize the flag manifold (i.e. the action corresponding to the `internal space'). To see how this works, let us first canonically quantize $\CP^{n-1}$, with action given in (\ref{flagpartaction}). Instead of working with normalized $u_i$, we first write the kinetic term of the Lagrangian as 
\bea
\mathcal{L}_0=i\,\sum\limits_{\alpha=1}^n\,\bar{z}^\alpha \circ \dot{z}^\alpha,
\eea
and impose the normalization constraint in the form
\bea
\sum\limits_{\alpha=1}^n\,\, |z^\alpha|^2=p\,.
\eea
Therefore the canonical momentum is $\pi^\alpha =\frac{\dd \mathcal{L}_0}{\dd \dot{z}^\alpha}=i \bar{z}^\alpha$, which leads to the algebra $\{z^{\alpha}, \bar{z}^\beta\}=p\,\delta^{\alpha\beta}$. This ultimately leads to the theory of Schwinger-Wigner quantization, which is a way of representing spin operators using creation-annihilation operators (for a review see, for example, \cite{Tsvelik}). In the present example it may be summarized as follows.

Suppose $\tau^a$ are a set of $\SU(n)$ generators in the fundamental representation. Introduce $n$ operators $a^\alpha$ and their conjugates $a^{\dagger,\alpha}$ with the canonical commutation relations
\bea\label{Weylalg}
[a^\alpha, a^{\dagger,\beta}]=\delta^{\alpha\beta}\, .
\eea
One can easily check that the operators
\bea
S^a = a^{\dagger,\alpha} \,\tau^a_{\alpha\beta}\,a^{\beta},
\eea
satisfy the commutation relations of $\mathfrak{su}(n)$, and $S^a$ act irreducibly on the subspace of the full Fock space specified by the condition
\bea
\sum\limits_{\alpha=1}^n\,a^{\dagger,\alpha} a^\alpha = p,
\eea
where $p$ is a positive integer representing the `number of particles'. For a given $p$ the representation one obtains is the $p$-th symmetric power of the fundamental representation. 

\vspace{0.3cm}\noindent

Now let us turn to a general flag manifold, with the kinetic term  $\int\,\theta$, where $d\theta=\Omega$ is the symplectic form (\ref{flagsymplform2}). Let us rewrite it as follows:
\bear\label{symplformzs}
\Omega= -{i\over 2}\,\sum\limits_{A,B}\,(p_A-p_B)\,u_A\circ d\bar{u}_B\wedge u_B\circ d\bar{u}_A=- i\,\sum\limits_{A, B}\,p_A\,u_A\circ d\bar{u}_B\wedge u_B\circ d\bar{u}_A=\\ \nonumber =i\,\sum\limits_{A, B}\,p_A\,du_A\circ \bar{u}_B\wedge u_B\circ d\bar{u}_A=
\textrm{using completeness}=d\left(i\,\sum\limits_{A}\,p_A\,u_A\circ d\bar{u}_A\right)
\eear
Defining $z_A = \sqrt{p_A} u_A$, we may therefore set
\bea
\theta=i\,\sum\limits_{A}\,z_A\circ d\bar{z}_A\,,\quad\quad \bar{z}_A\circ z_B=p_A\,\delta_{AB}.
\eea
Each $p_A$ corresponds to the `number of particles' of a particular species. The canonical quantization procedure then gives
\bea
[a_A^\alpha, \bar{a}_B^\beta]=\,\delta^{\alpha\beta}\,\delta_{AB}\,.
\eea
In other words, we introduce $n$ creation-annihilation operators $a^\alpha_A$ for each of the $p_A$. Then $p_A\in \mathbb{Z}_{\geq 0}$ is the occupation number of the $A$-th line of the Young tableau. The shift $p_A\to p_A+1$, which is inessential according to the above discussion, corresponds to adding a column to a Young diagram of full length. The differences $p_A-p_{A+1}$ are the Dynkin labels of the representation (which are the coefficients in the expansion of the highest weight in the highest weights of the fundamental representations). Whenever the lengths of two consecutive rows of the Young diagram coincide, the corresponding Dynkin label is zero, and the corresponding `symplectic form' degenerates, which signals that one should pass to a smaller flag manifold. This is consistent with our discussion in the previous subsection.

The next point is that the mutual orthogonality of the $z_A$s should be reflected in the operators $a_{A}^\alpha$ in some way. To illustrate this, let us consider the $\SU(3)$ adjoint representation. Let us label the six creation-anniliation operators as $a^\alpha,b^\alpha$ (three for each non-zero row), so that the $\mathfrak{su}(n)$ generators look as follows
\bea
S^a=a^{\dagger,\alpha} \tau^a_{\alpha\beta} a^\beta +b^{\dagger,\alpha} \tau^a_{\alpha\beta} b^\beta\, ,
\eea
where $\tau^a$ are the $n\times n$-generators in the defining representation. To model this representation on a subspace of the Fock space $F$, we build the operators
\bear
N_1=a^{\dagger,1} a^1+a^{\dagger,2} a^2+a^{\dagger,3} a^3,\quad
N_2=b^{\dagger,1} b^1+b^{\dagger,2} b^2+b^{\dagger,3} b^3\\
O_1=a^{\dagger,1} b^1+a^{\dagger,2} b^2+a^{\dagger,3} b^3
\eear 
and require the vectors $|\psi\rangle \in F$, on which the representation is built to satisfy
\bea
N_1 |\psi\rangle = 2 \,|\psi\rangle,\;N_2 |\psi\rangle = |\psi\rangle,\;O_1 |\psi\rangle=0.
\eea
 The values of $N_1$ and $N_2$ correspond to the number of boxes in the first and second rows of the Young diagram (i.e. they are `number operators' that count the number of particles of species $a$ and $b$, respectively). Notice that the classical condition $\bar{a}\circ b=0$ is translated to $O_1 |\psi\rangle=0$ with no counterpart $O_1^\dagger |\psi\rangle=0$. Indeed, the two equations would be incompatible, since $[O_1, O_1^\dagger]=N_1-N_2$ and $(N_1-N_2)\,|\psi\rangle=|\psi\rangle \neq 0$. This asymmetry is the same one that is already present in the Young diagram.
 
 Let us now explain how this generalizes to SU($n$). We introduce $n$ creation operators $a_A^{\dag,\alpha}$ for each row $A$ of the Young diagram ($A=1$ corresponds to the first row, i.e. the longest one), and impose the condition 
\bea\label{Okmconstr}
\mathcal{O}_{AB}|\psi\rangle\equiv a_A^\dagger\circ a_B \;|\psi\rangle=0 \quad \textrm{for}\quad A<B.
\eea
This is a compatible set of equations, since the operators $\mathcal{O}_{AB}$ satisfy the algebra
\bea
[\mathcal{O}_{AB}, \mathcal{O}_{CD}]=\delta_{BC} \mathcal{O}_{AD}-\delta_{AD}\mathcal{O}_{CB}\quad\textrm{where}\quad A<B,\;\;C<D.
\eea
The operators $\mathcal{O}_{AB}$ may be thus thought of as the positive roots of the Lie algebra $\mathfrak{su}(n)$. In Chapter 2, this algebra will reappear in the context of SU($n$) spin operators. 

\vspace{0.3cm}\noindent
The constraint (\ref{Okmconstr}) may be solved rather explicitly. More exactly, we are looking for the joint kernel of the operators $\mathcal{O}_{AB}$, $A<B$ acting on states in the $(N_1, N_2, \ldots)$-particle Fock space:
\bea\label{oscillstates}
|\Psi\rangle=A_{\alpha_1\ldots \alpha_{N_1}|\beta_1\ldots \beta_{N_2}|\ldots}\,a^{\dagger,\alpha_1}\ldots a^{\dagger,\alpha_{N_1}}\; b^{\dagger,\beta_1}\ldots b^{\dagger,\beta_{N_2}}\;\ldots\;|0\rangle\,.
\eea
The kernel is a linear space, and the basis in this space may be constructed as follows.
\begin{enumerate}
    \item{} Assign to each row of the Young diagram a letter $a, b, c, \cdots$. For example:
\[
\young(aaa,bb,c) 
\]
    \item{} For each column build antisymmetric combinations of the form 
    \bea \nonumber
    \sum\limits_{\sigma} \,(-)^\sigma\,a^{\dagger,\sigma(i)}\,b^{\dagger,\sigma(j)}\,c^{\dagger,\sigma(k)}\,d^{\dagger,\sigma(l)}\,,
    \eea
    where the number of letters participating is equal to the height of the column.
    \item{} Multiply these antisymmetric combinations (the number of `particles' of type $A$ will be precisely equal to the length of the $A$-th row in the Young diagram). To see that these are annihilated by operators $\mathcal{O}_{AB}, A<B$, note that the action of this operator removes the $B$-th letter and replaces it by the $A$-th letter, and since the $B$-th letter for $B>A$ always enters in skew-symmetric combinations with the $A$-th letter, the result will be zero.
\end{enumerate}

\subsubsection{Geometric quantization}\label{geomquantsec}
The mathematical counterpart of the procedure that we just described is called geometric quantization. One of the main statements of the subject -- the Borel-Weil-Bott theorem -- asserts that, given a representation $V$ of a group $G$, one can construct a holomorphic line bundle $\mathcal{L}_V$ over a suitable flag manifold of $G$ (in full generality one can take the manifold of complete flags), such that $V$ may be reconstructed as the space of holomorphic sections $H^0(\mathcal{L}_V)$ of $\mathcal{L}_V$. These holomorphic sections are polynomials, and indeed it is elementary to find a map from the space of states~(\ref{oscillstates}) to the space of polynomials -- this is essentially the Bargmann representation, as we review in Appendix~\ref{Bargmannapp}.  Given the background material accumulated to this point, we can somewhat specify what the line bundle in the Borel-Weil-Bott theorem is: it is characterized by its first Chern class that is  represented by the symplectic form $\Omega$~(\ref{symplformzs}), through which the kinetic term in the action standing in the path integral is defined. In other words, $[\Omega]=c_1(\mathcal{L}_V)\in H^2(\mathcal{F}, \mathbb{Z})$.

The flag manifold itself  that features in this construction is the manifold of `coherent states', which by definition are the states in the orbit of $G$ acting on the highest weight vector. This connection  becomes perhaps more transparent if one recalls  the discussion in section~\ref{appQuant}, where the integration of the symplectic form $\Omega$ over various two-cycles in the flag manifold was described. We may view the cycles $(\CP^1)_{{A_i},A_{i+1}}$ as the positive simple roots of $\mathfrak{su}(n)$, and $\Omega$ as a highest weight. It is a general theorem that highest weight orbits are K\"ahler manifolds~\cite{Kostant}. Coherent states are important for the construction of spin chain path integrals in Chapter 2, so we discuss them in more detail below in section~\ref{cohstates}. For a general discussion of geometric quantization we refer the reader to~\cite{KirillovReview} (see also~\cite{BykovHaldane2}).

\subsubsection{Simple examples of representations.} \label{sec:youngexample}

\vspace{0.3cm}\noindent
Let us present three example representations in SU($n$). We will return to these examples later on when we discuss coherent states. In all cases the states are built as polynomials in the creation operators $a^{\dagger,\alpha}$, $b^{\dagger,\alpha}$, etc., acting on the vacuum state $|0\rangle$.

\vspace{0.3cm}\noindent
\textbf{a)} 
\hspace{0.3cm}\young(aaaa)\hspace{0.5cm}
Symmetric powers of the fundamental representation $\Rightarrow$ Polynomials in $a^{\dagger,1},\dots, a^{\dagger,n}$ of degree~$4$.

\vspace{0.3cm}\noindent
\textbf{b)} 
\hspace{0.3cm}
\young(aa,b) \hspace{0.5cm}
In this case we have linear combinations of polynomials in $a^{\dagger,\alpha}$ and $b^{\dagger,\alpha}$ of the form $a^{\dagger,\alpha}\,(a^{\dagger,\beta} b^{\dagger,\gamma}-a^{\dagger,\gamma} b^{\dagger,\beta})$

\vspace{0.3cm}\noindent
\textbf{c)} 
\hspace{0.3cm}
\young(aaa,bb,c) \hspace{0.5cm}
Here we have linear combinations of polynomials in $a^{\dagger,\alpha}$, $b^{\dagger,\alpha}$ and $c^{\dagger,\alpha}$ of the form $a^{\dagger,\alpha}\,(a^{\dagger,\beta} b^{\dagger,\gamma}-a^{\dagger,\gamma} b^{\dagger,\beta})(a^{\dagger,\gamma} b^{\dagger,\delta} c^{\dagger,\lambda}-a^{\dagger,\gamma} b^{\dagger,\lambda} c^{\dagger,\delta}-a^{\dagger,\delta} b^{\dagger,\gamma} c^{\dagger,\lambda}-a^{\dagger,\lambda} b^{\dagger,\delta} c^{\dagger,\gamma}+a^{\dagger,\delta} b^{\dagger,\lambda} c^{\dagger,\gamma}+a^{\dagger,\lambda} b^{\dagger,\gamma} c^{\dagger,\delta})$.

\subsubsection{Example}

Apart from its aesthetic appeal, this construction offers certain calculational benefits. For instance, the calculation of values of the Casimir operators on various representations becomes a matter of simple harmonic oscillator algebra.
As an example we calculate the value of the quadratic Casimir of $\mathfrak{su}(n)$ in the representation described schematically by the following diagram:

\hspace{0.3cm}$\underbrace{\overbrace{\yng(4,3)}^{p_1}\!\!\!\!\!\!}_{p_2}$\hspace{0.5cm} where we assume there are $p_1$ boxes in the first row and $p_2$ boxes in the second one ($p_1\geqslant p_2$). We assign $n$ pairs of creation/annihilation operators $a, a^\dagger, b ,b^\dagger$ to each row. The rotation generators are
\bea\label{SchWign}
S^a=a^\dagger \circ T^a\circ a+b^\dagger \circ T^a\circ b.
\eea
The generators $T^a$ are unit-normalized: $\tr(T^a T^b)=\delta^{ab}$. Then $\sum\limits_a\;T^a \otimes T^a = P-{1\over n} \,I$, where $P$ is the permutation and $I$ the identity operator. Thus, for the Casimir one obtains (here for brevity we omit the state $|\psi\rangle$ on which these operators act, but its presence is implied)
\bear\nonumber 
&&\!\!\!\!\!\!\!\!\!\!\!\!\!\!\! C_2\equiv \sum\limits_a\;S^a\,S^a= \underbracket[0.6pt][0.6ex]{a^{\dagger,\alpha} \, a^{\beta}\;a^{\dagger,\beta} \, a^\alpha}_{=p_1^2+(n-1)p_1}+\underbracket[0.6pt][0.6ex]{a^{\dagger,\alpha} \, a^\beta \;b^{\dagger,\beta} \, b^\alpha}_{=-p_2}+\underbracket[0.6pt][0.6ex]{b^{\dagger,\alpha} \, b^\beta \, a^{\dagger,\beta} \, a^\alpha}_{=-p_2}+\underbracket[0.6pt][0.6ex]{b^{\dagger,\alpha} \, b^\beta \; b^{\dagger,\beta} \, b^\alpha}_{=p_2^2+(n-1)p_2}-\\ \nonumber \hspace{-0.5cm} &&\!\!\!\!\!\!\!\!\!\!\!\!\!\!\! -
{1\over n}\underbracket[0.6pt][0.6ex]{ (a^{\dagger,\alpha} \, a^\alpha+b^{\dagger,\alpha} \, b^\alpha)^2}_{=(p_1+p_2)^2}= p_1^2+(n-1)p_1+p_2^2+(n-1)p_2-{1\over n}(p_1+p_2)^2-2\,p_2
\eear

\vspace{0.3cm}
\noindent
This can be easily generalized to arbitrary representations of $\mathfrak{su}(n)$. Indeed, consider a Young diagram with $n$ rows (the maximal number for $\mathfrak{su}(n)$), the row lengths being $p_1\geq \ldots \geq p_{n-1}\geq p_n=0$. Introducing the variable $s_A=p_A-\frac{1}{n}\,\sum\limits_{B=1}^n\,p_B$ (so that $\sum\limits_{A=1}^n\,s_A=0$), the value of the second Casimir turns out to be
\bea
C_2=\sum\limits_{A=1}^n\,s_A(s_A-2A)\,,
\eea
in accordance with the result obtained long ago~\cite{PP2}.

\subsubsection{Coherent states}\label{cohstates}

Coherent states are a type of basis in a vector space on which a Lie group $G$ is represented. One takes a highest weight vector $|\psi \rangle$ and forms its $G$-orbit.  That is, one considers all vectors of the form $g\,|\psi \rangle$, where $g\in G$. This is a continuous basis, which is therefore overcomplete. In what follows we will be dealing solely with the case of compact $G=\tU(n)$, however we find it useful to remind the reader of how the definition just introduced fits into the familiar setup of quantum mechanics~(cf.~\cite{Klauder}). In this case one has a Heisenberg algebra $[a, a^\dagger]=\mathds{1}$ with a highest weight vector $|0\rangle$, which is annihilated by $a$ (and clearly fixed by the unit operator). The normalized coherent states are therefore given by the familiar formula
\bea
|v\rangle\equiv e^{-{1\over 2} |v|^2}\;e^{v\,a^\dagger}\,|0\rangle.
\eea

In this case coherent states are parametrized by complex numbers: $v\in \CC$. As we mentioned earlier, it is a general fact~\cite{Kostant} that the highest weight orbit in the projectivization $P(V)$ of an irreducible representation $V$ of a \emph{compact} Lie group is K\"ahler. For the coherent states we find below, $\nu$ will live in some flag manifold. 

In the case of $\SU(n)$ the coherent states can be expressed in terms of the creation-annihilation operators introduced via Schwinger-Wigner quantization above\footnote{A classic reference on coherent states for compact Lie groups, suitable for a mathematically inclined reader, is~\cite{Perelomov1}. A rather clear exposition of coherent states and geometric quantization can be also found in \cite{FH} and \cite{Saemann}. Some very explicit formulas for the coherent states of $\mathfrak{su}_3$ may be found in~\cite{MathurSen}. Another approach to the quantization of coadjoint orbits is developed in~\cite{AFS}.}. Having the bases at hand, in order to build the coherent states all one needs to do is to pick a particular state and form its orbit under $\SU(n)$.  For each of the three Young diagrams appearing in Section~\ref{sec:youngexample}, we build them explicitly; the general case should be clear from these examples.

\textbf{a)} The highest weight vector is $(a^{\dagger,1})^4|0\rangle$. Since $(g a^{\dagger,1} g^{-1})^4 = (\bar z_1 \circ a^{\dagger})^4$ for $z_1$ the first column\footnote{We use the same symbol for the matrix realization and the Fock space operator realization of a transformation $g\in \SU(n)$.} of $g \in \SU(n)$, we may parameterize the coherent states in this case as
\bea\label{CPNcohstates}
|v\rangle=(\bar{v}\circ a^\dagger)^4\,|0\rangle,\quad v\in \CP^{n-1} = \mathcal{F}_{1,n-1}
\eea

\textbf{b)} The highest weight vector is $a^{\dagger,1}\cdot(a^{\dagger,1} b^{\dagger,2}-a^{\dagger,2} b^{\dagger,1})|0\rangle$, and leads to\,\,$
|vw\rangle=(\bar{v}\circ a^\dagger)\cdot[(\bar{v}\circ a^\dagger)(\bar{w}\circ b^\dagger)-(\bar{w}\circ a^\dagger)(\bar{v}\circ b^\dagger)]|0\rangle,\quad \bar{w}\circ v=0$. Here $v$ and $w$ parametrize the partial flag manifold $\mathcal{F}_{1,1,n-2}$.

\textbf{c)} The highest weight vector $a^{\dagger,1}\cdot(a^{\dagger,1} b^{\dagger,2}-a^{\dagger,2} b^{\dagger,1})\cdot(a^{\dagger,1} b^{\dagger,2} c^{\dagger,3}-a^{\dagger,1} b^{\dagger,3} c^{\dagger,2}-a^{\dagger,2} b^{\dagger,1} c^{\dagger,3}-a^{\dagger,3} b^{\dagger,2} c^{\dagger,1}+a^{\dagger,2} b^{\dagger,3} c^{\dagger,1}+a^{\dagger,3} b^{\dagger,1} c^{\dagger,2} )|0\rangle$ leads to the coherent states
\bear
&
|uvw\rangle=(\bar{v}\circ a^\dagger)\cdot[(\bar{v}\circ a^\dagger)(\bar{w}\circ b^\dagger)-(\bar{w}\circ a^\dagger)(\bar{v}\circ b^\dagger)]\cdot \\ \nonumber &\cdot[(\bar{v}\circ a^\dagger)(\bar{w}\circ b^\dagger)(\bar{u}\circ c^\dagger)-(\bar{v}\circ a^\dagger)(\bar{u}\circ b^\dagger)(\bar{w}\circ c^\dagger)-(\bar{w}\circ a^\dagger)(\bar{v}\circ b^\dagger)(\bar{u}\circ c^\dagger)-\\ \nonumber
&-(\bar{u}\circ a^\dagger)(\bar{w}\circ b^\dagger)(\bar{v}\circ c^\dagger)+(\bar{w}\circ a^\dagger)(\bar{u}\circ b^\dagger)(\bar{v}\circ c^\dagger)+(\bar{u}\circ a^\dagger)(\bar{v}\circ b^\dagger)(\bar{w}\circ c^\dagger)]|0\rangle
\eear
with $\bar{w}\circ v=\bar{u}\circ w=\bar{u}\circ v=0$. These three variables parametrize $\mathcal{F}_{1,1,1,n-3}$.

It is easy to see that the above vectors are highest weight vectors. It follows from the representation~(\ref{SchWign}) (taking into account the obvious generalization to the case of three oscillators $a, b, c$) that those generators  $T^{a}$, which are upper-triangular, correspond to the following transformations of the operators $a^\dagger, b^\dagger, c^\dagger$:
\bear
\delta a^{\dagger,\alpha} = \sum\limits_{\beta<\alpha}\,\kappa_{\alpha\beta}\,a^{^\dagger,\beta}\,,\quad\quad\quad 
\delta b^{\dagger,\alpha} = \sum\limits_{\beta<\alpha}\,\kappa_{\alpha\beta}\,b^{\dagger,\beta}\,,\quad\quad\quad 
\delta c^{\dagger,\alpha} = \sum\limits_{\beta<\alpha}\,\kappa_{\alpha\beta}\,c^{\dagger,\beta}\,,
\eear
i.e. in the matrix $(a, b, c)$ the upper rows are added to the lower ones. Since the constructed states are defined through the  \emph{upper} minors of this matrix, they are invariant under such transformations, i.e. they are annihilated by all positive roots.

One of the central properties of coherent states is that they form an overcomplete basis. This is reflected in a fundamental identity -- the so-called `partition of unity'. For the case when the manifold of coherent states is $\CP^{n-1}$ (as in~(\ref{CPNcohstates})), which is the only case we will really be using, the identity takes the form
\bea\label{completeness}
\int\,d\mu(v, \bar{v})\;\frac{|v\rangle \langle v|}{\langle v|v \rangle}=\mathds{1}\,,
\eea
where $d\mu$ is the suitably normalized volume form on $\CP^{n-1}$. It is proportional to the top power of the Fubini-Study form, $d\mu\sim \omega_{\mathrm{FS}}^{n-1}$, and looks as follows when expressed in the inhomogeneous coordinates:
\bea
(d\mu)_{\CP^{n-1}}\sim\,\left(1+\sum\limits_{\alpha=1}^{n-1} \,v^\alpha \bar{v}^\alpha\right)^{-n}\, \prod\limits_{\alpha=1}^{n-1}\,(i\,dv^\alpha\wedge d\overline{v}^\alpha)\,.
\eea
For more complicated representations, where coherent states are labeled by more general flag manifolds than $\CP^{n-1}$, one would have to replace $d\mu$ with the corresponding volume form.

\subsection{Holstein-Primakoff and Dyson-Maleev representations}\label{HPDMsec}

In Section~\ref{SWbosons}, we demonstrated how Schwinger-Wigner oscillators arise from the canonical quantization of the flag manifold phase space in \emph{homogeneous} coordinates. We will now proceed to show that the famous Holstein-Primakoff representation corresponds to the quantization of the sphere -- the most elementary flag manifold -- in certain coordinates, related to the \emph{action-angle} and to the \emph{inhomogeneous} coordinates. A corresponding $\SU(n)$ flag manifold version can also be developed along the same lines. We start from the first-order Lagrangian
\bear\label{CP1inhomlagr}
&&\mathcal{L}_{\CP^1}=p{i\over 2}\frac{\bar{z}\dot{z}-z\dot{\bar{z}}}{1+z\bar{z}}=p\,\uprho^2\,d\varphi=p\,{i\over 2}(\bar{w}\dot{w}-w\dot{\bar{w}}),\\ \nonumber && \textrm{where}\quad\quad \begin{tabular}{c }
   $z=|z|\,e^{-i\varphi}$    \\
    $w= \uprho \,e^{-i\varphi}$   
\end{tabular}\,,\quad\quad \uprho^2=1-{1\over 1+z \bar{z}}\,.
\eear
As explained before, upon quantization $p\in \mathbb{Z}_+$ is a positive integer encoding the representation. 
We also need the expressions for the $\SU(2)$ charges. If we denote the vector $Z:=\left(\begin{tabular}{c}
    $1$  \\
    $z$ 
\end{tabular}\right)$, the spin variables are the moment maps
$ S^{a}=p\,\frac{Z^\dagger \tau^{a} Z}{\bar{Z}\circ Z}$, so that
\bea\label{HPspin}
S^+=p\,\frac{z}{1+|z|^2},\quad\quad S^-=p\,\frac{\bar{z}}{1+|z|^2},\quad\quad S^z=p\,\frac{1-|z|^2}{1+|z|^2}\,.
\eea
Using $z={|z|\over \uprho}\,w={w\over (1-|w|^2)^{1/2}}$, we find
\bea\label{HPspin2}
S^+=p\,w(1-|w|^2)^{1/2},\quad\quad S^-=p\,\bar{w}(1-|w|^2)^{1/2},\quad\quad S^z=p\,(1-2|w|^2)\,.
\eea
To canonically quantize the system~(\ref{CP1inhomlagr}), we denote $A:=\sqrt{p}\,w$\;, $A^\dagger:=\sqrt{p}\,\bar{w}$ and postulate the canonical commutation relations $[A, A^\dagger]=1$. Choosing the ordering compatible with the unitary relation $S^+=(S^-)^{\dagger}$, we find
\bea
S^+=(p-A^\dagger A)^{1/2}A,\quad\quad S^-=A^\dagger\,(p-A^\dagger A)^{1/2} ,\quad\quad S^z=(p-2A^\dagger A)\,,
\eea
which is the Holstein-Primakoff representation for the spin operators.

We have demonstrated that the Holstein-Primakoff realization arises from the quantization of the sphere $\CP^1$ which is the simplest example of a coadjoint orbit of a compact group. There is yet another well-known realization of the spin operators -- the so-called Dyson-Maleev realization -- whose advantage is that the resulting expressions for the spin operators are \emph{polynomial}. The reason why we wish to discuss this representation is that the corresponding setup is very similar to the one in which the integrable models will be formulated in Chapter~3. As we shall see there, these Dyson-Maleev variables may be used to demonstrate that the interactions in the sigma models are polynomial.

The Dyson-Maleev representation may as well be obtained in the framework of canonical quantization, however the primary objects in this case are the  orbits of the \emph{complexified} group $\text{SL}(n, \CC)$. In the mathematics literature\footnote{We wish to thank K.~Mkrtchyan for drawing our attention to this work and important discussions  on the subject. Some applications of the theory of `minimal' realizations of Lie algebras, as well as a list of related literature, may be found in~\cite{Karapet}.} this subject was initiated in~\cite{Joseph}. The question asked in that work was about constructing a representation of a given complex Lie algebra in terms of a minimal number of Weyl pairs (i.e. $q_j, p_j$-operators, such that $[q_j, p_k]=i \delta_{jk}$). As explained in~\cite{Joseph2}, the solution to this problem is in considering coadjoint orbits $\mathcal{O}$ of a minimal dimension of a corresponding Lie group. These are symplectic varieties, which may be naturally quantized in terms of $s$ Weyl pairs, where $s={1\over 2}\mathrm{dim}_{\CC}\,\mathcal{O}$. The classical limit of the Weyl pairs produces the Darboux coordinates on $\mathcal{O}$. It was also shown in~\cite{Joseph2} that, unless the Lie algebra in question is $\mathfrak{sl}(n)$, the minimal orbit is nilpotent, so typically this setup leads to the theory of nilpotent orbits. For $\mathfrak{sl}(n)$, which is our main case of interest, there is a continuum of semi-simple orbits, whose limiting point is a nilpotent orbit of the same (minimal) complex dimension $n-1$. 

Let us explain how this works for $\mathfrak{sl}(2)$. The semi-simple orbits may be labeled by the Cartan elements $\left(\begin{tabular}{c c}
  $a$   & $0$ \\
  $0$   & $-a$
\end{tabular}\right)$, where $a\in \CC\setminus \{0\}$ (the limit $a=0$ corresponds to the closure of the nilpotent orbit). The equation defining the orbit is ($M\in \mathfrak{sl}(2)$)
\bea\label{simpleorb}
M^2=a^2\,\mathds{1}_2\,.
\eea
Consider the following first-order Lagrangian (which should be viewed as the relevant counterpart of~(\ref{CP1inhomlagr})):
\begin{empheq}[box=\fbox]{align}
\label{DMlagr}
\hspace{1em}\vspace{5em}
\mathcal{L}=\sum\limits_{j=1}^2\,\left(V_i\cdot \mathcal{D}U_i+\bar{V_i}\cdot \bar{\mathcal{D} U_i}\right)+2\,(a\,\mathcal{A}+\bar{a}\,\bar{\mathcal{A}}),\quad\quad \mathcal{D}U_i=\dot{U}_i-\mathcal{A}\,U_i.
\end{empheq}
Here $U_i, V_i$ are the complex canonical variables, and the gauge field $\mathcal{A}$ is meant to generate the quotient by $\CC^\ast$. Just as before, the first term in the Lagrangian is a Poincar\'e-Liouville one-form corresponding to a certain (this time complex) symplectic form, and the introduction of a gauge field allows one to obtain the symplectic form on the orbit by means of a symplectic reduction. Here we will just take this fact for granted, but such representations are discussed in more detail in Chapter 3, in the context of integrable sigma models with flag manifold target spaces. The second term in the Lagrangian is a `Fayet-Iliopoulos term': under gauge transformations it shifts by a total derivative, but the action $\mathcal{S}=\int\,dt\,\mathcal{L}$ is invariant.

The group $\text{SL}(2, \CC)$ acts as $U\to g\circ U, V\to V\circ g^{-1}$, and from~(\ref{DMlagr}) one can derive the conserved charges corresponding to this action:
\bea
\mu=U\otimes V-{(V\circ U)\over 2}\,\mathds{1}_2
\eea
This is the moment map for the complex symplectic form $\omega=\sum\limits_{i=1}^2\,dV_i\wedge dU_i$, which is why we have denoted it by $\mu$. Varying the Lagrangian w.r.t. the gauge field, we obtain the constraint $V\circ U=2a$. As a result, $\mu$ satisfies the equation $\mu^2=a^2\,\mathds{1}_2$, so that $\mu$ belongs to the orbit~(\ref{simpleorb}). 

 Let us now choose  `inhomogeneous coordinates', i.e. we assume that at least one of $U_1, U_2$ is non-zero, say $U_1\neq 0$, in which case by a $\CC^\ast$-transformation we may set $U_1=1$. We also denote $U_2:=U$ and $V_2:=V$. The constraint $\sum\limits_{i=1}^2\,V_i \cdot U_i=2a$ may now be solved as $V_1=2a-V\cdot U$. The spin matrix $\mu$ has the following form in these variables:
\bea
\mu=\left(
    \begin{tabular}{c c}
       $a-V\cdot U$  & $V$  \\
       $U\cdot (2a-V\cdot U)$  & $V\cdot U-a$   
    \end{tabular}
    \right)
\eea
Quantization of~(\ref{DMlagr}) in the inhomogeneous coordinates $U, V$ amounts to imposing the canonical commutation relations $[U, V]=i$. In this case one has to deal with the ordering ambiguity (which is still much milder than the one in~(\ref{HPspin2}) and may easily be resolved by imposing the $\mathfrak{sl}(2)$ commutation relations), and as a result one arrives at the Dyson-Maleev representation
\bea
S^+=V,\quad\quad S^-=U\,(2a-U\,V),\quad\quad S^z=a-U\,V,\quad\quad \textrm{where}\quad\quad [U, V]=i\,.
\eea
By identifying $V=-i\,{\dd \over \dd U}$, we also obtain the well-known differential operator realization
\bea
S^+=-i\,{\dd \over \dd U},\quad\quad S^-=U\,\left(2a+i\,U\,{\dd \over \dd U}\right),\quad\quad S^z=a+i\,U\,{\dd \over \dd U},
\eea
which for $a=0$ is the standard form for the $\mathfrak{sl}(2)$-operators acting on the sphere $\CP^1$ with inhomogeneous coordinate $U$.

\pagebreak
\vspace{0.5cm}
\noindent
\rule{\textwidth}{1pt}
    \vspace{1ex}
\begin{center}
\vspace{-0.3cm}
{\Large     Chapter 2. From spin chains to sigma models}
\end{center}

\noindent
\vspace{-0.5ex}%
\rule{\textwidth}{1pt}

\addcontentsline{toc}{section}{\bfseries Chapter 2. From spin chains to sigma models}

\vspace{2cm}
In this chapter, we consider quantum spin systems in one spatial dimension. In their simplest form, these systems are described by the Heisenberg model, and are either ferromagnetic or antiferromagnetic, depending on the sign of the interaction term between neighboring spins on the chain. While the ferromagnet's ground state is the same for both classical and quantum chains (it is the state with all spins aligned along a common direction), this is not true for the antiferromagnet. Classically, the ground state is the so-called N\'eel state, with spins alternating between being aligned and antialigned along a common direction, but quantum mechanically the N\'eel state is no longer an eigenstate of the Heisenberg Hamiltonian. This fact can be understood from Coleman's theorem, which forbids the spontaneous ordering of a continuous symmetry in one spatial dimension~\cite{Coleman1973}.\begin{footnote}{Coleman's theorem is often confounded with the Mermin-Wagner-Hohenberg theorem, which forbids an ordered grounds state in two spatial dimensions at finite temperature, and applies equally well to both ferromagnets and antiferromagnets~\cite{MerminWagner1966,Hohenberg}.}\end{footnote}

The absence of an ordered ground state in the antiferromagnet has long been of interest to the physics community. Indeed, shortly after Heisenberg introduced his model of a ferromagnet in 1921, Bethe discovered an exact solution of the antiferromagnetic chain with spin $S=\frac{1}{2}$ at each site~\cite{bethe}. However, despite this initial progress, spin chains with $s>\frac{1}{2}$ were not amenable to such techniques, and fifty years would pass before their low energy properties could be characterized. In 1981, Duncan Haldane proposed a radical classification of antiferromagnetic chains: Those with integral spin $s$ have a finite energy gap above their quantum ground states, and exponentially decaying correlation functions. Meanwhile, those chains with half-odd integral spin have gapless excitations with algebraically decaying correlation functions~\cite{Haldane1983}.

Despite being consistent with Bethe's 1931 solution, Haldane's ``conjecture'' as it came to be known, was met with widespread skepticism~\cite{nobel}. This was likely due to the fact that spin-wave theory, a method that allows one to calculate the energy spectrum of antiferromagnets in higher dimensions, largely agreed with Bethe's one dimensional results. We now know this to be a coincidence, but at the time, this suggested to the community that spin wave results might be reliable in one dimension for all values of $S$. This would imply that all antiferromagnets would exhibit gapless excitations at low energies. Of course, this was in direct contradiction with Coleman's theorem, that invalidated spin wave theory in one dimension, but nonetheless, by the 1980s it was widely accepted that gapless excitations were universal among antiferromagnets.

In fact, Haldane's conjecture was met with surprise in other areas of physics as well. As we will demonstrate below, his argument hinges on a correspondence between antiferromagnets and the $\CP^1$ sigma model, a quantum field theory that was being used as a toy model for quantum chromodynamics at the time~\cite{Elitzur}. The role of the spin, $s$, manifests as a topological angle $\theta$ in the sigma model, so that integral $s$ translates to $\theta=0$ and half-odd-integral $s$ translates to $\theta=\pi$. Thus, Haldane's claim about antiferromagnets was also a claim about mass gaps in the $\CP^1$ sigma model. At that time, it was widely believed that a finite gap would exist for all values of $\theta$, and this was known exactly for $\theta=0$, and suggested numerically for small values of $\theta$~\cite{Zamolodchikov, BhanotNuclPhys1984,BhanotPRL1984}. It was shown 
in ~\cite{AffleckSUn1988} that  the $\CP^1$ model is gapless at $\theta =\pi$ but 
this is not true for $\CP^{n-1}$ with $n>2$.  In that case there is a first order 
transition at $\theta =\pi$ with the model remaining massive. This can be understood 
from the presence of relevant operators allowed by symmetry for $n>2$. The 
generalization of this behaviour to four-dimensional $\SU(n)$ Quantum Chromodynamics is 
a fascinating subject~\cite{Gaiotto_2015}. For large $n$ it has been established that the transition 
is first order with a finite mass~\cite{witten_largen,Witten_1998}. Whether or not this is true for $SU(3)$ is 
an open question.

Over the next few years, Haldane's conjecture would defy these skeptics, thanks to verification from multiple areas of research. Experimentally, neutron scattering on the organic nickel compound NENP, which is a quasi-one dimensional $s=1$ chain, detected a finite energy gap above the ground state~\cite{Buyers1986, Renard2003}.  Numerically, studies using exact diagonalization, Monte Carlo, and density matrix renormalization group methods were able to detect a finite gap in $S=1,2$ and 3~\cite{Botet1983,Nightingale1986,Kennedy1990, White1993, Schollwock1996, todo2001}. Very recently, this has been extended to $S=4$~\cite{Todo2019}. Meanwhile, in the $\CP^1$ sigma model, Monte Carlo methods were used to numerically verify the absence of a mass gap when $\theta=\pi$~\cite{Wiese1995,Azcoiti2003, Alles2008,Azcoiti2012, Wiese2012, Alles2014}, and a related integrable model was eventually discovered by the Zamolodchikov brothers~\cite{ZamolodchikovMassless}.

In many cases, the studies carried out in order to verify Haldane's claims were scientific breakthroughs in their own right. Indeed, the fields of density matrix renormalization group~\cite{white1,white2,ostlund}, and more generally tensor networks~\cite{vidal1, vidal2,vidal3}, as well as symmetry protected topological matter~\cite{spt1} all originated, in part, due to Haldane's conjecture. It is thus not a leap to claim that any generalization of Haldane's conjecture would be an impactful result to the physics community. And indeed, this is what led physicists, including Affleck, Read, Sachdev and others to extend Haldane's work to SU($n$) generalizations of spin chains in the late 1980s~\cite{Affleck1986, Affleck:1984ar,AffleckSUn1988, read}. At the time, these were purely hypothetical models with no experimental realization, but thanks to the correspondence between spin chains and sigma models, they were still interesting in their own right. Another motivation was a proposed relation between sigma models and the localization transition in the quantum Hall effect~\cite{Affleck1986, levine,evers}. And while this unsolved problem remains a motivator to study such models, recent advances from the cold atom community have revealed that SU($n$) chains (with $n\leq 10$) are now experimentally realizable, offering a much more physical motivation~\cite{wu2003, honerkamp2004, Cazalilla2009, gorshkov2010, bieri2012, scazza2014, taie2012, pagano2014, zhang2014, cazalilla2014, nonne2013, hofrichter,hideki2018}. This fact has led to a renewed theoretical interest in the field of SU($n$) spin chains. As a consequence, an SU($n$) version of Haldane's conjecture has recently been formulated~\cite{HaldaneSU3,Wamer2019,HaldaneSUN,Wamer2020}.

In this chapter, we review this recent effort of extending Haldane's conjecture from SU(2) to SU($n$). We begin in Section~\ref{section:ham} by introducing the SU($n$) Heisenberg chain, in the rank-$p$ symmetric representation. Unlike the familiar spin chains with SU(2) symmetry, for $n>2$, these symmetric representations form only a small subset of all possible irreducible representations. Near the end of this chapter, we return to this issue and analyze SU($n$) chains in other representations.

Next, in Section~\ref{section:exact}, we recall various exact results that exist for these SU($n$) Hamiltonians. Specifically, we discuss the Lieb-Shultz-Mattis Affleck theorem~\cite{LSM,AL}, and the Affleck-Kennedy-Lieb-Tasaki construction~\cite{aklt}.

In Section~\ref{section:fw}, we extend the familiar spin-wave theory to these SU($n$) chains, and obtain predictions for the velocities of low lying excitations. We observe that for $n>3$, there are multiple distinct velocities, which inhibit the automatic emergence of Lorentz invariance. 

Sections~\ref{section:derivation} through~\ref{symmrepsec} then provide a step-by-step derivation of a low-energy field theory description of the SU($n$) chain. This extends Haldane's original mapping of the spin chain to the $\CP^1$ model; now, the corresponding target space is the complete flag manifold, $\SU(n)/[\tU(1)]^{n-1}$. Thus, via these sections we establish a direct link from SU($n$) chains to the subject matter of Chapter 1. In Section~\ref{symmrepsec}, we also explain how the distinct flavour wave velocities flow to a common value upon renormalization.

The generalized Haldane conjecture is presented in Section~\ref{hooftanomsec}. This combines the exact results of Section~\ref{section:exact} with an analysis of mixed `t Hooft anomalies between the global symmetries of the chain. After quoting the results, we offer a detailed discussion of the mathematical structure behind these anomalies, which involves the concept of PSU($n$) bundles.

In Section~\ref{sec:massgen}, we reinterpret the SU($n$) Haldane conjecture in terms of fractional topological excitations, which generalize the notion of merons in SU(2)~\cite{meron}. Finally, in Section~\ref{linquadchains}, we explain how SU($n$) representations other than the rank-$p$ symmetric ones may admit a mapping to the same flag manifold target space, $\SU(n)/[\tU(1)]^{n-1}$. This leads us to non-Lagrangian embeddings of the flag manifold, resulting in the phenomenon that some low energy excitations have linear dispersion, while others have quadratic dispersion.

\section{Hamiltonian} \label{section:ham}

The familiar Heisenberg spin chain is characterized by a single integer, $2s$, which specifies the irreducible representation of SU(2) that appears on each site. In SU($n$), the most generic irrep is defined by $n-1$ integers, which give the lengths of the rows in its Young tableaux. In this chapter, we will mostly focus on the rank-$p$  symmetric irreps, which have Young tableaux
\bea
    \overbrace{\yng(10)}^p.
\eea

\begin{figure}[h]
\centering
\includegraphics[width = .7\textwidth]{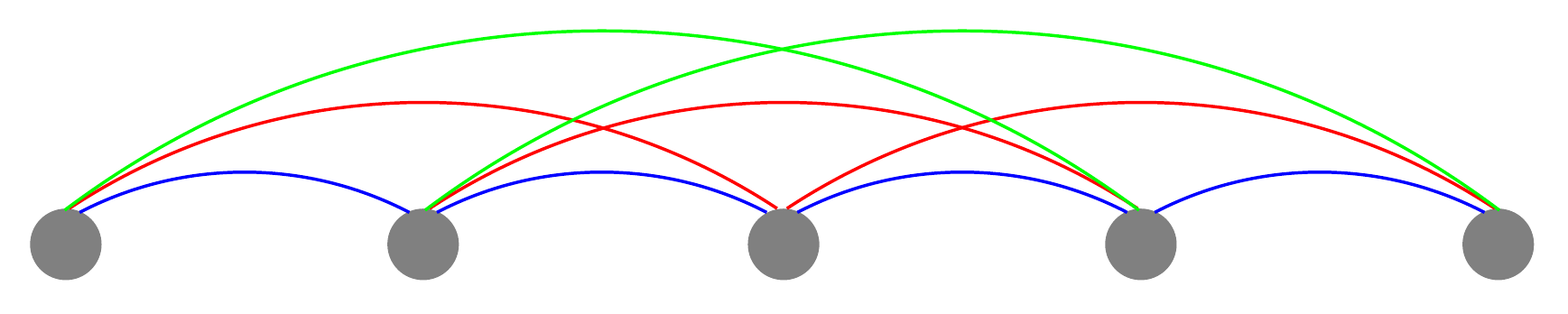}
\caption{Pictorial representation of the nearest (blue), next-nearest (red), and next-next-nearest (green) neighbour interactions occurring in (\ref{1:2}), for the case $n=4$.}
\label{cell}
\end{figure}

The simplest Hamiltonian one is tempted to write down is
\bea \label{1:1}
	H = J \sum_j \tr (S(j) S(j+1))
\eea
where $S(j)$ is an $n\times n$ Hermitian matrix with $\tr(S) =p$,\begin{footnote}{S(j) should be traceless; we have shifted it by a constant to simplify our calculations.}\end{footnote} whose entries correspond to the $n^2-1$ generators of SU($n$) and satisfy
\bea \label{1:3}
	[S_{\alpha\beta}, S_{\gamma \delta}] = \delta_{\alpha \delta} S_{\gamma \delta} - \delta_{\gamma \beta} S_{\alpha\delta}.
\eea
Indeed, in SU(2), $S_{\alpha \beta} = \vec{S} \cdot \vec{\sigma}_{\alpha \beta} + \frac{p}{2}\mathbb{I}$, and the Hamiltonian appearing in (\ref{1:1}) equals the Heisenberg model with spin $s = \frac{p}{2}$ (up to a constant). However, for $n>2$, this Hamiltonian possesses local zero mode excitations that destabilize the classical ground state and inhibit a low energy field theory description. To remedy this, we introduce an additional $n-2$ interaction terms, arriving at
\begin{empheq}[box=\fbox]{align}
\hspace{1em}\vspace{1em}\label{1:2}
H = \sum_j \sum_{r=1}^{n-1} J_r \tr(S(j)S(j+r))\;,\quad
\end{empheq}
where $J_1$ couples nearest-neighbours, $J_2$ couples next-nearest neighbours, and so on. See Figure~\ref{cell} for a pictorial representation of these interactions. This is the Hamiltonian that we will be studying throughout this chapter.

\subsection{Classical Ground State} \label{sub:class}

In the large-$p$ limit, the commutator (\ref{1:3}) is subleading in $p$, allowing us to replace $S$ by a matrix of classical numbers. To this order in $p$, the Casimir constraints of SU($n$) completely determine the eigenvalues of $S$. We have
\bea \label{1:1:2}
	S_{\alpha \beta} = p\bar u^\alpha u^\beta
\eea
for $u \in \mathbb{C}^n$ with $\overline u \circ u =1$. Note that $S_{\alpha \beta}$ are the components of the moment map $\mu$ from~(\ref{mugrass}), up to an additive constant term. The interaction terms appearing in (\ref{1:1}) reduce to 
\bea \label{eq:orthog}
	\tr(S(j)S(j+r)) = p^2 |\bar u(j) \circ u(j+r)|^2.
\eea
Since $u$ lives in $\mathbb{C}^n$, a classical ground state will posses local zero modes unless the Hamiltonian gives rise to $n-1$ constraints. This is the justification for our study of the modified Hamiltonian (\ref{1:2}), above, which removes any local zero modes by including longer range interactions. These interactions result in an $n$-site ordered classical ground state, which gives rise to a $\mathbb{Z}_n$ symmetry in their low energy field theory description. This $\mathbb{Z}_n$ symmetry is also present in the $p=1$ Bethe ansatz-solvable models~\cite{Sutherland, TsvelickWiegmann, AndreiFuruya}. In fact, it is expected that quantum fluctuations may produce an $n$-site unit cell through an ``order-by-disorder'' mechanism that generates effective additional couplings of order $p^{-1}$ that lift the local zero modes~\cite{HaldaneSU3, CorbozLajko}.

Since the classical ground state minimizing (\ref{1:2}) has an $n$-site order, it is characterized by $n$ normalized vectors that mutually minimize (\ref{eq:orthog}). That is, the classical ground state gives rise to an orthonormal basis of $\mathbb{C}^n$. As we recall from section~\ref{flaghomspacesec}, the space of $n$-tuples of mutually orthogonal vectors, defined up to a phase, is the complete flag manifold, which is the the mechanism how flag manifolds arise in the context of spin chains. Due to this $n$-fold structure, we rewrite the Hamiltonian (\ref{1:2}) as a sum over unit cells (indexed by~$j$):
\begin{equation} \label{1:1:1}
H = \sum_j \sum_{A=1}^n \sum_{r=1}^{n-1} J_r \tr(S(j_A)S(j_A+r)) \hspace{10mm} j_A := nj + (A-1).
\end{equation}

In the later sections of this chapter, we will expand about this classical ground state to characterize the low energy physics of (\ref{1:2}). But before this, we review some exact results that apply to SU($n$) Hamiltonians.

\section{Exact Results} \label{section:exact}

Haldane's original conjecture about SU(2) chains is supported by two rigorous results pertaining to Heisenberg Hamiltonians: the Lieb-Schultz-Mattis theorem~\cite{LSM}, and the Affleck-Kennedy-Lieb-Tasaki construction~\cite{aklt}. Similar results also exist for chains with SU($n$) symmetry, and this is what we review in this section.

\subsection{Lieb-Schultz-Mattis-Affleck Theorem (LSMA) Theorem}\label{LSMAsec}

The LSMA theorem is a rigorous statement about ground states in translationally invariant SU($n$) Hamiltonians~\cite{LSM,AL}:

\vspace{0.2cm}
\begin{center}

\fbox{\parbox{13.5cm}{
\centering Consider a translationally- and SU($n$)-invariant Hamiltonian of a spin chain with symmetric rank-$p$ representations at each site. If $p$ is not a multiple of $n$, then either the ground state is unique with gapless excitations, or there is a ground state degeneracy of at least $n\over \gcd(n,p)$.
}
}
\end{center}

\vspace{0.2cm}

\begin{figure}[h]
\centering
\includegraphics{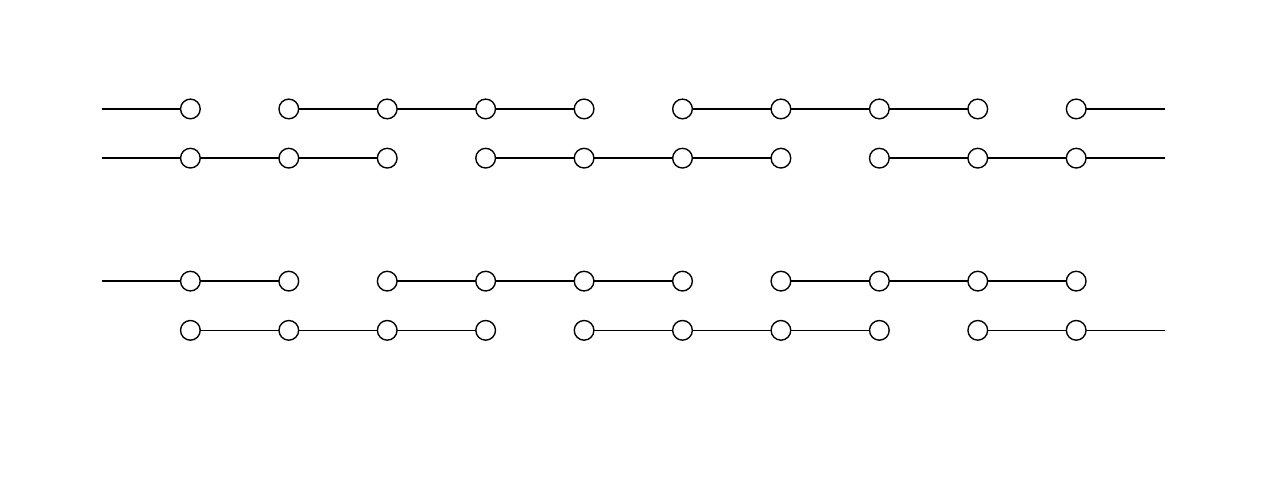}
\caption{A valence bond construction for the predicted two-fold degenerate ground state of SU(4) with $p=2$. Each node represents a fundamental $p=1$ irrep of SU(4). Each link represents an antisymmetrization between two nodes, and the antisymmetrization of four neighbouring nodes results in a singlet.}
\label{fig:su4aklt}
\end{figure}

\begin{figure}[h]
\centering
\includegraphics[width=.5\textwidth]{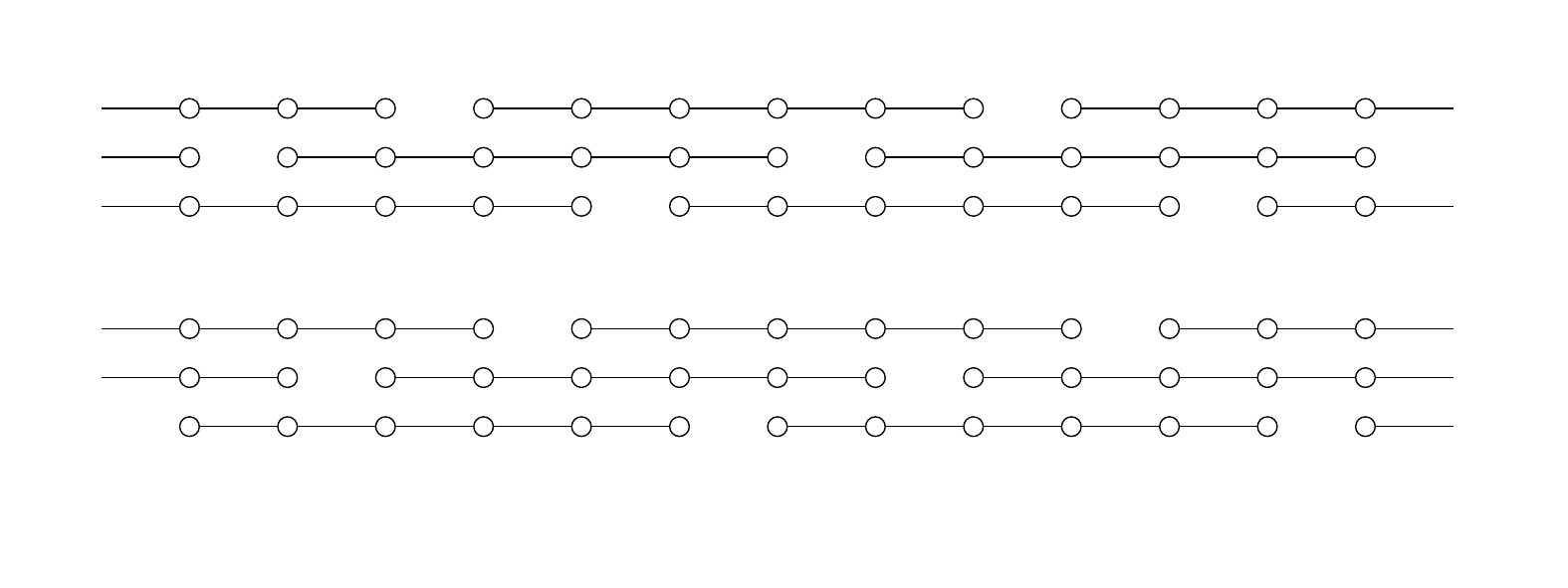}
\includegraphics[width=.49\textwidth]{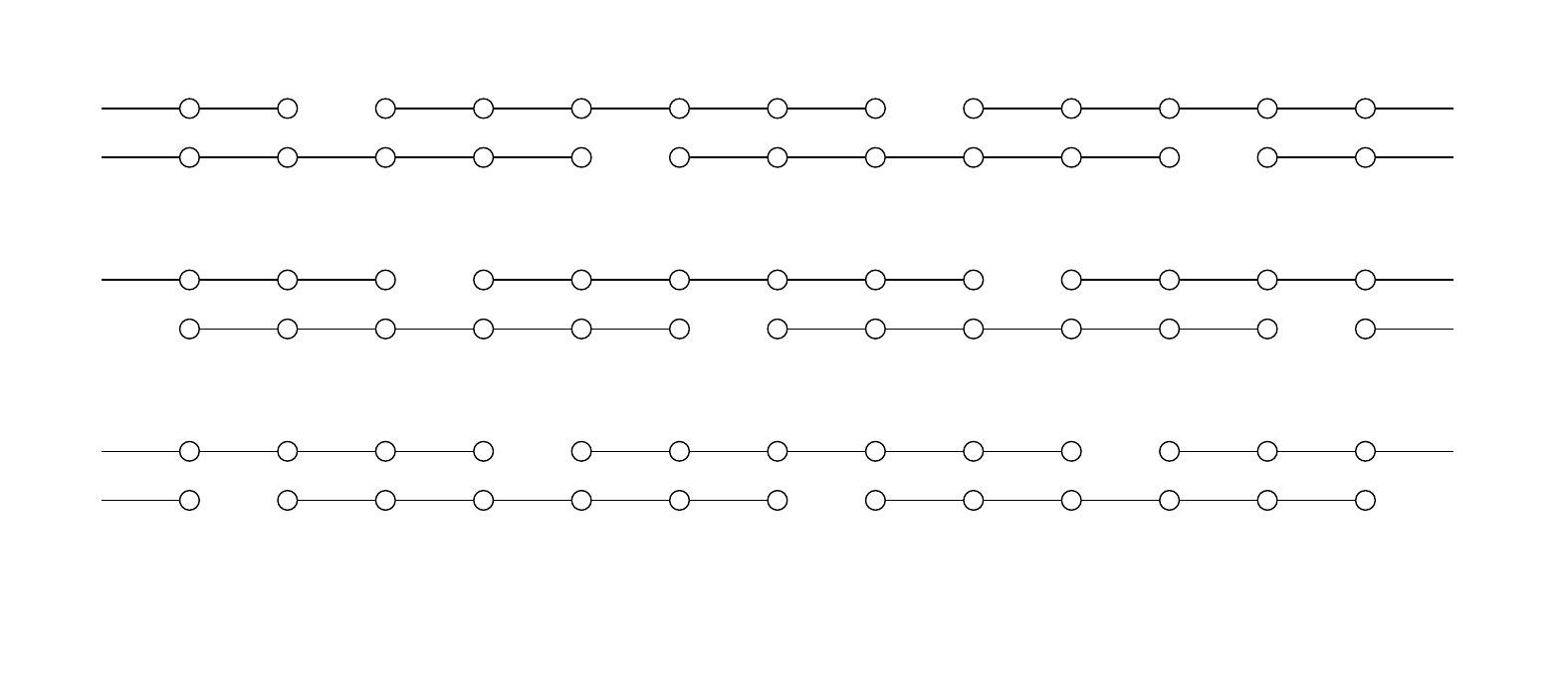}
\caption{Valence bond constructions for SU(6). The left subfigure corresponds to $p=3$, and has a 2-fold degenerate ground state. The right subfigure corresponds to $p=2$, and has a 3-fold degenerate ground state. Singlets are constructed out of 6 nodes, each of which represents a fundamental irrep in SU(6).} 
\label{fig:su6aklt}
\end{figure}

Let us show how the original proof in \cite{AL} can be extended to models with further range interactions. Explicitly, we consider the following Hamiltonian on a ring of $L$ sites:
\bea
	H = \sum_{r=1}^R H_r \hspace{10mm} H_r :=\sum_{j=1}^L J_r \tr(S(j)S(j+r))
\eea
where $S$ is defined as above. 
 We assume that $|\psi\rangle$ is the unique ground state of $H$, and is translationally invariant: $T|\psi\rangle = |\psi\rangle$. We then define a twist operator
\bea
	U = e^A \hspace{10mm} A:= \frac{2\pi i}{n L}\sum_{j=1}^L jQ(j)
\eea
with
\bea \label{la:2}
	Q = \sum_{A=1}^{n-1} S_{\alpha\alpha} - (n-1)S_{nn}
	= \tr(S) - nS_{nn} = p - nS_n^n.
\eea
Using the commutation relations (\ref{1:3}), it is easy to verify that 
\bea
	\Big[\tr(S(j)S(j+r)),Q(j)+Q(j+r)\Big] = 0
\eea
which then implies 
\bea
	U^\dag \tr(S(j)S(j+r)) U = e^{-\frac{r\pi i}{n L}(Q(j+r)-Q(j))} \tr (S(j)S(j+r))e^{\frac{r \pi i}{nL}(Q(j+r)-Q(j))}.
\eea
Using this, one can show that
\bea
	U^\dag H U =H+[ H,A]  + \fO(L^{-1})
\eea
so that $U|\psi\rangle$ has energy $\fO(L^{-1})$. Now, using the translational invariance of $|\psi\rangle$, we find 
\bea\label{TshiftLSMA}
	\langle \psi | U | \psi\rangle
	= \langle \psi | T^{-1} UT|\psi\rangle 
	= \langle \psi| U e^{\frac{2\pi i}{n}Q(1)} e^{-\frac{2\pi i}{nL}\sum_{j=1}^L Q(j)} |\psi\rangle.
\eea
Since $|\psi\rangle$ is a ground state of $H$, it is a SU($n$) singlet, and so must be left unchanged by the global SU($n$) transformation $e^{-\frac{2\pi i}{nL}\sum_{j=1}^L Q(j)} $. Moreover, using (\ref{la:2}), we have 
\bea \label{la:4}
	\langle \psi |U|\psi\rangle = e^{\frac{2\pi i p}{n} }\langle \psi | U e^{2\pi i S_n^n} |\psi\rangle.
\eea
As shown in section~\ref{SWbosons}, the matrices $S$ can be represented in terms of Schwinger bosons; the diagonal elements are then number operators for these bosons. Thus, $S_{nn}$ acting on $|\psi\rangle$ will always return an integer, and $e^{2\pi i S_{nn}}$ can be dropped. Thus, we find that so long as $p$ is not a multiple of $n$, 
\bea
	\langle \psi |U|\psi\rangle = 0
\eea
implying that $U|\psi\rangle$ is a distinct, low-lying state above $|\psi\rangle$. This completes the proof. Finally, we may also comment on the ground state degeneracy in the event that a gap exists above the ground state. Through the repeated application of (\ref{la:4}), we have 
\bea
	\langle \psi | U^{k} |\psi\rangle = e^{\frac{2\pi i pk}{n}} \langle \psi | U |\psi\rangle.
\eea
So long as $k < r:= n/\gcd(n,p)$, the family $\{ U^k |\psi\rangle\}$ is an orthogonal set of low lying states. If an energy gap is present, this suggests that the ground state is at least $r$-fold degenerate. See Figures \ref{fig:su4aklt} and \ref{fig:su6aklt} for a valence bond solid picture of these degeneracies in SU(4) and SU(6), respectively.

\subsection{Affleck-Kennedy-Lieb-Tasaki (AKLT) Constructions}\label{AKLTsec}

\begin{figure}[h]
\centering
\includegraphics[width = 0.7\textwidth]{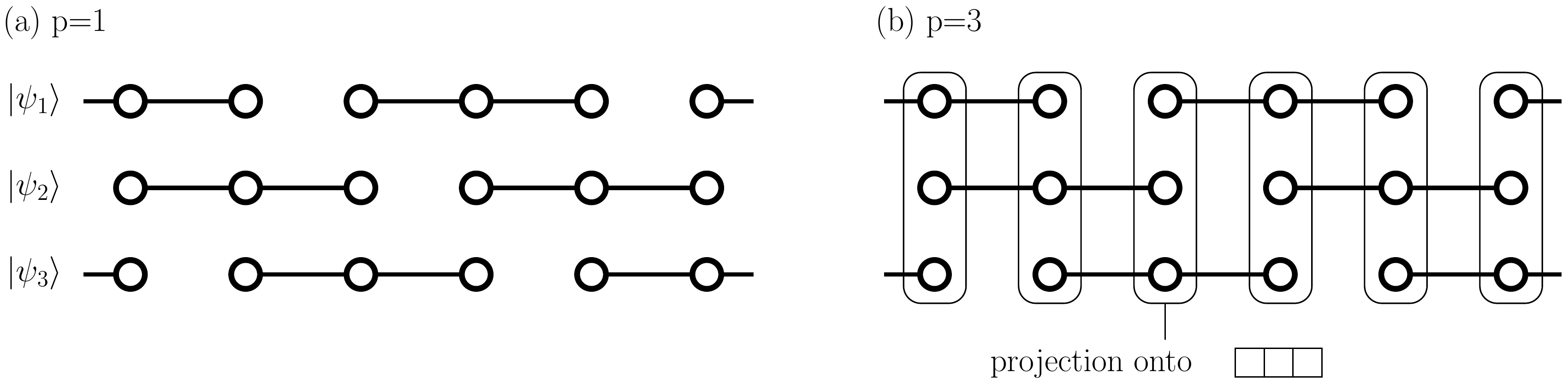}
\caption{AKLT constructions in SU(3). Left: When $p\neq n$, multiple valence bond solids can be formed. The ground state is not translationally invariant and degenerate. Right: When $p=n$, a unique, translationally invariant ground state can be constructed, by projecting on to the symmetric-$p$ representation at each site.}
\label{aklt}
\end{figure}

One of the first results that bolstered Haldane's conjecture was the discovery of the so-called AKLT model of a spin-1 chain, which exhibits a unique, translationally invariant ground state with a finite excitation gap~\cite{LSM,AL}. In this case, the number of boxes in the Young tableau is 2, and so the SU(2) version of the LSMA theorem does not apply. Recently, the AKLT construction has been generalized by various groups to SU($n$) chains~\cite{greiter2007, katsura2008, nonne2013, lechem2, furusaki, roy2018, gozel2}. Relevant to us are the symmetric representation AKLT Hamiltonians introduced in~\cite{greiter2007}. In particular, for $p$ a multiple of $n$, Hamiltonians are constructed that exhibit a unique, translationally invariant ground state. See Figure \ref{aklt} for the case $n=p=3$. Additionally, for $p$ not a multiple of $n$, with $ r:= n/ \gcd(n,p)$, Hamiltonians are constructed with $r$-fold degenerate ground states that are invariant under translations by $r$ sites (see Figures \ref{fig:su4aklt}, \ref{fig:su6aklt}). All of these models have short range correlations, and are expected to have gapped ground states, based on arguments of spinon confinement. The fact that the construction of a gapped, nondegenerate ground state is only possible when $p$ is a multiple of $n$ is consistent with the LSMA theorem presented above. 

\section{Flavour Wave Theory} \label{section:fw}

According to Coleman's theorem~\cite{Coleman1973}, we do not expect spontaneous symmetry breaking of the SU($n$) symmetry in the exact ground state of our Hamiltonian. Nonetheless, we may still expand about the classical (symmetry broken) ground state to predict the Goldstone mode velocities. If the theory is asymptotically free, then at sufficiently high energies the excitations may propagate with these velocities~\cite{gozel1}. In the familiar antiferromagnet, this procedure is known as spin wave theory; in SU($n$), it is called flavour wave theory~\cite{fw1,fw2}.

To begin, we introduce $n^2$ bosons in each unit cell to reproduce the commutation relations of the $S$ matrices: 
\bea
	S_{\alpha\beta}(j_A) = b^{\dag,\alpha}_A b^\beta_A. 
\eea
The counting is $n$ flavours of bosons for each of the $n$ sites of a unit cell. The condition $\tr(S) = p$ implies there are $p$ bosons at each site. The classical ground state involves only `diagonal' bosons of the type $b^A_A$ and $b^{\dag,A}_A$.  The `off-diagonal' bosons are Holstein-Primakoff bosons; in SU(2) they correspond to the operators $A, A^\dag$ introduced in Section~\ref{HPDMsec}. Flavour wave theory allows for a small number of Holstein-Primakoff bosons at each site, captured by
\[
	\nu(j_A) = \sum_{\alpha\not=A} b^{\dag,\alpha}_A
	 b^\alpha_A,
\]
 and writes the Hamiltonian (\ref{1:2}) in terms of these $n(n-1)$ bosons. In the large $p \gg \nu (j_A)$ limit, we expand 
\bear \nonumber
&& S^A_A(J_A) = p - \nu(j_A), \\ \nonumber
&& S^\alpha_{A}(j_A) \approx \sqrt{p}b^{\dag,\alpha}_A, \\ \nonumber
&& S^A_{\alpha}(j_A) \approx \sqrt{p}b^\alpha_A, 
\eear
to find
\bea
	\tr(S(j_A) S(j_B)) =
	p \left[ b^{\dag,A}_B b^A_B
	+ b^{\dag,A}_Bb^B_A
	+ b^{\dag,B}_Ab^{\dag, A}_B
	+ b^B_Ab^A_B \right]
	+\fO(p^{0}).
\eea
 In terms of these degrees of freedom, the Hamiltonian (\ref{1:2}) decomposes into a sum
\bea
	H = \sum_{A <B} H_{AB},
\eea
where $H_{AB}$ is a Hamiltonian involving only the two boson flavours $b^A_B$ and $b^B_A$. In momentum space, this gives $\frac{n(n-1)}{2}$ different $2\times 2$ matrices, each of which can be diagonalized by a Bogoliubov transformation:
\bea
	H_{A, A+t} = \text{const.} + \sum_k \omega_t(k) \sum_{m=1}^2 \left( d^{\dag,m}_t(k)d^{m}_t(k) + \frac{1}{2}\right)
\eea
where
\bea
	\omega_t(k) = 2p\sqrt{J_t J_{n-t}}\left| \sin \frac{ nka}{2}\right|.
\eea

Therefore, the corresponding flavour wave velocities are
\begin{equation}\label{flavourvel}
v_t = np \sqrt{J_t J_{n-t}} \hspace{10mm} t=1,2,\cdots, n-1
\end{equation}
When $n$ is odd, there are $n$ modes with each flavour wave velocity. When $n$ is even, this is true except for the velocity $v_{\frac{n}{2}}$, which has only $\frac{n}{2}$ modes. In each case, the number of modes adds up to $n(n-1)$. We note that for $n>3$, there is no longer a unique velocity, and the emergence of Lorentz invariance is absent. Only for a specific fine tuning of the couplings can Lorentz invariance be restored. These tuned models were the ones considered in~\cite{BykovHaldane1} and \cite{BykovHaldane2}.

\section{Derivation of the continuum theory} \label{section:derivation}

 In the present section our goal is to derive a path integral representation for spin chains with Hamiltonians of the type~(\ref{1:1:1}), using coherent states introduced in section~\ref{cohstates}. As a warm-up, we will start with a simpler example of a single SU(2)-spin (which may be thought of as a spin chain with one site) in Section~\ref{quantsph} (such systems were considered in~\cite{PerelomovBerry}, for example). The extension to a spin chain is rather straightforward and is discussed in Section~\ref{xxxchain}. There exist two different continuum limits of the spin chain: one based on the ferromagnetic ground state, which leads to a Landau-Lifschitz model with quadratic dispersion relations for the spin waves~(Section~\ref{ferro}), and one based on the anti-ferromagnetic ground state (the Haldane-type limit), which leads to linear dispersion and will be elaborated on in  Section~\ref{antifer}. Although these two situations are rather different, as we shall see in Section~\ref{linquadchains}, a general spin chain with equivalent (but arbitrary) representations at all sites leads to a continuum theory with both linear and quadratic dispersion modes, uniting the two cases.

\subsection{The quantum sphere $S^2$}\label{quantsph}

Let us consider in detail the case of SU(2)\footnote{General results along a similar line of reasoning were obtained in \cite{AFS}.}.  
We introduce the notation $A(\bar q,v)$ for the normalized matrix element between coherent states of any operator $\hat{A}$ (which for historical reasons is called the kernel of $\hat{A}$): $A(\bar{q}, v)=\frac{\langle q|\hat{A}|v\rangle}{\langle q|v\rangle}$. 

Once again we will consider the
rank-$p$ symmetric representations\footnote{For SU(2) all representations are of this type.}, and for such representations the scalar product of coherent states is
\bea\label{qvscalprod}
\langle q| v\rangle =(\bar{q}\circ v)^p\,.
\eea
This can be proven, for instance, by using the Fock space expressions of the type~(\ref{CPNcohstates}) for the coherent states. Viewing the Hilber space as a subspace of $(\CC^2)^{\otimes p}$, for the Hamiltonian acting on the quantum sphere we shall take 
\bea
\hat{T}_3=
\sum\limits_{s=1}^p 1\otimes ... \otimes \underset{\underset{\textrm{$s$-th position}}{\uparrow}}{\sigma_3} \otimes ... \otimes 1\,,
\eea
which may be interpreted as an external magnetic field $\vec{H}$ in the $z$-direction (under the assumption of a $\vec{S}\cdot \vec{H}$ coupling). The kernel of $\hat{T}_3$ is $T_3(\bar{q}, v)=p\, \frac{\bar{q}^1v^1-\bar{q}^2v^2}{\bar{q}^1v^1+\bar{q}^2v^2}$.

We will now present the derivation of the kernel of the ``evolution operator''
\bea
\hat{U}=e^{-i\alpha \hat{T}_3}\,.
\eea
Of course, in this finite-dimensional case this is merely a pedagogical exercise, since the action of $\hat{U}$ on a coherent state simply gives
$\hat{U}\,|v\rangle=|e^{-i\alpha \sigma_3}\circ v\rangle$,
and the corresponding matrix element is easily calculated:
\bea\label{evopans}
U(\bar{q}, v)=\left(\frac{e^{-\im\alpha}\bar{q}^1v^1+e^{\im\alpha} \bar{q}^2v^2}{\bar{q}^1v^1+\bar{q}^2v^2}\right)^p
\eea
As is standard in path integral calculations \cite{SF1}, in order to write a path integral representation for a matrix element $U(\bar{q}, v)$, one first needs to know the matrix elements of the generator~$\hat{T}_3$. Then one splits the ``time'' interval $\alpha$ into $k$ subintervals of length $\frac{\alpha}{k}$ and uses the formula $\hat{U}=\underset{k\to\infty}{\textrm{lim}} \left(1-\frac{\im\alpha}{k} \hat{T}_3\right)^k:=\underset{k\to\infty}{\textrm{lim}}\, \hat{\tau}^k$. Inserting the completeness relation (\ref{completeness}) between every two factors of $\hat{\tau}$, we arrive at the following expression for the kernel of the evolution operator:
\bear\label{Upath1}
&&U(\bar{q}, v)=\frac{\langle q|\hat{U}|v\rangle}{\langle q|v\rangle}
= \underset{k\to\infty}{\textrm{lim}} \int\;\prod\limits_{a=1}^{k-1} d\mu(z(a), \bar{z}(a))\times\;\;\\ \nonumber
&&  \times\;\tau(\bar{q}, z(k-1))\cdot \tau(\bar{z}(k-1), z(k-2))\cdots \tau(\bar{z}(2), z(1))\cdot  \tau(\bar{z}(1), v) \times \\ \nonumber && \;
\times\;\frac{    \langle q|z(1)\rangle \langle z(1)|z(2)\rangle           \,...\,       \langle z(k-2)|z(k-1)\rangle      \langle z(k-1)|v\rangle}{\langle q|v\rangle     \langle z(1)|z(1)\rangle     \,...\,       \langle z(k-1)|z(k-1)\rangle}
\eear
Here $z(a)$ is the coherent state of the $a$th `time slice'. To complete the derivation we use the explicit expression $
\tau(\bar{z}(a+1),z(a))=1-p\, \frac{\im\alpha}{k}\; \frac{\bar{z}(a)\circ \sigma_3\circ z(a+1)}{\bar{z}(a)\circ z(a+1)}$ for $\tau$ and~(\ref{qvscalprod}) for the scalar product of coherent states.  

We now want to ``take the limit'' in the formula (\ref{Upath1}), assuming that $z(a+1)-z(a) \sim \frac{1}{k}\dot{z}(a+1)$ (for a justification of this procedure see \cite{ZJ}). In order to do it we write the factors $\frac{\bar{z}(a)\circ z(a+1)}{\bar{z}(a+1)\circ z(a+1)}$ in the following form:
\bea\nonumber
\frac{\bar{z}(a)\circ z(a+1)}{\bar{z}(a+1)\circ z(a+1)}=\left(1-\frac{(\bar{z}(a+1)-\bar{z}(a))\circ  z(a+1)}{\bar{z}(a+1)\circ z(a+1)}\right)\simeq 1-{1\over k}\frac{\dot{\bar{z}}(a+1) \circ z(a+1)}{\bar{z}(a+1)\circ z(a+1) }
\eea
for $a=0,1, ..., k-2$. Then we obtain
\bea\label{evop}
U(\bar{q},v)=\!\!\int
\!\! \prod\limits_{t\in [0,1]} \!d\mu(z(t),\bar{z}(t)) \left(\frac{\bar{z}(1)\circ v}{\bar{q}\circ v}\right)^p \exp{\left(-p\!\int\limits_0^1 dt\,\frac{\dot{\bar{z}} \circ z}{\bar{z}\circ z}-p\,\im \,\alpha \int\limits_0^1 dt\, \frac{\bar{z}\circ \sigma_3\circ  z}{\bar{z}\circ z}\right)},
\eea
with boundary conditions $\bar{z}(0)=\bar{q}, z(1)=\nu$. The action in the exponent should be somewhat reminiscent of the action~(\ref{flagpartaction}) that we encountered in Chapter 1. Indeed, if in that formula we set $A_\mu\dot{x}^{\mu}=\alpha\,\sigma_3$, we would arrive exactly at the action in~(\ref{evop}), upon normalizing the coordinates as $|z|=1$ (in which case, following our convention, we relabel $z$ into $u$).

Before concluding this section, let us demonstrate that one can actually calculate the path integral~(\ref{evop}). To this end note that the equations of motion following from the action in the exponent of~(\ref{evop}) describe the rotation of the sphere around its $z$-axis (the one orthogonal to the plane of the stereographic projection). Passing to the inhomogeneous coordinate $z$ via $u^1={1\over \sqrt{1+|z|^2}}, u^2={z\over \sqrt{1+|z|^2}}$, we find that the e.o.m. are the equations of harmonic oscillations:
\bea \label{harmosc}
\im \dot{z}=2\alpha z ,\quad 
\im \dot{\bar{z}}=-2\alpha \bar{z} .
\eea
In fact with a particular choice of coordinates the Lagrangian standing in the exponent in~(\ref{evop}) may be turned exactly into the canonical Lagrangian of the harmonic oscillator, but this is not necessary for our purposes. Solving the equations with the prescribed initial conditions $\bar{z}(0)={\bar{q_2}\over \bar{q_1}}:=\bar{q}, \;z(1)={v_2\over v_1}:=v$, we obtain $\bar{z}(t)=\bar{q} e^{2\im \alpha t}$ and $z(t)=v e^{-2\im \alpha (t-1)}$. Plugging this into the exponent of the path integral (\ref{evop}), we get $e^{-p\,\im \alpha}$. The term $\frac{\bar{z}(1)\circ v}{\bar{q}\circ v}$ in front of the exponent produces $\frac{1+q\, \bar{y} \,e^{2\im \alpha}}{1+q \bar{y}}$, and altogether we get
\bea\label{evopfin}
U(\bar{q},v)=\left(\frac{e^{-\im \alpha}+\bar{q}\, v \,e^{\im \alpha}}{1+ \bar{q} v}\right)^p ,
\eea
which, as we know from (\ref{evopans}), is the right answer.

\subsection{Path integral for the spin chain}\label{xxxchain}

Similarly to what we did in (\ref{evop}), we now want to derive a path integral expression for the evolution operator of the spin chain $\hat{\mathbb{U}}=e^{i \alpha \hat{H}}$, $\hat H$ now being a spin chain Hamiltonian. At the same time we pass from the simple SU(2) case to the SU(3), or even SU($n$) model. We start with the Heisenberg Hamiltonian
\bea\label{xham}
\hat{H}_{\mathrm{Heis}}={1\over p}\sum\limits_{j=1}^L\; \left(\tr(S(j) S(j+1))+\textrm{const.}\right)\,,
\eea
where the spin operators are assumed to be in the symmetric powers of the fundamental representation, indexed by $p$ as before, and the constant may be chosen at our will. 
In order to build the path integral we first need to know the matrix elements of the Hamiltonian itself, which  amounts to knowing the matrix elements of $\mathbb{P}:={1\over p}\left(\tr(S(j)\cdot S(j+1))+\textrm{const.}\right)$. This operator acts in the tensor product \(\textrm{Sym}(\mathbb{C}^{n})^{\otimes p} \otimes \textrm{Sym}(\mathbb{C}^{n})^{\otimes p}\) and (for a suitable choice of the additive constant) is a restriction of the operator acting in $(\mathbb{C}^{n})^{\otimes p}\otimes (\mathbb{C}^{n})^{\otimes p}$ as a sum of permutations: $\mathbb{P}={1\over p}\sum\limits_{s, t=1}^p\,P_{s,t}$, where $P_{s,t}$ is the permutation of the $s$-th and $t$-th $\CC^n$-factors in the two copies of $(\mathbb{C}^{n})^{\otimes p}$. The tensor product of coherent states has the form  $(\bar{q}(j)\circ v(j))^p\;(\bar{q}(j+1)\circ v(j+1))^p$, and 

the matrix elements of $\hat{\mathbb{P}}$ is easily found to be
\bea
\mathbb{P}(\bar{q}_1,\bar{q}_2 ; v_1, v_2)=\frac{ \langle q_1, q_2 | \hat{\mathbb{P}} | v_1,v_2\rangle }{\langle q_1, q_2  | v_1,v_2\rangle}=p\;\frac{ ( \bar{q}_2\circ v_1)\,( \bar{q}_1\circ v_2) }{ (\bar{q}_1\circ v_1)\,(\bar{q}_2\circ v_2) }.
\eea
Now we can essentially repeat the steps from the previous section. The only difficulty is notational and it comes from the fact that in this case, as opposed to the previous example, we essentially have two ``space-time'' directions: one ``time'' or $\alpha$-direction, and a second ``spatial'' direction in which the spin chain is extended. As a consequence, our variables $z$ will now take two arguments: $z(a,j)$, where $a$ is the time index, and $j$ enumerates the sites of the spin chain. The integrand will again split into two terms: the first being a geometric phase term, and the second being the Hamiltonian:
\bea
U(\bar{q}, v) =  \underset{K\to\infty}{\textrm{lim}} \int\;\prod\limits_{a,j} d\mu(z(a,j), \bar{z}(a,j))\; \times I_{\text{geom}} \times I_{H}
\eea

The geometric term is local in the spin chain index $j$ and has a simplest (nearest-neighbor, or first-order) nonlocality in time, which is a general feature, since in the continuum limit it should lead to a one-form:
\bea
I_{\text{geom}}=\prod\limits_{a,j} \left(\frac{z(a,j)\circ \bar{z}(a+1,j)}{z(a+1,j)\circ \bar{z}(a+1,j)}\right)^p
\eea
On the other hand, the Hamiltonian term has a first-order nonlocality in the spin-chain direction, but also has a first-order nonlocality in the time direction, since the matrix elements of the Hamiltonian entering the integral are always of the form
$\frac{1}{k}\langle z(a+1,j)|\hat{H}|z(a,j)\rangle\,.
$ 
The latter nonlocality will not play a role, since the contribution of such matrix element always comes with a damping factor $\frac{1}{k}$, and the nonlocality being of order $\frac{1}{k}$ as well enters only subleading terms. In any case, the contribution of the Hamiltonian may be written as
\bea
I_{\mathrm{H}}=\prod\limits_{a} \left( 1+ p\,\frac{i \alpha}{k} \,\sum\limits_{j}\,\frac{ z(a,j)\circ \bar{z}(a+1,j+1)}{z(a,j)\circ \bar{z}(a+1,j)} \frac{z(a,j+1)\circ \bar{z}(a+1,j)}{z(a,j+1)\circ \bar{z}(a+1,j+1)} \right)
\eea
We may now exponentiate these expressions and take the limit $k \to \infty$, thus obtaining a continuous time variable $t$:
\bear\label{evopxxx}
&&\hspace{-2cm}U(\bar{q}, v) = \int\; \prod\limits_{t\in [0,1]}\prod_j d\mu(z(t,j),\bar{z}(t,j)) \left(\frac{\bar{z}(1,j)\circ z(1,j)}{\bar{z}(0,j) \circ z(1,j)}\right)^p \;\exp{(\im\mathcal{S})}, \; \\
\label{actionxxx}
&&\hspace{-2cm}\textrm{where}\quad \mathcal{S} = p\;\int\limits_0^1\,dt\, \sum\limits_j \left( i\, \frac{\dot{z}(j) \circ \bar{z}(j)}{|z(j)|^2} +
\alpha\, \frac{|z(j)\circ  \bar{z}(j+1)|^2}{|z(j)|^2\,|z(j+1)|^2}  \right)
\eear
with boundary conditions $z(0,j)=q(j),\;\bar{z}(1,j)=\bar{v}(j)$. We have suppressed the time argument in the second line above. The nontrivial question is how to take the continuum limit in the spin chain direction, indexed by ``$j$'', --- there are several inequivalent ways to do it. It is well-known that the isotropic (`XXX') spin chain has two ``vacua'', i.e. the states (or multiplets) with minimal and maximal energy. They also correspond to the extremal values of the spin: the vacuum with spin zero (or least possible spin in case the length of the chain does not allow for zero spin) is called antiferromagnetic, whereas the state with maximal spin (proportional to $L$ --- the length of the chain) is called ferromagnetic. Which one of these states is the true vacuum depends, of course, on the sign of the Hamiltonian.

\subsection{Ferromagnetic limit}\label{ferro}

The ferromagnetic limit is especially simple. It corresponds to the case where the $z$'s at the neighboring sites are very close to each other, that is $z(j+1)-z(j) \sim \frac{1}{L}$ ($L$ is the length of the spin chain, i.e. the number of sites). The first term in (\ref{actionxxx}) then simply produces
\bea\label{ferrcont1}
 \int\,dt\, \sum\limits_j  \im \frac{\dot{z}(j) \circ \bar{z}(j)}{z(j) \circ \bar{z}(j)}  \to L \int \, dt\,\int\limits_{-1/2}^{1/2}\,dx\; \im \,\frac{\dot{z}(t,x) \circ \bar{z}(t,x)}{z(t,x)\circ \bar{z}(t,x)},
\eea
whereas the expression in the second term can be rewritten in the same spirit:
\bear\nonumber
\sum\limits_j \frac{|z(j)\circ  \bar{z}(j+1)|^2}{|z(j)|^2\, |z(j+1)|^2} =1 -\sum\limits_j 
\frac{|\Delta z(j)|^2 |z(j)|^2- |z(j) \circ \Delta \bar{z}(j)|^2}{|z(j)|^4}+\cdots 
\eear
Upon taking the continuum limit and rescaling $x\to \frac{1}{L} x $ the full action acquires the form
\bea\label{LL}
\mathcal{S} = p\;\int\limits_0^1 \, dt\,\int\limits_\mathbb{R}\,dx\;\left[ \, \im \,\frac{\dot{z}(t,x) \circ \bar{z}(t,x)}{z(t,x)\circ \bar{z}(t,x)} - \, \left( \frac{\dx z \circ \dx \bar{z}}{z \circ \bar{z}} - \frac{(z \circ \dx \bar{z}) (\dx z \circ \bar{z})}{(z \circ \bar{z})^2} \right) \right]
\eea

Non-relativistic sigma-models of the type~(\ref{LL}) are known as Landau-Lifshitz models\footnote{The mathematical structures behind such models, in particular the connection with the geometry of loop groups, are discussed in~\cite{AH}.}. The target space of the model we have described is, obviously, $\CP^{n-1}$ (for example, in the second term in~(\ref{LL}) one immediately recognizes the Fubini-Study metric). The simplest example corresponds to $n=2$, i.e. when the target space is a usual 2-sphere. In this case the model is also known as the classical Heisenberg ferromagnet, and it is customary to use the unit three-vector $\vec{n}$ instead of the complex coordinates $z, \bar{z}$ (the two parametrizations are related via the stereographic projection: $n^1+\im n^2=\frac{2z}{1+z\bar{z}}, \; n^3=\frac{1-z\bar{z}}{1+z\bar{z}}$). Then the e.o.m., which follows from Lagrangian (\ref{LL}), is:
\bea\label{heisferr}
\frac{\dd \vec{n}}{\dd t}=\vec{n}\times \frac{\dd^2 \vec{n}}{\dd x^2} .
\eea
Expanding around a constant magnetization direction, $\vec{n}=\vec{n}_0+\delta\vec{n}$, one obtains the linear equation $\frac{\dd \delta\vec{n}}{\dd t}=\vec{n}_0\times \frac{\dd^2 \delta\vec{n}}{\dd x^2}$, which describes spin waves with quadratic dispersion.

\section{The antiferromagnetic limit}\label{anti}

The antiferromagnetic limit is much more involved. The main difference is that in this case the $z$-variables on neighboring sites are no longer close to each other. Let us first elaborate on the case of the sphere, that is $n=2$, which was for the first time explored in~\cite{Haldane1983}. In this case it is intuitively clear that the antiferromagnetic limit corresponds to the case where the spins on the neighboring sites have opposite directions, i.e. $\vec{n}(j+1)\simeq - \vec{n}(j)$. In terms of the complex coordinates used above this may be written as $z(j+1)\simeq -\frac{1}{\bar{z}(j)}$, or, using homogeneous coordinates, as $z^1(j+1) = \bar{z}^2(j), z^2(j+1) = - \bar{z}^1(j)$. Such a simple explanation is due to the fact that on the sphere there exists the antipodal involution, which in that case is also unique. However, this is no longer true for $\CP^{n-1}$ with $n \geq 3$.
This is the reason why it is not immediately obvious how one can extend the $\CP^1$ analysis to a higher-dimensional projective space. The answer crucially depends on the particular Hamiltonian at hand. The first model after the $\CP^1$-case to be successfully analyzed in~\cite{Affleck:1984ar} was the one of an alternating spin chain, so let us now recall how this was accomplished.

\subsection{Alternating representations}\label{aff1}

First of all let us consider the case of a spin chain with alternating representations: that is, on even sites one has some representation $\mathcal{R}$ and on odd sites the dual one $\bar{\mathcal{R}}$. In particular, this means that these representations can combine into a singlet and hence form an anti-ferromagnetic configuration. For the Hamiltonian one takes the Heisenberg Hamiltonian
\bea\label{heisham1}
H=\sum\limits_{j=1}^L\, \tr(S(j) S(j+1))\,,
\eea
where it is understood that $S(j)$ and $S(j+1)$ are in conjugate representations. For simplicity we assume in this section that either $R$ or $\bar{R}$ is the rank-$p$ symmetric representation.  The generalized Haldane limit for this kind of spin chain was constructed in~\cite{Affleck:1984ar}. In order to rephrase these results one should follow the steps of the previous section to obtain the following action in the $t$-continuum limit:
\bea\label{actionaff}
\mathcal{S} = p\;\int\limits_0^1\,dt\, \sum\limits_j \left( i\, \frac{\dot{z}(j) \circ \bar{z}(j)}{|z(j)|^2} +
\alpha\, \frac{|z(j)\circ  z(j+1)|^2} {|z(j)|^2\,|z(j+1)|^2}  \right).
\eea
The difference between the second terms in (\ref{actionxxx}) and (\ref{actionaff}) precisely reflects the difference between the representations at adjacent sites. 
The minimum of the Hamiltonian $\mathcal{H}=-\frac{|z(j)\circ  z(j+1)|^2} {|z(j)|^2\,|z(j+1)|^2} $ is clearly reached for $z(j+1)=\bar{z}(j)$. The important observation is that for such configurations the first term in (\ref{actionaff}) turns into a full derivative, since on every two neighboring sites $i \frac{\dot{z}(j) \circ \bar{z}(j)}{|z(j)|^2}+i \frac{\dot{z}(j+1) \circ \bar{z}(j+1)}{|z(j+1)|^2}=i\frac{d}{dt}(\log{|z(j)|^2})$. There is a simple but fundamental explanation of this fact. Consider the space $\CP^{n-1}\times \CP^{n-1}$, with the symplectic form on it being the sum of two Fubini-Study forms~(\ref{FSform}): $\Omega=\Omega_1 + \Omega_2$. We then have the following statement\footnote{The definition of Lagrangian submanifold was given in section~\ref{cohsec}.}:

\begin{empheq}[box=\fbox]{align}
\hspace{1em}\vspace{5em}
\textrm{The submanifold}\;\; \CP^{n-1}\subset\CP^{n-1}\times \CP^{n-1}\,,\; \textrm{def. by}\; z\to(z,\bar{z})\,,\; \textrm{is Lagrangian.}\quad
\end{empheq}

Indeed, since $\Omega=d\theta$, where $\theta$ is the one-form entering the first term in~(\ref{actionaff}), and the restriction of the symplectic form $\Omega\big|_{\CP^{n-1}}=0$ vanishes, it follows that $\theta\big|_{\CP^{n-1}}=df$ for some function~$f$, implying that the kinetic term is a total derivative.

Let us now expand the action (\ref{actionaff}) around the ``vacuum'' $z(j+1)=\bar{z}(j)$. The variables $z(j+1)$ and $z(j+2)$ are expressed in terms of $z(j)$ in the following fashion:
\bea
z(j+1)=\bar{z}(j)+\ol \bar{\tau}(j),\qquad z(j+2)=z(j) +\ol z(j)'
\eea
For convenience we introduce the projector $\Pi(j)=\mathds{1}-\frac{\bar{z}(j)\otimes z(j)}{|z(j)|^2}$ onto the subspace of~$\CC^{n}$ orthogonal to the vector $z(j)$. Then the terms in the Hamiltonian have the following expansions:
\begin{align}\nonumber 
\frac{|z(j) \circ z(j+1)|^2}{|z(j)|^2 |z(j+1)|^2}\simeq & \olsq \frac{\tau(j) \circ \Pi(j) \circ \bar{\tau}(j)}{|z(j)|^2},\; \\
\frac{|z(j+1) \circ z(j+2)|^2}{|z(j)|^2 |z(j+1)|^2}\simeq & \olsq \frac{\tilde{\tau}(j) \circ \Pi(j) \circ \overline{\tilde{\tau}}(j)}{{|z(j)|^2}}\,,
\end{align}
where $\tilde{\tau}(j)=\tau(j)-z(j)'$. The kinetic terms are expanded as follows:
\bea\label{kintermcpn}
i \frac{\dot{z}(j) \circ \bar{z}(j)}{|z(j)|^2}+i \frac{\dot{z}(j+1) \circ \bar{z}(j+1)}{|z(j+1)|^2}= i \ol \frac{ \tau(j) \circ \Pi(j) \circ \dot{\bar{z}}(j) - \bar{\tau}(j) \circ \Pi(j) \circ \dot{z}(j)}{|z(j)|^2}
+ \ldots 
\eea
where $\ldots$ denotes a full derivative. Thus, the action (\ref{actionaff}) acquires the following form:
\begin{align}
\mathcal{S} = p\;\int\limits_0^1\,dt\, \sum\limits_j \frac{1}{|z(j)|^2}\,\Big( & i \ol \left[ \tau(j) \circ \Pi(j) \circ \dot{\bar{z}}(j) - \bar{\tau}(j) \circ \Pi(j) \circ \dot{z}(j)\right]\;+ \\ \nonumber
& \hspace{-2cm}+\;{1\over L^2} \left[ \tau(j) \circ \Pi(j) \circ \bar{\tau}(j) +(\tau(j)-z(j)') \circ \Pi(j) \circ (\bar{\tau}(j)-\bar{z}(j)') \right]\Big).
\end{align}
Now we simply need to ``integrate out'' the fields $\tau, \bar{\tau}$. Upon setting $\tau, \bar{\tau}$ equal to their stationary values we also pass to the continuum limit with respect to the ``$j$'' index. This leads to the following expression:
\begin{align}\label{CPNaction}
\mathcal{S} = p \int\limits_0^1 \!dt\! \int\limits_\mathbb{-\infty}^{\infty} \!dx \Big[&{1\over 2} \partial_\mu z(x,t) \circ \frac{\Pi(z,\bar{z})}{|z(x,t)|^2}\circ \partial_\mu \bar{z}(x,t) \\
\nonumber &-{\im \over 2} \epsilon_{\mu\nu} \,\partial_\mu z(x,t) \circ \frac{\Pi(z,\bar{z})}{|z(x,t)|^2}\circ \partial_\nu \bar{z}(x,t)  \Big].
\end{align}
Clearly, the first term is the standard action of the $\CP^{n-1}$ sigma model, whereas the second term is the pull-back to the worldsheet of the K\"{a}hler form. The second term is topological and corresponds to the theta-angle $\theta = \pi p\;\;\textrm{mod}\;\;2\pi$.

\subsection{The large-$n$ limit}\label{largensec}

A useful method for the analysis of vector-like systems (such as the $\CP^{n-1}$ model, where the dynamical variable is the vector $z$) is the $1\over n$-expansion. The first step is in rewriting the model~(\ref{CPNaction}) as a gauged linear sigma model (GLSM). This is the first time we encounter such systems in this review, but later this point of view will be useful in the discussion of anomalies in Section~\ref{hooftanomsec}, and even essential in the analysis of the integrable models in Chapter 3. The GLSM action reads
\bea\label{CPNgauged}
\mathcal{S}=\int\,d^2x\,\left[\sum\limits_{\alpha=1}^n |D_\mu z^\alpha|^2-\lambda \left(\sum\limits_{\alpha=1}^n | z^\alpha|^2-{n\over g}\right)\right]+\frac{\theta}{2\pi}\,\int\,dA.
\eea
Here $z^\alpha$ are the components of $z$, which is normalized as $|z|^2={n\over g}$ (we have introduced the `t Hooft coupling constant $g$ of the sigma model), $D_\mu$ is a $\tU(1)$-covariant derivative, i.e. $D_\mu z^\alpha=\dd_\mu z^\alpha-i A_\mu\,z^\alpha$, and $\lambda$ is a Lagrange multiplier imposing the normalization constraint. The relation to~(\ref{CPNaction}) is as follows. The model~(\ref{CPNaction}) is invariant w.r.t. complex rescalings of the vector $z$, i.e. $z\to \uplambda\,z$ with $\uplambda \in \CC^\ast$, which is in accordance with the complex definition of the projective space. We have used this freedom to normalize the $z$ vector. Even more importantly, we have introduced a gauge field $A_\mu$. This gauge field does not have a kinetic term and enters the Lagrangian~(\ref{CPNgauged}) only algebraically. It can be eliminated via its e.o.m., which then leads one back to the system~(\ref{CPNaction}). 

In the model~(\ref{CPNaction}), the value of the topological angle is $\theta=p\,\pi$, however for the present discussion we prefer to leave it as a free parameter. The point of rewriting the action in the form~(\ref{CPNgauged}) is that it has become quadratic in the $z$ fields, so that they can be integrated out. The resulting action of the $\lambda$ and $A_\mu$ fields is
\bea
\mathcal{S}=n\,\left[i\,\mathrm{Tr\,Log}\left(-D_{\mu}^2-\lambda \right)+\int\,d^2x\,{\lambda\over g} \right]+\frac{\theta}{2\pi}\,\int\,dA.
\eea
Since this expression appears in the exponent in the integrand of the path integral, the large-$n$ limit corresponds to a stationary phase approximation. The critical point equation, obtained by varying w.r.t. $\lambda$, -- the so-called gap equation -- has the form (due to Lorentz and translational invariance one sets $A_{\mu}=0$ and $\lambda=\mathrm{const.}$ at the critical point)
\bea
{1\over g}-\int\,\frac{d^2k}{(2 \pi)^2}\,\frac{1}{k^2+\lambda}=0\,.
\eea
As the integral is UV-divergent, one imposes a cut-off $\Lambda$, and the solution is
\bea
\lambda=\Lambda^2\,e^{-\frac{4\pi}{g}}\,.
\eea
To get a qualitative picture of the phenomenon one may substitute this value into the original action~(\ref{CPNgauged}), arriving at a system of $n$ massive fields $z^\alpha$ (with mass $m^2=\lambda$) interacting with a gauge field $A_\mu$. One can show~\cite{DAdda1, AddaSUSY, WittenCPN} that the effect of the gauge field is to generate an attractive $\theta$-dependent potential between the `quarks' $z$ and `antiquarks' $\bar{z}$, which confines them for all values of $\theta$~\cite{WittenCPN}. The mass of the lowest bound state is $2m+\ldots$, where $\ldots$ are power-like corrections in $1\over n$ (which also depend on $\theta$). One concludes that, for large $n$, the model is massive for all values of $\theta$. For a review of $\theta$-dependence in sigma models and in gauge theories (as well as for the references on related lattice calculations) cf.~\cite{Vicari}.

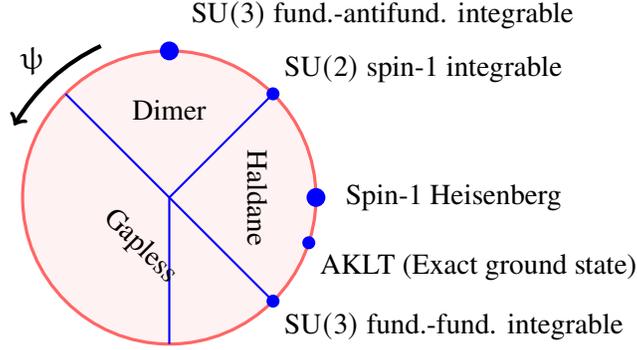
\begin{figure}
\centering
\bea\nonumber
\begin{tikzpicture}[
baseline=-\the\dimexpr\fontdimen22\textfont2\relax,scale=1.3]

\filldraw[color=red!60, fill=red!5, very thick](-1,0) circle (1.5);
\filldraw[color=blue, fill=blue, very thick](0.5,0) circle (0.08);
\filldraw[color=blue, fill=blue, very thick](-1,1.5) circle (0.08);
\node (Heis) at (0.5, 0) {};
\node (Dim) at (-1,1.5) {};
\draw[thick, color=blue,rotate around={45:(-1,0)}] (-1,0) --  (0.5,0) node (TB) {};
\draw[thick, color=blue,rotate around={-45:(-1,0)}] (-2.5,0) -- (0.5,0) node (LS) {};
\draw[thick, color=blue!100,rotate around={-18:(-1,0)}, opacity=0] (-2.5,0) -- (0.5,0) node (AKLT) {};
\draw[thick, color=blue] (-1,0) -- (-1,-1.5);
\filldraw[color=blue, fill=blue, very thick](TB) circle (0.05);
\filldraw[color=blue, fill=blue, very thick](LS) circle (0.05);
\filldraw[color=blue, fill=blue, very thick](AKLT) circle (0.05);
\draw[ultra thick, ->] (-1.7,1.55) arc (115:150:2);
\node[rotate=-90] at (-0.1,0) {\small Haldane};
\node at (-1,0.9) {\small Dimer};
\node[rotate=-45] at (-1.3,-0.5) {\small Gapless};
\node [below right=-0.15 and -0.15 of AKLT] {\small AKLT (Exact ground state)};
\node [below right=-0.15 and -0.15 of LS] {\small $\tSU(3)$ fund.-fund. integrable};
\node [above right=-0.15 and -0.15 of TB] {\small $\tSU(2)$ spin-1 integrable};
\node at (-2.4,1.4) {$\uppsi$};
\node [right=0.1 of Heis] {\small Spin-1 Heisenberg};
\node [above right=0 and 0 of Dim] {\small $\tSU(3)$ fund.-antifund. integrable};
\end{tikzpicture}
\eea
\caption{Phases of the bilinear-biquadratic spin chain.} \label{bilinbiquadpic}
\end{figure}

An interesting relation to spin chains may be obtained by considering the so-called bilinear-biquadratic spin chain~\cite{Affleck1986} defined by the Hamiltonian
\bea
H_{\textrm{bi}}=\frac{1}{2} \cos{\uppsi}\,\sum\limits_j\, \tr(S(j)S(j+1)) -\frac{1}{4}\sin{\uppsi}\,\sum\limits_j\,\left(\tr(S(j)S(j+1))\right)^2\,,
\eea
where $S$ are the matrices of spin-1 generators of $\mathfrak{su}(2)$. The phases of this chain as a function of $\uppsi$ are shown in Fig.~\ref{bilinbiquadpic}. For the discussion here only the Haldane and dimer phases are important; see~\cite{Fath, Mikeska, Andres} for the description of the full phase diagram and of the various phase transitions. Before explaining the relation to sigma models, let us comment on the five special points in the diagram.  They correspond to the following chains:
\begin{itemize}
    \item[$\circ$] Spin-1 Heisenberg, with Hamiltonian  $H=\sum\limits_j \tr( S(j)S(j+1)) $. By the original argument of Haldane it is gapped, and the field theory mapping results in an $S^2$ sigma model with $\theta=0$.
    \item[$\circ$] AKLT ($\mathrm{tan}(\uppsi)=-{1\over 3}$): this model lies in the Haldane phase as well. The ground state may be calculated exactly and is translationally invariant. See Section~\ref{AKLTsec}.
    \item[$\circ$] The critical $\tSU(2)$ spin-1 integrable point ($\uppsi={\pi\over 4}$). This is a higher-spin integrable extension of the spin-1/2 Heisenberg Hamiltonian~\cite{su2int1, su2int2}. The continuum limit is described by the $\tSU(2)_{k=2}$ WZNW model.
    \item[$\circ$] At $\uppsi=-{\pi\over 4}$ the symmetry is enhanced to $\tSU(3)$ and one has the $\tSU(3)$ extension~\cite{Sutherland} of the Heisenberg Hamiltonian $H=\sum\limits_{j} \tr( S(j)S(j+1))$, where $S$ now contain the generators of $\mathfrak{su}(3)$ in the fundamental representation at all sites. Again the spectrum is gapless, the critical theory described by $\tSU(3)_{k=1}$ WZNW model.
    \item[$\circ$] At $\uppsi={\pi\over 2}$ one again has an enhancement to $\tSU(3)$, this time with matrices $S$ whose entries generate alternating fundamental/anti-fundamental representations, i.e. $S_{\alpha\beta}(j+1)=\bar{S_{\alpha\beta}(j)}$. This model is integrable and gapped~\cite{Affleck1n, ChangAffleck, Parkinson, Batchelor, Klumper}, with a two-fold degenerate ground state and broken translational invariance (the `dimer'). 
\end{itemize}
The two large dots in the diagram, which correspond to the first and last points in the above list, are particularly important for us. As already mentioned, the $\uppsi=0$ case corresponds to the $S^2$ sigma model with vanishing $\theta$-angle. The $\uppsi={\pi\over 2}$ point corresponds to an alternating $\tSU(3)$ spin chain, exactly of the type considered in the previous section, so that the resulting field theory is a $\CP^2$ sigma model with $\theta=\pi$ ($p=1$ in the notation of the previous section). As we have mentioned, this spin chain has a gap in the spectrum, suggesting that the $\CP^{n-1}$-models with $n>2$ have a gap at $\theta=\pi$, which makes them  different from the $\CP^1$ model. At the same time this makes it consistent with the large-$n$ description above, which predicts a mass gap for the sigma models.

\subsection{Symmetric representations and the flag manifold as the space of N\'eel vacua: $\SU(3)$ case}\label{antifer}

We now want to move forward from the Hamiltonian (\ref{heisham1}) and find the sigma model which arises upon taking the continuum limit around the antiferromagnetic ``vacuum'' of the $\SU(3)$ spin chain with the Hamiltonian~(\ref{1:2}):
\bea\label{longrangehamsu3}
\mathcal{H}=\sum\limits_{i=1}^L\,\left(J_1\,\tr(S(j) S(j+1))+J_2\,\tr(S(j) S(j+2))\right)\,,
\eea
where $J_1>0, J_2>0$ are parameters that we leave free for the moment. 
First of all, completely parallel to the discussion of the isotropic spin chain in Section \ref{xxxchain} above, we can write a path integral expression for the evolution operator of the SU(3) chain (\ref{longrangehamsu3}). Similarly to (\ref{actionxxx}), the action appearing in the exponent in the integrand of the path integral has the following form:
\bear\label{actionnew}
&&\mathcal{S} = p\;\int\limits_0^1\,dt\, \sum\limits_j \left( i \,\frac{\dot{z}(j) \circ \bar{z}(j)}{|z(j)|^2} +
\alpha \cdot \mathcal{H}_j \right)\,,\quad\quad \textrm{where}\\ \nonumber
 &&\mathcal{H}_j= J_1\,\frac{|z(j) \circ  \bar{z}(j+1)|^2}{|z(j)|^2\,|z(j+1)|^2} +J_2\,
 \frac{|z(j)\circ  \bar{z}(j+2)|^2}{|z(j)|^2\,|z(j+2)|^2} \,.
\eear
In this formula each of the variables $z(j)$ has an additional (hidden) index, which takes three possible values corresponding to the fundamental representation of $\SU(3)$. 

We claim that in the case of (\ref{actionnew}) the antiferromagnetic vacuum configuration is when the $z$-vectors on any 3 neighboring sites are orthogonal to each other. First of all, this is consistent with what we had for the $\SU(2)$ case above, since the equation $1+ \bar{z}^1 z^2=0$ arising in that case (if one uses inhomogeneous coordinates) has a unique solution $z^2=-\frac{1}{\bar{z}^1}$, which is the antipodal involution discussed above. When $n=3$ we need to take three consecutive sites and impose orthogonality conditions on the three $z$-vectors $z_1, z_2, z_3$ sitting at these sites\footnote{Here we switch back to homogeneous coordinates.}:
\bea\label{2flag}
z_1 \circ \bar{z}_2=z_2 \circ \bar{z}_3=z_1\circ \bar{z}_3=0 .
\eea
The submanifold of $(\CP^2)^{\times 3}$ described by (\ref{2flag}) is the flag manifold $\mathcal{F}_3$. We're now going to elaborate on this simplest nontrivial example.

The first question is what will arise in the continuum limit from the kinetic term in~(\ref{actionnew}). The discussion in the previous section (see~(\ref{kintermcpn})) indicates that it is natural to first focus on  an arbitrary set of 3 consecutive sites. Then the kinetic term in the spin chain Lagrangian is the pull-back $\theta_t$ of the following one-form:
\bea\label{kinterm}
\theta= \im \frac{dz_1 \circ \bar{z}_1}{z_1 \circ \bar{z}_1}+\im \frac{dz_2 \circ \bar{z}_2}{z_2 \circ \bar{z}_2}+\im \frac{dz_3 \circ \bar{z}_3}{z_3 \circ \bar{z}_3}
\eea
This is the Poincar\'e-Liouville one-form for the product symplectic form $\Omega$ on $\CP^2\times \CP^2 \times \CP^2$,  so that 
$d\theta=\Omega$.  We claim that on the submanifold $\mathcal{F}_3$, described by (\ref{2flag}), this 2-form is zero. We may even formulate a slightly more general statement:

\vspace{0.3cm}\noindent \hspace{0.3cm}
\fbox{\parbox{14cm}{
\centering 
The submanifold $\mathcal{F}_3 \subset (\CP^2)^{\times 3}$, and more generally $\mathcal{F}_{n} \subset (\CP^{n-1})^{\times n}$,\\ is Lagrangian.
}
}

\vspace{0.3cm}
In fact, we already encountered a generalization of this statement in Section~\ref{cohsec}, however here we emphasize it due to its particular importance for the description of the antiferromagnetic interactions in spin chains. Let us focus on the consequences of this fact. It follows that
$\theta|_{\mathcal{F}_3}=df,$
where $f$ is a function (in fact, $f=i \log{(\epsilon_{\alpha\beta\gamma} \,z_1^\alpha \,z_2^\beta \,z_3^\gamma)}$),

so the integral $\int\limits_0^1 \theta_t\,dt = f(1)-f(0)$ reduces to the boundary term. We ignore this term in the present discussion.

Let us emphasize that the geometric setup discussed in the last two sections is general, and is key to understanding the target space of the sigma model that emerges in the continuum limit. The main conclusion is:

\vspace{0.3cm}\noindent \hspace{0.3cm}
\fbox{\parbox{14cm}{
\centering 
The target space of the sigma model is the `moduli space' of N\'eel vacua of the spin chain. It is a Lagrangian submanifold in the phase space of an elementary cell.
}
}

\vspace{0.3cm}
The first statement -- that  the expansion around the antiferromagnetic vacuum configuration leads to a sigma model whose target space is the manifold of vacua -- will be proven in the following sections. 

\section{The continuum limit}\label{cl}

The term $\mathcal{H}$ in (\ref{actionnew})
is equal to zero if we impose the background configuration (\ref{2flag}). Moreover, since $0\leq\mathcal{H}_j\leq J_1+J_2$, one immediately sees that the ferromagnetic and antiferromagnetic vacua saturate respectively the maximum and minimum of its possible values. In view of the fact that we will be building an expansion around the antiferromagnetic vacuum, from this observation we deduce an important consequence, namely that this expansion must start with a quadratic term, i.e. there is no linear term.

Let us assume that the number of sites of our spin chain is a factor of 3 (this is only needed for simplicity, and it does not play a big role for a sufficiently long spin chain). In this case we split the spin chain into $\hat{L}$ segments of length $3$ and focus for the moment on just one of these segments, which is the elementary cell number $k$ in the chain.

\subsection{The expansion around the ``vacuum'' configuration}\label{expansion}

On each of the three sites we have a three-dimensional complex vector $z_A$. Let us form a $3\times 3$ matrix of these vectors, which we denote by $Z$.

The antiferromagnetic configuration corresponds to the case where the three vectors are mutually orthogonal.

Now we need to take the fluctuations into account, and in order to build the sought for expansion we will employ the so-called $QR$ decomposition of a matrix. The $QR$ decomposition theorem says that an arbitrary matrix $Z$ may be decomposed into a product of a unitary matrix $U$ and an upper triangular one $B_+$:
\bea\label{QR}
Z= U\circ B_+
\eea
This statement is equivalent to the Gram-Schmidt orthogonalization theorem. Let us parametrize $B_+$ at link $k$ in the following way:
\bea
B_+(k)= \begin{pmatrix} 
  1 & \ol \xs(k) & \ol\ys(k) \\ 
  0 & 1 & \ol\zs(k) \\
  0 & 0 & 1 \\
\end{pmatrix} \begin{pmatrix} 
  \as(k) & 0 & 0 \\ 
  0 & \bs(k) & 0 \\
  0 & 0 & \cs(k) \\
\end{pmatrix}.
\eea

If we denote the columns of the matrix $U$ as $(u_1, u_2, u_3)$, the decomposition (\ref{QR}) says that
\bear\label{contin2}
&z_{1}(k) = \as(k)\, u_{1}(k),\quad z_{2}(k) = \bs(k)\,(u_{2}(k)+ \ol\xs(k)\, u_{1}(k)),\\ \nonumber &z_{3}(k)=\cs(k) \,(u_{3}(k)+\ol\ys(k)\, u_{1}(k)+ \ol\zs(k)\, u_{2}(k))\eear
Let us first of all write out the kinetic term (\ref{kinterm}) for three consecutive sites in these variables, to leading order in $\ol$. A simple calculation reveals that (suppressing the index $k$ for the moment)
\bear\label{kinetflag}
&& J_t =
\left( \im \frac{\dot{z}_1 \circ \bar{z}_1}{z_1 \circ \bar{z}_1}+\im \frac{\dot{z}_2 \circ \bar{z}_2}{z_2 \circ \bar{z}_2}+\im \frac{\dot{z}_3 \circ \bar{z}_3}{z_3 \circ \bar{z}_3}\right)_{\xs=\ys=\zs=0} \!\!\!-\\ && \nonumber-
\frac{\im}{L} \left( \,\xs\; u_1 \circ \dot{\bar{u}}_2 + \, \ys\;  u_1 \circ \dot{\bar{u}}_3 + \zs\; u_2 \circ \dot{\bar{u}}_3  - \textrm{c.c.}\right)\;+ ...
\eear
The hypothesis of the existence of a continuum limit implies that $u_{1}(k),\;u_{2}(k),\;u_{3}(k)$ vary mildly with $k$, in other words we may approximate
\bea\label{contin}
u_A(k+1)=u_A(k)+\ol u_A(k)'+ ...
\eea
\begin{center}
\vspace{-0.5cm}
\begin{figure}[h]
\centering
\includegraphics[width =0.9\textwidth]{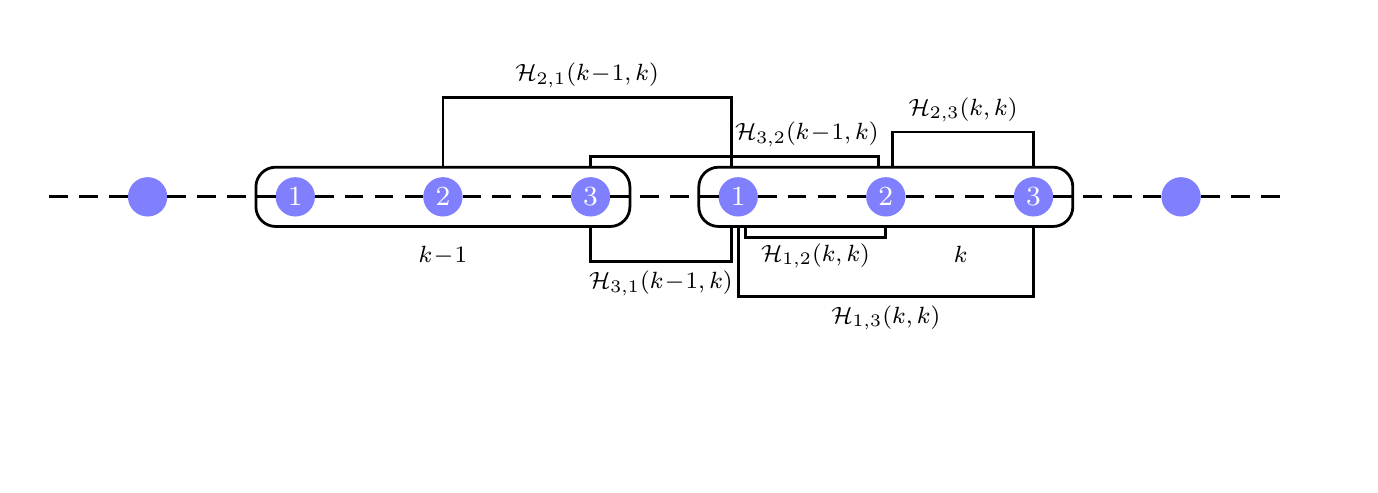}
\vspace{-1.5cm}
\caption{Explanation of the various terms calculated in (\ref{hamcontlimit}).}
\label{fig:chainpic}
\end{figure}
\end{center}
Let us introduce the quantity
\bea\label{hamperm2}
\mathcal{H}_{A,B}(k,k')=\frac{|z_A(k)\circ  \bar{z}_B(k')|^2}{|z_A(k)|^2 \; |z_B(k')|^2},
\eea
which is the density of the Hamiltonian $\mathcal{H}$ from (\ref{actionnew}), if the indices $A, B, k, k'$ change in a particular range. Indeed, we need to calculate $\mathcal{H}_{A,B}(k,k')$ for nearest- and next-to-nearest neighbor sites,
using the formulas (\ref{contin2})-(\ref{contin}) and keeping only the leading orders of $\olsq$ (see Fig.~\ref{fig:chainpic} for an explanation of what these terms stand for):
\begin{align} \nonumber
&\scalemath{0.95}{\mathcal{H}_{1,2}(k,k)=\frac{|z_{1}(k)\circ  \bar{z}_{2}(k)|^2}{|z_{1}(k)|^2 \; |z_{2}(k)|^2}\simeq \olsq |\xs(k)|^2,\quad
\mathcal{H}_{2,3}(k,k)=\frac{|z_{2}(k)\circ  \bar{z}_{3}(k)|^2}{|z_{2}(k)|^2 \; |z_{3}(k)|^2}\simeq \olsq |\zs(k)|^2}\\ \nonumber
&
\scalemath{0.95}{\mathcal{H}_{1,3}(k,k)=\frac{|z_{1}(k)\circ  \bar{z}_{3}(k)|^2}{|z_{1}(k)|^2 \; |z_{3}(k)|^2}\simeq \olsq |\ys(k)|^2} \\ \label{hamcontlimit}
&\mathcal{H}_{3,1}(k\!-\!1,k)=\frac{|z_{3}(k\!-\!1)\circ  \bar{z}_{1}(k)|^2}{|z_{3}(k\!-\!1)|^2 \; |z_{1}(k)|^2}
\simeq
\olsq |\!-\!u_{3}(k)' \!\circ \bar{u}_{1}(k)+\ys(k)|^2 \\ \nonumber 
&\mathcal{H}_{3,2}(k\!-\!1,k)=\frac{|z_{3}(k\!-\!1)\circ  \bar{z}_{2}(k)|^2}{|z_{3}(k\!-\!1)|^2 \; |z_{2}(k)|^2}
\simeq \olsq |-u_{3}'(k) \circ \bar{u}_{2}(k)+\zs(k)|^2\\ \nonumber
&\mathcal{H}_{2,1}(k\!-\!1,k)=\frac{|z_{2}(k\!-\!1)\circ  \bar{z}_{1}(k)|^2}{|z_{2}(k\!-\!1)|^2 \; |z_{1,k}|^2}
\simeq \olsq |-u_{2}'(k) \circ \bar{u}_{1}(k)+\xs(k)|^2
\end{align}

Substituting these values, together with the expression~(\ref{kinetflag}) for the kinetic term, into the action~(\ref{actionnew}), and eliminating the variables $x(k), y(k), z(k)$ that enter only quadratically, one obtains the Lagrangian (we set for simplicity $J_1=1, J_2=a$)\footnote{Note the convention for the epsilon-symbol: $\epsilon_{01}=1$.}

\bear\label{lagrf}
\mathcal{L}=
\frac{1}{1+a} (|u_{1}\circ\dot{\bar{u}}_{2}|^2-a\,\olsq |\bar{u}_{2}'\circ u_{1}|^2)-\frac{\im a}{1+a} \ol \epsilon_{\mu\nu} (u_{1} \circ \partial_\mu \bar{u}_{2})(\bar{u}_{1}\circ\partial_\nu u_{2})+\\ \nonumber +
\frac{1}{1+a}(|u_{1}\circ\dot{\bar{u}}_{3}|^2-a\,\olsq |\bar{u}_{1}\circ u'_{3}|^2)-\frac{\im }{1+a} \ol \epsilon_{\mu\nu} (u_{1} \circ \partial_\mu \bar{u}_{3})(\bar{u}_{1}\circ\partial_\nu u_{3})+ \\ \nonumber
+\frac{1}{1+a}(|u_{2}\circ\dot{\bar{u}}_{3}|^2-a\,\olsq |\bar{u}_{2}\circ u'_{3}|^2)-\frac{\im a}{1+a} \ol \epsilon_{\mu\nu} (u_{2} \circ \partial_\mu \bar{u}_{3})(\bar{u}_{2}\circ\partial_\nu u_{3})
\eear
Of course, each of the first terms in the three lines above can be brought to a canonical relativistic-invariant form by a rescaling of the space variable\footnote{It was noted in~\cite{BykovHaldane2} that there exists a canonical geometric expression for the metric arising in this way. Suppose $(\mathbf{\Phi}, \omega)$ is a symplectic manifold (in this case the phase space of an elementary cell), and $\mathcal{H}$ a function on it -- the classical Hamiltonian -- that attains a minimum on a Lagrangian submanifold $\mathcal{N}\mysub \mathbf{\Phi}$ (the target space of the sigma model). In this case one can define the (inverse) metric on $\mathcal{N}$ by the formula $g^{ij}=\omega^{ik} \left(\frac{\dd^2 \mathcal{H}}{\dd x^2}\right)_{kl}\omega^{lj}$.}. The fact that all the flavour wave velocities are equal in~(\ref{lagrf}) is really a coincidence that happens for $n=3$. For $n>3$ the flavor wave velocities are typically different and depend on the coupling constants of the spin chain Hamiltonian, see~(\ref{flavourvel}) and section~\ref{symmrepsec} for more details.

Let us analyze now the three epsilon-terms in the above expression. First of all, let us introduce notations for the corresponding three 2-forms:
\bear
&&u\equiv \im\,(u_1 \circ d \bar{u}_2)\wedge( \bar{u}_1 \circ d u_2)\\
&&v\equiv -\im\,(u_1 \circ d \bar{u}_3)\wedge( \bar{u}_1 \circ d u_3)\\
&&w\equiv \im\,(u_2 \circ d \bar{u}_3)\wedge( \bar{u}_2 \circ d u_3)
\eear
Then the three epsilon-terms in (\ref{lagrf}) are the pull-back of the following 2-form:
\bea
\omega = \frac{1}{1+a} \;(v-a\, u-a\, w)
\eea
The crucial fact is that $\omega$ may be split in two parts: a topological one (the $\theta$-term) and the non-topological one (the $B$-field, or Kalb-Ramond field, in sigma model terminology). The splitting may be achieved by noting that the above Lagrangian possesses a $\mathbb{Z}_3$ `quasi-symmetry', which acts on the vectors $(u_1, u_2, u_3)$ of the flag by cyclically permuting them:
\bea\label{Z3trans}
\mathbb{Z}_3:\quad\quad u_1\to u_2\to u_3\to u_1\,.
\eea
This symmetry has a transparent meaning: it arises because of the translational invariance of the Hamiltonian~(\ref{longrangehamsu3}), since the transformation~(\ref{Z3trans}) corresponds to shifting the elementary cell by one site. This has to be a symmetry at the level of the partition function, however for the Lagrangian~(\ref{lagrf}) this is only a `quasi-symmetry', meaning that under the action of $\mathbb{Z}_3$ it is shifted by an \emph{integral topological term}, i.e. an element of $H^2(\mathcal{F}, \mathbb{Z})$. As a result, the partition function, which is given by a path integral of the type 
\bea\label{partfuncshort}
\mathcal{Z}=\int\,e^{i\,\int\,dt\,dx\,\mathcal{L}}\,\prod\,dU\,,
\eea
is unaltered. This is the same argument that is used to prove that the path integral of Chern-Simons theory is well-defined~\cite{DeserJackiw, WittenJones}. Indeed, we will now show that $\omega$ may be split in a non-topological part that is invariant w.r.t. $\mathbb{Z}_3$ and a $\theta$-angle part that transforms non-trivially.

First of all, the two-forms transform as follows under $\mathbb{Z}_3$:
\bea
\mathbb{Z}_3:\quad\quad u\to w\to v\to u\,.
\eea
As a result, the only $\mathbb{Z}_3$-singlet is $u+v+w$, so that one may decompose
\bea\label{omega1}
\omega=\frac{1-2a}{3(1+a)}\left(u+v+w\right)-{1\over 3}\left(u- v+w-v\right)\,,
\eea
Let us now show that the second term is  topological. First of all, recall from section~\ref{cohsec} that every element of $H^2(\mathcal{F}_3, \mathbb{R})$ is  a linear combination of three forms $\Omega_A\; (A=1, 2, 3)$, which are the pull-backs to the flag manifold of the Fubini-Study forms corresponding to $u_1, u_2, u_3$. It is easy to relate these Fubini-Study forms to $u, v, w$. Recall that
\bea
\Omega_1 = du_1 \wedge d\bar{u}_1-(du_1 \circ \bar{u}_1)\wedge (d\bar{u}_1 \circ u_1)
\eea
and similar expressions hold for $\Omega_2$ and $\Omega_3$. Now let us use the identity $u_1\otimes \bar{u}_1+u_2\otimes \bar{u}_2+u_3\otimes \bar{u}_3=\mathds{1}_3$. Using this, we may rewrite the restriction of $\Omega_1$ to the flag manifold $\mathcal{F}_3$ in the following way:
\bea
\Omega_1|_{\mathcal{F}_3}=(du_1 \circ \bar{u}_2)\wedge (d\bar{u}_1 \circ u_2)+(du_1 \circ \bar{u}_3)\wedge (d\bar{u}_1 \circ u_3)=u-v.
\eea
External differentiation and restriction to a submanifold are commutative operations, therefore this restricted form is a closed 2-form on $\mathcal{F}_3$. Analogously $\Omega_2|_{\mathcal{F}_3}=w-u,\;\Omega_3|_{\mathcal{F}_3}=v-w$. In particular, we see that the sum
\bea\label{lagrembed1}
(\Omega_1+\Omega_2+\Omega_3)|_{\mathcal{F}_3}=0
\eea
is zero, as already discussed in section~\ref{cohsec}. It follows that the second term in~(\ref{omega1}) is topological, taking values in ${1\over 3}\,H^2(\mathcal{F}_3, \mathbb{Z})$. Keeping in mind~(\ref{lagrembed1}), we may write $\omega$ as
\bea\label{omegafullform}
\omega\simeq \frac{1-2a}{3(1+a)}\left(u+v+w\right)+\underbracket[0.6pt][0.6ex]{\Omega_1+2\Omega_2+3\Omega_3\over 3}_{:=\omega_{\mathrm{top}}}
\eea
Since the action stands in the exponent of the integrand in~(\ref{partfuncshort}), $\omega$ is defined modulo elements of $H^2(\mathcal{F}_3, \mathbb{Z})$. This is important, because under the action of $\mathbb{Z}_3$ the topological term $\omega_{\mathrm{top}}$ shifts precisely by such terms. Indeed, since the action of $\mathbb{Z}_3$ cyclically permutes $\Omega_1, \Omega_2, \Omega_3$, under its action one has
\bea\label{topformshift}
\mathbb{Z}_3:\quad\quad \omega_{\mathrm{top}}\to \omega_{\mathrm{top}}
-\underbracket[0.6pt][0.6ex]{\Omega_1+\Omega_2+\Omega_3\over 3}_{=0}+\underbracket[0.6pt][0.6ex]{\;\;\Omega_1\;\;}_{\in H^2(\mathcal{F}_3, \mathbb{Z})}\,\simeq \omega_{\mathrm{top}}\,,
\eea
where $\simeq$ means `up to an element of $H^2(\mathcal{F}_3, \mathbb{Z})$'. 
This property will be essential for the discussion of discrete 't Hooft anomalies in section~\ref{hooftanomsec} below.

\section{Symmetric representations: the general case}\label{symmrepsec}

In the previous section we considered the case of $\SU(3)$ spin chains. Next we discuss the generalization to the $\SU(n)$ case, where the Hamiltonian is given by~(\ref{1:2}). The N\'eel state in this case is given by a direct generalization of~(\ref{2flag}), namely requiring that the $n$ consecutive vectors $Z:=(z_1, \ldots z_n)$ are all pairwise orthogonal:
\bea\label{NeelSUN}
z_A\circ \bar{z}_B=0\quad\quad\textrm{for}\quad\quad A\neq B\,.
\eea
In order to derive the continuum theory, one follows the steps described in the previous section: one first introduces deviations from the N\'eel configuration~(\ref{NeelSUN}) and performs a factorization~(\ref{QR}) $Z=U\circ B_+$, where $U\in \tU(n)$ is unitary and $B_+$ is strictly upper-triangular. As before, the matrix $B_+$ describes the deviations from the N\'eel state in a single unit cell. One then expands the Lagrangian to quadratic order in the elements of $B_+$, as well as in the cell-to-cell variations, and integrates over the $B_+$ elements that enter algebraically. The calculation is rather tedious, and its details can be found either in~\cite{BykovHaldane2} or in~\cite{HaldaneSUN}.

\subsection{The flag manifold sigma model from an $\SU(n)$ spin chain}\label{flagchainsun}
Here we write out the resulting action of the field theory describing the SU($n$) chain in the rank-$p$ symmetric irrep:

\bea \label{3:3}
	S =  \sum_{1\leq A<B\leq n} \int dx d\tau \frac{1}{g_{|A-B|}} \left( v_{
|A-B|}|u_{A}\circ \dd_x\bar{u_{B}}|^2
	+ \frac{1}{v_{|A-B|}} |u_{A}\circ \dd_{\tau}\bar{u_{B}}|^2 \right)
\eea
\[
	  -\epsilon_{\mu\nu} \sum_{1\leq A<B\leq n} \lambda_{|A-B|} \int dx d\tau\;  (u_{A}\circ \dd_{\mu}\bar{u_{B}})(u_{B}\circ \dd_{\nu}\bar{u_{A}})  + S_{\text{top}},
\]
where $ v_{t} = np\sqrt{J_{t}J_{n-t}}$ is the flavour wave velocity associated with the pair of couplings $J_{t}$ and $J_{n - t}$ of the spin chain~(\ref{1:2}). The $\lambda$-terms are the generalizations of the non-topological contribution to $\omega$ discussed in the previous section for $n=3$ (the first term in~(\ref{omegafullform})), and $S_{\text{top}}=\int\,\omega_{\mathrm{top}}$ is a higher-$n$ generalization of the topological ($\theta$)-term, discussed in detail below. As we already mentioned earlier, the cases $n=2$ and $n=3$ may be thought of as being exceptional in that there is a single flavour wave velocity. The coupling constants are
\bea
	g_{t} = \frac{n}{v_{t}}(J_t + J_{n-t})
\eea
and
\bea \label{eq:lamform}
	\frac{ n \lambda_{t}}{p} = \frac{(n-t)J_{n-t} -tJ_{t}}{J_{t} + J_{n-t}}.
\eea
Since the coupling constants and velocities satisfy $g_t = g_{n-t}$ and $v_t = v_{n-t}$, we conclude that there are $\lfloor {n\over 2} \rfloor$ velocities and coupling constants, where
\bea \label{eq:floor}
	\lfloor {n\over 2} \rfloor = \begin{cases} \frac{n}{2} & \text{$n$ even} \\
	\frac{n-1}{2} & \text{$n$ odd} \\
	\end{cases}
\eea
The topological term is 
\bea
	S_{\text{top}} := \frac{2\pi i p}{n}\sum_{A=1}^{n} A \,Q_A ,\quad\quad\textrm{where}\quad\quad Q_A :=  \frac{1}{2\pi }\, 
\int 
\, \Omega_{A} 
	\label{eq:Stop}
\eea
is the integral of the Fubini-Study form over the worldsheet (as such, it is a quantized topological charge, see~(\ref{FSform})). Since $\sum_{A=1}^n\,\Omega_{A}=0$ (see section~\ref{cohsec}), one has
\bea\label{sumtopcharges}
\sum_{j=1}^n Q_A  =0\,,
\eea
so that
there are $n-1$ independent topological charges.

We note that the $\lambda$-terms appearing in (\ref{3:3}) are not quantized, despite the fact that they are pure imaginary in imaginary time. We give an interpretation of these terms below. In \cite{BykovHaldane2}, these $\lambda$-terms were absent as a result of the same fine-tuning that ensured a unique velocity. Indeed, the choice $J_t = \sqrt{\frac{n-t}{t}}$ ensures that $v_t \equiv \textrm{const.}$ for all $t$, and moreover that $\lambda_t = 0$ for all $t$.

\subsection{$\mathbb{Z}_n$ symmetry}\label{Znsymmsec}

Just as in the $\SU(3)$ case, we may introduce a discrete symmetry
\bea\label{Znsymm}
\mathbb{Z}_n:\quad\quad u_A\to u_{A+1}\,,\quad\quad u_{n+1}\equiv u_1\,.
\eea
It is easy to prove that the sum of $\lambda$-terms is invariant under this symmetry. Indeed, let us fix ${|A-B|=t\leq \lfloor {n\over 2} \rfloor}$, then every form 
\begin{equation}
B_t:=\left(\sum\limits_{\substack{1\leq A<B\leq n\\ |A-B|=t}}-\sum\limits_{\substack{1\leq A<B \leq n\\ |A-B|=n-t}}\right)\,(u_A\circ d\bar{u_B}\wedge u_B\circ d\bar{u_A})
\end{equation}
is $\mathbb{Z}_n$-invariant. Using the property $\lambda_{n-t}=-\lambda_t$ that follows from~(\ref{eq:lamform}), we may write the form entering the $\lambda$-terms as
\bea
\sum_{1\leq A< B \leq n} \lambda_{|A-B|} \,  (u_{A}\circ d\bar{u_{B}})\wedge (u_{B}\circ d\bar{u_{A}})=\sum\limits_{t=1}^{\lfloor {n\over 2} \rfloor}\,\lambda_t\,B_t\,,
\eea
proving that it is also invariant. In contrast, the topological part of the action is shifted under the $\mathbb{Z}_n$-transformation~(\ref{Znsymm}): $S_{\mathrm{top}}\to S_{\mathrm{top}}+2\pi i\,Q_1$, just as in the $\SU(3)$ case~(\ref{topformshift}). This is the ultimate reason that allows separating the topological terms from the non-topological $\lambda$-part of the skew-symmetric tensor field.

\subsection{Velocity Renormalization} \label{section:vrg}

The coupling constants $\{g_t\}$ and $\{\lambda_t\}$ in~(\ref{3:3}) correspond to the metric and torsion on the flag manifold, respectively~\cite{ohmori2019sigma}. However, a unique metric cannot be defined, since the theory (\ref{3:3}) lacks the Lorentz invariance that is often assumed for sigma models. Thus, we have a non-Lorentz invariant flag manifold sigma model (the same phenomenon was observed in~\cite{Wamer2019} where  SU(3) chains with self-conjugate representations  were considered). We will now use the renormalization group to show that at low enough energies, it is possible for the distinct velocities occurring  to flow to a single value, so that Lorentz invariance emerges.

The Lorentz invariant versions of the above flag manifold sigma model were studied in great detail in~\cite{ohmori2019sigma}. In particular, the renormalization group flow of both the $\{g_{t}\}$ and $\{\lambda_{t}\}$ were determined for general $n$. Moreover, field theoretic versions of the LSMA theorem were formulated, using the methods of 't Hooft anomaly matching (which we review below, in Section \ref{section:thooft}). We would like to apply these results to our SU($n$) chains which lack Lorentz invariance in general. First, it will be useful to introduce dimensionless velocities, $\upsilon_t$, defined according to $
	\upsilon_t := \frac{v_t}{\bar v}\,,\;  \bar v = \frac{1}{\lfloor \frac{n}{2} \rfloor}\sum_{t=1}^{\lfloor \frac{n}{2} \rfloor} v_t,$
and introduce new spacetime coordinates by means of a rescaling $	x \to  \frac{x}{\sqrt{\bar v}}\,,\;  \tau \to \sqrt{\bar v}  \tau.
$
We then consider the differences of velocities occurring in (\ref{3:3}), namely
\bea \label{delta}
	\Delta_{tt'} := \upsilon_t - \upsilon_{t'},
\eea
and ask how they behave at low energies. More precisely, we calculate the one-loop beta functions of these $\Delta_{tt'}$, to orders $\fO(g_{t})$ and $\fO(\lambda_{t})$. We will find that each of the $\Delta_{tt'}$ flows to zero under renormalization and we will show that this implies Lorentz invariance at our order of approximation. This is consistent with the fundamental SU($n$) models with $p=1$, where it is known by Bethe ansatz that Lorentz invariance is present~\cite{Sutherland, TsvelickWiegmann, AndreiFuruya}. There is in fact a similar phenomenon in 2+1 dimensional systems, where an interacting theory of bosons and Weyl fermions renormalizes to a Lorentz invariant model~\cite{lee2007emergence, grover2014emergent}.

The coefficients $\{g_t\}$ appearing in (\ref{3:3}) are dimensionless, and are all proportional to $\frac{1}{p}$. Since we've taken a large $p$ limit, we will expand all quantities in powers of the $\{g_t\}$. As we will see below, the coefficients $\{\lambda_t\}$ in (\ref{3:3}) do not enter into our one-loop calculations, and so we will neglect them throughout. Since we are interested in the low energy dynamics of these quantum field theories, we make the simplifying assumption that the matrices $U=(u_1, \ldots, u_n)$ are close to the identity matrix, and expand them in terms of the SU$(n$) generators.Recalling that the matrix $U$ is subject to the gauge transformations $U\to U\cdot D_g$, where $D_g=\mathrm{Diag}(e^{i\alpha_1}, \cdots, e^{i \alpha_n})$, we may fix this gauge invariance by the following parametrization\footnote{Throughout, repeated indices will be summed over.} of $U$:
\bea
U = \mathrm{Exp}\left(i\,\sum\limits_{\substack{ a\, \in\, \textrm{off-diagonal}\\ \textrm{generators}}}\sqrt{\tilde{g_a}\over 2}\,\phi_a T_a \right) = 1 + i\, \sqrt{\tilde{g_a}\over 2}\,\phi_a T_a - \frac{\tilde{g_a}}{4}\phi_a\phi_b T_a T_ b + \fO(\phi^3).
\eea

To explain how $\tilde{g_a}$ are related to $g_{|A-B|}$, we will assume that the following basis of off-diagonal  generators is chosen: $\{T_a\}=\{E_{AB}+E_{BA}, i(E_{AB}-E_{BA}),\;A<B\}$, where $E_{AB}$ are the elementary matrices with $1$ in $AB$-th position and $0$ elsewhere. Whenever the generator $T_a$ corresponds to one of these two generators, we set $\tilde{g_a}:=g_{|A-B|}$. As shown in~\cite{HaldaneSUN}, in this notation the expansion of the Lagrangian to quartic order in the $\phi$'s has the form

\bea \label{eq:2}
	\fL = \frac{1}{2}\left[ \frac{1}{\upsilon_a}(\partial_\tau \phi_a)^2 + \upsilon_a(\partial_x\phi_a)^2\right] 
	+\frac{\sqrt{\tilde{g_a} \tilde{g_b}\tilde{g_c}}}{\sqrt{2}}   \frac{h_a(\mu)}{\tilde{g_a}}  f_{bca}\partial_\mu\phi_a\partial_\mu\phi_b\phi_c
\eea
\[
\frac{\sqrt{\tilde{g_b}\tilde{g_c}\tilde{g_d}}}{4} \frac{  h_a(\mu)}{\tilde{g_a}}\Big[ \sqrt{\tilde{g_e}} \partial_\mu\phi_e \partial_\mu \phi_b \phi_c \phi_d f_{eca}f_{bda}
	+ \frac{4}{3} f_{bcE}f_{Eda}\sqrt{\tilde{g_a}}\partial_\mu\phi_{a}\partial_\mu \phi_b\phi_c\phi_d
	\Big] + \fO(\phi^5)\,.
\]
Here $f_{abc}, f_{abC}$ (the small-letter and capital-letter indices correspond to the off-diagonal generators and all generators respectively) are the structure constants defined by $[T_a,T_b] = 2if_{abC}T_C$. Also, $h_a(\mu)=\frac{1}{\upsilon_a}$ for  $\mu = \tau$ and $h_a(\mu)=\upsilon_a$ for $\mu = x$.

The calculations then follow the standard procedures of renormalization theory. We rewrite the free part of the above Lagrangian in `renormalized variables', i.e. $\mathcal{L}_0=\frac{1}{2}\left[ Z_a^\tau \frac{1}{\upsilon^r_a}(\partial_\tau \theta_a)^2 + \upsilon^r_aZ_a^x(\partial_x\theta_a)^2\right]$ (and the interaction terms accordingly). The bare and renormalized velocities are related as $\upsilon_a = \upsilon_a^{r} \sqrt{\frac{Z_a^x }{Z^\tau_a}}$, so that one can define the corresponding $\beta$-function $\beta_{\upsilon_a} := \frac{d \upsilon_a^r}{d\log M}$, where $M$ is the fixed energy scale.
\begin{figure}[h]
\centering
\includegraphics[width = .3\textwidth]{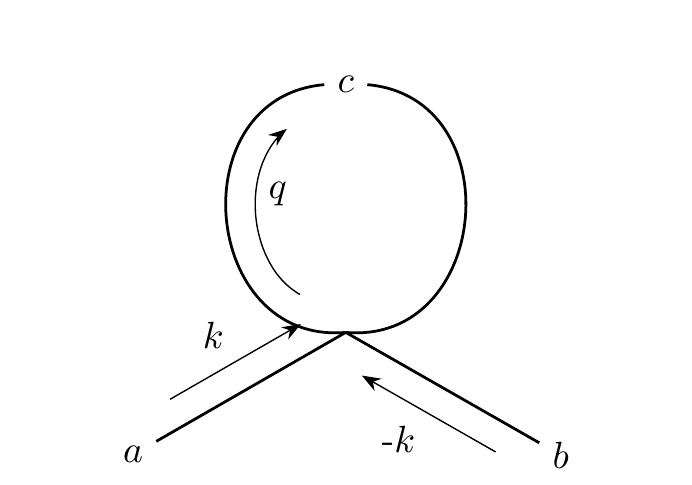}
\caption{The diagram contributing to velocity renormalization at one loop.}
\label{fig:feynman}
\end{figure}

One then has to choose the coefficients $Z_a^x, Z^\tau_a$ so as to cancel the one-loop divergences coming from the bubble graphs shown in Fig.~\ref{fig:feynman}. The details of the computation can be found in~\cite{HaldaneSUN}, the result being
\bea \label{beta:result}
	\beta_{\upsilon_t} =\frac{\upsilon_tg_t}{4\pi} \sum_{\substack{A=1 \\ A\not=t}}^{n-1} \frac{g_{|t-A|}}{g_A}\left[ \frac{\upsilon_t}{\upsilon_A} - \frac{\upsilon_A}{\upsilon_t}\right]\,,\quad\quad t=1,2, \cdots, q:= \lfloor \frac{n}{2}\rfloor\,. 
\eea
The equations for $\upsilon_t$ with $t>q$ may be obtained by using the identity $\upsilon_t = \upsilon_{n-t}$. As a result of~(\ref{beta:result}), the renormalization group flow equation $\frac{d \upsilon_t}{d\log M}=\beta_{\upsilon_t}$ (we drop the superscript $r$ to simplify the notation) is a non-linear system of ODE's for the functions $\upsilon_t(\log{M})$. One wishes to show that the `point' where all velocities are the same ($\upsilon_i=\upsilon_j$ for all $i, j$) is an attractor point of the system (clearly it is an equilibrium point). In general this might be a formidable task, so one can take a simpler step by linearizing the system of equations around the equilibrium and proving that the spectrum of the linearization operator is positive (i.e. that the equilibrium is stable). There are $q-1$ independent velocity differences that we will denote $\Delta^i := \Delta_{1,i+1}\;(i=1,2,\cdots, q-1).$ The linearized equation then takes the form
\bea \label{beta:matrix}
	\frac{d}{d\log M} \bDelta = R\bDelta\,,
\eea
where $R$ is a $(q-1)\times (q-1)$ matrix. The spectrum of $R$ will reveal the low energy behaviour of the $\Delta_{tt'}$: if the spectrum is strictly positive, one may conclude that all velocity differences flow to zero in the IR. In the highly symmetric case when all of the coupling constants $\{g_t\}$ are equal, one easily finds
\bea
		R = \frac{g}{2\pi}(n-1)\mathds{1}_{q-1}
\eea
showing that the spectrum of $R$ is strictly positive. In the non-symmetric case, for different choices of $n$ and values of the couplings, this has been checked numerically in~\cite{HaldaneSUN}, suggesting that the spectrum of $R$ is always positive.

Up to this conjecture, we have shown that the velocity differences $\Delta_{tt'}$ flow to zero at low energies. Another source of Lorentz non-invariance in the Lagrangian~(\ref{eq:2}) are the functions $h_a(\mu)$ in the interaction terms. These are however related to the velocities $\upsilon_a$ and therefore also flow to a common value, and thus Lorentz invariance of the entire model (\ref{3:3}) is possible if the velocities are initially close to each other.

\section{Generalized Haldane conjectures and $\mbox{'t}$ Hooft anomaly matching} \label{section:thooft}

Based on the renormalization group analysis in the previous section, we now argue that at low enough energies, the SU($n$) chains in the symmetric-$p$ irreps (without fine-tuning), may be described by a Lorentz invariant flag manifold sigma model
\begin{empheq}[box=\fbox]{align}
\hspace{1em}\vspace{5em}
\label{eq:rel1}
	&\fL = \sum_{A<B} \frac{1}{g_{|A-B|}} |u_{A}\circ \dd_\mu\bar{u_{B}}|^2
	- \epsilon_{\mu\nu}\sum_{A<B} \lambda_{|A-B|}\,(u_{A}\circ \dd_{\mu}\bar{u_{B}})(u_{B}\circ \dd_{\nu}\bar{u_{A}})  \\ \nonumber
	&\textrm{with topological theta-term}\\
\label{eq:fm1}
	&S_{\text{top}} = i\theta \sum_{A=1}^{n-1}A\, Q_A \hspace{10mm} \theta := \frac{2\pi p}{n}.
\end{empheq}
These sigma models have been studied in~\cite{BykovHaldane1, BykovHaldane2, ohmori2019sigma}. In \cite{ohmori2019sigma}, the renormalization group flows of the $\{\lambda_t\}$ and the $\{g_t\}$ were determined, and given a geometric interpretation. It was found that for $n>4$, the $\{g_t\}$ flow to a common value in the IR, and that for $n>6$, the $\{\lambda_t\}$ flow to zero in the IR. Even for $n<6$  Ohmori et al argue \cite{ohmori2019sigma} that a flow to the 
$\SU(n)_1$ WZNW model occurs. This is based on the observation that the $\lambda$ term doesn't induce any relevant operators in the WZNW model. 
Thus we may expect an $S_n$ (permutation group) symmetry to emerge at low enough energies, and for $n>6$. It is known that in these $S_n$-symmetric models, the unique coupling constant $g$ obeys~\cite{ohmori2019sigma}
\bea\label{normalmetrbeta}
	\frac{dg}{d\log M} = \frac{n +2}{4\pi}g.
\eea
and the theory is asymptotically free. The $S_n$-symmetric metric (with all $g_t$ equal) is known in the math literature as the `normal', or reductive, metric~\cite{FlagEinsteinMetrics}. This same metric will feature in the integrable models that we will describe in the next chapter. Interestingly, it is not K\"ahler (unless $n=2$) but it is Einstein, with cosmological constant proportional to $n+2$, which is the value in the r.h.s. of~(\ref{normalmetrbeta}) (a detailed discussion can be found in~\cite{BykovGN}).

\subsection{$\SU(n)$ Haldane conjectures}\label{HaldSUn}

The low-energy behavior of the sigma models~(\ref{eq:rel1})-(\ref{eq:fm1}) (and of the corresponding spin chains with symmetric rank-$p$ representations at each site) depends drastically on the values of the $\theta$-angles. In~\cite{HaldaneSU3, Tanizaki:2018xto, ohmori2019sigma, HaldaneSUN} the generalizations of Haldane's conjecture for this class of models were proposed. These are summarized in Table~\ref{Haldanetable}.

\begin{table}[h]
\centering
\begin{tabular}{c|c|c}
\textbf{Case}     & \textbf{Conjecture} & \textbf{Evidence}\\ \hline  
$p\neq 0\;\;\mathrm{mod}\;\;n$     & \begin{tabular}{c} Gapless  or\\ gapped with degenerate\\ ground states (proof) \end{tabular}  &
\begin{tabular}{l}
$\circ$ LSMA theorem  \cite{LSM, AL, HaldaneSUN} \, \\
\end{tabular}
\\ \hline
$\gcd(n,p)=1$ & Gapless, $\SU(n)_1$ CFT  & \begin{tabular}{l}$\circ$ `t Hooft anomalies: \\ \cite{Tanizaki:2018xto, ohmori2019sigma,lechemrg,Yao:2018kel}\\ $\circ$ Fractional instanton gas~\cite{meron, WamerMerons} \end{tabular} \\ \hline
$1<\gcd(n,p)<n$
& \begin{tabular}{c}
Gap \\ Degeneracy $d:={n\over \gcd(p,n)}$
\end{tabular}
& \begin{tabular}{l}$\circ$ No candidate CFT~\cite{lechemrg,Yao:2018kel}\,\,\,\,\,\, \\$\circ$ Fractional instanton gas\,\,\,\,
\end{tabular}
\\ \hline
$p=0\;\;\mathrm{mod}\;\;n$ & Gap & \begin{tabular}{l}
$\circ$ Numerics ($p=n=3$)~\cite{GozelMilaNumeric}\\
$\circ$ Absence of anomalies\\
$\circ$ Perturbations around the \\ integrable WZNW point~\cite{FromholzPerturb} \\
$\circ$ Fractional instanton gas \\
$\circ$ AKLT states:\\
\cite{greiter2007,katsura2008,nonne2013,furusaki,roy2018,gozel2}
\end{tabular}
\\ \hline
\end{tabular}
\bigskip
\par
\caption{Generalized Haldane conjectures for $\SU(n)$ spin chains with symmetric rank-$p$ representations.}
\label{Haldanetable}
\end{table}

Using the notion of 't Hooft anomaly matching (which we explain below in Sections~\ref{hooftanomsec}-\ref{Znanomsec}), both~\cite{ohmori2019sigma} and~\cite{Tanizaki:2018xto} were able to formulate a field-theoretic version of the LSMA theorem for SU($n$) chains. In short, the presence of an 't Hooft anomaly signifies nontrivial low energy physics. It was shown that in the flag manifold models, an 't Hooft anomaly is present so long as $p$ is not a multiple of $n$. In these cases the gapped phase must have spontaneously broken translation or PSU($n$) symmetry; the latter is ruled out by the Mermin-Wagner-Coleman theorem at any finite temperature. In the gapped phase, the ground state degeneracy is predicted to be $\frac{n}{\gcd(n,p)}$, which is consistent with the LSMA theorem presented in Section~\ref{LSMAsec} above. It is interesting to note that when the classical ground state has a different structure, as in the ground state of the two-site-ordered self-conjugate SU(3) chains~\cite{Wamer2019}, no anomaly occurs. This is consistent with the fact that the proof of the LSMA theorem also fails for such representations.

The authors of \cite{ohmori2019sigma} then argued that while an anomaly is present whenever $p \mod n \not=0$, an RG flow to an IR stable WZNW fixed point is possible only when $p$ and $n$ have no nontrivial common divisor (in Section~\ref{WZNWflags} below we review a simple relation between WZNW and flag manifold models). In this case, the flow is to SU($n)_1$. Otherwise, the candidate IR fixed point is SU($n)_q$, where $q=\gcd(n,p)$, however we don't expect $\SU(n)_k$ low energy theories, with $k>1$, to emerge without fine-tuning. This is because they contain relevant operators allowed by symmetry which destabilize 
them~\cite{TsvelickWiegmann,AndreiFuruya,AffleckSUn1988}. Integrable spin models are known which do exhibit $\SU(n)_k$ low energy theories but they require fine-tuned Hamiltonians~\cite{AndreiJohannessonPLA1984, JohannessonPLA1986,JohannessonNuclPhysB1986}. The most well-known example of this is 
the SU(2) case where integrable models of spin $s$ have low energy $\SU(2)_{2s}$ critical points~\cite{su2int1,su2int2,AffleckHaldane1987}. However, it has been established that these critical points are 
unstable against infinitesimal tuning of the spin Hamiltonian and one would require fine-tuning in order for the flag manifold sigma model to flow there. This can already be seen from Fig.~\ref{bilinbiquadpic} in the example of the spin-1 integrable chain, whose continuum limit is described by $\tSU(2)_{k=2}$ WZNW model. Any deformation of this chain would lead us to one of the two massive phases, either dimer or Haldane. There is also a general argument that no flow from the unstable $\tSU(n)_q$ theory to $\tSU(n)_1$ is possible, since this would violate the anomaly matching conditions derived in~\cite{lechemrg,Yao:2018kel} for generic SU($n$) WZNW models. For $\tSU(2)$ there is another anomaly-based argument of~\cite{Furuya}, which asserts that a flow between an $\tSU(2)_k$ and $\tSU(2)_{k'}$ theories is only possible if $k=k'\mod 2$.

\begin{figure}[h]
\centering
\includegraphics{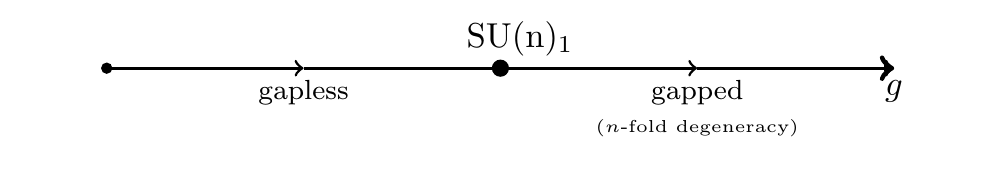}
\caption{A simplified phase diagram of the SU($n$) chains, as a function of coupling constant $g$ (a collective notion for the $\lfloor \frac{n}{2}\rfloor$ different coupling constants) when $p$ and $n$ are coprime.}
\label{fig:phasediag}
\end{figure}

Based on these anomaly arguments, we conclude that the rank-$p$ symmetric SU($n$) chains may flow to a SU($n)_1$ WZNW model if $p$ and $n$ do not have a common divisor. In this case, we expect gapless excitations to appear in the excitation spectrum. This prediction is a natural extension of the phase diagrams occurring in~\cite{AffleckHaldane1987} and~\cite{HaldaneSU3}. See Figure~\ref{fig:phasediag} for a simplified phase diagram of the SU($n$) chain in the case when $p$ and $n$ are coprime. 
Similar to the O(3) sigma model, we expect an RG flow from the flag manifold sigma model to the $\SU(n)_1$ WZNW model. This model has 
an SU(n)-invariant interaction term $\sum_aJ_L^aJ_R^a$ which is marginal. For one sign of this coupling, it is marginally irrelevant and flows 
to zero; for the other sign it flows to large values~\cite{lechemrg,HaldaneSUN}. As in the O(3) sigma model, we expect this coupling to have the irrelevant sign 
for sufficiently weak coupling in the flag-manifold sigma model.  If the coupling gets too large then this coupling constant changes sign and there is an RG flow to a gapped phase with broken translational symmetry, as occurs for SU(2) spin chains. 

We note that when $p$ and $n$ have a common divisor, at least one of the topological angles occurring in (\ref{eq:fm1}) is necessarily trivial. In the instanton gas picture of Haldane's conjecture (reviewed below in Section~\ref{meronsec}), each type of topological excitation must have a nontrivial topological angle in order to ensure total destructive interference in half odd integer spin chains~\cite{meron}.

\subsection{Derivation of the mixed 't Hooft Anomaly}\label{hooftanomsec}

One of the key tools in the analysis of the phase structure of the sigma model is the notion of `t Hooft anomaly matching. To introduce this concept, first one observes that the continuous global symmetry of the flag model~(\ref{eq:rel1})-(\ref{eq:fm1}) is (here $\mathbb{Z}_n\mysub \tSU(n)$ is the subgroup of the type $\omega^k\mathds{1}_n$, where $\omega$ is an $n$-th root of unity) \bea
\text{PSU}(n)=\SU(n)/\mathbb{Z}_n=\tU(n)/\tU(1)\,.
\eea
The reason is that the flag is described by $n$ vectors $U:=(u_1, \ldots, u_n)$, up to $\tU(1)^n$ phase transformations acting as $U\to U\cdot D\,,\;D=\mathrm{Diag}(e^{i\alpha_1},\cdots, e^{i \alpha_n})$. The global symmetry is given by the left action $U\to U_0 \cdot U$, and for $U_0\in \tU(n)$ the Lagrangian is invariant. On the other hand, the action of the center $\tU(1)$ can be compensated by a gauge transformation acting on the right. As a result, the faithfully acting symmetry is $\text{PSU}(n)$. Besides, the above models have a discrete $\mathbb{Z}_n$-symmetry~(\ref{Znsymm}) that acts by a cyclic permutation of the vectors $u_1, \ldots, u_n$. The claim~\cite{Tanizaki:2018xto, ohmori2019sigma} is that these two symmetries have a mixed anomaly, which we are about to describe. Overall our exposition in this section will be split into three main parts:
\begin{itemize}
    \item[$\circ$] Starting with a gauged linear representation for the flag manifold models, we derive the mixed $\tPSU(n)-\mathbb{Z}_n$ anomaly, following~\cite{Tanizaki:2018xto}.
    \item[$\circ$] We discuss how $\tPSU(n)$-bundles are related to the LSMA theorem and to fractional instantons
    \item[$\circ$] Following~\cite{ohmori2019sigma}, we explain how flag manifold models may be embedded in WZNW models. This serves to motivate the conjecture that flag manifold models flow in the IR to those conformal points for certain values of the $\theta$-angles
\end{itemize}

\subsubsection{Gauging the $\tPSU(n)$ global symmetry}

We start with the first point. On several occasions we already used the fact that the complete flag manifold $\tU(n)\over \tU(1)^n$ may be thought of as the space of orthonormal vectors $\bar{u}_A\circ u_B=\delta_{AB}$, each defined up to a phase: $u_A\sim e^{i\alpha_A}\,u_A$. It is of course standard in field theory to encode such equivalences by means of gauge fields, and in the present setup this leads to the so-called `gauged linear sigma model' formulation, which often simplifies the calculations (in the example of the $\CP^{n-1}$ model we already encountered it in section~\ref{largensec}). In the simplest case $g_t=\textrm{const.}, \lambda_t=\textrm{const.}$ the Lagrangian~(\ref{eq:rel1}) with the topological term~(\ref{eq:fm1}) may be rewritten as follows:
\bea\label{actionGLSManom}
S=\sum_{A=1}^{n}\int\,dx\,d\tau\, \left[-{1\over 2g}\left|(\diff+\im a_A) u_A\right|^2 +{\im \theta_A\over 2\pi}\diff a_A+{\lambda\over 2\pi} (\overline{u}_{A+1}\cdot \diff u_A)\wedge (u_{A+1}\cdot \diff \overline{u}_A)\right],
\eea
where $a_A$ are $\tU(1)$ gauge fields. 
As before, the first term is the usual kinetic term of the nonlinear sigma model, and the last term is the `non-topological' part of the  skew-symmetric field, as we explained in Sections~\ref{flagchainsun}-\ref{Znsymmsec}. It is linear both in space and time derivatives, but it is not topologically quantized to integers, and will not be important for the 't Hooft anomaly matching. The second term in the above action is the topological theta term of the two-dimensional $\tU(1)^n$ gauge theory. We may set $Q_A={1\over 2\pi}\int \diff a_A$, moreover the $Q_A$ so defined are equivalent to the topological charges that we encountered earlier, as one can see by eliminating the gauge fields through their e.o.m. Indeed, solving the e.o.m. of $a_A$, we find 
\bea\label{agaugefields}
a_A={\im\over 2}\left(\overline{u}_A \cdot \diff u_A -\diff \overline{u}_A\cdot u_A\right)=\im \overline{u}_A\cdot \diff u_A. 
\eea
We may interpret $da_A$ as the Fubini-Study form on the $A$-th copy of $\CP^{n-1}$ (see~(\ref{omegaFSunitary})). Moreover, as discussed in Section~\ref{cohsec} the flag manifold $\mathcal{F}_n$ is a Lagrangian submanifold of $(\CP^{n-1})^{\times n}$, which implies $\sum\limits_{A=1}^n \,da_A=0$. 
As a result, $
\sum\limits_{A=1}^n\,Q_A=0
$, which is of course the condition that we encountered many times before, cf.~(\ref{sumtopcharges}). As explained in Sections~\ref{flagchainsun}-\ref{Znsymmsec}, the above system~(\ref{actionGLSManom}) possesses a cyclic $\mathbb{Z}_n$ `quasi'-symmetry (i.e. a symmetry up to an integral of an element of~$H^2(\mathcal{F}, \mathbb{Z})$), if the $\theta$-angles are chosen as $\theta_A={2\pi p\,A\over n}$ for $p\in \mathbb{Z}$. 

Now we turn to the discussion of the mixed 't~Hooft anomaly between the $\tPSU(n)$ flavor symmetry and the $\mathbb{Z}_n$ permutation symmetry\footnote{The discussion in~\cite{Tanizaki:2018xto} contains also the case of a mixed $\tPSU(n)-\mathbb{Z}_{n'}$ anomaly, where $n'$ is a divisor of $n$, but we restrict here to the simpler case $n'=n$.}. The anomaly manifests itself in the fact that the partition function  of the system in a topologically non-trivial background  $\tPSU(n)$ gauge field is not invariant under the $\mathbb{Z}_n$ permutation, even at the point $\theta_\ell={2\pi p\,\ell\over n}$, where the system described by the action~(\ref{actionGLSManom}) is invariant. 

How do we introduce a background field for the $\tPSU(n)$ flavor symmetry? To answer this question, one should first clarify the difference between $\tSU(n)$ and $\tPSU(n)$ bundles. A $\tPSU(n)$ bundle $V_0$ over a worldsheet $\Sigma$   possesses an additional $\mathbb{Z}_n$-valued invariant, which is a member of the second cohomology $H^2(\Sigma,\pi_1(\tPSU(n)))$ -- the generalized Stiefel-Whitney class $w_2(V_0)$ (see, e.g.,~\cite{WittenSUSYindex, WittenLanglands}). Some examples of such bundles will be provided in the next section~\ref{PSUnex}, and for the moment we turn to a more formal definition of the invariant. First, the $\tPSU(n)$ bundle $V_0$ may be lifted, in a non-unique way, to a vector bundle $V$ over $\Sigma$, with structure group $\tU(n)$. The latter is characterized by its first Chern class $c_1(V)$ (its integral is the `abelian flux'). The non-uniqueness in choosing $V$  has to do with the fact that we could replace it with $V\otimes \mathcal{L}$, where $\mathcal{L}$ is any line bundle, since it would cancel out in the projective quotient anyway. Recalling that $c_1(V\otimes \mathcal{L})=c_1(V)+n\,c_1(\mathcal{L})$, we see that  $c_1(V)\;\mathrm{mod}\;n$ is a well-defined topological quantity. This $\mathrm{mod}\,n$-reduced class is called the generalized Stiefel-Whitney class $w_2(V_0)\in H^2(\Sigma, \mathbb{Z}_n)=\mathbb{Z}_n$, which characterizes the topologically non-trivial $\tPSU(n)$-bundles.

We will now convert this description into a relevant gauge theory formulation at the level of the Lagrangian. To mimic the description in terms of a vector bundle $V$ with structure group $\tU(n)$, we will introduce a $\tU(n)$ gauge field $\widetilde{A}$.  In order to implement the quotient, one has to postulate the additional gauge transformations
\bea\label{xishift}
a_A \mapsto a_A-\xi,\; \widetilde A\mapsto \widetilde A +\xi \mathds{1}_n\;,
\eea
where $\xi$ is a $\tU(1)$ gauge field, which simultaneously plays a role of gauge parameter here. Now, the point is that a $\tU(n)$ gauge field has an integer invariant -- the first Chern number -- that is expressed as follows: $\int\limits_{\Sigma} \,c_1(V)={1\over 2\pi}\,\int\,d\left(\mathrm{Tr}(\widetilde{A})\right)\in \mathbb{Z}$. Accordingly, since $\xi$ is a $\tU(1)$ gauge field, its curvature also has quantized periods, $\int\limits_{\Sigma} \,c_1(\mathcal{L})={1\over 2\pi}\,\int\,d\xi\in \mathbb{Z}$. Due to the shift symmetry~(\ref{xishift}) that we have imposed, the periods of $\widetilde{A}$ are shifted by multiples of $n$: $\int\limits_{\Sigma} \,c_1(V)\mapsto \int\limits_{\Sigma} \,c_1(V)+n\int\limits_{\Sigma} \,c_1(\mathcal{L})$, and as a result we get a $\mathbb{Z}_n\simeq \mathbb{Z}\big/ n\mathbb{Z}$ invariant in place of an integer invariant that one would have without the additional symmetry~(\ref{xishift}). As explained in~\cite{Kapustin:2014gua, Gaiotto:2014kfa, Aharony:2013hda}, this invariant may be encoded in a 2-form $\mathbb Z_n$  gauge field that we call $B$. In our notation it is simply $B:={1\over n}\,d\left(\mathrm{Tr}(\widetilde{A})\right)$. Under the shift~(\ref{xishift}) it changes as follows:
\bea
B\mapsto B+d\xi\,.
\eea
According to the above discussion, the integral
\bea\label{Bquant}
{1\over 2\pi}\int\limits_{\Sigma}\,B\in \mathbb{Z}_n
\eea
is a multiple of $1\over n$. To write a gauged version of the action~(\ref{actionGLSManom}), first of all we replace the covariant derivatives  $(\diff +\im a_A)u_A$  by the elongated derivatives  
$
(\diff+\im a_A + \im \widetilde{A}) u_A\;
$ 
-- a combination invariant under the shift~(\ref{xishift}). Besides, the curvatures $da_A$ are not invariant under the shift~(\ref{xishift}) but the combinations $da_A +B$ are. As a result, the effect of subjecting the system~(\ref{actionGLSManom}) to an external $\tPSU(n)$ gauge field is in the following modification of the action:
\bear\label{eq:PSU3_gauged}
S_{\mathrm{gauged}}=\sum_{A=1}^{3}\int_{M_2} &&\hspace{-0.5cm}\left[-{1\over 2g}\left|(\diff+\im a_A+\im \widetilde{A}) u_A\right|^2+{\im \theta_A\over 2\pi}(\diff a_A+B)\;+\right.\\ \nonumber &&\left.\hspace{-0.5cm}+\;{\lambda\over 2\pi} \{\overline{u}_{A+1}\cdot (\diff+\im\widetilde{A}) u_A\}\wedge \{u_{A+1}\cdot(\diff+\im \widetilde{A}) \overline{u}_A\}\right]. 
\eear
Notice that there is no need to write the $a_A$ gauge fields in the  $\lambda$-term, since their contributions vanish due to the orthogonality constraint between $u_A$. Performing the path integral,
\bea\label{partfuncpath}
Z[(A,B)]=\int \Diff a\Diff \overline{u}\Diff u \exp(S_{\mathrm{gauged}}), 
\eea
we obtain the partition function $Z[(A,B)]$ in the background $\tPSU(n)$  gauge field. $A$ is meant to represent the traceless part of the gauge field $\widetilde{A}$. One curious thing to notice about the partition function $Z$ is that it is no longer $2\pi$-periodic in the $\theta$-angles. Indeed, a shift of one of the angles, say $\theta_1\to\theta_1+2\pi$, produces a phase $Z[(A,B)]\mapsto Z[(A,B)]\cdot e^{i\int\,B}$.

\subsection{The $\mathbb{Z}_n$ anomaly in a $\tPSU(n)$ background}\label{Znanomsec}

It is instructive to notice that, in the presence of the $\tPSU(n)$ background gauge field, variation of the action w.r.t. $a_{\ell}$ no longer produces~(\ref{agaugefields}), and the following modified formula for the curvatures holds instead:
\bea\label{curvsumB}
\sum\limits_{A=1}^n\,(da_A+B)=0\,. 
\eea

If we consider the modified topological charges $\widetilde{Q}_A:={1\over 2\pi}\int\limits_{\Sigma}\,\left(da_A+B\right)$, which sum to zero, then according to~(\ref{Bquant}) they will be quantized in units of $1\over n$. We will encounter this phenomenon in the discussion of fractional instantons below and relate it to twisted boundary conditions, which in the present language are encoded in a non-trivial $\tPSU(n)$ gauge field.  Fractional instantons arising in the presence of twisted boundary conditions have also been discussed in the context of the resurgence program, cf.~\cite{DunneUnsal1, DunneUnsal2}.

Let us now show that the action~(\ref{eq:PSU3_gauged}) of the model in an external $\tPSU(n)$ gauge field is not invariant under the $\mathbb{Z}_n$ shift symmetry $u_A\to u_{A+1}$ (which should be obviously supplemented by $a_A\to a_{A+1}$). First of all, the metric and $\lambda$ terms are evidently invariant, so it suffices to compute the topological term. We recall that we have chosen our $\theta$-angles as $\theta_A=\frac{2\pi p A}{n}$, so that the topological term changes as follows under a $\mathbb{Z}_n$-shift (we set $a_{n+1}\equiv a_1$):
\bear\label{toptermshiftB}
S_{\mathrm{top}}=\im\,p\,\sum\limits_{A=1}^n{A\over n} \int\left(\diff a_A+B\right)\mapsto  \im\,p\,\sum\limits_{A=1}^n{A\over n} \int\left(\diff a_{A+1}+B\right)=\\ \nonumber =S_{\mathrm{top}}-{\im\,p\over n}\, \int\, \underbracket[0.6pt][0.6ex]{\sum\limits_{A=1}^n\left(\diff a_A+B\right)}_{=0\;\textrm{by}\; (\ref{curvsumB})}+\im\,p\, \int (\diff a_1+B). 
\eear
As $\int \diff a_1\in 2\pi\mathbb{Z}$, this term drops off in the path-integral. However, since $\int B\in {2\pi\over n}\mathbb{Z}$, the $B$-term  contributes a phase, so we have
\bea
Z[(A,B)]\mapsto Z[(A,B)]\exp\left(\im\,p\,\int B\right)
\label{partfuncshiftB}
\eea
under the $\mathbb{Z}_n$ permutation. 
This is the mixed 't~Hooft anomaly between $\tPSU(n)$ and $\mathbb{Z}_n$. There is no local counter term that can eliminate the generation of the $B$-term under the $\mathbb Z_n$ exchange symmetry. Indeed the only counter-terms allowed are  $\im q \int B$ where $q\in \mathbb Z\bmod n$, and these are invariant under the $\mathbb Z_n$ symmetry. 

In the cases of continuous global symmetries it was argued by 't Hooft long ago~\cite{HooftAnom} that the anomalies should match between the UV and IR limits of the theory: even if the effective theory in the IR looks drastically different from the original UV theory, both theories should exhibit the same anomalies. This is typically used to derive constraints on the IR dynamics, which might be otherwise difficult to deduce directly from the UV theory. Originally this idea was developed for the study of chiral symmetries in QCD, however it is believed that the same property holds for discrete symmetries or mixed continuous-discrete symmetries as in our example here. In~\cite{Tanizaki} the UV/IR matching of anomalies was related to the so-called `adiabatic continuity' of the theory, i.e. to the smooth dependence of the physical properties of a theory compactified on a circle on the radius of the circle (cf.~\cite{SulejmanpasicVolume, Asorey, DunneUnsal1} for examples of when this does or does not hold).

By the anomaly matching argument, the ground state at the $\mathbb{Z}_n$ invariant point, $\theta_A={2\pi pA\over n}$, with $p$ not a multiple of~$n$, cannot be trivially gapped. According to~\cite{PhysRevB.83.035107}, in $1+1$ dimension intrinsic topological order is ruled out, so the system must either have spontaneous symmetry breaking or  conformal behavior in the low-energy limit. The same statement is obtained by the LSMA theorem for the lattice $\SU(n)$ chain (see section~\ref{LSMAsec}), and the argument reviewed here provides its field-theoretic counterpart.

\subsection{Examples of $\tPSU(n)$-bundles}\label{PSUnex}

An attentive reader might have noticed that, in the above discussion, at least two of the steps were reminiscent of what we already encountered in other contexts. First of all, the calculation of the shift in the topological terms~(\ref{toptermshiftB}) is very similar to the shift calculated in~(\ref{topformshift}), albeit with an important distinction that in the latter case the shift was by a $2\pi$-quantized term, which is immaterial in the path integral. Secondly, at the level of the partition function~(\ref{partfuncshiftB}) the $\mathbb{Z}_n$-transformation is a change of variables in the path integral~(\ref{partfuncpath}), so whenever $\exp{\left(\im p \int B\right)}\neq 1$ this really means that the partition function vanishes. This argument is very similar to the insertion of the  translation operator $T$ in~(\ref{TshiftLSMA}) in the proof of the LSMA theorem, which leads to the vanishing of a certain matrix element. Both of these similarities are not coincidences.

\subsubsection{Flat bundles as twisted boundary conditions}

We start with a somewhat more intuitive explanation of the background $\tPSU(n)$ gauge field. In fact, in several cases the results of the previous section may well be formulated in terms of \emph{flat} background gauge fields, $dA-A\wedge A=0$. We will explain this on two examples: those of a torus $\mathbb{T}^2$ and of a sphere $S^2$.

The torus example is somewhat easier, as there is a general statement (reviewed in~\cite{WittenSUSYindex}) that, for a \emph{simple} group $G$, every $G$-bundle over $\mathbb{T}^2$ admits a flat connection\footnote{The fact that $G$ is simple is crucial here. For example, for a line bundle $\mathcal{L}$ with gauge group $\tU(1)$ one would have an additional invariant -- the first Chern number $\int\limits_{\Sigma} c_1(\mathcal{L})={1\over 2\pi}\,\int\limits_{\Sigma} F$, expressed through the curvature $F$ of the connection.}. This is also true for the topologically non-trivial $\tPSU(n)$ bundles of the previous section. To calculate the corresponding invariant $w_2$, one views the gauge field as an $\SU(n)$ gauge field and computes its holonomies $a$ and $b$ along two meridians of the torus. Since $\pi_1(\mathbb{T}^2)=\mathbb{Z}^2$, in $\tPSU(n)$ the holonomies would satisfy $ab=ba$, which in $\SU(n)$ is relaxed to $aba^{-1}b^{-1}=\omega\in w_2$. Note that these holonomies $a$ and $b$ may well be non-trivial even for a flat gauge field $A$, which is the reason that it suffices to consider flat connections.

Now, the ultimate use of flat connections is that they may be completely eliminated at the expense of imposing twisted boundary conditions on the fields. Indeed, a flat connection has the form $A=-g^{-1}dg$, where $g$ is locally a function on the worldsheet, which, when viewed globally, encodes the holonomies $a$ and $b$. If $x$ and $y$ are the coordinates along the meridians of the torus, one has $g(x+2\pi, y)=a\circ g(x,y)$ and $g(x, y+2\pi)=b\circ g(x,y)$. Accordingly, if $u_A$ are the matter fields of the model charged under the $\tPSU(n)$ gauge group (like the unit vector fields of the flag models), we can now perform a gauge transformation $u_A \to g\circ u_{A}$, which completely eliminates the gauge field at the expense of imposing twisted boundary conditions $u_A(x+2\pi, y)\sim a\circ u_A(x,y)$, $u_A(x, y+2\pi)\sim b\circ u_A(x,y)$ ($\sim$ means `up to a phase', since the $u_A$ take values in a projective space).

\begin{figure}[h]
\centering
\begin{tikzonimage}[width=.45\textwidth]{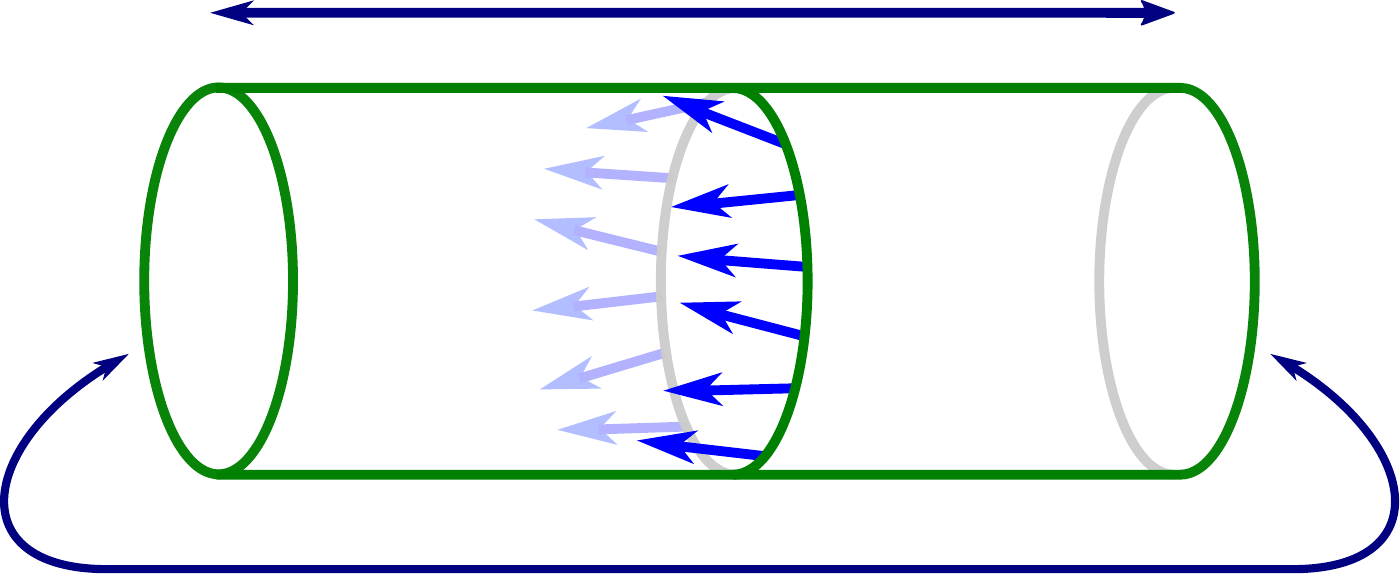}
\node at (0.5, 1.1) {$\beta$};
\node at (0.5, -0.2) {Gluing with twist\; $U$};
\end{tikzonimage}`
\caption{A schematic depiction showing that the twisted partition function $\mathrm{Tr}(U e^{-\beta H})$ of the spin chain leads, in the continuum limit, to a $\tPSU(n)$ bundle over a torus. This bundle is described by the twist operator $U$ entering the LSMA theorem.} 
\label{figcyl}
\end{figure}

In the language used before, eliminating the gauge field amounts to setting $A=B=0$ and imposing the twisted boundary conditions. This allows us to make a connection between the shifts in the topological terms: (\ref{toptermshiftB}) in the presence of the $A, B$ gauge fields and (\ref{topformshift}) without them. The point is that, with the twisted boundary conditions, the shift $\int \,\Omega_1$ in~(\ref{topformshift}) is no longer quantized as an integer times $2\pi$, but rather as an integer times $2\pi\over n$, thus reproducing the shift by the $B$-field in~(\ref{toptermshiftB}). One way to see this is to observe that, with the twisted boundary conditions,  the  fluxes of the gauge fields $a_A$ are quantized in multiples of $2\pi\over n$, and since $\Omega_1=da_1$, this leads to the corresponding statement for $\int \,\Omega_1$. Indeed, the twists have the form $u_A(x+2\pi, y)\sim a\circ u_A(x,y)$, where $\sim$ means that actually $a\in \tPSU(n)$ and is only defined up to a power of the root of unity $\omega$. In the formulation with $a_A$ gauge fields, to undo this ambiguity one performs a  gauge transformation 
$
u_A \rightarrow e^{i\varphi_A(x,  y)}\cdot u_A
$, where the gauge parameter $\varphi_A(x, y)$ has periodicity $\varphi_A(x+2\pi, y)-\varphi_A(x, y)={2\pi\over n}$.  
Since the gauge transformation affects the gauge fields $a_A\rightarrow a_A-\diff \varphi_A$, we conclude that the fluxes of $a_A$ will be quantized as multiples of $2\pi/n$. As we recall, $Q_A={1\over 2\pi}\int\,da_A$ are the topological charges, so we come to the conclusion that these topological charges are quantized in multiples of~$1\over n$. This means that the shift~(\ref{topformshift}) that was immaterial for periodic boundary conditions (or, in general, for maps from a closed Riemann surface $\Sigma$), now produces a non-vanishing contribution, equal to the one of the $B$-field in~(\ref{toptermshiftB}).

In fact, we have already encountered an example of a non-trivial $\tPSU(n)$-bundle over a torus (albeit in a discretized form) in the proof of the LSMA theorem in Section~\ref{LSMAsec}. There we defined the so-called twist operator acting on a spin chain of length $L$:
\bea
	U = e^A \quad\quad A:= \frac{2\pi i}{n L}\sum_{j=1}^L jQ_j,\quad\quad Q = \sum_{\alpha=1}^{n-1} S_\alpha^\alpha - (n-1)S^n_n\in \mathfrak{su}(n)\,.
\eea
Clearly, in the continuum limit we get
\bea\label{gxtwist}
U=\prod\limits_{x\in [0,2\pi)}\,g(x),\quad\quad g(x)=e^{i{x\over n}Q_x}\,.
\eea
In the framework of flag manifold sigma models that arise in the continuum limit, the setup of the LSMA theorem is as follows: we wish to compute the partition function $\mathrm{Tr}(U e^{-\beta H})$ with the insertion of the twist operator. From the worldsheet point of view this means that we take the theory on a cylinder and glue the fields at the two of its ends by the twist function~$U$, as shown in Fig.~\ref{figcyl}. Topologically this produces a $\tPSU(n)$-bundle over a torus, whose $w_2$-invariant is characterized by the periodicity property of $g(x)$: $g(x+2\pi)=\upxi\cdot g(x)$, where $\upxi\in \mathbb{Z}_n$. Looking back at~(\ref{gxtwist}), we find $\upxi=e^{\frac{2\pi i}{n} Q}$. If one deals with rank-$p$ symmetric representations, as in the LSMA theorem of section~\ref{LSMAsec}, the invariant is $\upxi=e^{\frac{2\pi i p}{n}}$. This is the `anomaly factor' that appears in~(\ref{la:4}) upon inserting the translation operator in~(\ref{TshiftLSMA}), which should be seen as parallel to making a cyclic permutation $u_A\to u_{A+1}$ in the path integral~(\ref{partfuncpath}) and obtaining a factor~(\ref{partfuncshiftB}) as a result.

\subsubsection{Fractional instantons as sections of $\tPSU(n)$ bundles}

As a next step, we consider the worldsheet $S^2$. We wish to explain that examples of sections of topologically non-trivial $\tPSU(n)$ bundles over a sphere $S^2$ are provided by the so-called `fractional instantons'  that we will introduce shortly. For a recent discussion of these fractional isntantons in the context of resurgence, see the recent paper~\cite{unsal2020}.

The fibers of the relevant bundles are the target spaces of the sigma model, i.e. the flag manifolds. We recall that a section of a topologically trivial bundle is simply a map $S^2\to \mathcal{F}$ from the worldsheet to the flag manifold target space. One can alternatively think of it as a map $\mathbb{R}^2\to \mathcal{F}$ with `decay conditions at infinity', meaning that the infinity of $\mathbb{R}^2$ is mapped to a single point in $\mathcal{F}$. In general, fiber bundles over $S^2$ with structure group $\tPSU(n)$ may be defined using a patching function $S^1\to \tPSU(n)$, where $S^1=U_+\cap U_-$ is the equator -- the intersection of the two patches $U_\pm$ on $S^2$ (the northern and southern hemispheres). Topologically the bundles are determined by the class of the patching map $S^1\to \tPSU(n)$ in the homotopy group $\pi_1(\tPSU(n))\simeq \mathbb{Z}_n$. Now, suppose we want to construct a section of such a bundle $\pi: E\to S^2$, with fiber $\mathcal{F}$ and structure group $\tPSU(n)$. Over either $U_+$ or $U_-$ one can trivialize the bundle, i.e. one identifies $\pi^{-1}(U_\pm)\simeq U_\pm\times \mathcal{F}$. Constructing a section of $E$ is the same as specifying two maps $f_\pm: U_\pm\to \mathcal{F}$ patched across the equator, i.e. $f_+\big|_{S^1}=g\circ f_-\big|_{S^1}$, where $g: S^1\to \tPSU(n)$ is the patching function.

Let us now explain how this construction may be used for the description of fractional instantons. The latter, by definition, are maps $\mathbb{R}^2\to \mathcal{F}$ with the following behavior at infinity:
\bea \label{meronsinfin}
	u_A^\alpha = \frac{1}{\sqrt{n}}\omega^{A \alpha} e^{\im w_\alpha \varphi}\,, \hspace{5mm} \omega=e^{2\pi \im\over n}\,.
\eea
Here $\varphi$ is the angle around a `circle at infinity', and $w_\alpha\in \mathbb{Z}$ are the winding numbers. Unless all $w_\alpha$ have the same value, such maps do not satisfy decay conditions at infinity (if $w_\alpha=\textrm{const.}$, the winding is undone by a gauge transformation). We will view $\mathbb{R}^2$, together with the circle at infinity, as the upper hemisphere $U_+$, and the fractional instanton will serve to define the map $f_+: U_+\to \mathcal{F}$. Now, on $U_-$ we will define a constant map, given by $\tilde{u}_A^\alpha = \frac{1}{\sqrt{n}}\omega^{A \alpha} $. Clearly, along the equator the two are related by the patching map $g=e^{-{\im\over n}\,(\sum_{\alpha=1}^n w_\alpha)\,\varphi}\,\mathrm{Diag}(e^{\im w_1 \varphi}, \cdots, e^{\im w_n \varphi})\in \tPSU(n)$. The topology of the bundle is characterized by $[\sum_{\alpha=1}^n w_\alpha\;\;\mathrm{mod}\;\;n]\in \mathbb{Z}_n=\pi_1(\tPSU(n))$. If $w_\alpha=\textrm{const.}$, the corresponding invariant vanishes, in line with the discussion above.

\subsection{Relation to the WZNW model}\label{WZNWflags}

In Section~\ref{Znanomsec} we described the mixed $\tPSU(n)-\mathbb{Z}_n$ anomalies that arise for flag manifold sigma models. In the case when such anomalies are present, one possibility is that the renormalization group flow leads to a conformal field theory in the IR. Moreover, as it is recorded in Table~\ref{Haldanetable} and discussed in section~\ref{HaldSUn}, one conjectures that the resulting conformal field theory is the $\SU(n)_{1}$ WZNW model. To motivate this relation, we recall what is perhaps the most vivid way to see a connection between the flag sigma model and the WZNW model. The idea  is to embed the former into the latter~\cite{Tanizaki:2018xto, ohmori2019sigma}. One starts with the WZNW Lagrangian, defined as follows (here $G\in \SU(n)$):
\begin{align}\label{UVWZNW}
S_{\mathrm{WZNW}}:= \frac{R^2}{2} \int_{M_2}\text{Tr}(\dd_\mu G\dd^\mu G^\dagger)
+{\im \over 12\pi} k  \int_{M_3} \text{Tr}((G^\dagger d G)^3) \,,
\end{align}
where $M_2=\Sigma$ is the two dimensional spacetime and $M_3$ is a three-manifold whose boundary is $M_2$, i.e.\ $\partial M_3 = M_2$.   The coefficient $k$ is quantized to be a positive integer.  In the UV the radius $R$ of the target space is large and the theory comprises $n^2-1$ `asymptotically free' bosons.  The renormalization group flow then interpolates between this free point and a conformal fixed point at $R^2={k\over 4\pi}$ in the IR~\cite{Witten:1983ar}.

The main statement is that the action~(\ref{UVWZNW}), when restricted to unitary matrices $G$ \emph{with a fixed spectrum}, produces the action of a flag manifold sigma model, whose $\theta$-angles are dictated by the spectrum of $G$. In other words, we will be considering matrices $G$ of the form
\begin{align}\label{calU}
\begin{split}
&G  = U\, \Omega_0 \,U^\dagger\,,\quad\quad \textrm{where}\quad\quad
\Omega_0 = \text{diag}(e^{\im \varphi_1},e^{\im \varphi_2},\cdots,e^{\im \varphi_n})\,,\quad\quad U\in \SU(n)\,.
\end{split}
\end{align}

For simplicity we assume $\varphi_A\neq \varphi_B$ for $A\neq B$. In this case the matrix $U$ is defined up to right multiplication by a diagonal matrix $D$, $U\sim U\cdot D$, so that $U$ defines a point in a flag manifold. Using \eqref{calU}, the kinetic term of the WZNW action \eqref{UVWZNW} can be easily computed as:
\begin{align}\label{kinetic}
\frac{R^2}{2}\text{Tr}(\partial_\mu G\partial_\mu G^\dagger)
=
R^2\sum_{A}\partial_\mu u_A\circ  \partial_\mu \bar{u}_A
-R^2\sum_{A, B}
e^{\im(\varphi_A-\varphi_B)}
|u_A\circ \partial_\mu\bar{u}_B|^2\,.
\end{align}

To compute the WZ term one takes $M_3= M_2 \times {\cal I}$ where $ {\cal I}= [0,1]$ is an interval with coordinate $y$.  To make sure that $M_3$ has a single boundary $M_2$, one effectively compactifies the second boundary $M_2\times \{y=1\}$ by requiring $G\big|_{y=1}=\mathds{1}$. Since the WZ action (after exponentiation) doesn't depend on the extension of the fields to the bulk of $M_3$, one may choose 
\bear\label{extension}
&&G(z,\bar z ,y) =  U(z,\bar z) \Omega(y) U(z,\bar z)^\dagger\,,\\ \nonumber
&&\Omega(y) =  \text{Diag}(e^{\im\varphi_1(y) },e^{\im\varphi_2(y)}, \cdots, e^{\im\varphi_{n}(y)}),\quad\quad \textrm{where}\quad\quad 
\varphi_A(0) = \varphi_A\,,\,\varphi_A(1) =0\,.
\eear
We may now substitute \eqref{extension} into the second term of \eqref{UVWZNW},   
and one finds that it splits into a sum of two: $\int\limits_{M_3} \text{Tr}((G^\dagger d G)^3 )=\int\limits_{M_2}\,\left(\Omega_{\mathrm{top}}+\Omega_{\lambda}\right)$. 

The first one produces the $\theta$-terms of the flag model:
\begin{align}\label{WZ2}
\int\limits_{M_2}\,\Omega_{\mathrm{top}}
= {k\over 2\pi} \int_{M_2} \sum_{A}   \varphi_A \,d\bar{u}_A\wedge \!\circ\, du_A\,,
\end{align}
whereas the second one is the non-topological part of the $B$-field (the `$\lambda$-term', as we referred to it earlier):
\begin{align}\label{WZB}
\int\limits_{M_2}\,\Omega_{\lambda}=- {k \over 4\pi}\sum_{A\neq B}\sin(\varphi_A-\varphi_B)  \,
\int_{M_2} (  u_B\circ d\bar{u}_A )\wedge (\bar{u}_B\circ du_A) \,.
\end{align}
As was shown in section~\ref{Znsymmsec}, the $\mathbb{Z}_n$-invariant values correspond to the choice of angles (with an overall subtraction so that $\mathrm{det}\,\Omega_0=1$)
\bea\label{phivalues}
\varphi_A=\frac{2\pi\,A}{n}-{2\pi (n+1)\over 2 n}\,.
\eea

One also needs to have a mechanism to restrict the spectrum of the matrix $G$ in~(\ref{UVWZNW}) to be of the form $e^{\im\, \varphi_A}$, with the values~(\ref{phivalues}). The paper~\cite{ohmori2019sigma} proposes the following scenario: one adds to~(\ref{UVWZNW}) a potential $
V=\sum _{j=1}^{\lfloor {n\over 2}\rfloor}  \, g_j \, \text{Tr} (U^j)\text{Tr}((U^\dagger)^j)\,.
$ In the limit when all $g_n\to \infty$ one restricts to the locus $\text{Tr} (U^j)=0, j=1 \ldots \lfloor {n\over 2}\rfloor$, which can be shown to imply a spectrum of the form~(\ref{phivalues}).

\section{A gas of fractional instantons}\label{meronsec}

In the previous section the generalized Haldane conjectures for an $\SU(n)$ spin chain with symmetric rank-$p$ representations at each site were formulated. Following~\cite{WamerMerons}, we will now recall an intuitive explanation for these conjectures based on fractional topological excitations. This is a generalization of an older work in SU(2)~\cite{meron}, which explains the generation of the Haldane gap in terms of merons in the $S^2$ nonlinear sigma model. 

In the case of the $S^2$ target space the idea was to arrive at the O(2) model in a special limit, when a large potential is added to restrict the field $\vec{n}$ to the XY plane. In the absence of the $\theta$-angles, it is well known that a mass gap is generated in the O(2) model, due to vortex proliferation~\cite{Kosterlitz_1973}. In the case of  the $S^2$ model with a large potential this mass gap is still generated when the potential is weakened, and $\vec{n}$ can lift off the plane. The resulting nonplanar vortices are known as merons~\cite{meron}.\begin{footnote}{`Meron' means half-instanton, and refers to the fact that these configurations have half-integer quantized topological charges. The word `instanton' is used, in place of vortex or soliton, because one of the two dimensions in the O(2) model corresponds to Euclidean time.}\end{footnote} This argument was used to identify merons as the mass-gap generating mechanism in the $S^2$ sigma model which corresponds to the purely isotropic case of $m=0$. We will now review a generalization of this argument that suggests that a mass gap is present in the SU($n)/[\mbox{U}(1)]^{n-1}$ flag manifold sigma model (without topological terms), and that it is generated by topological excitations.

\subsection{Squashing to the $XY$-model}

Following \cite{meron}, the strategy is to break the symmetry of the flag manifold down to U(1), where a phase transition is well understood in terms of vortex proliferation. One starts by adding to the Lagrangian an anisotropic potential $V_1$ that breaks the SU($n$) symmetry down to $[\tU(1)]^{n-1}$:
\bea
V_1 = m\sum_{A=1}^n \sum_{\alpha<\beta} \big( |(u_A^\alpha|^2 - |(u_B^\beta|^2\big)^2\,.
\eea

It is the SU($n$) generalization of adding the term $\sum_j {S}_z(j){S}_z(j)$ to the SU(2) Hamiltonian. In the limit $m\to \infty$, the potential $V_1$ restricts all of the components $u_A^\alpha$ of $u_A$ to be of equal absolute value (equal to $1\over \sqrt{n}$ due to normalization), with arbitrary phases. Taking into account the gauge group $\tU(1)^{n-1}$, this gives $n(n-1)$ real parameters. The number of (real) orthogonality constraints is formally also $n(n-1)$, however one should take into account the remaining $\tU(1)^{n-1}$ \emph{global} symmetry acting as $u_A^\alpha\to e^{i\theta_{\alpha}}\,u_A^\alpha$. As a result, the configuration minimizing the potential $V_1$ is (up to permutations of the vectors $u_1, \cdots, u_n$) 
\bea \label{eq:typical}
	u_A^\alpha = \frac{1}{\sqrt{n}}\omega^{A \alpha} e^{\im\sigma_\alpha}\,, \hspace{5mm} \omega=e^{2\pi \im\over n}\,, \hspace{5mm} \sigma_\alpha \in [0,2\pi].
\eea
Orthonormality of these states follows from the identity $
	\sum_{\rho=1}^n \omega^{\rho j} =0$ for $j\not= 0 \mod n$. The $\mathbb{Z}_n$ symmetry $u_A\to u_{A+1}$ is represented on~(\ref{eq:typical}) by a shift
	\bea\label{sigmaZn}
	\sigma_\alpha\to \sigma_\alpha+\frac{2\pi \alpha}{n}\,.
	\eea

The formula~(\ref{eq:typical}) defines an embedding $(S^1)^{n-1}\mysub \mathcal{F}$, and in the present setup this is the torus of asymptotic vortex configurations away from the core. Substituting~(\ref{eq:typical}) into the Lagrangian~(\ref{eq:rel1}), one obtains a generalized XY-model with $n-1$ $S^1$-valued fields, coupled to each other. Although this model could perhaps be studied in full generality, we would like to make use of the known results for the standard XY-model, and to this end we will add an additional potential $V_2=m\sum_{A=1}^n\sum_{\alpha=2}^{n-1} \Big(\mathrm{Im}[(u_A^1\bar{u}_A^\alpha)^n] \Big)^2$ that will suppress all but one fields. The potentials $V_1$ and $V_2$ have common minima (zero locus), as one can see by rewriting $V_2$ in terms of (\ref{eq:typical}): $
	V_2= 4m\sum_{A=1}^n\sum_{\alpha=2}^{n-1}\sin^2(n(\sigma_1-\sigma_\alpha)).
$
Due to the factor of $n$ the set of minima of $V_2$ is invariant under the $\mathbb{Z}_n$ symmetry~(\ref{sigmaZn}), which corresponds to translational invariance in the underlying lattice model.
It is clear that the effect of $V_2$ is to equate all but one of the U(1) fields (up to the shift~(\ref{sigmaZn})). Due to residual gauge symmetry, everything then depends only on one variable $\sigma:=\sigma_n-\sigma_1$. This is equivalent to the O(2) model of a vector $\vec{n}\in\mathbb{R}^3$ restricted to the XY plane. By inserting this restricted form of $u_A$ into (\ref{eq:rel1}), it is easy to show that the $\lambda_{A,B}$ terms vanish, and the resulting O(2) coupling constant is
\bea\label{eq:gg}
g^{-1} = \sum_{A,B} g_{A,B}^{-1}\,.
\eea

\subsection{Fractional Instantons}

For $n=2$ the potential $V_2$ vanishes and the perturbation $V_1$ is equivalent to adding a mass term $m(n_3)^2$ to the $S^2$ model Lagrangian, restricting $\vec{n}$ to lie in the XY plane in the large $m$ limit. This follows from the equivalence $n^i = u_1^\dag \sigma^i u_1$ (we already used it earlier in~(\ref{evop}), where $n_3=\frac{\bar{z}\circ \sigma^3 \circ z}{\bar{z}\circ z}$ was written in stereographic coordinates).

The vortices of the model, in order to be well-defined at the core, must become non-planar. They are called merons and have topological charge $Q= \pm \frac{1}{2}$ (the sign depending on whether $n_3 = \pm 1$ at their core), as compared to $Q=\pm 1$ for the more familiar instantons and antiinstantons of the $S^2$ model. More exactly, to use the notation developed for flag manifold models, we will be thinking of the topological charge as a pair of charges $\vec{Q}=(Q_1, Q_2)$, such that $Q_1+Q_2=0$ (cf.~(\ref{eq:Stop}) above), and one may set $Q=Q_1$. As explained earlier (see section~\ref{cohsec}, for example), this corresponds to the embedding $S^2\mysub (S^2)_1\times (S^2)_2$ mapping $\vec{n}\to (\vec{n}, -\vec{n})$, in which case $Q_i={1\over 2\pi}\int\limits_{\Sigma}\,f^\ast \Omega_i$, where $\Omega_1, \Omega_2$ are the two Fubini-Study forms, subject to $(\Omega_1+\Omega_2)\big|_{S^2}=0$, and $f: \Sigma\to S^2\mysub (S^2)_1\times (S^2)_2$ is a map from a worldsheet~$\Sigma$.
In this simplest case $Q_1$ is the area on $S^2$ of the image of $f$, multiplied by the number of times a typical point is covered. If $\Sigma$ is a closed Riemann surface, such as $S^2$, $Q_1$ and $Q_2$ are integers. In particular, a map $f: S^2 \to S^2$ may be thought of as a map $\mathbb{R}^2 \to S^2$ with a fixed asymptotic value at infinity: $\vec{n}(\infty)=\vec{n}_0$. In the case of a meron, on the contrary, the asymptotic behavior at infinity is such that $f|_{\infty}: S^1\to S^1\subset S^2$, where $S^1\subset S^2$ is the equator $n_3=0$. As a result, $f$ maps $D\to S^2$, where $\Sigma=D$ is a disc, with the condition that the boundary of the disc is glued to the equator in a prescribed way. The simplest situation is when the disc covers once the upper or lower hemisphere, in which case the corresponding area $Q_1=\pm{1\over 2}$ (this corresponds also to a single winding of the boundary map $f|_{\infty}$).

The setup can be generalized to the SU($n$) case as follows. As discussed earlier, the minimum of $V_1$ is achieved at the configuration~(\ref{eq:typical}), which defines an embedding $(S^1)^{n-1}\mysub \mathcal{F}$. We wish to compute the topological charges $Q_{A}={1\over 2\pi}\int\limits_D \,f^\ast \Omega_{A}$ for a map $f: \Sigma=D\to \mathcal{F}$, such that $f\big|_{\infty}: (S^1)_{\mathrm{WS}} \to (S^1)^{n-1}$
is a map with a fixed set of winding numbers $\vec{w}\in \pi_1((S^1)^{n-1})\simeq \mathbb{Z}^{n-1}$. From~(\ref{eq:typical}) it is natural to think of $\vec{w}$ as being the set of windings of $n$ angles $\sigma_j$ modulo the winding vector $(1, \cdots, 1)$ that can be removed by an overall $\tU(1)$ gauge transformation, i.e. $\vec{w}=(w_1, \cdots, w_n)\;\mathrm{mod}\;(1, \cdots, 1)$.

To compute the topological numbers, we recall the flag manifold embedding $\mathcal{F}\mysub (\CP^{n-1})^{\times n}$ (see Section~\ref{cohsec}) and denote $\pi_A$ the projection on the $A$-th projective space. Given a map $f: \Sigma\to \mathcal{F}$, we construct a map $f_{A}=\pi_A\circ f$ to $\CP^{n-1}$. If $(Z_1 : \cdots: Z_n)$ are the homogeneous coordinates on the projective space, we choose the standard $n$ patches $\{U_{A}\}_{A=1, \ldots, n}$, each one given by the condition $Z_A\neq 0$ for some $A$. Let us now consider the special maps $f_0$ (the `elementary fractional instantons') characterized by the fact that $(f_0)_{A}(\Sigma)\mysub U_{s(A)}$, $s$ being a permutation. This means that the image of each $(f_0)_{A}$ lies in a single patch $U_{s(A)}$. In each patch $U_A$ we set $Z_A=1$ and write the two-form $\Omega\big|_{U_A}=d\theta_A$\,,where $\theta_A=\frac{i\,\sum\limits_{B}\,Z_B d\bar{Z}_B}{1+\sum\limits_{B\neq A}|Z_A|}$ is a well-defined Poincar\'e-Liouville one-form, so that by Stokes theorem \bea (Q_0)_{A}={1\over 2\pi}\int_{S^1_{\mathrm{WS}}}\,f_0^\ast \theta_{s(A)}={1\over n}\sum_B\,\left(w_B-w_{s(A)}\right)\,,
\eea
where we have substituted the asymptotic values~(\ref{eq:typical}). Clearly $\sum_{A} (Q_0)_{A}=0$, as required. A general fractional instanton may be thought of as a collection of instantons `on top' of an elementary fractional instanton, resulting in the topological charge $\vec{Q}_0+(s_1, \cdots, s_n)$, where $s_A\in \mathbb{Z}$ are integers, $\sum_A s_A=0$.

Earlier we introduced the potential $V_2$, and so we would like to restrict to the configurations that asymptotically minimize this potential, i.e. to the special maps $D\to \mathcal{F}$, such that $(S^1)_{\mathrm{WS}}=\dd D\to S^1\mysub (S^1)^{n-1}$. In this case we may set $w_1=\cdots = w_{n-1}=0$. Apart from that, let us restrict to a single winding,  $w_n=1$. The topological charges of an elementary fractional instanton then are ($s(k)=n$)
\bea\label{Qmeron}
\vec{Q}_0=\left({1\over n}, \cdots,\!\!\!\underbracket[0.6pt][0.6ex]{\;{1\over n}-1\;}_{\;\;\textrm{position} \;k\;\;}\!\!\!,\cdots, {1\over n}\right)
\eea

\subsection{Mass generation}\label{sec:massgen}

As we have seen, there are several topological sectors. While the number of configurations has increased, the original argument from SU(2) for mass generation carries over to this more general case: for each species of topological excitation in this $n$-fold family there is a species of particle in the Coulomb gas formalism~\cite{meron}. That is, each particle has a partition function that is represented (at large distances) by a sine-Gordon (sG) model,
\bea \label{eq:SG}
	\fL_{SG} = \frac{1}{2} (\partial_\mu \sigma)^2 + \gamma \cos \left(\frac{2\pi}{g}  \sigma\right),
\eea
in the limit of large $\gamma$, which represents the fugacity, or density, of the fractional instanton gas. This expression is derived in detail in \cite{Affleck_1989} and relies on the fact that all higher-loop corrections to the (fractional) instanton gas are IR finite in the sG model~\cite{Kosterlitz_1973}. Formally speaking, expanding the partition function of the sG model in $\gamma$, i.e. in the cosine interaction, produces a multi-vortex Coulomb gas partition function of the XY-model (a similar trick has been used in Liouville theory, cf.~\cite{Teschner}).

In~(\ref{eq:SG}) $g$ is the O(2) coupling constant in (\ref{eq:gg}), and plays the role of temperature in the sG model. Since all of the $n$ species arise from the same action, each will have the same fugacity and critical $g$, so that the above model (\ref{eq:SG}) is merely copied $n$ times, and the SU(2) analysis from \cite{meron} can be applied directly: for large $m$, the fractional instantons are dilute and we are in a massless boson phase. As $m$ is lowered, the effective critical temperature is increased until the topological excitations condense and a mass gap is produced.\begin{footnote}{It is also worth noting that the exact critical exponents for the sG model are well known, and are reviewed in \cite{Affleck_1989}.}\end{footnote} Thus, we conclude that fractional instantons are responsible for generating a mass gap in the flag manifold sigma model (\ref{eq:rel1}), in the absence of topological angles.

\subsection{Destructive Interference in the Presence of Topological Angles} \label{sec:top}

We now restore the topological angles $\theta_\alpha$, and study how the mass generating mechanism changes. For large $m$, we are in the O(2) model and the $\theta$-terms do not play a role. However, as $m$ is lowered towards zero, the fugacity $\gamma$ in the sine-Gordon model is modified to
\bea \label{eq:fug}
	\gamma\sum_{A=1}^n e^{i\vec{\theta}\cdot\vec{Q}^A},
\eea
where the sum is over the $n$ species of fractional instanton, and $\theta_A = A \theta$, with $\theta = \frac{2\pi p}{n}$. Using (\ref{Qmeron}), one easily finds that this sum equals $\gamma\sum_A \zeta^{pA}$. So long as $p$ is not a multiple of $n$, this sum vanishes, and the Coulomb gas is in its massless phase.

 At first glance, this appears to be inconsistent with the conjecture discussed in section~\ref{section:thooft}, which also predicts a gap when $p$ is not a multiple of $n$, but has a nontrivial shared divisor with $n$. This discrepancy is resolved by considering higher order topological excitations. This is summarized in the following table:
 
 \vspace{0.5cm}
\begin{minipage}{14cm}
\centering
\begin{tabular}{c|c|c|c}
\textbf{Case}     & \textbf{Winding} & \textbf{Fugacity} & \textbf{Conclusion}\\ \hline  
$p= 0\;\;\mathrm{mod}\;\;n$     & 1  &
$n \gamma$ & Mass generation
\\ \hline
\begin{tabular}{c}
$p$ and $n$ coprime\\ (no common divisor)
\end{tabular}& \begin{tabular}{c}
$1, \cdots, n-1$\\ $n$
\end{tabular}  & \begin{tabular}{c}
$0$\\ $n\gamma$
\end{tabular} & Massless \\ \hline
\begin{tabular}{c}
$\gcd(p,n)  \not= 1, n$\\ $\Rightarrow 1< d< n$
\end{tabular}& \begin{tabular}{c}
$1, \cdots, d-1$\\ $d$
\end{tabular} & \begin{tabular}{c}
$0$\\ $n\gamma$
\end{tabular} & Mass generation \\ \hline
\end{tabular}
\par
\bigskip
Here $d:={n\over \gcd(p,n)}$. Winding number $w$ refers to the map $(S^1)_{\mathrm{WS}}\to (S^1)_{\mathrm{target}}$. For simplest fractional instanton configurations the topological charge is $w\,\vec{Q}_0$, and the fugacity is $\gamma \sum_A\zeta^{pAw}$.
\end{minipage}

 \vspace{0.5cm}
 While objects that have winding number greater than $\pm 1$ have larger action, they too must be considered, and do not necessarily lead to a vanishing fugacity.
This is also true in the case of SU(2), where so-called ``double merons'' do not having cancelling contributions~\cite{meron}. However these events are not strong enough to open a gap at the isotropic point $m=0$, and it is conjectured in~\cite{WamerMerons} that this holds for general $n$. When $p$ and $n$ have a nontrivial common divisor different from $n$, configurations with a smaller value of the action begin to contribute to mass generation, starting with objects that have winding number $d$. As a result, the critical value $m$ is larger than at $\theta = \frac{2\pi}{n}$ (although still lower than at $\theta=0$).

\section{More general representations: linear and quadratic dispersion}\label{linquadchains}

As we already discussed earlier, depending on the sign of the coupling constant $J$ the ground state of the Heisenberg chain is either ferro- or anti-ferromagnetic. The continuum limits around these two states lead to rather different models: in the ferromagnetic case this is the Landau-Lifschitz model (section~\ref{ferro}) that describes spin waves with quadratic dispersion, whereas in the anti-ferromagnetic case one obtains a relativistic sigma model (section~\ref{anti}), which in the gapless case describes excitations with linear dispersion.

In the present section, following~\cite{Wamer2020}, we will consider spin chains of the following type: at each site we will place spins in an arbitrary representation $\mathcal{R}$ of $\SU(n)$, with the condition that this representation is the same for all sites. The representation will be characterized by the lengths of the rows in the Young diagram, $p_1\geq p_2\geq \cdots \geq p_{n-1}\geq 0$. Among such models we will pick out those that lead to sigma models with the flag manifold target space~$\tU(n)\over \tU(1)^n$ in the continuum limit. One restriction that is imposed by this setup is that all the nonvanishing $p_i$'s will be distinct (if some of the rows of the Young diagram were of the same length, one would obtain a partial flag manifold $\tU(n)\over \tU(n_1)\times \cdots \times \tU(n_m)$ as the target space, cf.~\cite{BykovHaldane2}).  The curious feature of the general situation is that,  although we will take anti-ferromagnetic couplings between the spins, some of the modes will have quadratic dispersion relation, just as in the ferromagnetic case (although others will have linear dispersion). One can then work out conditions on the representation $\mathcal{R}$ that ensure that only linear modes remain. For such representations we will deduce the topological angles of the resulting models, which, as we saw earlier, are to a large extent responsible for the phase structure of these models.

To start with, we recall that in the semiclassical (large spin) limit the spin operators $S(j)$ of the chain should be replaced by the corresponding moment maps $\mu(j)=U^\dag \text{diag}(p_1,\cdots,p_n) U$, where $U$ is a unitary matrix, and $p_1 \cdots p_n$ are the lengths of the rows in the Young diagram of the representation $\mathcal{R}$ ($p_n=0$). In this case the spin-spin interaction between sites $i$ and $j$ becomes $\tr(S(i) S(j)) \to \sum_{A, B} p_A p_B 
	|\bar{u}_A(i)\circ u_B(j)|^2$. As a result, the simplest SU($n$) chain Hamiltonian, namely the nearest-neighbour model, becomes
\bea \label{eq:101}
	H = J \sum_j \sum_{A, B=1}^{n-1}p_A p_B |\bar{u}_A(j)\circ u_B(j+1)|^2,\hspace{10mm} J>0.
\eea
The sums over $A$ and $B$ stop at $n-1$, since $p_n=0$ by definition. This nearest-neighbour model is the logical starting point for any SU($n$) generalization of the antiferromagnetic spin chain. However, in most cases, we will be required to consider Hamiltonians with longer range interactions if we hope to map to the flag manifold sigma model. This should already be clear from the discussion of the rank-$p$ symmetric representations in sections~\ref{antifer} and~\ref{symmrepsec} above. Since the complete flag manifold is the space of $n$-tuples of mutually orthogonal fields taking values in $\CP^{n-1}$, one must add  $(n-1)$-neighbour interactions in order to impose orthogonality on the  $n$ fields. Instead, if one couples less than $n$ sites of the chain together, there will be leftover degrees of freedom, which manifest as local zero modes, ultimately prohibiting any field theory mapping. 

We will now explain how this construction generalizes as we increase the number $k$ of nonzero $p_i$.

For the purposes of presentation we will consider three main examples: $k=1$ (the symmetric representations), $k=n-1$ (which in the case of self-conjugate representations, $R\simeq \bar{R}$, is a generalization of the construction described in section~\ref{aff1}) and the case $n=\lambda k$. We refer the reader to~\cite{Wamer2020} for  more general situations.

\subsection{Spin chain ground states}

We will be using some graphical notation for describing classical spin configurations  of SU($n$) chains. First, let $\{\vec{e^i} \}$ be an orthonormal basis of $\mathbb{C}^n$. 

\begin{figure}[h]
\centering
\vspace{-.5cm}
\includegraphics[width = .9\linewidth]{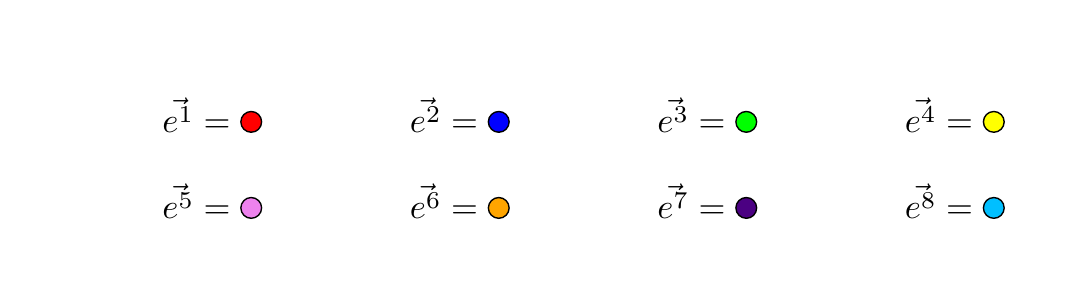} 
\caption{Colour dictionary for the first eight basis elements in $\mathbb{C}^n$. These coloured circles will be used to pictorially reprsent classical states of the chain.}
\label{fig:colour:l}
\end{figure}

We will use coloured circles to represent the first few elements of this basis, as shown in~Fig~\ref{fig:colour:l}.

When drawing a classical ground state, we will arrange the same-site vectors into a single column, and use a white space to separate neighbouring chain sites. For example, the N\'eel state of the SU(2) antiferromagnet is 
\bea 
	\cir{e1} \hspace{5mm}\cir{e2}\hspace{5mm}\cir{e1}\hspace{5mm}\cir{e2}\hspace{5mm}\cir{e1}\hspace{5mm}\cir{e2}\hspace{5mm}\cir{e1}\hspace{5mm} \cir{e2} \hspace{5mm}.
\eea

The benefit of these ground state pictures is that it makes it easy to read off the energy cost of a term $\tr(S(i)S(j)) = \sum_{A, B} p_A p_B |\bar{u}_A(i)\circ u_B(j)|^2.$ 
The right hand side of this expression vanishes unless one of the complex unit vectors (i.e. one of the colours) at site $i$ equals one of the complex unit vectors at site $j$. In this case, the r.h.s. equals $p_{A_0} p_{B_0}$, where $A_0$ and $B_0$ are the respective positions of the unit vector/colour in column $i$ and column $j$.

With this notation in place, we will now describe the ground state structure of  SU($n$) chains with some  sufficiently simple representations at each site.

\subsubsection{Case 1: $k=1$.} \label{k=1}

We begin with recalling what occurs for the symmetric representations of SU($n$), with Young tableaux that have a single row of length $p_1$ (see sections~\ref{antifer}-\ref{symmrepsec} above). For a nearest-neighbour SU($n$) Heisenberg Hamiltonian any configuration that has no energy cost per bond will be a classical ground state. Since $k=1$, and only a single node is present at each site, the N\'eel state shown above is such an example. However, for $n>2$, the basis at each site is larger than 2 (i.e. there are other colours available), and this leads to an infinite number of other ground states, resulting in a zero-energy mode that destabilizes any candidate ground state above which we would like to derive a quantum field theory. As a consequence, the nearest-neighbour Hamiltonian must be modified by longer-range interactions\footnote{These interactions may be dynamically generated from the nearest-neighbour model \cite{CorbozLajko}.}. Since there are $n$ possible colours, we require up to $(n-1)$-neighbour interactions, all of which are taken to be antiferromagnetic, in order to remove the zero modes. For example, in SU(5), with interactions up to 4th neighbour, one such ground state is
\bea \label{eq:sym1}
	\cir{e1} \hspace{5mm}\cir{e2}\hspace{5mm}\cir{e3}\hspace{5mm}\cir{e4}\hspace{5mm}\cir{e5}\hspace{5mm}\cir{e1}\hspace{5mm}\cir{e2}\hspace{5mm} \cir{e3} \hspace{5mm} \cir{e4} \hspace{5mm}\cir{e5} \hspace{5mm}.
\eea
This is a colourful depiction of the configuration~(\ref{NeelSUN}). As we recall from sections~\ref{anti}-\ref{symmrepsec}, the flag manifold $\tU(n)\over \tU(1)^n$ appears here as the space of $n$-tuples of pairwise orthogonal complex vectors, each vector coming from a copy of $\CP^{n-1}$ at one site of the chain.

\subsubsection{Case 2: $k=n-1$.} \label{sub:case2}

The second class of representations that we consider have Young tableaux with $n-1$ nonzero rows, and are arguably simpler than the symmetric representations considered above. Since in this case, according to the theory of geometric quantization explained in section~\ref{appQuant}, the on-site representation $\mathcal{R}$ already corresponds to the complete flag manifold, a nearest-neighbour Heisenberg interaction is sufficient to derive the associated sigma model. Let us first demonstrate this for the case of  SU(4). The interaction term
	$\tr(S(j)S(j+1)) = \sum_{A, B=1}^3 p_A p_B |\bar{u}_A(j)\circ u_B(j+1)|^2$ 
is never zero for two adjacent sites, which requires choosing the colour for six nodes. 
Using the inequality $p_2^2+p_3^2 \geq 2p_2p_3$, after a moment's thought one finds that the ground states have the following form:
\bea\label{kn1ground}
	\cir{e1} \hspace{5mm}\cir{e2}\hspace{5mm}\cir{e1}\hspace{5mm}\cir{e2}\hspace{5mm}\cir{e1}\hspace{5mm}\cir{e2}\hspace{5mm}\cir{e1}\hspace{5mm}
\eea
\[
	\cir{e3} \hspace{5mm}\cir{e4}\hspace{5mm}\cir{e3}\hspace{5mm}\cir{e4}\hspace{5mm}\cir{e3}\hspace{5mm}\cir{e4}\hspace{5mm}\cir{e3}\hspace{5mm}
\]
\[
	\cir{e4} \hspace{5mm}\cir{e3}\hspace{5mm}\cir{e4}\hspace{5mm}\cir{e3}\hspace{5mm}\cir{e4}\hspace{5mm}\cir{e3}\hspace{5mm}\cir{e4}\hspace{5mm}
\]
This pattern extends to general $n$: the first row of nodes establishes a N\'eel-like state, while the remaining $n-2$ rows have a ``reverse-ordered'' pattern: the colour ordering along a column switches direction between even and odd sites. For these representations, the unit cell is always 2 sites in length, which leads to a $\mathbb{Z}_2$ translation symmetry in the sigma model.

\subsubsection{Case 3: $n = \lambda k$}

To construct the ground state, one partitions $n$ colors into $\lambda$ sets, with $k$ colors in each set. We would like to place each set at one of the consecutive $\lambda$ sites, so to this end we add up $(\lambda-1)$-neighbour interactions (always with antiferromagnetic couplings) which make sure that the $k$-planes at the consecutive sites are orthogonal to each other. On top of that, in order to obtain the complete flag manifold, one still needs to orthogonalize $k$ vectors inside each $k$-plane, which can be achieved by adding a weaker $\lambda$-neighbour interaction that serves to reverse order within each set of the partition. For example, in SU(6) with $k=2$, the Hamiltonian we should consider is
\bea \label{eq:nk2}
	H = \sum_j \Big( J_1 \tr(S(j)S(j+1)) +J_2 \tr(S(j)S(j+2)) + J_3 \tr(S(j)S(j+3))\Big)
\eea
with $J_1>J_2 \gg J_3 >0$, which has, for example, the following ground state: 
\bea \label{nlambdakground}
	\cir{e1} \hspace{5mm}\cir{e3}\hspace{5mm}\cir{e5}\hspace{5mm}\cir{e2}\hspace{5mm}\cir{e4}\hspace{5mm}\cir{e6}\hspace{5mm}\cir{e1}\hspace{5mm}\cir{e3} \hspace{5mm} \cir{e5} \hspace{5mm}
\eea
\[
	\cir{e2} \hspace{5mm}\cir{e4}\hspace{5mm}\cir{e6}\hspace{5mm}\cir{e1}\hspace{5mm}\cir{e3}\hspace{5mm}\cir{e5}\hspace{5mm}\cir{e2}\hspace{5mm} \cir{e4} \hspace{5mm}\cir{e6} \hspace{5mm}
\]
The $J_1$ and $J_2$ terms serve to partition the colours into three sets (the `$2$-planes'): $\{\cir{e1}, \cir{e2} \}$, $\{ \cir{e3}, \cir{e4}\}$, $\{\cir{e5},\cir{e6}\}$, and the $J_3$ terms serve to reverse order within each of these three sets. Clearly  the unit-cell has size $2\lambda$ for these representations. 

\subsection{Conditions for linear dispersion and topological angles}

In the previous section we constructed spin chain Hamiltonians, whose classical minima lie on a complete flag manifold. This defines an embedding
\bea
i:\quad \mathcal{F}\mysub \mathcal{M}\,,
\eea
where $\mathcal{M}=\prod_{A=1}^d\,\mathcal{M}_A$ is the phase space of an elementary cell of length $d$. As we recall from sections~\ref{ferro}-\ref{anti}, the next step in deriving a continuum theory is in evaluating the restriction of the symplectic form 
\bea
\Omega_{\mathcal{M}}=\sum\limits_{A=1}^d\,\Omega_A\,,
\eea
which comes from the kinetic term in the Lagrangian (the `Berry phase'), to the space of minima of the Hamiltonian, i.e. to the flag manifold. In the ferromagnetic situation of Section~\ref{ferro} this restriction is non-degenerate. On the other hand, in the anti-ferromagnetic situation of section~\ref{anti} the restriction is identically zero (i.e. the flag manifold is a Lagrangian submanifold). The general situation is intermediate (the restricted form is degenerate but not exactly vanishing), and the relevant characteristic is the rank of the restriction $\Omega_{\mathcal{M}}\big|_{\mathcal{F}}$, which defines the number of fields with quadratic dispersion. By tuning the values of the integers $p_i$, i.e. by suitably choosing the representation $\mathcal{R}$, one can reduce the rank down to zero, in which case $\mathcal{F}\mysub \mathcal{M}$ is an isotropic submanifold. 

\begin{table}[h]
\centering
\begin{tabular}{| c  | c  r |} \hline
Representation &  Conditions & \\
\hline \hline 
$k=1$    & none &  \\ \hline

\multirow{2}{*}{\vspace{-10mm} $k=n-1$}  & 
$p_{i} + p_{n-i+1} = p_1 \hspace{5mm}  $& $n$ even; $i=2,\cdots, \frac{n}{2}$ \\
\cline{2-3}
 & \parbox{1cm}{
 \begin{align*}\vspace{-3mm}
p_{i} + p_{n-i+1} = p_1  \hspace{5mm}  \\
2p_{\frac{n+1}{2}} = p_1 \hspace{5mm}  \\
\end{align*}}
\vspace{-5mm}  & $n$ odd; $i=2,\cdots, \frac{n-1}{2}$ \\
\hline

\multirow{2}{*}{\vspace{-10mm} $n=k\lambda$}   & $ p_{i} + p_{k+1-i} =p_1+p_k \hspace{5mm} $ &  $k$ even; $i=2,\cdots,\frac{k}{2}$ \\ \cline{2-3}
&   \parbox{1cm}{\begin{align*}
p_{i} + p_{k+1-i} &= p_1 +p_k \hspace{5mm}  \\
2p_{\frac{k+1}{2}} &= p_1+p_k \hspace{5mm}  \\
\end{align*}}   & $k$ odd; $i=2,\cdots, \frac{k-1}{2}$   \\  \hline 
\end{tabular}
\caption{Rank reduction conditions (elimination of modes with  quadratic dispersion) for the two-form $\Omega_{\mathcal{M}}\big|_{\mathcal{F}}$.}\label{result2}
\end{table}

At least in the linearly dispersing case (when $\Omega_{\mathcal{M}}\big|_{\mathcal{F}}=0$), following the algorithm described in the previous sections, one can proceed to evaluate the topological angles. According to~\cite{Wamer2020}, the two-form entering the topological term is\footnote{For the case of spins with rectangular Young tableau at each site, when the resulting flag manifold is the manifold of partial flags, the same expression was obtained in~\cite{BykovHaldane2}.}
\bea\label{toptermgen}
\omega_{\mathrm{top}}={1\over d}\,\sum\limits_{A=1}^d\,A\cdot\Omega_A\big|_{\mathcal{F}}\,.
\eea
We recall (cf.~(\ref{symplformzs})) that in the general case each form $\Omega_A$ may be written as $\Omega_A=i\,\sum\limits_{k=1}^n\,p_k\,du_A^{(k)}\wedge \circ \bar{du}_A^{(k)}$, where $\{u_A^{(k)}\}_{k=1}^n$ are mutually orthogonal vectors at the $A$-th site of the unit cell. These vectors are represented by the circles in a given column of the colour diagram, such as~(\ref{eq:sym1}), (\ref{kn1ground}) or~(\ref{nlambdakground}), and the `restriction to $\mathcal{F}$' in~(\ref{toptermgen}) means replacing the given vector $u_A^{(k)}$ by the vector of the flag corresponding to the indicated colour.

Let us demonstrate how this works for $k=n-1$. 
The elementary cell consists of two sites, $d=2$, so that $\Omega_{\mathcal{M}}=i\,\sum\limits_{j=1}^2\sum\limits_{A=1}^n\,p_A\,du_A(j)\wedge \circ \bar{du_A(j)}$. According to the pattern of ground states (see (\ref{kn1ground})), two of the colours occur once (in the first position of the column), and the remaining $n-2$ colours occur twice, with reverse ordering. Therefore the restriction of the symplectic  form is
\bea
	{1\over i}\,\Omega_{\mathcal{M}}\big|_{\mathcal{F}} = p_1 \,(du_1\wedge \circ \bar{du_1} +du_2\wedge \circ \bar{du_2})
	+
	\sum_{A=3}^{n} (p_{A-1} +p_{n-A+2}) du_A \wedge \circ \bar{du_A}.
\eea
Here $u_A$ without the site label in brackets are meant to represent the $n$ orthogonal vectors of the embedded flag manifold $\mathcal{F}$. We recall that $\sum_{A=1}^{n} \, du_A \wedge \circ \bar{du_A}\big|_{\mathcal{F}}=0$, so that the expression can be simplified: 
\bea\label{OmegaMrestr}
	{1\over i}\, \Omega_{\mathcal{M}}\big|_{\mathcal{F}} = 
	\sum_{A=3}^{n} (p_{A-1} +p_{n-A+2} - p_1) \,du_A \wedge \circ \bar{du_A}.
\eea
Now we have up to $(n-2)$ fields with quadratic dispersion. The exact number will depend on how many of the conditions $p_{A-1}+p_{n-A+2}-p_1=0$ are satisfied. 
The representations that satisfy every constraint, and thus give rise to sigma models with purely linear dispersion, correspond to the self-conjugate representations $R \simeq \bar{R}$ of SU($n$). The case of SU(3) with $2p_2=p_1$ was considered in detail in~\cite{Wamer2019}. Similar constraints can be derived for other representations, see Table~\ref{result2} and ref.~\cite{Wamer2020}. Restricting in an analogous way the two-form $\omega_{\mathrm{top}}$~(\ref{toptermgen}), one obtains the topological term. Similarly to~(\ref{OmegaMrestr}), it may be expanded as 
\bea
\omega_{\mathrm{top}}=i\,\sum\limits_{A=1}^n\,\theta_A\,\,du_A \wedge \circ \bar{du_A}\,.
\eea
For the simple representations that we have discussed here the values of the $\theta$-angles are recorded in Table~\ref{result4}. The discrete symmetry in the general case is $\mathbb{Z}_d$: it acts on $\omega_{\mathrm{top}}$ by shifting $\omega_{\mathrm{top}}\to \omega_{\mathrm{top}}+\Omega_1$ (compare with~(\ref{topformshift})). One could in principle derive the mixed $\text{PSU}(n)-\mathbb{Z}_d$ anomalies in this case as well, which would provide a generalization of Haldane-type conjectures to this type of representations.

\begin{table}[!h]
\centering 
\begin{tabular}{l | l c } 
Representation &  Topological Angles  \\ \hline \hline
$k=1$ & $\theta_A = \frac{2\pi p_1}{n}(A-1)$ \vspace{1mm} & $i=1,2,\cdots, n$ \\ \hline
$k=n-1$ & $\theta_A = \pi p_A$ \vspace{1mm} & 
$i=1,2,\cdots,n$ \\ \hline
$k=\frac{n}{\lambda}$ & $\theta_{A,B} = \frac{\pi(p_A + p_{k+1-B})}{\lambda}(B-1) + \pi p_{k+1-A}$ \vspace{1mm} &
$A=1,\cdots, k; \;B=1, \cdots, \lambda$ \\ \hline
\end{tabular}
\caption{Possible topological angles for some representations of SU($n$) chains. In the last row the index is split as $A\to (A, B)$.}
\label{result4}
\end{table}

\pagebreak
\vspace{0.5cm}
\noindent
\rule{\textwidth}{1pt}
    \vspace{1ex}
\begin{center}
\vspace{-0.3cm}
{\Large     Chapter 3. Integrable flag manifold sigma models and beyond}
\end{center}

\noindent
\vspace{-0.5ex}%
\rule{\textwidth}{1pt}

\addcontentsline{toc}{section}{\bfseries Chapter 3. Integrable  flag manifold sigma models and beyond}

\vspace{2cm}
In the present chapter we pass to the subject of integrable sigma models with flag manifold target spaces, as well as some more general models. Recall that the integrability of the $S^2$-model~\cite{Zamolodchikov}, which predicted massive excitations over the vacuum state, was one of the motivations for Haldane's proposal that $\SU(2)$ integer-spin chains have a gap in the spectrum. It was subsequently shown~\cite{ZamolodchikovMassless} that the $\theta=\pi$ model is soluble as well, this time with a massless spectrum, in line with Haldane's treatment of the half-integer-spin chains.  In the case of $\SU(n)$ chains the resulting flag manifold sigma models described in the previous chapter are apparently not integrable, and their integrable counterparts discussed below feature a very special metric and  $B$-field. One striking parallel between the two types of models is the important role played by the $\mathbb{Z}_n$ symmetry, as we explain below in Section~\ref{Zmsymmintmod}. Another important feature of the proposed integrable models is their relation to nilpotent orbits, which also featured in our discussion of the Dyson-Maleev representation in Section~\ref{HPDMsec} above.

Sections~\ref{compstructzerocurvsec},~\ref{CS4Dsec} and~\ref{PCMrelsec} are dedicated to various aspects of the classical theory of integrable sigma models. The reason why we discuss this in great detail is that, when the target space of the model is not symmetric, constructing even a classical integrable theory is a significant challenge.  In section~\ref{GNsec} we will argue that the integrable flag manifold sigma models are in fact equivalent to (generalized) chiral Gross-Neveu models. This relation allows one to take a glimpse in the quantum realm of these models, at least in the one-loop approximation. For example, the analysis of the one-loop $\beta$-function in section~\ref{betafuncsec} gives rather important insights in the structure of these models. Another quantum aspect of the problem is the subject of chiral anomalies that we touch upon in section~\ref{superquiversec}.  Besides, rather surprisingly, the formulation of sigma models as Gross-Neveu models implies that the interactions in these sigma models are \emph{polynomial}. This fact is based on, or perhaps partially explained by, two seemingly unrelated observations. One is that the Dyson-Maleev variables provide a polynomial parametrization for the spin operators. The other is that, at least in the simplest cases~\cite{Zagermann}, the integrable models of the relevant class may be obtained by dimensional reductions of 4D gravity, expressed in Ashtekar variables, which are known to make the interactions in gravity polynomial. These fascinating inter-relations are explained in section~\ref{polintsec}.

Before we describe the theory in full generality, let us provide an example of the relation between sigma models and Gross-Neveu models. Consider the \emph{bosonic} Thirring model. In terms of a Dirac spinor $\Psi=\begin{pmatrix} U\\ \bar{V}\end{pmatrix}$ the two-dimensional Thirring Lagrangian reads:
\bea\label{ThirringLagr}
\mathscr{L}=\bar{\Psi}\slashed{\dd}\Psi+{1\over 2}\,(\bar{\Psi}\gamma_\mu \Psi)^2=V \bd U+\bar{U} \dd \bar{V} +|U|^2 |V|^2.
\eea
To obtain the bosonic Thirring model, we now regard the variables $U$ and $V$ as bosonic. Eliminating $V, \bar{V}$, we obtain the sigma model form of the system:
$\mathscr{L}=\frac{\bd U\,\dd\bar{U}}{U \bar{U}}$. The target space is a cylinder with multiplicative coordinate $U$. At the quantum level, the elimination of $V, \bar{V}$ means we have to integrate over these variables in the path integral. As a result, one should take into account the corresponding determinant, which is the source of an emerging dilaton. In this case the dilaton $\Phi\sim \log{|U|^2}$ is linear along the cylinder. This is the system describing the asymptotic region of Witten's cigar~\cite{WittenCigar}. As we shall see,  a wide class of sigma models may be seen to arise by a very similar procedure from \emph{chiral gauged} Gross-Neveu models, which are natural extensions of~(\ref{ThirringLagr}).

\section{The models and the zero-curvature representation}\label{compstructzerocurvsec}

We start with a more conventional formulation of sigma models by describing their metric and $B$-field in Lie-algebraic terms. In this chapter we will always assume that the worldsheet $\Sigma$ is a two-dimensional Riemannian manifold. As for the target-space $\mathcal{M}$, in full generality we will not require it to be a flag manifold but rather a manifold with the following properties\footnote{Generalizations to non-simple groups $G$ are also possible.}:
\begin{align} \nonumber
&\circ\quad \mathcal{M}\, \textrm{is a homogeneous space}\; \;G/H, \;\;G\; \textrm{semi-simple and compact} \\ \label{prop}
&\circ\quad \mathcal{M}\, \textrm{has an integrable}\; G\textrm{-invariant complex structure}\; \mathscr{J} \\
\nonumber
&\circ\quad \textrm{The Killing metric}\; \mathbb{G}\; \textrm{on}\; \mathcal{M}\; \textrm{is Hermitian w.r.t.}\; \mathscr{J} 
\end{align}
Let us explain what we mean by `Killing metric' on a homogeneous space. To this end, we decompose the Lie algebra $\mathfrak{g}$ of the Lie group $G$ as 
\bea\label{ghm}
\mathfrak{g}=\mathfrak{h}\oplus \mathfrak{m}\,,
\eea
where $\mathfrak{m}$ is the orthogonal complement to $\mathfrak{h}$ with respect to the Killing metric on $\mathfrak{g}$. We may accordingly decompose the Maurer-Cartan current $J=g^{-1}dg=J_{\mathfrak{h}}\oplus J_{\mathfrak{m}}$. The `Killing metric' $\mathbb{G}$ on $G\over H$ is defined by the line element
\bea
ds^2=-\mathrm{Tr}(J_{\mathfrak{m}}^2)\,.
\eea
The corresponding metric on $\mathfrak{m}$ will be called Killing as well. In the case of trivial $\mathfrak{h}$ this would then reduce to the canonical Killing metric, hence the name.

For a target space $\mathcal{M}$ with the properties~(\ref{prop}), one can define a \mbox{sigma model,} whose equations of motion may be rewritten as the flatness condition for a one-parameter family of connections $A_u, u\in \CC^\ast$. This flatness condition is an extension to this broader class of target spaces of a property that is encountered in sigma models with symmetric target spaces \cite{Pohlmeyer, Forger, Zakharov1}. In the latter case, this property is an important sign of integrability of the model: it may be used to find B\"acklund transformations \cite{Uhlenbeck, Devchand}, and it is a starting point for the construction of classical solutions of the models \cite{HitchinTori}.

Complex simply-connected homogeneous manifolds $G/H$ with $G$ semi-simple were classified long ago \cite{Wang}. They are given by the following theorem: any such manifold $G/H$ corresponds to a subgroup $H$, whose semi-simple part coincides with the semi-simple part of the centralizer of a toric subgroup of $G$. For the case of $G=\SU(n)$, for example, invariant complex structures exist on those of the  manifolds
\bea\label{complhom}
\mathcal{M}_{n_1, \ldots, n_m | n}=\frac{\SU(n)}{S(\tU(n_1)\times \ldots \times \tU(n_m))},\quad\quad m\geq 0, \;\;n_i> 0,\;\;\sum\limits_{i=1}^m \,n_i\leq n\,,
\eea
that are even-dimensional. If $\sum_{i=1}^m \,n_i= n$, the manifold in (\ref{complhom}) is a flag manifold. Otherwise, it is a toric bundle over a flag manifold. The fiber $\tU(1)^{2s}$ ($2s=n-\sum_{i=1}^m \,n_i$) of the toric bundle is even-dimensional, since the flag manifold itself is even-dimensional.

The models, which will be of interest for us in the present paper, are defined by the following action:
\bea\label{action}
\mathcal{S}[\mathbb{G}, \mathscr{J}]:=\int_\Sigma\,d^2 z\,\|\dd X\|^2_\mathbb{G}+\int_\Sigma\,X^\ast \omega,
\eea
where $\omega$ is the fundamental Hermitian form corresponding to the pair $(\mathbb{G}, \mathscr{J})$, defined as
\bea\label{kahform}
\omega=\mathbb{G}\circ \mathscr{J}\,.
\eea
In general, the Killing metric $\mathbb{G}$ is not K\"ahler, i.e. the fundamental Hermitian form is not closed: $d\omega \neq 0$. Even if the manifold $\mathcal{M}$ admits a K\"ahler metric, it is in general different from $\mathbb{G}$. As an example of such a phenomenon one can consider the flag manifold $\SU(3)\over S(\tU(1)^3)$. The sigma model (\ref{action}) for this flag manifold was investigated in detail in \cite{BykovFlag1, BykovFlag2}. Other examples of models of the class (\ref{prop}) are provided by Hermitian symmetric spaces -- symmetric spaces with a complex structure. These manifolds are K\"ahler, and the invariant metric is essentially unique (up to scale), thus leading to the closedness of $\omega$: $d\omega=0$. We will discuss this special case in Section \ref{symmspace}. For the moment let us note the following  equivalent rewriting of the action~(\ref{action}):
\begin{empheq}[box=\fbox]{align}
\hspace{1em}\vspace{1em}\label{actioncompl}
\mathcal{S}[\mathbb{G}, \mathscr{J}]:=
\int_\Sigma\,d^2 z\; \mathbb{G}_{j\bar{k}}\,\dd U^j \bar{\dd U^k}\,,\quad
\end{empheq}
where we have introduced complex coordinates $U^j$ on $\mathcal{M}$. Curiously, models with the $B$-field of the form~(\ref{kahform}) appeared in~\cite{WittenTop} in the context of topological sigma models, and gauged Wess-Zumino-Novikov-Witten theories with this feature were studied in~\cite{Gawedzki, ShataBoer}.

\subsection{The zero-curvature representation}\label{complzerocurv}

Let us now formulate the requirements~(\ref{prop}) on the target space $\mathcal{M}={G\over H}$ in Lie algebraic terms, and prove that the e.o.m. that follow from the action~(\ref{actioncompl}) admit a zero-curvature representation.

We will assume that the quotient space $G/H$ possesses a $G$-invariant almost complex structure $\mathscr{J}$. We are not postulating that $\mathscr{J}$ be integrable -- this will rather follow from the requirement of the existence of a Lax connection. The almost complex structure acts on $\mathfrak{m}$ (the subspace featuring in the decomposition~(\ref{ghm})) and may be diagonalized, its eigenvalues being $\pm i$ (see the following section for details). We denote the $\pm i$-eigenspaces by $\mathfrak{m}_{\pm}\mysub \mathfrak{m}_{\CC}$:
\bea\label{liealgdec}
\mathfrak{g}_{\CC}=\mathfrak{h}_{\CC}\oplus \mathfrak{m}_+\oplus \mathfrak{m}_-,\quad\quad \mathscr{J}\circ \mathfrak{m}_\pm=\pm i\,\mathfrak{m}_\pm\,.
\eea
$G$-invariance of the almost complex structure implies that $[\mathfrak{h}, \mathfrak{m}_\pm]\subset \mathfrak{m}_\pm$. We introduce the current
\bea\label{currcompldecomp}
J=g^{-1}dg=J_0+J_++J_-,\quad\quad J_0\in\mathfrak{h},\;\;\;J_\pm\in \mathfrak{m}_\pm\,.
\eea
It takes values in the Lie algebra $\mathfrak{g}$, and we have decomposed it according to the decomposition (\ref{liealgdec}) of the Lie algebra. In these terms the action (\ref{action}) may be rewritten as follows (henceforth we will be using bracket notation for the scalar product of two elements $\alpha, \beta \in \mathfrak{g}$ in the Killing metric):
\bea\label{actionLiealg}
\mathcal{S}[\mathbb{G}, \mathscr{J}]:=\int_\Sigma\,d^2 z\,\;\boldlangle (J_+)_z,\,(J_-)_{\bar{z}}\boldrangle\;.
\eea

\begin{figure}[h]
    \centering
    \includegraphics[width=0.3\textwidth]{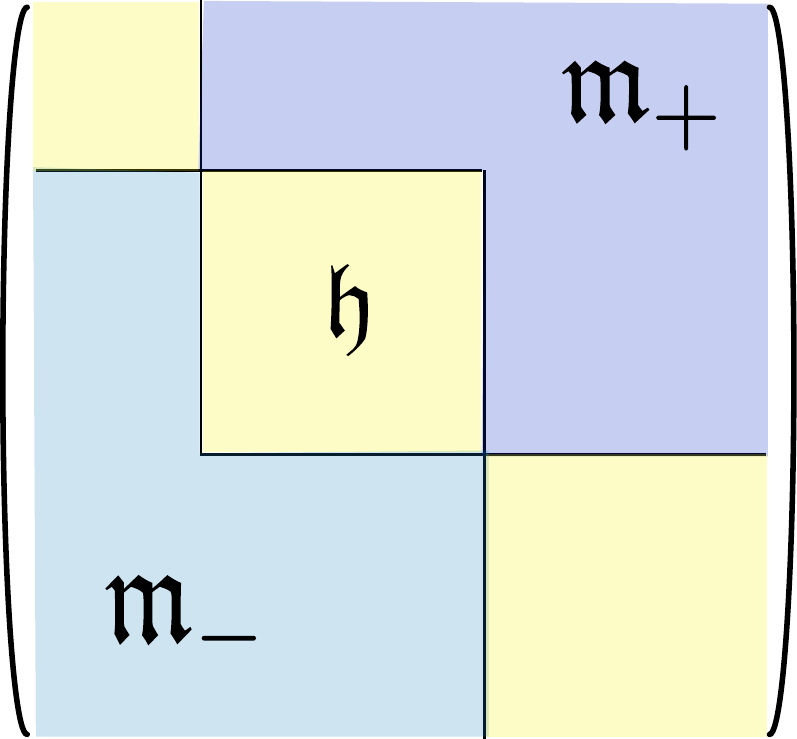}
    \caption{The decomposition (\ref{liealgdec}) of the Lie algebra.}
    \label{figcompstr}
\end{figure}

\emph{Example.} Let us consider the flag manifolds of the group $G=\tSU(n)$, which are the main subject of this review. A typical integrable complex structure on the flag manifold defines the holomorphic/anti-holomorphic subspaces $\mathfrak{m}_\pm$ shown in Fig.~\ref{figcompstr}.
It is useful do decompose $J_{\pm}$ in the irreducible representations $V_{AB}$  of the stabilizer $\mathfrak{h}$, see~(\ref{mirreps}). For this purpose we parametrize the unitary matrix $g$ as follows:
\bea\label{gtaumatr}
g=\{\tau_1, \tau_2, \ldots, \tau_{m-1}, \tau_m\}\,,
\eea
where $\tau_1 \ldots \tau_{m}$ are groups of $n_1 \ldots n_{m}$ orthonormal vectors, each group parametrizing a plane of the corresponding dimension in~$\CC^n$. The projection of $J_\pm$ on $V_{AB}$ is given by $J_{AB}:=\tau_A^\dagger d\tau_B$, and the full action~(\ref{actionLiealg}) takes the form
\bea\label{currcompaction}
\mathcal{S}[\mathbb{G}, \mathscr{J}]:=\int_\Sigma\,d^2 z\,\;\sum\limits_{A<B}\,\mathrm{Tr}\left((J_{BA})_{\bar{z}} (J_{AB})_z\right)\,.
\eea
These are the integrable models with flag manifold target spaces that we wish to study.

We return to the action~(\ref{actionLiealg}). The Noether current, constructed using the above action, will be denoted by $K$. It is derived by taking an infinitesimal $(z, \bar{z})$-dependent variation $g\to e^{\epsilon(z, \bar{z})}\circ g$ in the above action, which leads to

\bea\label{Scurr}
K=g\cdot\;\underbracket[0.6pt][0.6ex]{2\big((J_+)_z dz+(J_-)_{\bar{z}} d\bar{z}\big)}_{:=S} \;\cdot g^{-1}\,=g S g^{-1}
\eea
Since the target space $\mathcal{M}=G/H$ is homogeneous, the equations of motion of the model are equivalent to the conservation of $K$:
\bea\label{curcons}
d\ast K=0.
\eea
Here, $\ast$ denotes the Hodge star operator, whose action on one-forms is defined by $\ast dz=i\,dz, \;\ast d\bar{z}=-i\,d\bar{z}$. . In order to be able to build a family of flat connections we require that $K$ be flat (This will be used in (\ref{connection})-(\ref{flatnessAu}) below.):
\bea \label{curflat}
dK-K\wedge K=0.
\eea
We have to show, of course, that it is possible to satisfy this relation. Equations (\ref{curcons})-(\ref{curflat}) may be rewritten in terms of $S$ (introduced in (\ref{Scurr})) as follows:

\bear \label{eqcons}
&&d\ast S+\{J, \ast \,S\}=\\ \nonumber&&=-2i dz\wedge d\bar{z}\;\;\big(\bar{\mathscr{D}} (J_+)_z-[(J_+)_z, (J_+)_{\bar{z}}]+ \mathscr{D} (J_-)_{\bar{z}}+[(J_-)_z, (J_-)_{\bar{z}}]\big)=0\\ \label{eqflat}
&&dS+\{J-{1\over 2} S, S\}=\\ \nonumber&&=-2dz\wedge d\bar{z}\;\; \big( \bar{\mathscr{D}} (J_+)_z-[(J_+)_z, (J_+)_{\bar{z}}]-\mathscr{D} (J_-)_{\bar{z}}-[(J_-)_z, (J_-)_{\bar{z}}]\big)=0
\eear
Here, $\mathscr{D}$ is the covariant derivative for the gauge group $H$: $\mathscr{D}_jM_k:=\dd_jM_k+[(J_0)_j, M_k]$ ($j, k= z, \bar{z}$). The conditions (\ref{eqcons})-(\ref{eqflat}) are equivalent, if
\bea\label{algcond}
[\mathfrak{m}_+, \mathfrak{m}_+]\subset \mathfrak{m}_+,\quad\quad [\mathfrak{m}_-, \mathfrak{m}_-]\subset \mathfrak{m}_-\,.
\eea

We will discuss below in Section~\ref{compstructsec0} that if the metric $\mathbb{G}$ is Hermitian w.r.t. the chosen almost complex structure $\mathscr{J}$,
this requirement is equivalent to the integrability of $\mathscr{J}$.

Consider now the following family of connections $A_u$, indexed by a parameter $u\in \CC^\ast$:
\bea\label{connection}
A_u={1-u \over 2}\,K_z dz+{1-u^{-1}\over 2} \,K_{\bar{z}}d\bar{z}\;.
\eea

Conservation and flatness of the Noether current $K$, eqs. (\ref{curcons})-(\ref{curflat}), imply that $A_u$ is flat for all $u$~\cite{Pohlmeyer}:
\bea\label{flatnessAu}
dA_u-A_u\wedge A_u=0\quad\textrm{for all}\quad u\in \CC^\ast\;.
\eea
This completes the derivation of the zero-curvature representation for the class of models~(\ref{actionLiealg}), which includes the flag manifold models~(\ref{currcompaction}).

\subsection{Complex structures on flag manifolds}\label{compstructsec0}

We now turn to the general theory of complex structures on flag manifolds, which will play an important role throughout this chapter.

A very detailed treatment of complex structures on homogeneous spaces was given as early as in the classic work~\cite{BorelHirzebruch}, so here we mostly present an adaptation of some of these statements to our needs. To start with, on the manifold $\tU(n)\over \tU(1)^n$ of complete flags in $\CC^n$ there are $2^{\frac{n(n-1)}{2}}$ invariant almost complex structures, with $n!\leq 2^{\frac{n(n-1)}{2}}$ of them being integrable.\footnote{
We note that for large $n$, according to Stirling's formula,  $e^{n\,\log(n)}<e^{\log(2)\,n^2\over 2}$.}

As we already saw in~(\ref{liealgdec}), the complex structure $\mathscr{J}$ induces a decomposition $\mathfrak{m}_{\CC}=\mathfrak{m}_+\oplus \mathfrak{m}_-$, 
where $\mathfrak{m}_\pm$ play the role of holomorphic tangent spaces to $G/H$, i.e. $\mathscr{J}\circ a=\pm i\,a$ for $a\in \mathfrak{m}_\pm$. In section~\ref{genmetr}~(formula~(\ref{mirreps})) we have already decomposed  $\mathfrak{m}_{\CC}$ into irreducible components. Using this decomposition, we may define an almost complex structure on $\mathcal{F}$ by defining the action of  $\mathscr{J}$ as follows:
\bea
\mathscr{J}\circ V_{AB}=\pm\,i\,V_{AB}\quad\quad \textrm{for}\quad\quad 1\leq A<B\leq n\,.
\eea
As a result, one has exactly $2^{\frac{n(n-1)}{2}}$ possibilities. There are several equivalent definitions of integrability of a complex structure:
\begin{itemize}
\item[$\circ$] Vanishing of the Nijenhuis tensor: \bea\label{nijenhuis}
[\mathscr{J}\circ X, \mathscr{J}\circ Y]-\mathscr{J}\circ([\mathscr{J}\circ X, Y]+[X, \mathscr{J}\circ Y])-[X, Y]=0
\eea
for arbitrary vector fields $X, Y$.
\item[$\circ$] Using vector fields: the commutator of two holomorphic vector fields should be holomorphic, i.e.
\bea \label{integrdist1}
(1-i\,\mathscr{J})\,[(1+i\,\mathscr{J})X, (1+i\,\mathscr{J})Y]=0\,.
\eea
(The property (\ref{integrdist1}) may also be stated as the condition that the distribution of holomorphic vector fields is integrable.) This is easily seen to be equivalent to~(\ref{nijenhuis}).
\item[$\circ$] Using forms: the holomorphic forms should constitute a differential ideal in the algebra of forms, i.e. the following condition should be satisfied: $d(J_-)_a\sim \sum\limits_b \,R_{ab}\wedge (J_-)_b$ for some one-forms $R_{ab}$. 
\end{itemize}
If the restriction to $\mathfrak{m}$ of the adjoint-invariant metric $\langle\bullet, \bullet\rangle$ on $\mathfrak{u}(n)$  is Hermitian w.r.t. the chosen almost complex structure $\mathscr{J}$,  the last definition implies
\bea\label{algcond0}
[\mathfrak{m}_+, \mathfrak{m}_+]\subset \mathfrak{m}_+,\quad\quad [\mathfrak{m}_-, \mathfrak{m}_-]\subset \mathfrak{m}_-\,.
\eea
This is proven in Appendix~\ref{complstructapp}. The latter will serve us as a working definition of an integrable complex structure.
  
 \vspace{0.3cm}
 
 On the complete flag manifold $\mathcal{F}_n$, one can define an almost complex structure by choosing $\frac{n(n-1)}{2}$ mutually non-conjugate forms $J_{A_1 B_1}, \ldots, J_{A_{\frac{n(n-1)}{2}} B_{\frac{n(n-1)}{2}}}$ and declaring them holomorphic. The remaining $\frac{n(n-1)}{2}$ forms will be therefore anti-holomorphic. To determine which of these complex structures are integrable, it is useful to use a diagrammatic representation. We draw $n$ vertices, as well as arrows from the node $A_1$ to the node $B_1$, from $A_2$ to $B_2$ and so on, so that all pairs of nodes are connected (such diagrams are called~`tournaments', see~\cite{SalamonTourn}). As we shall now prove, the integrability of the almost complex structure, defined in this way, is equivalent to the acyclicity of the graph (the condition that it should not contain closed cycles).

Let us start with $\mathcal{F}_3$, and let $e_A, \;A=1, 2, 3$ be the standard unit vectors with components $(e_A)_\alpha=\delta_{A\alpha}$ ($\alpha=1, 2, 3$). To the holomorphic one-forms one can associate a subspace $\mathfrak{m}_+$ of the Lie algebra $(\mathfrak{su}(3))_\CC=\mathfrak{sl}(3)$ as follows:
\bea
\mathfrak{m}_+=\mathrm{Span}(E_{A_1 B_1}, E_{A_2 B_2}, E_{A_3 B_3}),\quad \textrm{where}\quad E_{AB}=e_A\otimes e_B
\eea
Integrability of the complex structure is equivalent to the requirement that $\mathfrak{m}_+$ is a subalgebra: $[\mathfrak{m}_+, \mathfrak{m}_+]\subset \mathfrak{m}_+$. On the other hand, the matrices $E_{mn}$ have the commutation relations
\bea
[E_{AB}, E_{CD}]=\delta_{BC}E_{AD}-\delta_{AD} E_{CB}
\eea
In the tournament diagram, $E_{AB}$ is represented by an arrow from $A$ to $B$; thus one sees that the closedness of $\mathfrak{m}_+$ under commutation is equivalent to the following statement:
\bear\label{property}
&&\textrm{For any two consecutive arrows $A \to B$ and $B \to C$}\\ \nonumber
&&\textrm{their `shortcut' segment $(A, C)$ has the arrow $A\to C$}
\eear
For the diagram with three vertices, i.e. for the $\mathfrak{su}(3)$ case under consideration, it is clear that the cyclic quivers are the only ones that do not lead to integrability. 

\begin{figure}
    \centering 
    \includegraphics[width=0.3\textwidth]{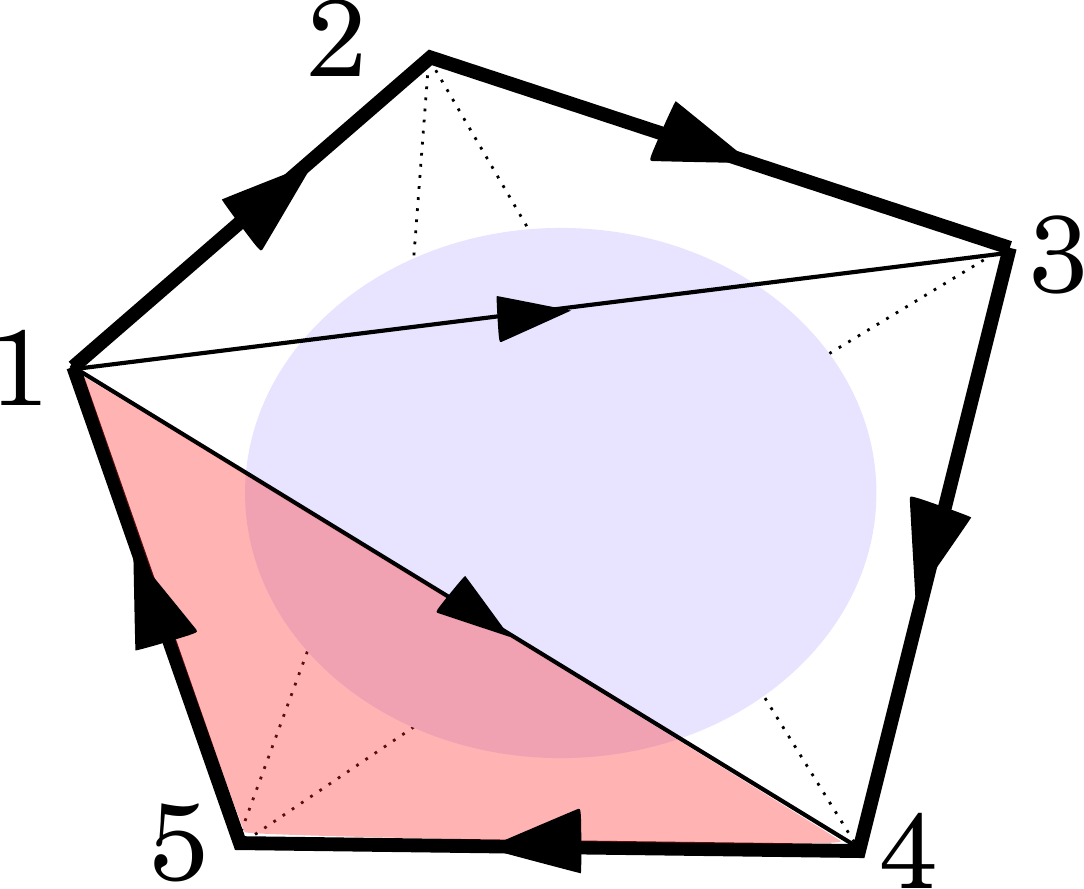}
  \caption{The procedure showing that a cycle $(1,2,3,4,5)$ in a graph leads to the violation of condition (\ref{property}). Using (\ref{property}), we replace the pair of segments $(1,2), (2,3)$ by $(1,3)$, i.e. cut a corner. Then we replace $(1,3)$, $(3,4)$ by $(1,4)$, arriving at the cyclic red triangle, which violates (\ref{property}).}
  \label{cut}
\end{figure}

In the general case of the flag manifold $\mathcal{F}_n$, suppose we have $n$ pairwise-connected vertices, and the graph is acyclic. Then the requirement (\ref{property}) is satisfied, since otherwise there would be a cycle with three vertices. Reversely, suppose the graph has a cycle. Then, using (\ref{property}), one can `cut corners' to reduce again to the cycle with three vertices, which is prohibited (see Fig. \ref{cut}).

One can then establish that there are exactly $n!$ acyclic diagrams. They correspond to the total orderings of the set of $n$ vertices. This is proven in Appendix~\ref{complstructapp}. By the logic explained above this means that there are $n!$ complex structures on a complete flag manifold $\tU(n)\over \tU(1)^n$. Analogously there are $m!$ complex structures on a partial flag manifold $\tU(n)\over \tU(n_1)\times \cdots \times \tU(n_m)$. The number of complex structures may be interpreted as follows. Choosing a complex structure is equivalent to choosing a complex quotient space representation~(\ref{flagdefhol}). In order to construct such a representation, one should choose a sequence of embedded linear spaces of the type~(\ref{embeddedspaces}), and the dimensions of these spaces are given by the partial sums of the integers $n_A$. These dimensions are therefore determined by an ordering of the set $\{n_A\}$, and there are $m!$ such orderings.

Now that we have described all invariant complex structures on an arbitrary flag manifold, we can take a fresh glance at the spaces of closed two-forms shown in~(\ref{2formsdiag}). As discussed in Chapter 1, the space of closed two-forms has real dimension $m-1$. Using the description of the cohomology~(\ref{H2quot}) based on the embedding of the flag manifold into a product of Grassmannians~(\ref{flaggenembed}), one can describe the space of closed two-forms as a hyperplane
\bea\label{hyperplaneclosed}
x_1+\ldots+ x_m=0\quad\quad \textrm{in}\quad\quad \mathbb{R}^m\,.
\eea
Consider a typical point in this vector space, where all $x_i$'s are distinct. This corresponds to a non-degenerate two-form, i.e. a symplectic form~$\Omega$. Given any such form, one can show that there is a unique complex structure $\mathscr{J}$, such that the corresponding symmetric tensor $\mathbb{G}:=-\Omega\circ \mathscr{J}$ is positive-definite. This is tantamount to saying that  $\mathbb{G}$  defines a K\"ahler metric on the flag manifold. In the simplest case of a Grassmannian, when $m=2$, the closed two-form is a multiple $\alpha\cdot \Omega_{\mathrm{FS}}$ of the generalized Fubini-Study form. There are also two invariant complex structures: $\mathscr{J}$ and $-\mathscr{J}$. As a result, the real line of invariant closed two-forms parametrized by $\alpha=x_2-x_1$ is divided into two rays $\alpha\gtrless 0$, and on each of these rays one picks a suitable complex structure $\pm \mathscr{J}$ to define a metric. The two rays are interchanged by the action of $S_2$: $x_2\leftrightarrow x_1$.

Returning to the general case, one finds that the hyperplane~(\ref{hyperplaneclosed}) is divided into $m!$ chambers, such that the points in the interior of each chamber may be thought of as K\"ahler forms corresponding to the same invariant complex structure. The chambers are interchanged by the action of the permutation group on $x_1, \ldots, x_m$, which is free provided that $x_i$'s are all distinct. This action is clearly synchronized with the action of $S_m$ that interchanges the complex structures. The boundaries between the chambers correspond to the case when several $x_i$'s coincide, which leads to the degeneration of the two-forms. As already mentioned at the end of section~\ref{appQuant}, in this case we may find a suitable smaller flag manifold, on which the two-form $\Omega$ is non-degenerate. The smaller flag manifold is obtained from the original one by a forgetful projection. By induction, more and more severe degenerations will correspond to forgetting more and more structure of the flag, and the extreme case when all $x_i=0$ corresponds to the flag manifold collapsing to a point. From this perspective, the space~(\ref{hyperplaneclosed}) of closed two-forms on the original flag manifold encodes the symplectic forms on the original manifold and on various smaller flag manifolds alike. 

\subsection{Symmetric spaces}\label{symmspace}

As a first example we consider the case of Hermitian symmetric spaces $\mathcal{M}$, i.e. symmetric spaces that admit a complex structure. First, we recall that in terms of the decomposition~(\ref{ghm}) symmetric spaces are characterized by the property $[\mathfrak{m}, \mathfrak{m}]\mysub \mathfrak{h}$. The Hermitian symmetric spaces are, in turn, characterized by the relation $[\mathfrak{m}_+, \mathfrak{m}_+]=0$. Indeed, this follows from the symmetric space property $[\mathfrak{m}, \mathfrak{m}]\mysub \mathfrak{h}$ and the integrability of the complex structure $[\mathfrak{m}_+, \mathfrak{m}_+]\mysub \mathfrak{m}_+$. Conversely, if $[\mathfrak{m}_+, \mathfrak{m}_+]=0$, one shows, using ad-invariance of the Killing metric on $\mathfrak{g}$, that $[\mathfrak{m}_+, \mathfrak{m}_-]$ is orthogonal to $\mathfrak{m}_\pm$, and hence $[\mathfrak{m}_+, \mathfrak{m}_-]\mysub \mathfrak{h}$.

The case of a symmetric target space is special in that the form $\omega$ is closed: $d\omega=0$. In other words, the Killing metric $\mathbb{G}$ is K\"ahler. Moreover, it is the only case when this is so:

\vspace{0.2cm}
\begin{center}
\fbox{\parbox{13.5cm}{
\centering The Killing metric on $\mathcal{F} = \mathcal{F}_{n_1, ... ,n_m}$ is K\"ahler if and only if $\mathcal{F}$ is a symmetric space, i.e. $m=2$ and $\mathcal{F} = Gr_{n_1, n}$ is a Grassmannian.
}
}
\end{center}

\vspace{0.2cm}

To prove this, we note that the components $J_\pm$ of the current $J$ (see the decomposition (\ref{currcompldecomp})) represent holomorphic/anti-holomorphic one-forms. The Killing metric on $\mathcal{F}$, which is $ds^2=-2 \,\mathrm{Tr}\,(J_+J_-)\,,
$
is therefore Hermitian. The K\"ahler form is, accordingly,
\bea
\omega=i\,\mathrm{Tr}\,(J_+\wedge J_-)\,.
\eea
In calculating the exterior derivative of $\omega$, we will be using the flatness equation $dJ-J\wedge J=0$ and the properties $\mathfrak{h}\perp  \mathfrak{m}_\pm, \mathfrak{m}_\pm\perp  \mathfrak{m}_\pm$ (isotropy of $\mathfrak{m}_\pm$). Simplifying the resulting expression, one gets 
\bea
d\omega=i\left( \mathrm{Tr}\,(J_+\wedge J_- \wedge J_-)-\mathrm{Tr}\,(J_-\wedge J_+ \wedge J_+)\right)\,.
\eea
The three-forms in the r.h.s. are of type $(1, 2)$ and $(2, 1)$ respectively (they are complex conjugate to each other). Therefore
$
d\omega=0$ if and only if  $\mathrm{Tr}\,(J_+\wedge J_- \wedge J_-)=0\,.
$
Due to the non-degeneracy of the Killing metric, this can only hold if
\bea
[\mathfrak{m}_-, \mathfrak{m}_-]=0\,.
\eea
It is easy to see that this holds if and only if $m=2$ (see Fig.~\ref{figcompstr}, for example). 

As a result, whenever the target space is a Grassmannian, i.e. $m=2$, the second term in the action (\ref{action}) is in fact topological and therefore does not affect the equations of motion. In this case we return to the well-known theory of integrable sigma models with symmetric target spaces. However the canonical Lax connection in this case is different from the one in (\ref{connection}). Indeed, the connection usually employed in the analysis of sigma models with symmetric target spaces has the form
\bea
\tilde{A}_\lambda={1-\lambda \over 2}\,\tilde{K}_z dz+{1-\lambda^{-1}\over 2} \,\tilde{K}_{\bar{z}}d\bar{z},\quad\quad\textrm{where}\quad\quad \tilde{K}=2\,g\cdot\big[g^{-1}dg\big]_{\mathfrak{m}}\cdot g^{-1}
\eea
 is the Noether current derived using the canonical action
\bea\label{actioncan}
\mathcal{S}[\mathbb{G}]=\int_\Sigma\,d^2 z\,\|\dd X\|^2_{\mathbb{G}}\;.
\eea
In the case of a Hermitian symmetric target space the difference between the two actions, (\ref{action}) and (\ref{actioncan}), is a topological term:
\bear
&&\mathcal{S}[\mathbb{G}, \mathscr{J}]-\mathcal{S}[\mathbb{G}]=\int_\Sigma\,X^\ast \omega,\\ \nonumber
&&\textrm{where} \quad d\omega=0\quad\textrm{if}\;\;\mathcal{M}\;\textrm{is symmetric.}
\eear
Therefore the two actions lead to the same equations of motion. Nevertheless, the Noether currents $K$ and $\tilde{K}$ are different, although both are flat. For the current $K$ this was shown in~(\ref{eqcons})-(\ref{algcond}), whereas the flatness

$d\tilde{K}-\tilde{K}\wedge \tilde{K}=0$ of $\tilde{K}$ does not, in fact, require using the equations of motion -- it is purely a consequence of the structure of the Lie algebra of the symmetric space (in particular, the fact that $[\mathfrak{m}, \mathfrak{m}]\subset \mathfrak{h}$). Moreover, the flatness condition may be solved, in this case, in a local fashion\footnote{`Local' means that $\widehat{g}$ is a local function of the fields of the model.}:
\bea\label{Cartancurr}
\tilde{K}=-\widehat{g}^{-1}d\widehat{g},\quad \textrm{where}\quad \widehat{g}=\sigma(g) g^{-1},
\eea
$\sigma$ being Cartan's involution on the Lie group $G$. By definition, $\sigma$ is a group homomorphism, $\sigma(g_1 g_2)=\sigma(g_1)\sigma(g_2)$, and $\sigma(h)=h$ for $h\in H$. The formula $\widehat{g}=\sigma(g) g^{-1}$, viewed as a map $g\in G/H \to \widehat{g}\in G$, describes the Cartan embedding\footnote{Cartan's embedding is known to be totally geodesic. By definition, this means that the second fundamental form of $\hat{\sigma}({G\over H})\subset G$ vanishes: $(\nabla_X Y)^\perp=0$ for any two vectors $X, Y\in T(\hat{\sigma}({G\over H}))$. It is easy to check that if $\hat{\sigma}: \mathcal{M}\subset \mathcal{N}$ is a totally geodesic submanifold, and $X: \Sigma \to \mathcal{M}$ is a harmonic map (i.e. a solution to the sigma model e.o.m.), then $\hat{\sigma} \circ X: \Sigma \to \mathcal{N}$ is also harmonic. This means that the classical solutions of the symmetric space $G/H$ model are a subset of solutions of the principal chiral model.}
\bea\label{Cartembed}
G/H \hookrightarrow G\;.
\eea

Flatness and conservation of the current $\tilde{K}$ lead to the flatness of the family $\tilde{A}_\lambda$. A question naturally arises of what the relation between $A_u$ and $\tilde{A}_\lambda$ is. The answer is that the connections $A_u$ and $\tilde{A}_\lambda$ are gauge-equivalent, if one makes the following identification of spectral parameters:
\bea\label{spectrpar}
\lambda=u^{1/2}\;.
\eea

The gauge transformation $\mathcal{G}$ relating $A_u$ and $\tilde{A}_\lambda$, 
\bea
\tilde{A}_\lambda=\mathcal{G} A_u \mathcal{G}^{-1}-\mathcal{G} d\mathcal{G}^{-1},
\eea
may be constructed explicitly. The following formula holds for the case when the target-space is the Grassmannian $Gr_{n_1, n_1+n_2}:={\SU(n_1+n_2)\over S(\tU(n_1)\times \tU(n_2))}$ and the complex structure is chosen so that it splits $\mathfrak{m}$ as $\mathfrak{m}=\left(\begin{smallmatrix}
 0 & \mathfrak{m}_+\\\mathfrak{m}_- & 0 \end{smallmatrix}\right)$:
\bea
\mathcal{G}=g\Lambda g^{-1}, \quad \textrm{where}\quad \Lambda=\lambda^{{1\over 2}{n_1-n_2\over  n_1+n_2}}\;\mathrm{Diag}(\underbracket[0.6pt][0.6ex]{\lambda^{-1/2}, \ldots, \lambda^{-1/2}}_{n_1}, \underbracket[0.6pt][0.6ex]{\lambda^{1/2}, \ldots, \lambda^{1/2}}_{n_2})\;.
\eea

Although the flag manifold~(\ref{flagunitary}) in general is not a symmetric space, it is a so-called $\mathbb{Z}_m$-graded space. Perhaps most well-known are the $\mathbb{Z}_3$-graded spaces, examples of which are provided by twistor spaces of symmetric spaces~\cite{Salamon} and nearly K\"ahler homogeneous spaces~\cite{Butruille} (the latter also appear in the context of string compactifications, cf.~\cite{Lust1, Lust2, Lechtenfeld1}). In this language the ordinary symmetric spaces are $2$-symmetric spaces. Similarly to what happens for symmetric spaces, the e.o.m. of a certain class of  sigma models with $\mathbb{Z}_m$-graded target-spaces may be rewritten as flatness conditions for a one-parameter family of connections. These models were introduced in \cite{Young}, and the construction of Lax connections for these models was elaborated in~\cite{BeisertLucker}. The relation to the Lax connections of Section~\ref{complzerocurv} has been recently established in~\cite{DelducKameyama} (this is an extension to $\mathbb{Z}_m$ of our discussion above regarding symmetric spaces). The fact that the integrals of motion of the models are in involution was proven, for instance, in~\cite{LacroixPhD}.

To summarize, there are certain relations (that we recall in Appendix~\ref{gradedsec}) between the models based on the $\mathbb{Z}_m$-graded spaces and the models discussed so far in this chapter. In general, however, we view the approach based on complex structures as rather different from the one based on $\mathbb{Z}_m$-gradings. This will be emphasized in the next sections, where  complex structures will be shown to play a key role through $\beta\gamma$-systems, as well as in the formulation of the integrable sigma models as gauged chiral Gross-Neveu models.

\subsection{Dependence on the complex structure: $\mathbb{Z}_m$-symmetry of the models}\label{Zmsymmintmod}

As we have emphasized, the models studied in the present Chapter depend explicitly on the complex structure $\mathscr{J}$ on the target space. It turns out, however, that the action~(\ref{action}), albeit depending on the complex structure, might produce the same equations of motion even for different choices of complex structure. This is due to the fact, that for certain complex structures, which we denote by $\mathscr{J}_1$ and $\mathscr{J}_2$, the difference in the two actions may just be a topological term:
\bea\label{toptermdiff}
\mathcal{S}[\mathscr{J}_1]-\mathcal{S}[\mathscr{J}_2]=\int\limits_\Sigma\,\mathscr{O}_{12},\quad\quad d\mathscr{O}_{12}=0\,.
\eea

Let us describe precisely the situation when this happens. To this end we recall that, as was established in Section~\ref{compstructsec0}, the complex structures are in a one-to-one correspondence with an ordering of the mutually orthogonal spaces $\CC^{n_1}, \ldots \CC^{n_m}$ constituting a flag, i.e. a point in a flag manifold $\tU(n)\over \tU(n_1)\times \cdots \tU(n_m)$. The statement is then as follows: 

\vspace{0.2cm}
\begin{center}
\fbox{\parbox{13.5cm}{
\centering The actions $\mathcal{S}[\mathscr{J}_1]$ and $\mathcal{S}[\mathscr{J}_2]$ differ by a topological term, as in~(\ref{toptermdiff}), if and only if the corresponding sequences of spaces $\{\CC^{n_1}, \ldots \CC^{n_m}\}$ differ by a \emph{cyclic} permutation.
}
}
\end{center}

\vspace{0.2cm}
This was proven in~\cite{BykovAnom}, and for the sake of completeness we recall the proof in Appendix~\ref{integrcyclsymm}. The important point is that this $\mathbb{Z}_m$ `symmetry' is very parallel to the $\mathbb{Z}_n$-symmetry of sigma models arising from spin chains, which was ultimately a reflection of the translational invariance of the latter and whose importance was emphasized in Sections~\ref{Znsymmsec} and \ref{section:thooft}. We write `symmetry' in quotation marks, because it is really a symmetry of the theory only in the case $n_1=\cdots =n_m$, when it can be realized by a cyclic permutation of the groups of vectors $\{u_1^{(1)}, \cdots , u_{n_1}^{(1)}\}$, \ldots , $\{u_1^{(m)}, \cdots , u_{n_m}^{(m)}\}$. In all other cases this should be seen as the equivalence of different theories, defined by the action functionals $\mathcal{S}[\mathscr{J}]$ for different complex structures $\mathscr{J}$. The same issue arises in the case of spin chains, when $1$-site translational invariance (leading to $\mathbb{Z}_m$-symmetry) is only present when the representations at each site are equivalent, which again leads to the condition $n_1=\cdots =n_m$. On the other hand, continuum limits of spin chains with different representations at different sites may still be described by partial flag manifold sigma models~\cite{BykovHaldane2}.

\section{Relation to 4D Chern-Simons theory}\label{CS4Dsec}

In the recent paper~\cite{CYa}, a novel approach to the construction of (at least classically)  integrable sigma models has been proposed. The flag manifold models of the previous sections, as well as their deformations, may as well be obtained within this framework. Besides, as we shall see in Section~\ref{GNsec}, when combined with the gauged linear sigma model approach, this construction provides a novel formulation of sigma models as gauged Gross-Neveu models. Deformed models appear naturally in this formalism through the introduction in the Lagrangian of the classical $r$-matrix. This is a very well-known object in integrable theories, but for completeness we shall start by recalling its definition and providing the simplest examples that we will use later on.

\subsubsection[The classical $r$-matrix]{The classical $r$-matrix.}\label{CYBEsec}

The classical $r$-matrix $r(u)$ takes values in $\mathfrak{g}\otimes\mathfrak{g}$, where $\mathfrak{g}$ is a semi-simple or, more generally, reductive Lie algebra and $u$ is a parameter taking values in a complex abelian group ($\CC$, $\CC^\ast$ or the elliptic curve $E_\tau$, depending on whether one deals with the rational/trigonometric/elliptic case respectively). The $r$-matrix satisfies the classical Yang-Baxter equation (CYBE), which takes values in $\mathfrak{g}\otimes\mathfrak{g}\otimes\mathfrak{g}$ and has the following form:
\bea
[r_{12}(u), r_{13}(u\cdot v)]+[r_{12}(u), r_{23}(v)]+[r_{13}(u\cdot v), r_{23}(v)]=0\,.
\eea
Since we mostly have the trigonometric case in mind, we write the equation in multiplicative form, that is to say $u, v \in \CC^\ast$. The notation $r_{12}(u)$ means $r_{12}(u)=r(u)\otimes \mathds{1}$, and analogously for other pairs of indices. Solutions to the above equation have been extensively studied in the classic paper~\cite{BelavinDrinfeld}.

\vspace{0.3cm}\noindent
For the purposes of the present paper it is more convenient to think of the $r$-matrix as a map $r(u): \mathfrak{g} \to \mathfrak{g}$, or equivalently $r(u)\in \mathrm{End}(\mathfrak{g})\simeq \mathfrak{g}\otimes \mathfrak{g}^\ast$. In this case we will write $r_u(a)\in \mathfrak{g}$ for the $r$-matrix acting on a Lie algebra element $a\in \mathfrak{g}$. One also often assumes the so-called `unitarity' property of the $r$-matrix: 
\bea\label{rmunit}
\mathrm{Tr}(r_u(A)\,B)=-\mathrm{Tr}(A\,r_{u^{-1}}(B))\,.
\eea
As we will see shortly, for our purposes it will be useful to weaken this condition slightly. In the new notations the CYBE looks as follows:
\bea
[r_u(a), r_{uv}(b)]+r_u([r_v(b), a])+r_{uv}([b, r_{v^{-1}}(a)])=0\,.
\eea

\vspace{0.3cm}\noindent
The solution of interest has the form (for now we assume $\mathfrak{g}\simeq \mathfrak{su}(n)$)
\bear\label{trigrmatrix}
&&r_u=\upalpha_u\,\pi_++\upbeta_u \,\pi_-+\upgamma_u\,\pi_0\,,\\ \label{abg}
&&\upalpha_u=\frac{u}{1-u},\quad\quad \upbeta_u=\frac{1}{1-u},\quad\quad \upgamma_u={1\over 2}\,\frac{1+u}{1-u}\,,
\eear
where $\pi_\pm$ are projections on the upper/lower-triangular matrices, and $\pi_0$ is the projection on the diagonal. The rational limit is achieved by setting $u=e^{-\upepsilon}$ and taking the limit $\upepsilon\to 0$, in which case $r_u\to \frac{1}{\upepsilon}\,\mathds{1}$.

The ansatz
\begin{empheq}[box=\fbox]{align}
\hspace{1em}\vspace{1em} \label{rR}
r_u=\frac{1}{2}\,\frac{1+u}{1-u}\,\mathrm{Id}+{i\over 2}\,\mathcal{R}\quad
\end{empheq}
transforms the CYBE to an equation on $\mathcal{R}$, which does not depend on the spectral parameter:
\bea\label{mCYBE}
[\Rc(a), \Rc(b)]+\Rc([\Rc(b), a]+[b, \Rc(a)])-[a, b]=0\,.
\eea
It is known in the literature as the `classical modified Yang-Baxter equation'. The solution~(\ref{trigrmatrix}) corresponds to\footnote{Another option is taking an $\mathcal{R}$-matrix induced by a complex structure on the Lie group $G$ with Lie algebra~$\mathfrak{g}$~\cite{BykovDeform}. In this case~(\ref{mCYBE}) is the condition of vanishing of the Nijenhuis tensor~(\ref{nijenhuis}).} $\mathcal{R}=i\,(\pi_+-\pi_-)$.

\subsection{The `semi-holomorphic' 4D Chern-Simons theory}\label{4DCSsec}

Having the right tools in place, we proceed to explain the construction of~\cite{CYa}, which is based on a certain `semi-holomorphic' 4D Chern-Simons theory that we will now describe.  In this case the four-dimensional `spacetime' is a product  $\Sigma\times \mathscr{C}$, where $\Sigma$ is called the `topological plane' and is endowed with coordinates $z, \bar{z}$ -- this will  eventually be the worldsheet of the sigma model, -- and $\mathscr{C}$  is a  complex curve with coordinates $w, \bar{w}$ (this is the spectral parameter curve). The latter is required to admit a nowhere-vanishing holomorphic differential $\omega=dw\neq 0$, which means that its canonical class is trivial: $K_{\mathscr{C}}=0$. As a result, the curve is either the complex plane, a cylinder or an elliptic curve (a torus):  $\mathscr{C}\simeq \CC, \CC^\ast, E_{\tau}$. The Chern-Simons action of the model is
\bea\label{complexCS}
S_{\mathrm{CS}}={1\over \hbar}\int\limits_{\Sigma\times \mathscr{C}}\,\omega\wedge\mathrm{Tr}\left(A\wedge (dA+{2\over 3} A\wedge A)\right)\,,
\eea
where $A=A_{z} dz+A_{\bar{z}}d\bar{z}+A_{\bar{w}}d\bar{w}$ is a gauge field corresponding to a (semi-simple) gauge group $G$. One couples this theory to certain two-dimensional systems of a very particular sort, called $\beta\gamma$ systems. These are defined for complex symplectic target spaces, their action in local Darboux coordinates $(p, q)$ being $S_{\beta\gamma}=\int\limits_{\Sigma}\,d^2z \,p_i \bd q^i$. In the context of~\cite{CYa} one considers target spaces of the form $T^\ast \mathcal{M}$, where $\mathcal{M}$ is a complex manifold endowed with a holomorphic action of the group $G$, and writes down a sum of two $\beta\gamma$-system actions, one holomorphic and the other anti-holomorphic:
\begin{figure}[h]
\centering
   \includegraphics[width=0.2\linewidth]{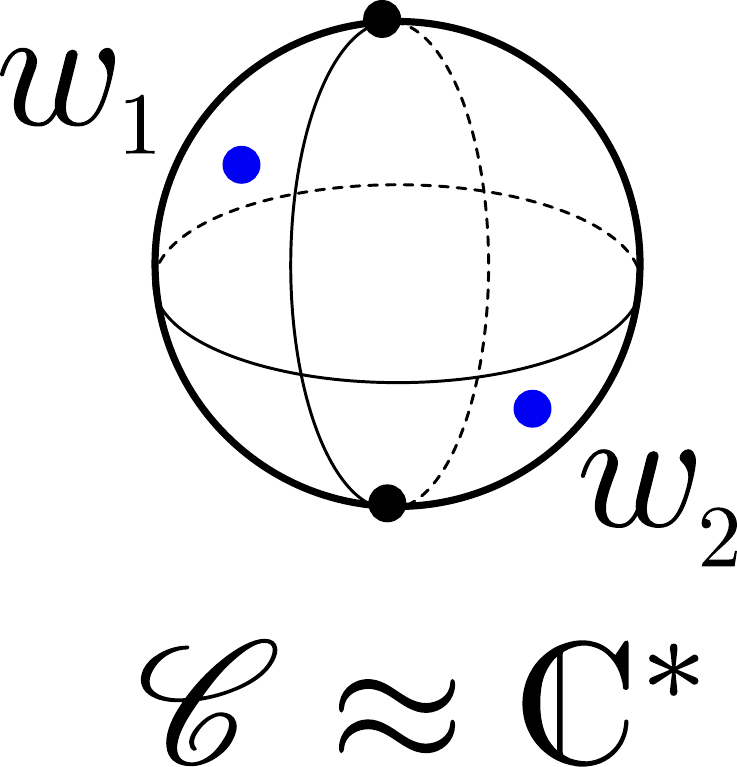}
   \caption{Two $\beta\gamma$-defects located at points $w_1, w_2$ on the spectral parameter curve $\mathscr{C}$.}
   \label{fig:sphereCstar}
\end{figure}
\bear\label{defectaction}
&&S_{\mathrm{def}}=\int\limits_{\Sigma}\,d^2z \,\left(p_i \bar{D}^{(w_1)}q^i+\bar{p}_i D^{(w_2)}\bar{q}^i\right)\,, \\ \nonumber
&&\textrm{where}\quad\quad \bar{D}^{(w_1)}q^i=\bar{\dd} q^i-\sum\limits_a \left(A_{\bar{z}}^{(w_1)}\right)_a v_a^i
\eear
and $v_a$ are the holomorphic vector fields on $\mathcal{M}$ generating the action of $G$. The full action functional is the sum of two:~(\ref{complexCS}) and~(\ref{defectaction}). The next step is to impose the `light-cone' gauge $A_{\bar{w}}=0$, in which case two of the e.o.m. become linear in $A_{z}, A_{\bar{z}}$, and the third one is the zero-curvature constraint:
\bear
&&\bar{\dd}A_z-\dd A_{\bar{z}}+[A_z, A_{\bar{z}}]=0,\\ \label{Azconn}
&&\dd_{\bar{w}} A_z=\delta^{(2)}(w-w_1) \sum_{a } p_i v_a^i\, \tau_a\\ \label{Abarzconn}
&&\dd_{\bar{w}} A_{\bar{z}}=\delta^{(2)}(w-w_2) \sum_{a } \bar{p}_{\bar{i}} v_a^{\bar{i}}\, \tau_a\,.
\eear
The delta-functions in the r.h.s. of the latter equations mean that $A_z(w), A_{\bar{z}}(w)$  depend meromorphically on $w$, and the first equation is then the zero-curvature equation (on the worldsheet $\Sigma$) for the family of connections $A=A_z dz+A_{\bar{z}}d\bar{z}$ depending on the parameter $w$.

In order to solve the equations~(\ref{Azconn})-(\ref{Abarzconn}), one needs to invert the operator $\dd_{\bar{w}}$. One of the key observations in~\cite{CYa} is that, with suitable boundary conditions, the Green's function $\bar{\dd}_{\bar{w}}^{-1}$ is the classical $r$-matrix~\cite{BelavinDrinfeld}, viewed as an element of $ \mathrm{End}(\mathfrak{g})$. In the rational case the Green's function is simply the Cauchy kernel, enhanced with additional matrix structure, i.e.
\bea\label{rmatrat}
r_w=\frac{\mathrm{Id}}{w}\in \mathrm{End}(\mathfrak{g})\,.
\eea
As we reviewed in Section~\ref{CYBEsec}, the $r$-matrix is sometimes written as an element of $\mathfrak{g}\otimes \mathfrak{g}$, and the two definitions are simply related by raising/lowering an index, using the Killing metric on $\mathfrak{g}$. Accordingly, the more conventional  representation for~(\ref{rmatrat}) would be $r(w)=\frac{\sum \tau_a \otimes \tau_a}{w}\in \mathfrak{g}\otimes \mathfrak{g}$, where $\tau_a$ are the generators of $\mathfrak{g}$. In the trigonometric case one needs to impose boundary conditions at the ends of the cylinder, i.e. at the two punctures on  $\mathscr{C}=\CC^\ast\simeq \CP^1\setminus \{0, \infty\}$ (these are shown as black dots in Fig.~\ref{fig:sphereCstar}). As explained in~\cite{CYa} at length, the relevant boundary conditions amount to picking a decomposition\footnote{Such decomposition is the same as picking a complex structure on $\mathfrak{g}$, compatible with the metric. It is also known in the literature as a Manin triple.} of the Lie algebra $\mathfrak{g}_{\CC}=\mathfrak{g}_+\oplus \mathfrak{g}_-$, where $\mathfrak{g}_\pm$ are two isotropic subspaces of $\mathfrak{g}$, and requiring that $A_z\in \mathfrak{g}_+$ at $w=0$ but $A_z\in \mathfrak{g}_-$ at $w=\infty$. In that case the $r$-matrix has the form
\bea
r_w=\frac{\Pi_+}{1-w}-\frac{\Pi_-}{1-w^{-1}}\in \mathrm{End}(\mathfrak{g})\,,
\eea
where $\Pi_\pm$ are the projectors on $\mathfrak{g}_\pm$, or alternatively 
$r(w)=\frac{\sum \tau_a^+ \otimes \tau_a^-}{1-w}-\frac{\sum \tau_a^- \otimes \tau_a^+}{1-w^{-1}}\in \mathfrak{g}\otimes \mathfrak{g}$.

Once we know the $r$-matrix, the solution to~(\ref{Azconn}) is $A_z=r_{w_1}\left(\sum_{a } p_i v_a^i\, \tau_a\right)$, and analogously for~(\ref{Abarzconn}). Substituting the solution back into the full action $S=S_{\mathrm{CS}}+S_{\mathrm{def}}$ (using the fact that the Chern-Simons action is quadratic in the $A$-fields in the gauge $A_{\bar{w}}=0$), we obtain the action that only depends on the $p, q$-variables:
\bear\label{rdefaction}
&&S=\int\,d^2z\,\left(p_i \bar{\dd}q^i+\bar{p}_i \dd\bar{q}^i+r_{w_1-w_2}\left(p_i v_a^i\, \tau_a,  \bar{p}_{\bar{i}} v_a^{\bar{i}}\, \tau_a)\right)\right)=\\ \nonumber
&&=\int\,d^2z\,\left(p_i \bar{\dd}q^i+\bar{p}_i \dd\bar{q}^i+{1\over w_1-w_2}\sum |p_i v_a^i|^2\right)\,,
\eear
where in passing to the second line we have restricted ourselves to the rational $r$-matrix. 
One sees that the action is quadratic in the $p$-variables, which are the coordinates in the fiber of the cotangent bundle~$T^\ast \mathcal{M}$. Integrating out these variables as well, we get the sigma model form of the action:
\bea\label{metrsigmaaction}
S\sim\int\,d^2z\,\left( G_{i\bar{j}} \bar{\dd}q^i \dd\bar{q}^{\bar{j}}\right)\,,\quad\quad \textrm{where}\quad\quad G_{i\bar{j}}=\left(\sum\limits_a\,v_a^i v_a^{\bar{j}}\right)^{-1}
\eea
is the metric on the target space.  Note that in order for the expression for the metric $G_{i\bar{j}}$ to make sense  the matrix $\sum\limits_a\,v_a^i v_a^{\bar{j}}$ has to be invertible. This is equivalent to the requirement that $\mathcal{M}$ is a homogeneous space. 

The model~(\ref{metrsigmaaction}) is clearly of the same type as the general class of models~(\ref{actioncompl}) introduced earlier. 
In the next section, following~\cite{BykovNilp}, we will prove directly that, in the case when $\mathcal{M}$ is a flag manifold, the two models are equivalent (meaning that the metric $G_{i\bar{j}}$ coincides with the Killing metric discussed at the beginning of  Section~\ref{compstructzerocurvsec}). In proving this, it will turn out extremely useful to introduce a gauged linear sigma model approach to the models in question, which will ultimately lead us to the formulation of sigma models as generalized Gross-Neveu models in section~\ref{GNsec}. We will also see that the formalism  described here, especially when combined with the GLSM-presentation, makes it very easy to construct integrable deformations (trigonometric, and possibly even elliptic) of the sigma models by picking the corresponding $r$-matrices in~(\ref{rdefaction}).  

\subsection{The gauged linear sigma model and the $\beta\gamma$-systems}
\label{GLSMbetagamma}

In the present section we will prove that the flag manifold models obtained from the coupling of two $\beta\gamma$-systems through a four-dimensional Chern-Simons field are -- in the rational case -- equivalent to the models that we described earlier in Section~\ref{compstructzerocurvsec}. To this end we therefore effectively set the $r$-matrix to be the identity operator: $r=\mathrm{Id}$. Our main tool in identifying the two types of models will be the gauged linear sigma model representation that was developed in~\cite{BykovGLSM1, BykovGLSM2} for flag models of type~(\ref{currcompaction}). In the case when the target space is a Grassmannian, the metric $\mathbb{G}$ is K\"ahler, and this representation is equivalent to the K\"ahler quotient $Gr_{k, n} \simeq \mathrm{Hom}(\CC^k, \CC^n)\sslash \tU(k)$. In the general case our construction leads to a quotient w.r.t. a non-reductive group and to the `Killing' metric $\mathbb{G}$, which is not K\"ahler in general.

The construction is as follows. We introduce the field $U\in \mathrm{Hom}(\CC^M, \CC^n)$, satisfying the orthonormality condition $U^\dagger U=\mathds{1}_M$, as well as the ``gauge'' field $\widetilde{\mathcal{A}}=\mathcal{A} \,dz+\bar{\mathcal{A}} \,d\bar{z}$ of the  special form shown in Fig.~\ref{fig:gaugefieldGLSM}.
\begin{figure}[h]
\centering
   \bea\label{colormatr}
\mathcal{A}=
\begin{tikzpicture}[
baseline=-\the\dimexpr\fontdimen22\textfont2\relax,scale=1]
\matrix[matrix of math nodes,left delimiter=(,right delimiter=),ampersand replacement=\&] (matr) {
\node (X0) {\ast};\&\ast\&\ast\&\ast\&\ast\&\node (X00) {\ast};\\
\node (X1) {\ast};\& \node (X2) {\ast};\&\ast\&\ast\&\ast\&\node (X11) {\ast};\\
\&\&\ast\&\ast\&\ast\&\ast\\
\&\&\node (X3) {\ast};\&\node (X4) {\ast};\&\ast\&\node (X22) {\ast};\\
\&\&\&\&\ast\&\ast\\
\&\&\&\&\node (X5) {\ast};\&\node (X6) {\ast};\\
};
\node[fill=blue!10, fit=(X0.north west) (X11.south east) ] {};
\node[fill=blue!10, fit=(X2.south east) (X22.south east) ] {};
\node[fill=blue!10, fit=(X4.south east) (X6.south east) ] {};
\draw[blue!50, line width=1pt, rounded corners] ([yshift=-3pt,xshift=-3pt]X1.south west)  -- ([yshift=-3pt,xshift=-2.5pt]X2.south east) node[right,black] {};
\draw[blue!50, line width=1pt, rounded corners] ([yshift=-3pt,xshift=-3pt]X2.south east)  -- ([yshift=-3pt,xshift=-3pt]X3.south west) node[right,black] {};
\draw[blue!50, line width=1pt, rounded corners] ([yshift=-3pt,xshift=-3pt]X3.south west)  -- ([yshift=-3pt,xshift=-2.5pt]X4.south east) node[right,black] {};
\draw[blue!50, line width=1pt, rounded corners] ([yshift=-3pt,xshift=-3pt]X4.south east)  -- ([yshift=-3pt,xshift=-3pt]X5.south west) node[right,black] {};
\draw[blue!50, dotted, line width=1pt] ([yshift=-3pt,xshift=-3pt]X3.south west)  -- ([yshift=-40pt,xshift=-3pt]X3.south west) node[right,black] {};
\draw[blue!50, dotted, line width=1pt] ([yshift=-3pt,xshift=-3pt]X5.south west)  -- ([yshift=-10pt,xshift=-3pt]X5.south west) node[right,black] {};
\draw[latex-latex] ([yshift=-6pt,xshift=-3pt]X5.south west) --  node[above, yshift=0pt] {{\scriptsize$ d_1$}} ([yshift=-6pt,xshift=3pt]X6.south east);
\draw[latex-latex] ([yshift=-37pt,xshift=-3pt]X3.south west) --  node[below, yshift=2pt] {{\scriptsize$ d_2$}} ([yshift=-37pt,xshift=58pt]X3.south west);
\draw[latex-latex] ([yshift=6pt,xshift=-3pt]X0.north west) --  node[above, yshift=-2pt] {{\scriptsize$ d_{m-1}$}} ([yshift=6pt,xshift=3pt]X00.north east);
\end{tikzpicture},\quad\quad\quad \bar{\mathcal{A}}=(\mathcal{A})^\dagger\,.
\eea
   \caption{The gauge field entering the GLSM description~(\ref{lagr2}) of the flag manifold sigma models. Here $d_1, \cdots, d_{m-1}:=M$ are the dimensions of the complex spaces in the flag~(\ref{complexflag}).}
   \label{fig:gaugefieldGLSM}
\end{figure}
The Lagrangian reads
\bea\label{lagr2}
\mathscr{L}=\mathrm{Tr}\left(\|\bar{D} U\|^2\right)\,,\quad\quad\textrm{where}\quad\quad \bar{D} U=\bar{\dd}U+i\, U\bar{\mathcal{A}}\,.
\eea
This Lagrangian is equivalent to~(\ref{currcompaction}), as can be shown by eliminating the field $\mathcal{A}$. Indeed, varying w.r.t. $\bar{\mathcal{A}}$ one obtains $(D_z U^{\dagger}\circ U)_{\mathfrak{p}_M}=0$, where $\mathfrak{p}_M$ is the space of upper-block-triangular matrices of size $M\times M$. Let us parametrize $U$ as \bea
U=\{\tau_1, \tau_2, \ldots, \tau_{m-1}\}\,,
\eea
where $\tau_1 \ldots \tau_{m-1}$ are groups of $n_1 \ldots n_{m-1}$ orthonormal vectors. Adding the last group $\tau_m$ of $n_m$ vectors orthogonal to all the rest, we obtain the matrix $g$ from~(\ref{gtaumatr}). One then has the orthonormality condition 
$\tau_A^\dagger \tau_B=\delta_{AB}\; (A, B=1,\ldots, m)$ and the completeness relation
\bea\label{complete2}
\sum\limits_{A=1}^m\,\tau_A \tau_A^\dagger=\mathds{1}_{n}\,,
\eea
which is equivalent to $g g^\dagger=\mathds{1}_{n}$. In this notation  components of the gauge field $\mathcal{A}$ have the form (see Fig.~\ref{colormatr})
\bear\nonumber
&&[\mathcal{A}]_{AB}=i\,\tau_A^\dagger \dd \tau_B\quad\textrm{for}\quad A\leq B\,;\\ \nonumber&& [\mathcal{A}]_{AB}=0\quad\textrm{for}\quad A> B\,.\quad A, B=1\ldots m-1\,.
\eear
In order to compute the Lagrangian~(\ref{lagr2}), it is useful first to calculate
\bea
\bar{D} \tau_A=\bar{\dd} \tau_A-\sum\limits_{B=A}^{m} \tau_B \tau_B^\dagger \, \bar{\dd}\tau_A=\textrm{using}\;(\ref{complete2})=\sum\limits_{B=1}^{A-1} \tau_B \tau_B^\dagger\, \bar{\dd}\tau_A\quad\quad A=1,\ldots, m-1\,.
\eea
Substituting into (\ref{lagr2}), we obtain the final expression for the Lagrangian in the `non-linear form'
\bea\label{flaglagrfin}
\mathcal{L}=\sum\limits_{A<B}\;\mathrm{Tr}\left((J_{BA})_z^\dagger \,(J_{BA})_z\right),\quad\quad\textrm{where}\quad\quad (J_{BA})_z=\tau_B^\dagger\,\dd\tau_A\,.
\eea
This is clearly the same as~(\ref{currcompaction}), up to an exchange of complex structure $\mathscr{J}\to -\mathscr{J}$.

We return to the GLSM~(\ref{lagr2}). Due to the orthonormality condition $U^\dagger U=\mathds{1}_M$, the gauge group of the model is $\tU(n_1)\times \cdots \times \tU(n_m)$. A natural question is whether one can instead use a quotient w.r.t. the complex group of upper/lower-block-triangular matrices. To answer this, we give up the orthonormality condition and assume that $U\in \mathrm{Hom}(\CC^M, \CC^n)$ is an arbitrary complex matrix of rank $M$. We then write down the following Lagrangian:
\bea\label{lagr3}
\mathscr{L}=\mathrm{Tr}\left((\bar{D} U)^\dagger\,\bar{D} U\,\frac{1}{U^\dagger U}\right)\,.
\eea
It is easy to see that it is invariant w.r.t. complex gauge transformations  $U\to U g$, where  $g\in P_{d_1, \ldots, d_{m-1}}$ (a parabolic subgroup of $\text{GL}(M,\mathbb{C})$. The Gram-Schmidt orthogonalization procedure brings the Lagrangian~(\ref{lagr3}) to the form~(\ref{lagr2}), but for a number of reasons the complex form is preferable. In order to get rid of the denominator in the Lagrangian, we introduce an auxiliary field $V\in \mathrm{Hom}(\CC^n, \CC^M)$ and write down a new Lagrangian
\bea\label{lagr4}
\mathscr{L}=\mathrm{Tr}\left(V \bar{D} U\right)+\mathrm{Tr}\left(\bar{U} D \bar{V}\right)-\mathrm{Tr}\left(VV^\dagger U^\dagger U\right)\,,
\eea
that turns into the original one upon elimination of the field $V$. This is therefore a far-reaching generalization of the elementary example considered at the very start of this Chapter, where using a similar procedure we obtained a sigma model with target space the cylinder $\CC^\ast$.  One should also keep in mind that in the process of integration over the $V$-variables, a non-trivial one-loop determinant typically arises, which leads to a non-zero dilaton, in exactly the same way as it happens in the context of Buscher rules for $T$-duality~\cite{Buscher, TseytlinDil, SchwarzDil} (this is particularly important for the deformed models discussed in section~\ref{classintsec}). Next we perform yet another quadratic transformation, in order to eliminate the quartic interaction. To this end we introduce the complex matrix field $\Phi\in \mathrm{End}(\CC^n)$ and its Hermitian conjugate: $\bar{\Phi}=(\Phi)^\dagger$. We write one more Lagrangian
\bea\label{lagr5}
\mathscr{L}=\mathrm{Tr}\left(V \bar{\mathscr{D}} U\right)+\mathrm{Tr}\left(\bar{U} \mathscr{D} \bar{V}\right)+\mathrm{Tr}\left(\Phi \bar{\Phi}\right)\,,
\eea
where $\bar{\mathscr{D}}$ is the ``elongated'' covariant derivative
\bea
\bar{\mathscr{D}} U=\bar{\dd} U+i \,U \bar{\mathcal{A}}+i\,\bar{\Phi} U\,.
\eea
Let us clarify the geometric meaning of the Lagrangian~(\ref{lagr5}). The first two terms correspond to a sum of the so-called $\beta\gamma$-systems on the flag manifold $\mathcal{F}$, in a background field~$\Phi$~\cite{Nekrasov, WittenBeta}. In the terminology of the previous section our field $\bar{\Phi}$ should be viewed as the component $A_{\bar{z}}$ of the Chern-Simons gauge field along the ``topological plane'' (i.e. the worldsheet~$\Sigma$). The quadratic form in the interaction term  $\mathrm{Tr}\left(\Phi \bar{\Phi}\right)$ in~(\ref{lagr5}) is in this context the inverse propagator of the field $A_{\bar{z}}$, which in the present (rational) case is proportional to the identity matrix.

By definition, such a system may be defined for an arbitrary complex manifold $\mathcal{M}$ ($\mathrm{dim}_\CC \mathcal{M}=D$) with the help of a complex fundamental $(1,0)$-form $\theta=\sum\limits_{i=1}^D\,p_i\,dq_i$ (the complex analogue of the Poincar\'e-Liouville one-form) on $T^\ast \mathcal{M}$. Here $q_i$ are the complex coordinates on $\mathcal{M}$ and  $p_i$ are the complex coordinates in the fiber of the  holomorphic cotangent bundle. The action of the $\beta\gamma$-system is then simply $S=\int\limits_\Sigma\,d^2z\,\sum\limits_{i=1}^D\,p_i\,\bar{\dd}q_i$. In the case of the flag manifold this action can be most conveniently written, using two matrices $U\in \mathrm{Hom}(\CC^M, \CC^n), V\in \mathrm{Hom}(\CC^n, \CC^M)$ and the gauge field $\mathcal{A}$. Indeed, it will be shown in the next section that the fundamental $(1,0)$-form can be written as $\theta=\mathrm{Tr}\left(V dU\right)\big|_{\mu_{\CC}=0}$, where $(U, V)$ satisfy the condition
\bea\label{mommapzero}
\mu_{\CC}=VU\big|_{\mathfrak{k}^\ast}=0\,, \quad\quad\textrm{and}\quad\quad \mathfrak{k}=\mathrm{Lie}(P_{d_1, \ldots, d_{m-1}})
\eea
is the Lie algebra of the corresponding parabolic subgroup of $\text{GL}(M, \CC)$. It is also assumed that the space of matrices, satisfying this condition, is factorized w.r.t. the action of $P_{d_1, \ldots, d_{m-1}}$, i.e. one has a complex symplectic reduction. The condition~(\ref{mommapzero}) is precisely the condition of vanishing of the moment map $\mu_\CC=0$ for the action of the parabolic group $P_{d_1, \ldots, d_{m-1}}$ on the space of matrices $(U, V)$ endowed with the symplectic form $\omega_0=\mathrm{Tr}(dU\wedge dV)$. As a result,
\bea\label{omegared}
d\theta=\omega_{\mathrm{red}}
\eea
is the complex symplectic form, arising after the reduction w.r.t. the parabolic group. In order to ensure the condition~(\ref{mommapzero}) at the level of the Lagrangian of the model, one needs the gauge field $\bar{\mathcal{A}}\in \mathrm{Lie}(P_{d_1, \ldots, d_{m-1}})$. Indeed, differentiating the Lagrangian~(\ref{lagr5}) w.r.t. $\bar{\mathcal{A}}$, one arrives at the condition~(\ref{mommapzero}).

\subsection{Relation to the quiver formulation}

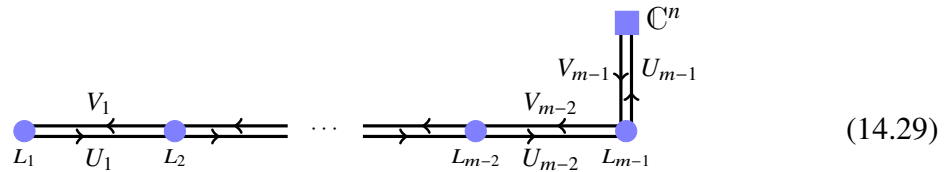
\begin{figure}[h]
\centering
   \bea
\begin{tikzpicture}[
baseline=-\the\dimexpr\fontdimen22\textfont2\relax,scale=1]
\begin{scope}[very thick,decoration={
    markings,
    mark=at position 0.4 with {\arrow{>}}}
    ] 
\draw[postaction={decorate}] ([yshift=-2pt,xshift=0pt]0,0) --  node [below, yshift=0pt] {\footnotesize $U_1$} ([yshift=-2pt,xshift=0pt]2,0);
\draw[postaction={decorate}] ([yshift=-2pt,xshift=0pt]2,0) --    ([yshift=-2pt,xshift=0pt]3.5,0);
\draw[postaction={decorate}] ([yshift=-2pt,xshift=0pt]4.5,0) --    ([yshift=-2pt,xshift=0pt]6,0);
\draw[postaction={decorate}] ([yshift=-2pt,xshift=0pt]6,0) --  node[below, yshift=0pt] {{\footnotesize$ U_{m-2}$}} ([yshift=-2pt,xshift=0pt]8,0);
\end{scope}
\begin{scope}[very thick,decoration={
    markings,
    mark=at position 0.6 with {\arrow{<}}}
    ] 
\draw[postaction={decorate}] ([yshift=2pt,xshift=0pt]0,0) --  node[above, yshift=0pt] {{\footnotesize $ V_1$}} ([yshift=2pt,xshift=0pt]2,0);
\draw[postaction={decorate}] ([yshift=2pt,xshift=0pt]2,0) --  ([yshift=2pt,xshift=0pt]3.5,0);
\draw[postaction={decorate}] ([yshift=2pt,xshift=0pt]4.5,0) --  ([yshift=2pt,xshift=0pt]6,0);
\draw[postaction={decorate}] ([yshift=2pt,xshift=0pt]6,0) --  node[above, yshift=0pt] {{\footnotesize$ V_{m-2}$}} ([yshift=2pt,xshift=0pt]8,0);
\end{scope}
\begin{scope}[very thick,decoration={
    markings,
    mark=at position 0.6 with {\arrow{<}}}
    ] 
\draw[postaction={decorate}] ([yshift=0pt,xshift=-2pt]8,0) --  node[right, xshift=3pt, yshift=5pt] {{\footnotesize$ U_{m-1}$}} ([yshift=0pt,xshift=-2pt]8,1.3);
\end{scope}
\begin{scope}[very thick,decoration={
    markings,
    mark=at position 0.4 with {\arrow{>}}}
    ] 
\draw[postaction={decorate}] ([yshift=0pt,xshift=2pt]8,0) --  node[left, xshift=-3pt, yshift=5pt] {{\footnotesize$ V_{m-1}$}} ([yshift=0pt,xshift=2pt]8,1.3);
\filldraw[blue!50] (7.87,1.3) rectangle ++(8pt,8pt);
\end{scope}
\filldraw[blue!50] (0,0) circle (4pt); \filldraw[blue!50] (2,0) circle (4pt); \filldraw[blue!50] (6,0) circle (4pt); \filldraw[blue!50] (8,0) circle (4pt);
\node at (8.5,1.5) {$\CC^n$};
\node at (0,-0.35) {\scriptsize $L_1$};
\node at (2,-0.35) {\scriptsize $L_2$};
\node at (6,-0.35) {\scriptsize $L_{m-2}$};
\node at (8,-0.35) {\scriptsize $L_{m-1}$};
\node at (4,0) {\scriptsize $\cdots$};

\end{tikzpicture}
\eea
   \caption{Nakajima quiver for the cotangent bundle to the flag manifold $T^\ast \mathcal{F}$.}
   \label{fig:Nakajimaquiv}
\end{figure}

Before passing to further topics, let us clarify the relation between the complex symplectic form, as discussed in the previous section (constructed using the symplectic quotient with respect to a parabolic subgroup), and the symplectic form that arises as a result of a reductive quotient, defined by the so-called quiver. We recall that $T^\ast \mathcal{F}$ is a hyper-K\"ahler manifold that may be constructed by a hyper-K\"ahler quotient of flat space (though we stress that the real symplectic form -- the K\"ahler form -- will not concern us here). This quotient is based on a linear quiver diagram~\cite{Nakajima} shown in Fig.~\ref{fig:Nakajimaquiv}.
This is essentially the same quiver that we encountered in section~\ref{kahquiv} (Fig.~\ref{flagquivpic}), but this time with a doubled set of arrows, which is related to the fact that this time we have the cotangent bundle $T^\ast \mathcal{F}$ rather than the flag manifold itself. In each node there is a vector space $L_k\simeq \CC^{d_k}$, and to each arrow from node $i$ to node $j$ corresponds a field, taking values in $\mathrm{Hom}(L_i, L_j)$. The full space of fields is therefore
\bea
\mathscr{W}_0:=\oplus_{i=1}^{m-1}\,\left(\mathrm{Hom}(L_i, L_{i+1})\oplus \mathrm{Hom}(L_{i+1}, L_i)\right)\,.
\eea
In each node there is an action of a gauge group $\tGL(L_i)$. We then consider the GIT-quotient $\mathscr{W}_f:=\mathscr{W}/\mathscr{G}$ of the stable subset $\mathscr{W}\subset \mathscr{W}_0$ w.r.t. the group $\mathrm{G}:=\prod\limits_{i=1}^{m-1}\,\tGL(L_i)$. In $\mathscr{W}_f$ we define a submanifold given by the vanishing conditions for the moment maps ($U_0=0, V_0=0$):
\bea\label{mommapi}
\mathcal{F}:= \{\mu_i=U_{i-1}V_{i-1}-V_iU_i=0\,,\quad\quad i=1, \ldots, m-1\,\}\subset \mathscr{W}_f\,.
\eea
The (well-known) statement is that the resulting space is the flag manifold~(\ref{flagdefhol})-(\ref{flagunitary}), which is why we have denoted it by $\mathcal{F}$. On $\mathscr{W}_0$ there is a natural complex {symplectic~form}
\bea
\Omega=\sum\limits_{i=1}^{m-1}\,\mathrm{Tr}(dU_i \wedge dV_i)\,.
\eea
The construction just described may be interpreted as the symplectic quotient w.r.t. the complex group $\mathrm{G}$, and it endows $\mathcal{F}$ with a certain symplectic form $\Omega_\mathcal{F}$. We prove the following statement:

\vspace{0.3cm}
\fbox{\parbox{13.5cm}{
\centering 
$\Omega_\mathcal{F}=\omega_{\mathrm{red}}$,  where  $\omega_{\mathrm{red}}$ is the symplectic form~(\ref{omegared}) that arises as a result of the reduction w.r.t. a parabolic subgroup of $\tGL(M, \CC)$.
}
}

\vspace{0.3cm}
\noindent To prove this, consider the fields $\{U_i\}$. $U_i$ is a matrix with $d_i$ columns and $d_{i+1}$ rows. By the action of $\tGL(d_{i+1}, \CC)$ one can bring $U_i$ to the form where the first $d_{i+1}-d_i$ rows are zero and the last $d_i$ rows represent a unit matrix. The stabilizer of this canonical form w.r.t. the joint (left-right) action of $\text{GL}(d_{i+1}, \CC)\times \text{GL}(d_i, \CC)$ is the subgroup $P_{d_i, d_{i+1}} \subset P_{d_i, d_{i+1}}\times \text{GL}(d_{i}, \CC)$, embedded according to the rule $g\to (g, \pi_i(g))$, where $\pi_i(g)$ is the projection on the block of size $d_i\times d_i$. Iterating this procedure, i.e. bringing all matrices  $U_i$ ($i=1, \ldots, m-2$) to canonical form, we arrive at the situation, when one is left with a single non-trivial matrix $U_{m-1}:=U$, and the resulting symmetry group is precisely $P_{d_1, \ldots, d_{m-1}}$. We also denote $V_{m-1}:=V$. Now, let $a\in \mathfrak{k}=\mathrm{Lie}(P_{d_1, \ldots, d_{m-1}})$. By definition of the stabilizer $a U_{m-2}=U_{m-2}\pi_{m-2}(a)$, therefore $\mathrm{Tr}(aU_{m-2}V_{m-2})=\mathrm{Tr}(\pi_{m-2}(a)V_{m-2}U_{m-2})=\mathrm{Tr}(\pi_{m-2}(a)U_{m-3}V_{m-3})$, where in the second equality we have used the equation~(\ref{mommapi}). Since $\pi_{m-2}(a)\in \mathrm{Stab}(U_{m-3})$, we can iterate this procedure, and at the end we will obtain $\mathrm{Tr}(aU_{m-2}V_{m-2})=0$. Due to the equation $U_{m-2}V_{m-2}-VU=0$ we get $VU\big|_{\mathfrak{k}^\ast}=0$, which coincides with~(\ref{mommapzero}). Besides, since the matrices $U_i$ ($i=1, \ldots, m-2$) are constant, the restriction of the symplectic form $\Omega$ coincides with $\mathrm{Tr}(dU_{m-1}\wedge dV_{m-1})=\mathrm{Tr}(dU\wedge dV)$. 

\vspace{0.5cm}
Let us clarify the role of the field $\Phi$. Differentiating the Lagrangian~(\ref{lagr5}) w.r.t. $\bar{\Phi}$, we obtain
\bea\label{Phiz}
\Phi=-i\,UV\,.
\eea
This coincides with the expression for the $z$-component of the Noether current for the action of the group  $\text{GL}(n, \CC)$ on the space of matrices $(U, V)$. For the $\beta\gamma$-system written above $\Phi$ is nothing but the moment map for the action of this group.

\section{Relation to the principal chiral model}\label{PCMrelsec}

We recall that the principal chiral model is a sigma model with target space a compact Lie group, such as $\tSU(n)$. In the present section, following~\cite{BykovNilp}, we describe the relation between flag manifold models and the principal chiral model.

\subsection{Nilpotent orbits}

\noindent
Our starting point will be the formulas~(\ref{mommapzero})-(\ref{Phiz}):
\bea\label{PhizUV}
\Phi=-i\,UV\,,\quad\quad \mu_{\CC}=VU\big|_{\mathfrak{k}^\ast}=0\,,
\eea
where $\mathfrak{k}=\mathrm{Lie}(P_{d_1, \ldots, d_{m-1}})$.

\subsubsection{Grassmannian}
As a warm-up we consider the case of a Grassmannian, i.e. $m=2$. Then the vanishing of the moment map is simply $VU=0$. Therefore $\Phi^2=0$, which means that $\Phi$ belongs to a nilpotent orbit of the group $\tGL(n, \CC)$. From the expression for $\Phi$ it also follows that $\mathrm{Im}(\Phi)\subset\mathrm{Im}(U)\subset \mathrm{Ker}(\Phi)$. As is well-known,
\bea\label{cotgrass}
\{(U, \Phi): \mathrm{rk}(U)=M, \;\Phi^2=0, \;\;\mathrm{Im}(\Phi)\subset\mathrm{Im}(U)\subset \mathrm{Ker}(\Phi)\} \simeq T^\ast \text{Gr}_{M, n}
\eea
is the cotangent bundle to a Grassmannian (notice the rank condition!), and the forgetful map 
\bea
T^\ast \text{Gr}_{M, n}\to \{\Phi: \Phi^2=0\}
\eea
provides a resolution of singularities of the nilpotent orbit in the r.h.s. (the Springer resolution). The conditions in the l.h.s. of~(\ref{cotgrass}) imply the factorization~(\ref{PhizUV}) for $\Phi$, and the non-uniqueness in this factorization corresponds exactly to the gauge symmetry $U\to U g, V\to g^{-1} V$, where $g\in \text{GL}(M, \CC)$.

Let us now derive the equations of motion for the field $\Phi$. First of all, the Lagrangian~(\ref{lagr5}) implies the equations of motion $\bar{\mathscr{D}} U=0,\,\bar{\mathscr{D}} V=0
$ for the fields $U$ and $V$. Therefore $\bar{\mathscr{D}} \Phi=0$, i.e.
\bea\label{dPhi}
\bar{\dd}\Phi +i\,[\bar{\Phi}, \Phi]=0\,.
\eea
This equation is nothing but the equation of motion of the principal chiral field. Indeed, introduce a 1-form $j=i(\Phi\,dz+\bar{\Phi}\,d\bar{z})$ with values in the Lie algebra $\mathfrak{u}(n)$. In this case~(\ref{dPhi}) together with the Hermitian conjugate equation may be written in the form of two conditions
\bea
d\ast j=0,\quad\quad dj-j\wedge j=0\,,
\eea
which are the e.o.m. of the principal chiral field. This is consistent with the fact, reviewed in section~\ref{complzerocurv}, that the Noether current of the model~(\ref{action}) is flat.

The condition $\Phi^2=0$ means that the Jordan structure of  $\Phi$ consists of $m_2$ cells of sizes $2\times 2$ and $m_1$ cells of sizes $1\times 1$. In this case $n=2m_2+m_1$ and $\mathrm{dim \;Ker}(\Phi)=m_1+m_2$. Since $\mathrm{Im}(U)\subset \mathrm{Ker}(\Phi)$ and $\mathrm{rk}(U)=M$, we get the condition $M\leq m_1+m_2$. This easily leads to\footnote{The inequalities are saturated in the case $\mathrm{Ker}(\Phi)\simeq \mathrm{Im}(U)$, when the number of $2\times 2$ cells is maximal and equal to $n-M$, and the number of cells of size $1\times 1$ is $2M-n$. Note that this is only possible in the case $M\geq {n\over 2}$. Reduction of the number of cells of type $2\times 2$ corresponds to the degeneration of the matrix $\Phi$.} $m_2\leq n-M$, $m_1\geq 2M-n$. 

The dynamical equation~(\ref{dPhi}) imposes severe constraints on the way in which the Jordan structure of the matrix  $\Phi$ can change as one varies the point $z, \bar{z}$ on the worldsheet. Indeed, it implies that $\Phi=kQ(z)k^{-1}$, where $Q(z)$ is a matrix that depends holomorphically on $z$. The Jordan structure of the matrix $Q(z)$ is the same as that of $\Phi$, and the vanishing of the Jordan blocks occurs holomorphically in~$z$. In particular, the Jordan structure changes only at ``special points'' -- isolated points on the worldsheet. As a result, ``almost everywhere'' the dimension of the kernel $\mathrm{dim\;Ker}(\Phi):=\widetilde{M}$ is the same, and the map $\lambda: (z, \bar{z})\to \mathrm{Ker}(\Phi)$ is a map to the Grassmannian $Gr_{\widetilde{M}, n}$. A more careful analysis of the behavior of $\Phi$ at a special point would show that $\lambda$ may be extended to these points. This may be summarized as follows: let $g(z, \bar{z})$ be a solution of the principal chiral model, i.e. a harmonic map to the group $G$, satisfying the condition $\Phi^2=0$, where $\Phi:=g^{-1}\dd g$ is a component of the Noether current, and let the dimension of the kernel of  $\Phi$ at a typical point of the worlsheet be $\widetilde{M}$. Then one can construct a harmonic map to the Grassmannian $Gr_{\widetilde{M}, n}$ by the rule $(z, \bar{z}) \to \mathrm{Ker}(\Phi)$.

\subsubsection{The partial flag manifold}

Let us extend the results of the previous section to more general flag manifolds. We return first to the equation for~$\Phi$:
\bea\label{dPhi1}
\bar{\dd}\Phi +i\,[\bar{\Phi}, \Phi]=0\,,
\eea
but this time we assume that the matrix $\Phi$ satisfies, in a typical point $(z, \bar{z})\in \Sigma$, the condition
\bea\label{phinilp}
\Phi^m=0 \quad\quad \textrm{and}\quad\quad \Phi^{m-1}\neq0\;.
\eea
The matrix $\Phi$ naturally defines a flag
\bea\label{kerflag}
f:=\;\;\{0\subset\mathrm{Ker}(\Phi)\subset \mathrm{Ker}(\Phi^2)\subset \cdots \subset \mathrm{Ker}(\Phi^m)\simeq \CC^n\}
\eea

The relation between the principal chiral model equations written above and the flag manifold models is summarized by the following assertion:

\vspace{0.3cm}
\fbox{\parbox{13.5cm}{
\centering 
Given a matrix $\Phi$ satisfying~(\ref{dPhi1})-(\ref{phinilp}), the map $(z, \bar{z}) \to f$ is a solution to the e.o.m. of the flag manifold sigma model~(\ref{currcompaction}).
}
}

\vspace{0.3cm}
To prove this, consider a matrix $U$ of the form $U=(U_{m-1}|\cdots|U_{1})$, where $U_i$ is a matrix, whose columns are the linearly independent vectors from $\mathrm{Ker}(\Phi^i)/\mathrm{Ker}(\Phi^{i-1})$. Let us relate the dimensions of these spaces to the dimensions of the Jordan cells of the matrix $\Phi$. To this end we bring $\Phi$ to the Jordan form
\bea
\Phi^{(0)}=\mathrm{Diag}\{J_{s_1}, \ldots, J_{s_\ell}\},\quad\quad \sum\limits_{j=1}^{\ell} s_j=n\,,
\eea
where $J_{s}$ is a Jordan cell of size $s\times s$. We have chosen the ordering $s_1\geq \ldots \geq s_\ell$, where, according to the supposition~(\ref{phinilp}), $s_1=m$.  We denote by $\kappa_i$ the number of Jordan cells of size at least $i$ ($\kappa_1=\ell$). The following two properties are obvious:
\begin{itemize}
\item $\kappa_{i+1}\leq \kappa_i$, i.e. $\kappa_1, \ldots, \kappa_m$ is a non-increasing sequence.
\item $\mathrm{dim \,Ker}(\Phi)=\kappa_1$, $\mathrm{dim \,Ker}(\Phi^2)=\kappa_1+\kappa_2$ etc., \\therefore $\mathrm{dim \,Ker}(\Phi^i)/\mathrm{Ker}(\Phi^{i-1})=\kappa_i$.
\end{itemize}
It follows that $U_i\in \mathrm{Hom}(\CC^{\kappa_i}, \CC^n)$ and $U\in \mathrm{Hom}(\CC^{M}, \CC^n)$, where $M=\sum\limits\limits_{i=1}^{m-1} \kappa_i$.

Since, by construction, $\mathrm{Im}(U)\simeq \mathrm{Ker}(\Phi^{m-1})$ and $\mathrm{Im}(\Phi)\subset \mathrm{Ker}(\Phi^{m-1})$, we have $\mathrm{Im}(\Phi)\subset \mathrm{Im}(U)$, i.e. there exists a matrix $V\in \mathrm{Hom}(\CC^n, \CC^M)$, such that $\Phi=-i\,UV$. Let us now derive the equations of motion for the matrices $U$ and~$V$. Since $\Phi^k U_k=0$ and $\bar{D}\Phi=0$, one has $\Phi^k \bar{D} U_k=0$. The columns of the matrix $(U_k|\cdots | U_1)$ span the kernel of $\Phi^k$, hence $\bar{D} U_k=-i\,\sum\limits_{j\leq k}\,U_j \bar{\mathcal{A}}_{jk}$, where $\bar{\mathcal{A}}_{jk}$ are matrices of relevant sizes. Out of the matrices $\bar{\mathcal{A}}_{jk}$ ($1\leq j\leq k\leq m-1$) we form a single matrix $\bar{\mathcal{A}}$, which schematically looks as in~(\ref{colormatr}). Then, clearly, the following equation is satisfied:
\bea
\bar{\mathscr{D}} U=\bar{\dd} U+i \,U \bar{\mathcal{A}}+i\,\bar{\Phi} U=0\,.
\eea
As $\Phi=-i\,UV$, from the non-degeneracy of $U$ it follows that $\bar{\mathscr{D}} V=0$. Because $U_1, \ldots, U_k \subset \mathrm{Ker}(\Phi^k)$, in the matrix $\Phi^k U\sim U(VU)^k$ the last $\sum\limits_{i=1}^k \kappa_i$ columns vanish, therefore the matrix $(VU)^k$ is strictly-lower-triangular and has zeros on the first $k$ block diagonals (the main diagonal is counted as the first one). We denote by $\mathfrak{k}$ the parabolic subalgebra of $\mathfrak{gl}(M)$ that stabilizes the subflag of~(\ref{kerflag}) with the last element omitted. We have proven that $VU\big|_{\mathfrak{k}^\ast}=0$. Therefore a solution $\Phi(z, \bar{z})$ of the system~(\ref{dPhi1})-(\ref{phinilp}) produces a solution $(U, V)$ to the equations of motion of the sigma model with target space the flag manifold
\bear
&&\!\!\!\!\!\!\!\!\!\!\!\!\!\!\!{\tU(n)\over \tU(\kappa_1)\times \cdots \times \tU(\kappa_m)}, \quad\quad \textrm{where}\\
\nonumber &&\!\!\!\!\!\!\!\!\!\!\!\!\!\!\!\kappa_j=\mathrm{dim \,Ker}(\Phi^j)/\mathrm{Ker}(\Phi^{j-1}) \quad \quad \textrm{is a non-increasing sequence.}
\eear
The complex structure on the flag is uniquely determined by the structure of the complex flag~(\ref{kerflag}).

\section{Sigma models as generalized Gross-Neveu models}\label{GNsec}

In the previous section we showed that the integrable flag manifold models that we formulated at the start of Section~\ref{compstructzerocurvsec} may equivalently be written in the form~(\ref{lagr5}) (which was motivated by the relation to 4D Chern-Simons theory) or in the form~(\ref{lagr4}), if one eliminates the auxiliary fields $\Phi, \bar{\Phi}$.  We have also seen in Section~\ref{4DCSsec} that, in principle, there is a way of constructing the (trigonometric and elliptic) deformations of these models by appropriately inserting the $r$-matrix into the Lagrangian. The purpose of this section is to emphasize, following~\cite{BykovGN}, that all of these systems -- either deformed or undeformed -- are really examples of \emph{chiral gauged bosonic} Gross-Neveu models. This way of formulating sigma models is not merely a simple reformulation, but offers substantial calculational benefits. For example, as we shall show, these methods allow to solve the renormalization group (generalized Ricci flow) equations and arrive at a beautiful universal one-loop solution. At the quantum level these models have chiral anomalies, which may be cancelled by adding fermions. We will show that this naturally leads to the notion of super-quiver-varieties and allows, among other things, to arrive at a novel formulation of supersymmetric models. 

\subsection{The bosonic chiral Gross-Neveu model}

We start by taking a closer look at the system~(\ref{lagr4}). It turns out very fruitful to rewrite it in Dirac form. We introduce $n$ `Dirac bosons'
\bea
\Psi_a=\begin{pmatrix} U_a\\ \bar{V}_a\end{pmatrix}\,,\quad\quad a=1, \ldots, n\,.
\eea
The Lagrangian~(\ref{lagr4}) is, in this notation,
\begin{empheq}[box=\fbox]{align}\label{psilagr}
\hspace{1em}\vspace{1em}
\mathcal{L}=\bar{\Psi_a} \slashed{D} \Psi_a + \left(\bar{\Psi_a}{1+\gamma_5\over 2}\Psi_a\right) \,\left(\bar{\Psi_b}{1-\gamma_5\over 2}\Psi_b\right)\,.\quad
\end{empheq}
The Dirac notations are standard: $\sigma_{1, 2}$ are the Pauli matrices,  $\slashed{\dd}:=\sum\limits_{i=1}^2\, \sigma_i \,\dd_i$ and $\gamma_5:=i\,\sigma_1\sigma_2$. The Lagrangian~(\ref{psilagr}) is the bosonic incarnation of the so-called chiral Gross-Neveu model (equivalently the $\SU(n)$ Thirring model~\cite{WittenThirring}) interacting with a gauge field. As in~(\ref{lagr4}) and in the example at the very start of this Chapter, the equivalence with the sigma model formulation (with a metric, $B$-field and possibly dilaton) is established through the elimination of $V$ and $\bar{V}$.

The model~(\ref{psilagr}) is `chiral',  meaning that there is a symmetry\footnote{In Minkowski signature this would have been the usual $\tU(1)$ chiral symmetry. This difference in chiral transformations has been observed in~\cite{Zumino, Mehta}.}
\bea
U\to \uplambda U, \quad V\to \uplambda^{-1} V\,, \quad\quad \textrm{where}\quad\quad \uplambda\in \CC^\times\,.
\eea
A general $\SU(n)$-invariant Lagrangian would only retain a $\tU(1)$-symmetry, $|\uplambda|=1$, and is not invariant under the full $\CC^*$, which arises in~(\ref{psilagr}) due to the chiral projectors. In other words, for a Euclidean worldsheet signature, chiral symmetry is equivalent to the complexification of the original (non-chiral) symmetry. This chiral symmetry is of extreme importance at least for two reasons:
\begin{itemize}
\item[$\circ$] Chiral symmetry ensures that the quartic interaction terms are quadratic in the $V$-variables, which are the `momenta' conjugate to $U$. This allows integrating them out and arriving at a metric form of the sigma model.
\item[$\circ$] We will also be interested in sigma models on projective spaces, Grasmannians etc., and these may be obtained by taking the quotient w.r.t. $\CC^*$, i.e. by gauging the chiral symmetry. In doing so, one needs to verify that it is free of anomalies.
\end{itemize}

\subsection[The deformed Gross-Neveu models]{The deformed Gross-Neveu models.}\label{classintsec}

As promised earlier, we proceed to discuss the deformations of the Gross-Neveu systems~(\ref{psilagr}), or equivalently of the original sigma models. As we shall see, the Gross-Neveu form of the model is particularly useful in this case. For example, the one loop renormalization group flow is described in this case by a couple of elementary Feynman diagrams, which should be contrasted with the highly non-trivial generalized Ricci flow equations that one obtains in the geometric formulation of the sigma model. This will also serve as our first step towards a definition and solution of these models at the quantum level.

We will first show how one can construct a deformation of the Lagrangian~(\ref{lagr5}). This will serve to relate our construction to the formulation in terms of the 4D Chern-Simons theory described in section~\ref{4DCSsec}.

Using the  three matrices $U \in \mathrm{Hom}(\CC^{M}, \CC^n), V\in \mathrm{Hom}(\CC^n, \CC^{M}), \Phi\in \mathrm{End}(\CC^{n})$ (we will always assume $M\leq n$) that we already encountered in Section~\ref{GLSMbetagamma}, we write down the deformed Lagrangian

\bea\label{lagr55}
\mathcal{L}=\mathrm{Tr}\left(V \bar{\mathscr{D}} U\right)+\mathrm{Tr}\left(\bar{U} \mathscr{D} \bar{V}\right)+\mathrm{Tr}\left(r_s^{-1}(\Phi) \bar{\Phi} \right)\,,
\eea
where $r_s$ is the classical $r$-matrix, depending on the deformation parameter $s$, that we encountered in Section~\ref{CYBEsec}. The covariant derivative is ${\bar{\mathscr{D}}U=\bd U+i\,\bar{\Phi} U+i \,U \bar{\mathcal{A}}}$, where $\mathcal{A}$ is a gauge field, whose structure depends on the actual target space under consideration, and $\Phi, \bar{\Phi}$ are auxiliary fields. We recall that, from the perspective of~\cite{CYa}, $\Phi, \bar{\Phi}$ are the components of the four-dimensional Chern-Simons gauge field along the worldsheet, and the quadratic term in $\Phi, \bar{\Phi}$ contains $r_s^{-1}$, because, as discussed in section~\ref{4DCSsec}, the Green's function of the gauge field is effectively the classical $r$-matrix. One can eliminate these auxiliary fields,  since they enter the Lagrangian quadratically, arriving at the following expression:
\bea \label{lagr6}
\mathcal{L}=\mathrm{Tr}\left(V \bar{D} U\right)+\mathrm{Tr}\left(\bar{U} D \bar{V}\right)+\mathrm{Tr}\left(r_s(U V) (U V)^\dagger\right)\,,\quad
\eea
where $\bar{D}U=\bd U+i \,U \bar{\mathcal{A}}$. Clearly, in Dirac notation this leads to the deformed Gross-Neveu model of the following form:
\begin{empheq}[box=\fbox]{align}
\hspace{1em}\vspace{1em}\label{lagr7}
\mathcal{L}=\bar{\Psi_a} \slashed{D} \Psi_a + (r_s)_{ab}^{cd}\left(\bar{\Psi_a}{1+\gamma_5\over 2}\Psi_c\right) \,\left(\bar{\Psi_d}{1-\gamma_5\over 2}\Psi_b\right)\,.
\end{empheq}
To summarize, we have obtained a chiral gauged bosonic Gross-Neveu model, where the deformation is encoded in the classical $r$-matrix that defines the quartic vertex.

\subsubsection[The zero curvature representation]{The zero curvature representation.}

The main property of the system~(\ref{lagr6})-(\ref{lagr7}) is that its e.o.m. admit a zero-curvature representation. To write  it down, we observe that in the undeformed case, when $r_s$ is proportional to the identity operator, the above Lagrangians have an $\SU(n)$ global symmetry and a corresponding Noether current one-form
$
\mathds{K}=K\,dz+\bar{K}\,d\bar{z}=U\, V\,dz+\bar{V} \,\bar{U}\,d\bar{z}\,.
$
Using this one-form, we define a family of connections, following~\cite{CYa}: 
\bea\label{rmatrflatconn}
\mathscr{A}=r_{\kappa_1}(K)\,dz-r_{\kappa_2}(\bar{K})\,d\bar{z}\,.
\eea
Here $\kappa_1, \kappa_2$ are complex parameters that will be related below. We wish to prove that the connection $\mathscr{A}$ is flat:
\bea\label{flatA}
d\mathscr{A}+\mathscr{A}\wedge \mathscr{A}=-dz\wedge d\bar{z}\,\left(r_{\kappa_2}(\dd \bar{K})+r_{\kappa_1}(\bd K)+[r_{\kappa_1}(K), r_{\kappa_2}(\bar{K})]\right)\overset{?}{=}0\,.
\eea
To this end, we will use the equations of motion of the model~(\ref{lagr6}). To write them out, we define the `conjugate' operator $\hat{r}$ by the relation $\mathrm{Tr}(r_s(A)\,B)=-\mathrm{Tr}(A\,\hat{r}_{s^{-1}}(B))$. When the unitarity relation~(\ref{rmunit}) holds, $\hat{r}=r$. The e.o.m. for the $U$ and $V$ variables may be shown to imply 

the following concise equations for the `Noether current' $K$:
\bear\label{noethcurrdefflat}
\bd K=[\hat{r}_{s^{-1}}(\bar{K}), K]\,,\quad\quad
\dd \bar{K}=[\bar{K}, r_s(K)]\,.
\eear
These are the deformed versions of the equations~(\ref{dPhi1}) that we encountered earlier (since in eliminating $\Phi$ from (\ref{lagr55}) one easily sees that in the undeformed case $\Phi\sim K$). Substituting in the equation~(\ref{flatA}), we see that it is satisfied if the matrix $r$ obeys the equation
\bea\label{CYBE1}
r_{\kappa_2}([\bar{K}, r_s(K)])+r_{\kappa_1}([\hat{r}_{s^{-1}}(\bar{K}), K])+[r_{\kappa_1}(K), r_{\kappa_2}(\bar{K})]=0\,.
\eea
The reason why in our case $\hat{r}$ is not necessarily equal to $r$ is that we will mostly be dealing with a non-simple Lie algebra $\mathfrak{gl}(n)=\mathfrak{sl}(n)\oplus \CC$ ($\CC\mysub \mathfrak{gl}(n)$ corresponds to matrices proportional to the unit matrix). We will assume a block-diagonal $r$-matrix, acting as follows:
\bea\label{glnr}
r_s=(r_s)_{\mathfrak{sl}(n)}+(r_s)_{\CC}\,,\quad\quad (r_s)_{\mathfrak{sl}(n)}\in \mathrm{End}(\mathfrak{sl}(n)),\quad\quad (r_s)_{\CC}:=b(s)\,\mathrm{Tr}.
\eea
Here $(r_s)_{\mathfrak{sl}(n)}$ acts on traceless matrices as in~(\ref{rR}), and $(r_s)_{\CC}$ acts on matrices of the type $\alpha\cdot\mathds{1}$ as multiplication by  $n \,b(s)$. In this case $\hat{r}_s=r_s-(b(s^{-1})+b(s))\,\mathrm{Tr}$. If the unitarity relation is satisfied, the mismatch vanishes. However, in either case $b(s)$ completely drops out from the equation~(\ref{CYBE1}), so we will prefer allowing an arbitrary function $b(s)$ for the moment.

Postulating the relations $\kappa_1=u,\; \kappa_2=uv,\; v=s^{-1}$ (implying $\kappa_1=\kappa_2\,s\,$) between the parameters, we identify~(\ref{CYBE1}) with the classical Yang-Baxter equation for $\mathfrak{g}=\mathfrak{sl}(n)$ from Section~\ref{CYBEsec}.

\subsection[The $\beta$-function and the Ricci flow]{The $\beta$-function and the Ricci flow.}\label{betafuncsec}
\begin{figure}
\centering
\bea\nonumber
\begin{tikzpicture}[
baseline=-\the\dimexpr\fontdimen22\textfont2\relax,scale=1.3]
\draw[-stealth, blue!50, line width=2pt, rounded corners] (-1,0)  -- (0,0) node[right,black] {};
\draw[blue!50, line width=2pt, rounded corners] (-0.5,0)  -- (1,0) node[right,black] {};
\draw[-stealth, red!50, line width=2pt, rounded corners] (2,0)  -- (3,0) node[right,black] {};
\draw[red!50, line width=2pt, rounded corners] (2.5,0)  -- (4,0) node[right,black] {};
\draw[-stealth, blue!50, line width=2pt, rounded corners] (5,0)  -- (5.6,0) node[right,black] {};
\draw[-stealth, blue!50, line width=2pt, rounded corners] (5.3,0)  -- (6.7,0) node[right,black] {};
\draw[blue!50, line width=2pt, rounded corners] (6.3,0)  -- (7,0) node[right,black] {};
\draw[-stealth, red!50, line width=2pt, rounded corners] (6,1)  -- (6,0.4) node[right,black] {};
\draw[-stealth, red!50, line width=2pt, rounded corners] (6,0.7)  -- (6,-0.7) node[right,black] {};
\draw[red!50, line width=2pt, rounded corners] (6,-0.4)  -- (6,-1) node[right,black] {};
\node at (-1,-0.3) {\footnotesize $z_1$};
\node at (1,-0.3) {\footnotesize  $z_2$};
\node at (0,0.5) {${1\over z_2-z_1}$};
\node at (0,-0.4) {\footnotesize $i$};
\node at (2,-0.3) {\footnotesize  $z_1$};
\node at (4,-0.3) {\footnotesize  $z_2$};
\node at (3,0.5) {$-{1\over \bar{z_2}-\bar{z_1}}$};
\node at (3,-0.4) {\footnotesize $j$};
\node at (5.3,-0.3) {\footnotesize $i$};
\node at (6.8,-0.3) {\footnotesize $j$};
\node at (6.25,0.7) {\footnotesize $k$};
\node at (6.25,-0.7) {\footnotesize $l$};
\node at (7.3,0.7) { $-(r_s)_{ij}^{kl}$};
\end{tikzpicture}
\eea
\caption{Feynman rules of the deformed model~(\ref{lagr6})-(\ref{lagr7}), with $\mathcal{A}=0$.} \label{fig1}
\end{figure}
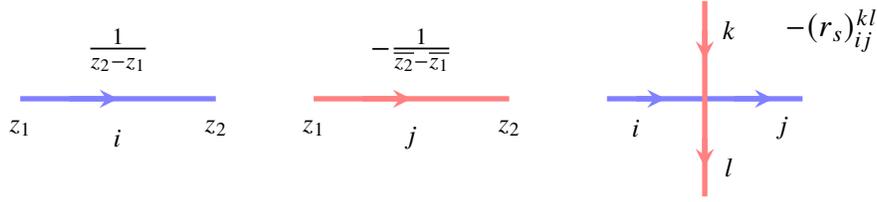

\begin{figure}
\centering
\bea\nonumber
\begin{tikzpicture}[
baseline=-\the\dimexpr\fontdimen22\textfont2\relax,scale=1.2]
\draw [-stealth, red!50, line width=2pt] (2,0) arc [radius=1, start angle=100, end angle= 120];
\draw [-stealth, red!50, line width=2pt] (2,0) arc [radius=1, start angle=100, end angle= 180];
\draw [-stealth, red!50, line width=2pt] (2,0) arc [radius=1, start angle=100, end angle= 255];
\draw [red!50, line width=2pt] (2,0) arc [radius=1, start angle=100, end angle= 260];
\draw [-stealth, blue!50, line width=2pt] (1,0) arc [radius=1, start angle=80, end angle= 60];
\draw [-stealth, blue!50, line width=2pt] (1,0) arc [radius=1, start angle=80, end angle= 0];
\draw [-stealth, blue!50, line width=2pt] (1,0) arc [radius=1, start angle=80, end angle= -75];
\draw [blue!50, line width=2pt] (1,0) arc [radius=1, start angle=80, end angle= -80];
\node at (2.2,0) {\footnotesize $k$};
\node at (0.8,0) {\footnotesize $i$};
\node at (0.8,-2) {\footnotesize $j$};
\node at (2.2,-2) {\footnotesize $l$};
\draw [-stealth, red!50, line width=2pt] (5,-1.95) arc [radius=1, start angle=260, end angle= 240];
\draw [-stealth, red!50, line width=2pt] (5,-1.95) arc [radius=1, start angle=260, end angle= 180];
\draw [-stealth, red!50, line width=2pt] (5,-1.95) arc [radius=1, start angle=260, end angle= 110];
\draw [red!50, line width=2pt] (5,-1.95) arc [radius=1, start angle=260, end angle= 100];
\draw [-stealth, blue!50, line width=2pt] (4,0) arc [radius=1, start angle=80, end angle= 60];
\draw [-stealth, blue!50, line width=2pt] (4,0) arc [radius=1, start angle=80, end angle= 0];
\draw [-stealth, blue!50, line width=2pt] (4,0) arc [radius=1, start angle=80, end angle= -75];
\draw [blue!50, line width=2pt] (4,0) arc [radius=1, start angle=80, end angle= -80];
\node at (5.2,0) {\footnotesize $l$};
\node at (3.8,0) {\footnotesize $i$};
\node at (3.8,-2) {\footnotesize $j$};
\node at (5.2,-2) {\footnotesize $k$};
\end{tikzpicture}
\eea
\caption{Diagrams contributing to the $\beta$-function at one loop.} \label{fig2}
\end{figure}
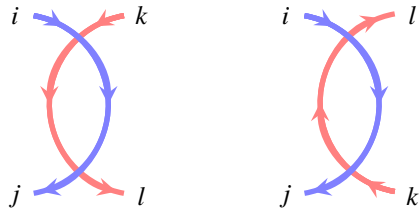

In this section we turn to the analysis of the elementary quantum properties of the theory defined by the Lagrangian~(\ref{lagr6})-(\ref{lagr7}), in the ungauged case $\mathcal{A}=0$. The main question we pose is whether this Lagrangian preserves its form after renormalization, at least to one loop order -- in other words, whether it is sufficient to renormalize the parameters of the $r$-matrix. To this end we write out the Feynman rules of the system in Fig.~\ref{fig1}.
At one loop the two diagrams contributing to the renormalization of the quartic vertex are shown in Fig.~\ref{fig2}.
There is a relative sign between the two diagrams, due to the different directions of the lines in the loops. Otherwise, the type of the divergence is the same -- it is logarithmic\footnote{Here we are talking about UV divergences.}, proportional to~$\int\frac{d^2z}{z \bar{z}}$. As a result, the one-loop $\beta$-function is

\begin{empheq}[box=\fbox]{align}\label{betafunc}
\hspace{1em}\vspace{1em}
\beta_{ij}^{kl}=\sum\limits_{p, q=1}^{n}\,\left((r_s)_{ip}^{kq} (r_s)_{pj}^{ql}-(r_s)_{ip}^{ql} (r_s)_{pj}^{kq}\right)\quad
\end{empheq}
As already discussed earlier, we will assume a block-diagonal $r$-matrix~(\ref{glnr}), where $(r_s)_{\mathfrak{sl}(n)}$ acts as in~(\ref{rR}), and $(r_s)_{\CC}$ acts on a unit matrix as multiplication by $n\cdot b(s)$. This is translated into the four-index notation as follows:
\bea\label{rsu1}
(r_s)_{ij}^{kl}=
a_{kl}(s)\,\left(\delta_i^k \delta_j^l-{1\over n}\delta_{ij}\delta^{kl}\right)
+b(s)\,\delta_{ij}\delta^{kl}\,.
\eea

The coefficients $a_{kl}(s)$ are defined by the action of~(\ref{rR}) in the standard basis: $r_s(e_k\otimes e_l)=a_{kl}(s)\,e_k\otimes e_l$. Concretely (see~(\ref{abg})),
\bea\label{acoefs}
a_{ij}=\begin{cases}
\quad \frac{s}{1-s}=\upalpha,\quad\quad \;\;i<j\\
\quad\frac{1}{1-s}=\upbeta,\quad\quad \;\;i>j\\
\quad{1\over 2}\frac{1+s}{1-s}=\upgamma,\quad\quad i=j\,.
\end{cases}
\eea
Substituting~(\ref{rsu1}) in~(\ref{betafunc}) and doing the summations, we obtain

\bea\label{betafunc2}
\beta_{ij}^{kl}=\left[\frac{ns}{(1-s)^2}+(i-j)a_{ij}\right]\,
\underbracket[0.6pt][0.6ex]{
\left(\delta_i^k \delta_j^l-{1\over n}\delta_{ij}\delta^{kl}\right)
}_{:=\Pi_{ij}^{kl}}\,.
\eea
Since $a_{ij}={1\over 2}\frac{1+s}{1-s}+{i\over 2}\,\mathcal{R}_{ij}$, the one-loop result is not of the form~(\ref{rsu1}) (due to the term proportional to $i-j$). For this reason the straightforward Ricci flow equation ${d r_s\over d\tau}=\beta$ for $s(\tau)$ does not have a solution. However, this can be easily remedied by allowing reparametrizations of coordinates along the flow. Let us reparametrize
\bea\label{rescaling}
U\to \kappa U, \quad V\to V \kappa^{-1}\,,\quad\quad \textrm{where} \quad\quad \kappa=\mathrm{Diag}\{\kappa_1, \ldots , \kappa_n\}\,.
\eea
The kinetic term in~(\ref{lagr6}) is invariant, so the only effect is in the effective replacement of the $r$-matrix by $\tilde{r}$, where $(\tilde{r}_s)_{ij}^{kl}=\frac{\kappa_i \bar{\kappa_k}}{\kappa_j \bar{\kappa_l}}(r_s)_{ij}^{kl}$. One has to conjugate the $\beta$-function tensor analogously, and the equation we will aim to solve is ${d \tilde{r}_s\over d\tau}=\tilde{\beta}$. It may be rewritten as an equation for the original $r$-matrix as follows:
\bear
&&{d\over d\tau}(r_s)_{ij}^{kl}=\beta_{ij}^{kl}-{d\over d\tau}\left(\log{\left(\frac{\kappa_i \bar{\kappa_k}}{\kappa_j \bar{\kappa_l}}\right)}\right)\,(r_s)_{ij}^{kl}:=\widehat{\beta}_{ij}^{kl}\,\\ \nonumber
&&\textrm{where}\quad\quad \widehat{\beta}_{ij}^{kl}=\left[\frac{ns}{(1-s)^2}+(i-j)a_{ij}-{d\over d\tau}\left(\log{\left(\big|{\kappa_i\over\kappa_j}\big|^2\right)}\right) a_{ij}\right]\,\Pi_{ij}^{kl}
\eear
The unwanted term may now be canceled by the simple substitution
\bea\label{coordflow}
\kappa_j=e^{{\tau\over 2}\,j}\,.
\eea
Recalling again the expression for $a_{ij}(s)$, we find that the remaining equations may be written as $\dot{b}=0$, ${d\over d\tau}\left({1\over 2}{1+s\over 1-s}\right)={ns\over (1-s)^2}$, or $\dot{s}=ns$. Therefore
\begin{empheq}[box=\fbox]{align}\label{sflow}
\hspace{1em}
b=\mathrm{const.},\quad s=e^{n\,\tau}\quad 
\end{empheq}
In the geometric formulation (i.e. for the sigma model defined by a metric, $B$-field and dilaton) the RG-flow equations are the generalized Ricci flow equations. They look as follows~\cite{Curci, Polchinski}:
\bear
&&-\dot{g_{ij}}=R_{ij}+{1\over 4}H_{imn}H_{jm'n'}g^{mm'}g^{nn'}+2\,\nabla_i\nabla_j \Phi\,,\quad\\ \nonumber
&&-\dot{B_{ij}}=-{1\over 2}\,\nabla^k\,H_{kij}+\nabla^k\Phi\,H_{kij}\,,\quad\\ \nonumber
&&-\dot{\Phi}={\mathrm{const.}}-{1\over 2}\,\nabla^k\nabla_k\,\Phi+\nabla^k\Phi \nabla_k \Phi+{1\over 24}\,H_{kmn}H^{kmn}\,,
\eear
where $g$ is the metric, $B$ is the skew-symmetric field with $H$ its `curvature', and $\Phi$ is the dilaton. We arrive at the important conclusion that the trigonometrically deformed system~(\ref{lagr6})-(\ref{lagr7}) with $\mathcal{A}=0$ provides a solution to these Ricci flow equations. Moreover, one might appreciate the simplicity of the solution via the Gross-Neveu formulation. For more on this we refer the reader to~\cite{BykovGN}.

\subsubsection{$\beta$-functions of homogeneous models}

The calculation that we presented above was performed for the ungauged model ($\mathcal{A}=0$). It turns out, however, that the introduction of the gauge field does not alter the $\beta$-function. Moreover, one can easily see that, even if we replace the size-$n$ vectors $U$ and $V$ by $M\times n$-matrices, which would be necessary for considering Grassmannian $Gr_{M, n}$ or flag manifold target spaces, the calculation again leads to exactly the same answer, since the additional matrix index is simply a spectator index for the diagrams in Fig.~\ref{fig2}. This phenomenon is rather remarkable and has implications for the undeformed (i.e. homogeneous) models as well. In the geometric formulation this means that the metric and $B$-field of \emph{all} homogeneous models satisfy the generalized Einstein condition with the same `cosmological constant', equal to $n$:
\bea\label{genEinstein}
R_{ij}+{1\over 4}H_{imn}H_{jm'n'}g_{\mathrm{hom.}}^{mm'}g_{\mathrm{hom.}}^{nn'}=n \,(g_{\mathrm{hom.}})_{ij}\,.
\eea
Here $g_{\mathrm{hom.}}$ is the homogeneous metric, and $H$ is the curvature of the $B$-field equal to the fundamental Hermitian form of the metric. As we recall from the discussion in Section~\ref{symmspace}, in the case of Grassmannians the metric $g_{\mathrm{hom.}}$ is K\"ahler, so that the fundamental Hermitian form is closed and $H=0$. In that case the equation~(\ref{genEinstein}) translates into the usual Einstein equation $R_{ij}=n \,(g_{\mathrm{hom.}})_{ij}$. This is the well-known fact that the one-loop $\beta$-function for symmetric space models is equal to the dual Coxeter number of the symmetry group $G$ (in particular, it is independent of the denominator $H$ for symmetric spaces $G\over H$), cf.~\cite{ZJBig} as a general reference and~\cite{Morozov} for the case of Hermitian symmetric spaces, as well as the lectures~\cite{ZaremboLect}. In the case of K\"ahler symmetric spaces -- such as Grassmannians -- there is an alternative explanation: since the Ricci form represents the first Chern class of the manifold, one can attribute the cosmological constant $n$ to the fact that $c_1(Gr_{M, n})=n\,[\mathscr{C}]$, where $\mathscr{C}$ is a generator of $H^2(Gr_{M, n}, \mathbb{Z})$. For non-symmetric spaces -- such as the general flag manifolds -- this logic does not work, and there is no direct relation between the first Chern class and the dual Coxeter number. A related fact is that for non-symmetric spaces the metric in our sigma models is not K\"ahler, and one instead has to take into account the non-zero field $H$ in~(\ref{genEinstein}). The effect of including this field is that the two terms in the $\beta$-function -- the curvature and the $H$-field term -- sum up in such a way that the $\beta$-function is again proportional to $n$.

\subsubsection{The `sausage' example}

Before concluding this section, let us describe a simplest example, namely the deformation of the sphere $S^2$ -- the so-called `sausage'~\cite{FOZ}. Already on this example one can see all the salient features of the general solution described above. 
Let us derive this solution, starting from the Lagrangian~(\ref{lagr6}) of the deformed model and making the rescalings~(\ref{rescaling}), (\ref{coordflow}):
\begin{eqnarray} \label{lagr66}
\mathcal{L}  &=  V\circ \bar{D} U+  \bar{U}\circ D\bar{V} +\mathrm{Tr}\left(r_s(\kappa U\otimes V \kappa^{-1}) (\kappa U\otimes V \kappa^{-1})^\dagger \right)\\ \nonumber
 &=  V\circ \bar{D} U- D \bar{U}\circ \bar{V} +\upgamma (|U_1|^2  |V_1|^2+|U_2|^2  |V_2|^2) \\
 \nonumber  &+ \upalpha\, e^{-\tau}\, |U_1|^2\cdot |V_2|^2+\upbeta\, e^{\tau}\, |U_2|^2\cdot |V_1|^2.
\end{eqnarray}
Next we pass to the inhomogeneous gauge $U_1=1$ and relabel $U_2:=W, V_2:=V$. Variation w.r.t. the gauge field gives the constraint $V\circ U=0$, which is solved by $V_1=-W\cdot V$. The Lagrangian acquires the following form in these coordinates:
\bea
\mathcal{L}=V \cdot \bar{D} W- \bar{V}\cdot  D \bar{W}   +  \left(2 \upgamma |W|^2 +\upalpha\, e^{-\tau}\, +\upbeta\, e^{\tau}\, |W|^4\right) |V|^2
\eea
Eliminating the fields $V, \bar{V}$ and  using the expressions~(\ref{acoefs}) for $\upalpha, \upbeta, \upgamma$ with $s=e^{2\tau}$, we arrive at 
\bea\label{sausagelagr}
\mathcal{L}=\frac{\left(e^{-\tau}-e^{\tau}\right)|\bar{D} W|^2}{\left(e^{\tau}+|W|^2\right)\left(e^{-\tau}+|W|^2\right)}\,,
\eea
which corresponds to the `sausage' metric. As all models of the type~(\ref{actioncompl}), the Lagrangian~(\ref{sausagelagr}) features a $B$-field equal to the fundamental Hermitian form of the metric, however since this is a manifold of complex dimension $1$, the $B$-field is closed, so that $H=0$. One can also show that in this exceptional case the dilaton $\Phi$ is constant (for details see~\cite{BykovGN}), so that the Ricci flow equation is especially simple:
\bea
-{d g_{ij}\over d\tau}=R_{ij}\,.
\eea
The range of the Ricci time variable is $\tau\in (-\infty, 0)$ (accordingly $s=e^{2\tau}\in(0,1)$), and the corresponding solution is called `ancient' in the terminology used in Ricci flow literature. Geometrically the `sausage' looks exactly as its name suggests:
\bea
\begin{tikzpicture}[
baseline=-\the\dimexpr\fontdimen22\textfont2\relax,scale=1]
\draw[mygreen1, line width=2pt] (-2,0.5) --   (2,0.5);
\draw[mygreen1, line width=2pt] (-2,-0.5) --   (2,-0.5);
\draw [mygreen1, line width=2pt] (2,0.5) arc [radius=0.5, start angle=90, end angle= -90];
\draw [mygreen1, line width=2pt] (-2,0.5) arc [radius=0.5, start angle=90, end angle= 270];
\node at (5.5,0) {$0<s<1$};
\node at (0,0) {Length $\sim |\log{s}|$};
\end{tikzpicture}
\eea
The solution has two characteristic regimes. The first one is  $s\to 0$, in which case  $ds^2\to\frac{|dW|^2}{|W|^2}$, so that one obtains an infinitely long cylinder. Since the cylinder is flat, this should be interpreted as the UV  limit with asymptotic freedom. The opposite regime is the IR limit $s\to 1$, where one obtains a round metric on $\CP^1$, albeit with a vanishing radius, which is a sign of an IR singularity. The same behavior persists qualitatively in the case of $\CP^{n-1}$~\cite{BykovGN}, where Ricci flow interpolates between a cylinder $(\CC^\ast)^{n-1}$ in the UV (asymptotic freedom) and a `round' projective space of vanishing radius in the IR (one has a similar behavior in the case of the so-called $\eta$-deformed $\CP^{n-1}$, as shown in~\cite{BykovLust}).

\subsection{Sigma models with polynomial interactions}\label{polintsec}

In the previous sections we have claimed that the sigma models of a wide class, including the familiar $\CP^{n-1}$, Grassmannian, flag, etc. models, are equivalent to chiral Gross-Neveu models. One remarkable consequence of this fact is that the corresponding sigma models are therefore models with polynomial interactions, and all non-linear constraints that are usually present in conventional formulations can be bypassed.

The model~(\ref{lagr6})-(\ref{lagr7}) with $\mathcal{A}=0$ is a Gross-Neveu model with purely quartic interactions. One might wonder, what happens when one includes the gauge field, for example as in the $\CP^{n-1}$-model. This system is described by~(\ref{lagr6}), where $U$ is a column vector and $V$ a row vector, both of length $n$, and additionally one has a $\CC^\ast$ gauge field $\mathcal{A}$. It turns out the gauge field may be completely eliminated by passing to the inhomogeneous $\CC^\ast$-gauge
\bea\label{inhomgauge}
U_n=1\,.
\eea
Variation of the Lagrangian~(\ref{lagr6}) w.r.t. $\bar{\mathcal{A}}$ gives $V\circ U=0$, so that 
\bea\label{Vinhom}
V_n=-\sum\limits_{k=1}^{n-1}\,V_k U_k\,.
\eea
This removes the gauge field at the expense of modifying the Feynman rules of the theory. Dropping fermionic fields for the moment, we write the Lagrangian of the model~(\ref{lagr6}) in this gauge:
\bear\label{inhomgaugelagr}
&&\mathcal{L}=\sum\limits_{k=1}^{n-1}\,\left(V_k \bd U_k-\bar{V}_k \dd \bar{U}_k+\upbeta |V_k|^2\right)+\\ \nonumber
&&\underbracket[0.6pt][0.6ex]{+\sum\limits_{l, m=1}^{n-1}\,a_{lm}\,|U_l|^2 |V_m|^2+\upgamma\,\big|\sum\limits_{p=1}^{n-1}\,U_p V_p\big|^2}_{\textrm{quartic vertices}}+\underbracket[0.6pt][0.6ex]{\upalpha \,\left(\sum\limits_{k=1}^{n-1}\,|U_k|^2\right)\,\big|\sum\limits_{p=1}^{n-1}\,U_p V_p\big|^2 }_{\textrm{sextic vertices}}
\eear
We see that the propagators and vertices are modified, and on top of that sextic vertices have appeared. A fascinating feature of the Lagrangian~(\ref{inhomgaugelagr}) is that its interaction terms are again polynomial in the $(U, V)$-variables. In other words, instead of a nonlinear $\sigma$-model we have arrived at a different nonlinear theory -- the theory of several bosonic fields (albeit with fermionic propagators) with polynomial interactions. 
One might also notice that the procedure of going to inhomogeneous coordinates that we have just described is in fact reminiscent of what we did at the end of section ~\ref{HPDMsec} while introducing the so called Dyson-Maleev variables. In other words, the generalized Dyson-Maleev variables allow to eliminate the gauge fields and turn the sigma models into models of multiple bosonic fields with polynomial interactions.

Another fascinating parallel that this discussion invokes is that with Ashtekar variables in 4D general relativity. It is well-known that dimensional reduction of general relativity, possibly with additional matter fields, along two commuting Killing vectors leads to integrable sigma models~\cite{Geroch, Belinsky, MaisonIntegr} (see~\cite{Nicolai} for a review). The target space depends on the particular gravitational system that one started with~\cite{GibbonsMaison, BreitenlohnerMaison}. For example, in the case of pure gravity one gets $SL(2, \mathbb{R})\over SO(2)$, whereas gravity with $n-1$ vector fields leads to $\tSU(n, 1)\over \text{S}(\tU(n)\times \tU(1))$ -- the hyperbolic analogue of projective space. These are all complex (Hermitian) symmetric spaces, albeit of Minkowski signature, which makes it slightly different from what we encountered in most of this article, but the general structure of the models is the same.  An interesting consequence comes from the salient property of Ashtekar variables in general relativity, namely that they make the interactions polynomial, cf.~\cite{Ashtekar}. This of course hints on the relation to the polynomiality of interactions in the sigma models that we have just discussed. A more careful analysis~\cite{Zagermann} shows that the Noether currents $K$ of the sigma model are bilinear combinations of Ashtekar's canonical variables $(A, E)$: $K\sim A\otimes E$. On the other hand, these same Noether currents, when calculated from the Lagrangian~(\ref{psilagr}) or~(\ref{lagr6}), have the form $K\sim U\otimes V$, so that the canonical variables $(U, V)$ may be naturally interpreted as the dimensional reductions of Ashtekar variables.

Apart from the polynomiality  of interactions, the latter observation has yet another important consequence. Since the $(A, E)$ variables are canonical, one has the Poisson brackets $\{A(x), E(y)\}\sim \delta(x-y)$, and as a result the Noether currents $K$ have local Poisson brackets as well, schematically of the form $\{K(x), K(y)\}\sim K(x)\,\delta(x-y)$. This of course also immediately follows from the analogous Poisson structure of the $(U, V)$ variables, which can be  seen from the first-order  Lagrangians~(\ref{psilagr}), or~(\ref{lagr6}). Since the flat connection~(\ref{rmatrflatconn}) is linear in the $K$-variables, its components also have ultralocal Poisson brackets of the standard form~\cite{DelducKameyama} (for background see~\cite{ReshFad, STSh, Kulish}). As simple as it may look from this perspective, it is a rather exceptional property for sigma models, where the Poisson brackets of Lax operators typically produce non-ultralocal terms~\cite{ReshFad2, Maillet1, Maillet2}, proportional to $\delta'(x-y)$. The latter cause significant difficulties in the discretization of such systems, which may therefore be overcome for the integrable models discussed in this Chapter.

\subsection{Integrable models related to quiver varieties}\label{superquiversec}

The formulation of sigma models in terms of Gross-Neveu models  suggests a natural, but rather far-reaching, generalization~\cite{BykovSUSYCPN}. The first step towards this generalization is to realize that at the quantum level one is forced to supplement the purely bosonic models described above with fermions. This is necessary because in general gauged models of the type~(\ref{psilagr}) or (\ref{lagr6})-(\ref{lagr7}) suffer from gauge anomalies. These are in fact a property of the kinetic term in the Lagrangian~(\ref{psilagr}), so that to this end we may omit the interaction term. The one-loop determinant of the matter fields leads us to Schwinger's calculation~\cite{Schwinger} of the effective action of the gauge fields $\mathcal{A}$:
\bea\label{effaction}
\mathcal{S}_{\mathrm{eff.}}={\upxi\over 2} \int\,dz\,d\bar{z}\,F_{z\bar{z}}{1\over \triangle}F_{z\bar{z}}\,,\quad\quad F_{z\bar{z}}=i\,(\dd \bar{\mathcal{A}}-\bd \mathcal{A})\,.
\eea
The coefficient $\upxi$ collects some numerical factors and is proportional to the number of matter fields we have integrated over. The action is invariant w.r.t. the gauge transformations of the original $\tU(1)$, $\mathcal{A}\to \mathcal{A}+\dd \alpha$,  $\bar{\mathcal{A}}\to \bar{\mathcal{A}}+\bd \alpha$ ($\alpha\in \mathbb{R}$), but not w.r.t. the complexified $\CC^\ast$ gauge transformations $\mathcal{A}\to \mathcal{A}+\dd \alpha$,  $\bar{\mathcal{A}}\to \bar{\mathcal{A}}+\bd \bar{\alpha}$ ($\alpha\in \CC$). The non-Abelian analogue of this calculation leads to the WZNW action, as shown in~\cite{PW} and discussed in~\cite{Nair1, Nair2} in the Euclidean case, but even the simple abelian effective action~(\ref{effaction}) suffices for most purposes. For example, it is clear that the anomaly may be canceled by including fermions symmetrically with the bosons:
\bea\label{cpnferm0}
\mathcal{L}\to \bar{\Psi_a} \slashed{D} \Psi_a + \bar{\Theta_a} \slashed{D} \Theta_a\,,
\eea
where $\Psi$ are the bosons and $\Theta$ the fermions. In this case the respective determinants  cancel. We emphasize that such a simple mechanism is possible because we have rewritten the bosonic part of the theory in fermionic form in the first place.  

In~(\ref{cpnferm0}) we have dropped the interaction terms to emphasize that the anomaly is a property of the kinetic term in the Lagrangian. As it turns out,  the kinetic term has a clear geometric meaning and defines what may be called the `super-phase space' of the model. Moreover, the whole theory of generalized integrable Gross-Neveu models of the type~(\ref{psilagr}) can be cast in pure differential-geometric terms.

The relevant geometric context is as follows. Suppose we have a super phase space~$\mathbf{\Phi}$, which is a complex symplectic (quiver) supervariety. There is a gauge (super)-group $\mathbf{G}_{\textrm{gauge}}$ acting in the nodes of the quiver, and matter fields $U\in \mathsf{W}$, $V\in \bar{\mathsf{W}}$ are in representations $\mathsf{W}\oplus \bar{\mathsf{W}}$ of $\mathbf{G}_{\textrm{gauge}}$. We assume that the quiver is `doubled', meaning that every representation arises together with its dual (Nakajima quivers have this property~\cite{Nakajima, NakajimaReview}). Apart from the gauge nodes, the quiver will typically have some global nodes with an action of a complex global symmetry  (super)-group~$\mathbf{G}_{\textrm{global}}$. We can therefore define the complex moment map $\mu$ for the action of $\mathbf{G}_{\textrm{global}}\circlearrowright\mathbf{\Phi}$. In this setup one can, quite naturally, define the following Lagrangian:
 
\vspace{-0.2cm}
\begin{empheq}[box=\fbox]{alignat=3}
\hspace{1em}\vspace{1em} \label{SUSYlagrGen}
&\quad \mathcal{L}=\left(V\cdot \bar{\mathcal{D}} U+\bar{U}\cdot \mathcal{D} \bar{V}\right)+\vkappa\,\mathrm{STr}(\mu \,\bar{\mu})\,.\quad
\end{empheq}
The Gross-Neveu system~(\ref{psilagr}) is a special case. The kinetic term in~(\ref{SUSYlagrGen}) corresponds to the $\beta\gamma$-systems -- it is a pull-back of the canonical Poincar\'e-Liouville one-form corresponding to the complex symplectic form of the quiver. The second term provides a coupling between the holomorphic and (anti)-holomorphic $\beta\gamma$-systems and comes with an arbitrary coefficient $\vkappa$ that should be seen as a coupling constant (in the sigma model setup this is the inverse squared radius of the target space). One can directly show that the moment map $\mu$ satisfies the e.o.m.
\bea\label{mueq0}
\bd \mu=\vkappa\,[\bar{\mu}, \mu]\,,
\eea
which is the e.o.m. of the principal chiral model, thus once again pointing at the relation between the models, which has already been emphasized in section~\ref{PCMrelsec}. In particular, we already encountered the above equation in~(\ref{dPhi1}) and its deformed version in~(\ref{noethcurrdefflat}) (for models with kinetic term given by the Poincar\'e-Liouville one-form the components of the Noether current coincide with the moment map variables).

One also needs to impose the chiral anomaly cancellation conditions that in the general setup have the form
\bea\label{anomcancelGen}
\mathrm{Str}_{\mathsf{W}}(T_a T_b)=0\,,\quad \textrm{where} \quad  T_a, T_b\in \mathfrak{g}_{\textrm{gauge}}\,.
\eea
These are in fact the Euclidean analogues of the standard anomaly cancellation conditions in the WZNW models~\cite{WittenFactor}. We expect that in most cases the condition $\mathrm{Str}_{\mathsf{W}}(T_a)=0$ holds as well (this is so in all known examples, at least). One conjectures that the Lagrangian~(\ref{SUSYlagrGen}), supplemented with the conditions~(\ref{anomcancelGen}), defines a quantum integrable model. In particular, the well-known quantum anomalies~\cite{AbdallaAnomaly, AbdallaCancel} in the Yangian charges of the model, or in other words in the so-called L\"uscher's non-local charge~\cite{LuscherNonlocal, Bernard1, Bernard2} (these results are reviewed in the book~\cite{AbdallaBook}), should cancel as a result of the cancellation of the chiral anomalies.

Let us specify what this formal setup brings in the case of the $\CP^{n-1}$-model with fermions. All known models, whose bosonic part is the $\CP^{n-1}$ sigma model, can be attributed to one of the two cases. In both of these cases the phase spaces are complex symplectic  quotients of the form
\bea\label{compsympred}
\mathbf{\Phi}=(T^\ast \CC^{n|n})\!\sslash\! \mathbf{G}_{\textrm{gauge}}\,,
\eea
where $\mathbf{G}_{\textrm{gauge}}$ is a subgroup of $\text{GL}(1|1)\mysub \text{GL}(n|n)$, the latter being the symmetry group of $\mathsf{T^\ast} \CC^{n|n}$. In this language, the two cases are distinguished by the choice of $\mathbf{G}_{\textrm{gauge}}$:
\begin{itemize}
    \item[$\circ$] The `minimal fermions' phase space: $\mathbf{G}_{\textrm{gauge}}=\CC^\ast=\left\{\; g\in \text{SL}(1|1)\,:\; g=\begin{pmatrix} 
      \uplambda & 0  \\
      0 &  \uplambda
   \end{pmatrix}\; \right\}$
    \item[$\circ$] The `supersymmetric' phase space: $\mathbf{G}_{\textrm{gauge}}=\left\{\; g\in \text{SL}(1|1)\,:\; g=\begin{pmatrix} 
      \uplambda & 0  \\
      \upxi &  \uplambda
   \end{pmatrix}\; \right\}$. In this case $\uplambda\in \CC^\ast$ is a bosonic element, and $\upxi\in \CC$ is a fermionic element. As a result, here the quotient~(\ref{compsympred}) is a genuine super-symplectic reduction.
\end{itemize}
Both situations correspond to the following elementary quiver:
\bea
\begin{tikzpicture}[
baseline=-\the\dimexpr\fontdimen22\textfont2\relax,scale=1]
\begin{scope}[very thick,decoration={
    markings,
    mark=at position 0.4 with {\arrow{>}}}
    ] 
\draw[postaction={decorate}] ([yshift=-2pt,xshift=0pt]0,0) --  node [below, yshift=0pt] {\footnotesize $U$} ([yshift=-2pt,xshift=0pt]2,0);
\end{scope}
\begin{scope}[very thick,decoration={
    markings,
    mark=at position 0.6 with {\arrow{<}}}
    ] 
\draw[postaction={decorate}] ([yshift=2pt,xshift=0pt]0,0) --  node[above, yshift=0pt] {{\footnotesize $ V$}} ([yshift=2pt,xshift=0pt]2,0);
\end{scope}
\filldraw[blue!50] (1.85,-0.15) rectangle ++(8pt,8pt);
\filldraw[blue!50] (0,0) circle (4pt); 
\node at (2.5,0.5) {$\CC^n$};
\node at (-0.5,-0.5) { $\CC^{1|1}$};
\end{tikzpicture}
\eea
Here $U\in \mathrm{Hom}(\CC^{1|1}, \CC^n)$ and $V \in \mathrm{Hom}(\CC^n, \CC^{1|1})$. The difference comes from the action of the gauge group $\mathbf{G}_{\textrm{gauge}}$ on $\CC^{1|1}$. In fact, one can also identify the configuration spaces $M$ of the models, since in both cases $\mathbf{\Phi}=T^\ast M$:
\bea
M_{\mathrm{min}}=\CP^{n-1|n},\quad\quad\quad M_{\mathrm{SUSY}}=\Pi T (\CP^{n-1})\,.
\eea
Here $\Pi T$ stands for the `fermionic tangent bundle', i.e. the tangent bundle where the fibers are assumed fermionic. Accordingly the super-projective space $\CP^{n-1|n}$ may be seen as the total space of the super-vector bundle $\Pi(\mathcal{O}(1)\oplus \cdots \oplus \mathcal{O}(1))$ over $\CP^{n-1}$, which puts the two configuration spaces on par with each other.

Specifying the phase or configuration space is of course not sufficient to formulate the theory. One additionally needs to choose the global symmetry group $G_{\mathrm{global}}$, which in turn defines the Hamiltonian in~(\ref{SUSYlagrGen}) via the complex moment map $\mu$. For $M_{\mathrm{min}}$ there are two choices that lead to well-known models:
\begin{itemize}
    \item[$\circ$] $G_{\mathrm{global}}=SL(n, \CC)$. In this case~(\ref{SUSYlagrGen}) defines the $\CP^{n-1}$ model with `minimally coupled fermions'~\cite{AbdallaCancel} (this is also the reason for the name of the phase space $M_{\mathrm{min}}$). 
    \item[$\circ$] $G_{\mathrm{global}}=PSL(n|n, \CC)$. This is the sigma model with target space $\CP^{n-1|n}$ that has been widely studied in the literature, cf.~\cite{ReadSaleur, SaleurSchomerus}. In the case $n=4$ this leads to the so-called  twistor string~\cite{WittenTwistor}.
\end{itemize}

Finally, it is very instructive to look at the `supersymmetric configuration space' $M_{\mathrm{SUSY}}$. As proven in detail in~\cite{BykovSUSYCPN}, in order to obtain an interacting $(2, 2)$-supersymmetric $\CP^{n-1}$ sigma model, in this case one should choose~$G_{\mathrm{global}}=SL(n, \CC)$. This provides a new approach to constructing worldsheet-supersymmetric models by starting from a model with target space supersymmetry (in this case $\tGL(n|n, \CC)$) and gauging part of the supergroup in such a way that the target space supersymmetry disappears and gives way to worldsheet supersymmetry. This approach does not rely on superspace methods, and one might say that worldsheet supersymmetry is \emph{emergent} in this case.

\pagebreak

\vspace{0.5cm}
\noindent
\rule{\textwidth}{1pt}
    \vspace{1ex}
\begin{center}
\vspace{-0.3cm}
{\Large    Conclusion}
\end{center}

\noindent
\vspace{-0.5ex}%
\rule{\textwidth}{1pt}

\addcontentsline{toc}{section}{\bfseries Conclusion}

\vspace{2cm}
Despite the ubiquity of flag manifolds in mathematics, they might not be equally familiar to the physics audience. One of the goals of this review was to fill this gap in the physics literature and to introduce these rich objects, explaining that they are useful and in certain cases inevitable in  physics applications.

We started in Chapter 1 by defining what flag manifolds are, and by describing their differential-geometric structures. As a first application of these methods, we considered an `almost textbook' example of a mechanical particle interacting with a non-Abelian gauge field. As we explained, the isotopic `spin' degrees of freedom take values in a suitable flag manifold. The symplectic form on the flag manifold is the classical analogue of the concept of representation of the gauge group, w.r.t. which the particle is charged. The `Berry phase', which is formulated in terms of the chosen symplectic form, serves as a kinetic term for spin motion and is nothing but a one-dimensional version of the (originally two-dimensional) WZNW-term. Exactly the same argument as in WZNW theory leads to the quantization of the parameters entering the symplectic form. The resulting `quantum numbers' have a transparent interpretation as the lengths of the rows in the Young diagram corresponding to the representation of the particle.

The approach that we developed on the example of the mechanical particle is in fact rather universal and is colloquially known as `geometric quantization'. In the rest of Chapter~1 we explained how the Berry phase action can be quantized, and that this leads to various representations of spin operators well-known in condensed matter physics, such as the Schwinger-Wigner, Holstein-Primakoff and Dyson-Maleev representations. We also emphasized that the very same flag manifolds can also be understood as the manifolds of coherent states for the relevant representations. We subsequently used these coherent states to construct path integrals for spin chains in Chapter 2.

Overall, in Chapters 2 and 3 we attempted to cover two major topics related to sigma models with  a two-dimensional worldsheet and a flag manifold target space. The first topic is how such sigma models arise in the continuum limits of spin chains, and the second one is the description of integrable flag manifold models. 

In Chapter 2, we reviewed SU($n$) spin chains in various representations, and discussed at length how these representations give rise to sigma models with different flag manifold target spaces. This can be understood by considering how the Young tableau parameters $p_\alpha$, which generalize the notion of spin in the antiferromagnet, determine the target space of the chain's matrix degree of freedom, $S$. Mathematically speaking, $S$ is a moment map, and different $p_\alpha$s  define different co-adjoint orbits. For SU(2), the target space of $S$ is always $S^2$, and leads to the familiar $\CP^1$ model, but for $n>2$, the 2-sphere is promoted to some flag manifold of SU($n$).

For most of the chapter, we focused on the totally symmetric representations of SU($n$), which corresponded to $p_1=p$, and $p_\alpha=0$ for $\alpha>1$. This defines a coadjoint orbit isomorphic to $\CP^{n-1}$ at each site of the chain. However, we did not end up deriving a sigma model with this projecitve target space. Instead, by considering Hamiltonians with longer range interaction terms, that have classical ground states with $n$-site order, we obtained the complete $\tU(n)/[\tU(1)]^n$ flag manifold sigma model. Loosely speaking, the longer-range interaction terms served to couple the different $\CP^{n-1}$ sectors together, as well as impose orthogonality. Related to this, we found that these sigma models possessed $n$ topological terms, each of which corresponded to the pull-back to the complete flag manifold of the Fubini-Study form on $\CP^{n-1}$. The corresponding topological angles were found to be $\theta_A = \frac{2\pi p A}{n}$.

Unless the various interaction terms in the Heisenberg Hamiltonian were tuned to special values, we learned that these flag manifold sigma models lack Lorentz invariance. This is due to the fact that multiple velocities exist in the most general case. However, in Section~\ref{section:vrg} we reviewed how these velocities flow to common value under renormalization, thus establishing that Lorentz invariance does indeed emerge at low energies. This fact led to an SU($n$) generalization of Haldane's conjecture: when $p$ and $n$ are coprime, the corresponding SU($n$) chain will be in a gapless phase at low energies; otherwise, a finite energy gap will persist, with ground-state degeneracy equal to $n/\gcd(n,p)$. This conjecture was supported by various exact results, including the LSMA theorem and AKLT constructions, as well as by 't Hooft anomaly matching conditions. In short, 't Hooft anomalies are present in the flag manifold sigma model for all $p$ not a multiple of $n$, but only when $\gcd(n,p)=1$ can the theory flow to a stable conformal field theory (which in this case is the SU($n)_1$ WZNW model). These anomalies are mixed between the PSU($n$) symmetry of the model, and a global $\mathbb{Z}_n$ symmetry, which derives from the underlying $n$-site order of SU($n$) chain.

Finally, we concluded Chapter 2 by reinterpreting this generalized Haldane conjecture in terms of fractional topological excitations in the sigma model. These correspond to nontrivial sections in a PSU($n$) bundle, and have topological charges that are multiple of~$1/n$. We explained how these excitations give rise to an energy gap via a Coulomb gas mechanism, similar to the Kosterlitz-Thouless phase transition in the classical XY model. When $\gcd(n,p)=1$, these excitations interfere, resulting in an effective fugacity of zero, and lead to a gapless phase in the sigma model. More general representations of SU($n$), and how their chains lead to sigma models with both linear and quadratic dispersion, were also reviewed at the end of Chapter 2.

The subsequent narrative was centered around a slightly different circle of questions related to flag manifold sigma models. More exactly, in Chapter 3 we described a wide class of \emph{integrable} sigma models with complex homogeneous target spaces and deformations thereof. This class includes the flag manifold sigma models as rather representative examples. It has long been known that the construction of integrable models with target spaces that are not symmetric (even if homogeneous) is a significant challenge already in classical theory. For this reason we started in Chapter 3 by describing from various angles the classically integrable models with flag manifold target spaces. Technically the key new ingredient that needs to be included to make such models integrable is a non-topological $B$-field of a special kind. We explained that these models can be obtained by one of the three approaches: in a more conventional way by constructing Noether currents satisfying zero-curvature equations, using a remarkable relation to the principal chiral model via nilpotent orbits, or by the novel techniques related to four-dimensional Chern-Simons theory. Besides, we showed that these sigma models are \emph{exactly and explicitly} equivalent to chiral gauged Gross-Neveu systems, whose integrability properties have been known since the 1970's.

In the latter part of Chapter 3 we concentrated on studying the proposed bosonic Gross-Neveu models, as well as their fermionic completions. It turned out that this perspective makes the analysis of the underlying sigma models substantially easier than in the pure geometric formulation with a metric, $B$-field and dilaton. In support of this opinion we provided a calculation of the one-loop $\beta$-function of a wide class of trigonometrically deformed sigma models. This $\beta$-function is common for all of these models and provides a far-reaching generalization of the so-called `sausage' solution that corresponds to the $S^2$ target space. We also explained that the general solution explains some puzzles about the undeformed, homogeneous models: for example, it provides an explanation of why the $\beta$-functions of symmetric space models depend only on the dual Coxeter number of the symmetry group and extends this result to the (non-symmetric) flag manifold models. Finally, we showed that the purely bosonic gauged models suffer from chiral anomalies, which may be canceled by adding fermions. More generally, we formulated a broad differential-geometric setup for sigma models whose phase spaces are  quiver super-varieties satisfying anomaly cancellation conditions. We demonstrated how this setup may be applied to the $\CP^{n-1}$ model with fermions, yielding all known quantum integrable models with bosonic core $\CP^{n-1}$, and emphasized that this approach provides a new way of constructing models with worldsheet supersymmetry by gauging models with target space supersymmetry.

\vspace{2cm}

\textbf{Acknowledgments.} We would like to thank 
Yu.~Amari, I.~Ya.~Aref'eva, G.~Arutyunov, A.~Bourget, R.~Donagi, S.~Frolov, A.~Hanany, E.A.Ivanov, S.~Ketov,  C.~Klim\v{c}\'{\i}k, M. Lajko, G.~Lopes Cardoso, D.~L\"ust, A.~Ya.~Maltsev, T.~McLoughlin, F. Mila, K.~Mkrtchyan, H.~Nicolai, M.~Nitta, V.~Pestun, N.~Sawado, N.~Seiberg, A.G.Sergeev, V.~Schomerus, E.~Sharpe, S.~Shatashvili, A.A.Slavnov, T.~Sulejmanpasic, J.~Teschner, S.~Theisen, A.~Tseytlin, K.~Zarembo and P.~Zinn-Justin for 
helpful discussions and N.~Seiberg, K.~Zarembo for comments on the manuscript. DB is especially grateful to A.~A.~Slavnov for long-term support, and to E.~A.~Ivanov for proposing  the idea of writing a review article on the subject of flag manifold sigma models. DB would also like to thank the Max-Planck-Institut für Physik in Munich (Germany), where part of this work was done, for hospitality. The research of I.~Affleck and K.~Wamer was supported by NSERC Discovery Grant 04033-2016, as well as by scholarships from NSERC and the Stewart Blusson Quantum Matter Institute. The work of D.~Bykov was performed at the Steklov International Mathematical Center and supported by the Ministry of Science and Higher Education of the Russian Federation (agreement no. 075-15-2019-1614). 

\vspace{2cm}

{\setstretch{0.8}
\setlength\bibitemsep{5pt}
\printbibliography
}
\appendix

\section{Kähler potential from the quiver quotient formulation}\label{kahpotapp}

We showed in section~\ref{kahlstructsec} that there at least two ways to derive invariant K\"ahler metrics on flag manifolds: using the so-called quasipotentials and also using the Nakajima-type quiver shown in Fig.~\ref{flagquivpic}. In this section we prove the equivalence of the two approaches.

The space of matrices $\{U_A\}$ shown in Fig.~\ref{flagquivpic} is endowed with the standard symplectic form $\Omega_A=i\,\mathrm{Tr}(dU_A\wedge dU_A^\dagger)$ and, accordingly, a metric $(ds^2)_A=\mathrm{Tr}(dU_A  dU^\dagger_A)$. The full symplectic form is then
\bea
\Omega_0=\sum\limits_{A=1}^{m-1}\,i\,\mathrm{Tr}(dU_A\wedge dU_A^\dagger)\,.
\eea
At each circular node $j$ one has the action of a gauge group $\tU(L_A):=\tU(d_A)\mysub \tGL(d_A, \CC)$ that preserves the symplectic form. Accordingly one can define the moment maps for this action: $\mu_A=U_{A}^\dagger U_{A}-U_{A-1}U_{A-1}^\dagger$. The main statement is that the flag manifold may be defined as a quotient:
\bea\label{flagsymplquot}
\mathcal{F}=\{\mu_A=\zeta_A\,\mathds{1}_{d_A},\quad A=1, \ldots, m-1\}\big/ \tU(L_1)\times \cdots \times \tU(L_{m-1})\,,
\eea
where $\zeta_A>0$ are positive constants (in the supersymmetric setup~\cite{Donagi:2007hi} they are called Fayet-Iliopoulos parameters). Notice that there are $m-1$ such constants, consistent with our previous discussion that all three spaces in~(\ref{2formsdiag}) have the same dimension.  The conditions in~(\ref{flagsymplquot}) ensure that each of the matrices $U_{A}^\dagger U_{A}$ is non-degenerate, which implies $\mathrm{rk}(U_A)=d_A$. The linear spaces $L_A$ of the flag may be obtained as $\mathrm{Im}(U_{m-1}\cdots U_{A+1} U_{A})\subset \CC^n$: the matrix $U_{m-1}\cdots U_{A+1} U_{A}$ has rank $d_A$, so that it defines $d_A$ vectors in $\CC^n$, and the quotient w.r.t.~$\tU(L_A)$ amounts to considering the linear space spanned by these vectors (compare with the example~(\ref{L2})-(\ref{L1}), depicted in Fig.~\ref{flagparam}). Clearly, the $L_A$ so defined are nested in each other: $L_{A-1}\subset L_A$. 

The apparatus of symplectic quotient provides a symplectic form on $\mathcal{F}$ by restricting the original symplectic form $\Omega_0$ to the level set of the moment maps $\Omega=\Omega_0\big|_{\mu=\zeta}$. Since the whole setup is K\"ahler, so that there is a complex structure and metric involved, the reduction also provides a K\"ahler metric on the flag manifold, which should coincide with the metric given by the K\"ahler potential~(\ref{KahlerPotFlagMfds}).

Let us see how this happens. The strategy, known from the general theory of K\"ahler quotients~(cf.~\cite{HKLR}), is as follows: one considers generic matrices $U_A, U_A^\dagger$, not necessarily satisfying the moment map constraints, and one needs to find the complexified symmetry transformation  $g_1\times \cdots \times g_{m-1}\subset \tGL(L_1, \CC)\times \cdots \times \tGL(L_{m-1}, \CC)$, such that the transformed variables would satisfy the constraints. Introducing $M_A=g_A^\dagger g_A$, it is easy to see that we may rewrite this requirement as
\bea\label{complgroup1}
U_A^\dagger M_{A+1} U_A-M_A U_{A-1} M_{A-1}^{-1} U_{A-1}^\dagger M_A=\zeta_A \,M_A\,,\quad\quad M_m=\mathds{1}\,.
\eea
Given a solution $M_1, \ldots, M_{m-1}$, we obtain the K\"ahler potential of the quotient manifold as follows:
\bea\label{kahpotquot1}
\mathcal{K}=\sum\limits_{A=1}^{m-1}\,\zeta_A\,\tr(\log{M_A})=\sum\limits_{A=1}^{m-1}\,\zeta_A\,\log{(\det M_A)}\,.
\eea
We proceed to compute the determinants of the matrices $M_A$. Denoting $y_A:=M_{A-1}^{-1} U_{A-1}^\dagger M_A$ we may rewrite~(\ref{complgroup1}) in two equivalent forms:
\bear\label{complgroup2}
&&U_A^\dagger M_{A+1} U_A=M_A(\zeta_A+U_{A-1} y_A)\,.\\ \label{yUeq}
&&y_{A+1} U_A-U_{A-1}y_A=\zeta_A\,.
\eear
Multiplying~(\ref{complgroup2}) by $U_{A-1}^\dagger$ from the left and by $U_{A-1}$ from the right and using~(\ref{yUeq}), we find
\bear\label{Uinduct}
U_{A-1}^\dagger U_A^\dagger M_{A+1} U_A U_{A-1}&=&M_{A-1}(\zeta_{A-1}+U_{A-2} y_{A-1} )(\zeta_A+ y_A U_{A-1})=\\ \nonumber
&=& M_{A-1}(\zeta_{A-1}+U_{A-2} y_{A-1} )(\zeta_A+\zeta_{A-1}+U_{A-2}y_{A-1})\,.
\eear
Next we introduce the matrix $W_{B}:=U_{m-1}\cdots U_{B}$. As discussed above, $\mathrm{Im}(W_B)=L_B$. Recalling that $M_m=\mathds{1}$, we may continue~(\ref{Uinduct}) by induction to demonstrate that
\bea\label{WMP}
W_{B}^\dagger W_{B}=M_{B}\cdot P_B,
\eea
where $P_B$ is a product of  matrices of the type $a+U_{B-1}y_{B}$ ($a$ are constants). It turns out that the latter matrices are triangular in a certain basis, their diagonal blocks being constant. Indeed, it follows from~(\ref{yUeq}) that $(a+U_{B-1} y_B)U_{B-1}=(a+\zeta_{B-1})U_{B-1}+U_{B-1}U_{B-2}y_{B-1}$ (to be continued by induction), so that the matrix $a+U_{B-1} y_B=D_B+N_B$, where $D_B$ is diagonal with eigenvalues $a, a+\zeta_{B-1}, \ldots$ and $N_B$ is strictly triangular, in the sense that it maps $\mathrm{Im}(U_{B-1}U_{B-2}\cdots U_{B-C})$ to $\mathrm{Im}(U_{B-1}U_{B-2}\cdots U_{B-B} U_{B-C-1})$ for all $C$. It follows that $\mathrm{det}\,P_B=\mathrm{const.}\neq 0$, so that~(\ref{WMP}) implies $\mathrm{det}(M_{B})\sim \mathrm{det}(W_{B}^\dagger W_{B})$, up to a constant coefficient. Substituting into~(\ref{kahpotquot1}) and identifying $\zeta_{B}=\gamma_{B}$, we find agreement with~(\ref{Zmat})-(\ref{KahlerPotFlagMfds}).

\section{Symplectic forms on coadjoint orbits}\label{adjorbsec}

We saw in section~\ref{genmetr} that the most general invariant two-form on a flag manifold ${G\over H}={SU(n)\over S(U(n_1)\times \cdots \times U(n_m))}$ is
\bea
\Omega=\sum\limits_{A<B}\,a_{AB}\,\mathrm{Tr}(j_{AB}\wedge j_{BA})\,.
\eea
Here we wish to prove that the requirement of it being closed leads to the Kirillov-Kostant form~(\ref{symplformflag}). To check, in which case the above two-form is closed, we will take advantage of the flatness of the Maurer-Cartan current, $
dj-j\wedge j=0\,.
$
It follows that $
\mathscr{D}j_{AB}=\sum\limits_{C\neq (A, B)}\,j_{AC}\wedge j_{CB}\,,$
where $\mathscr{D}$ is the $H$-covariant derivative, defined as follows:
$
\mathscr{D}j_{AB}:=dj_{AB}-\{j_{\mathfrak{h}}, j_{AB}\}\,.
$
From the condition that $\Omega$ is closed it follows that
\bea\label{closed1}
a_{AB}+a_{BC}+a_{CA}=0\quad\quad \textrm{for all pairwise different}\quad\quad (A, B, C)\,.
\eea
The general solution to this equation is
\bea\label{closed2}
a_{AB}=p_A-p_B\,.
\eea
Therefore we have a family of homogeneous symplectic forms with $m-1$ real parameters. These forms may be compactly written as follows:
\bea\label{symplformflag1}
\Omega =\mathrm{Tr}(p\,j\wedge j)\,,\quad\quad \textrm{where}\quad\quad p=\mathrm{Diag}(p_1\,\mathds{1}_{n_1}, \ldots, p_m \mathds{1}_{n_m})\,.
\eea
The element $p$  may be normalized to be traceless: $\mathrm{Tr}(p)=0$. The stabilizer $H$ may now be thought of as the stabilizer of the matrix $p\in \mathfrak{u}_n$, and the flag manifold itself -- as an adjoint orbit:
\bea
\mathcal{F}_{d_1, \ldots, d_m}=\{g\,p\,g^{-1},\quad g\in \SU(n)\}\,.
\eea

\section{Coherent states as polynomials}\label{Bargmannapp}

In section~\ref{cohstates} we described the coherent states of $SU(n)$ using the Schwinger-Wigner representation in Fock space. In place of the Fock space generated by the creation operators acting on a vacuum state $|0\rangle$ we may equivalently use the space of polynomials with a Gaussian inner product -- this is the celebrated Bargmann representation~\cite{Bargmann}. The map is simple:
\bea\label{aadagpol}
(a_1^\dagger)^{q_1}\cdots (a_n^\dagger)^{q_n}|0\rangle \quad \mapsto \quad z_1^{q_1}\cdots z_n^{q_n}
\eea
This map is a Hilbert space isomorphism, meaning that the scalar product is preserved, if one picks the Gaussian scalar product on the space of polynomials (here $\hat{f}$ and $\hat{g}$ are two polynomials):
\bea\label{Bargmannscal}
(\hat{f}, \hat{g})_{\mathrm{Bargmann}}=\int\,\overline{\hat{f}(z)} \; \hat{g}(z)\,e^{-\sum\limits_{j=1}^n|z_j|^2}\,\prod\limits_{\alpha=1}^n\,(i\,dz^\alpha\wedge d\overline{z^\alpha})
\eea
Let us also discuss the relation to the definition of states as \emph{in}homogeneous polynomials used in the classical work~\cite{Berezin}, where coherent states were used to describe the quantization of a sphere $S^2\sim \CP^1$ --- the simplest homogeneous K\"{a}hler (symplectic) manifold\footnote{Coherent states, written in inhomogeneous coordinates, are also discussed in~\cite{Perelomov1}.}. To start with, we observe that, since in our applications to representation theory the number of oscillators is fixed,  we may introduce a new  Hilbert space, isomorphic to the one of homogeneous polynomials~(\ref{aadagpol}). Suppose we have a rank-$p$ symmetric representation, so that we are dealing with homogeneous polynomials of degree~$p$. It is an elementary fact that the following two spaces are isomorphic:
\bear\nonumber 
&\textrm{Homogeneous polynomials of degree}\;p\; \textrm{in}\;n\;\textrm{variables}\quad \leftrightarrow \\ \nonumber & \textrm{Polynomials of degree}\;\leq p\; \textrm{in}\;n-1\;\textrm{variables}
\eear
In order to pass from the first to the second definition one sets one of the variables equal to unity, say $z_1=1$. This is the counterpart of passing to inhomogeneous coordinates on a projective space (see the very beginning of Chapter~1). Going backwards amounts to  homogenizing the polynomial. To find the correct integration measure on the space of polynomials of degree $\leq p$, we first start with the homogeneous polynomials $\hat{f}$ and $\hat{g}$ of degree~$p$ and make the change of variables $\{z_1\to \lambda, z_2\to \lambda z_2, \cdots, z_n\to \lambda z_{n}\}$. In this case, clearly, $\hat{f}=\lambda^p\,f(z)$ and analogously for $g$, where $f(z), g(z)$ are now inhomogeneous polynomials of $n-1$ complex variables. As a result of the change of variables in the integral~(\ref{Bargmannscal}) we obtain
\bear \nonumber
&&\scalemath{0.95}{(\hat{f}, \hat{g})_{\mathrm{Bargmann}}=\int\,e^{-|\lambda|^2\left(1+\sum\limits_{\alpha=1}^{n-1}|z^\alpha|^2\right)}\,\left(i\,|\lambda|^{2(p+n-1)}\,d\lambda\wedge d\bar{\lambda}\right)\;\,\overline{f(z)} \; g(z)\;\prod\limits_{\alpha=1}^{n-1}\,(i\,dz^\alpha\wedge d\overline{z^\alpha})=}\\ \nonumber
&&\scalemath{0.95}{=\quad \textrm{integrating over}\,\lambda, \bar{\lambda}\quad \sim\quad  \int \frac{\overline{f(z)} \; g(z)}{\left(1+\sum\limits_{\alpha=1}^{n-1}|z^\alpha|^2\right)^{p+n}}\;\prod\limits_{\alpha=1}^{n-1}\,(i\,dz^\alpha\wedge d\overline{z^\alpha})}
\eear
In other words, if $\hat{f}, \hat{g}$ are homogeneous polynomials of degree $p$ and $f, g$ are their inhomogeneous counterparts (obtained by setting $z_1=1$), then $(\hat{f}, \hat{g})_{\mathrm{Bargmann}}=(f, g)$ provided we define the scalar product in the space of inhomogeneous polynomials of degree $\leq p$ as follows\footnote{From a mathematical standpoint (which uses the Borel-Weil-Bott theorem briefly mentioned in section~\ref{geomquantsec}) $f$ and $g$ are sections of the line bundle $\mathcal{O}(p)$ over $\CP^{n-1}$. The integrand may be understood as a scalar product in the fiber at a given point $z$ on the base. In that case $(1+\sum_{\alpha=1}^{n-1}\,|z^\alpha|^2)^{-p}$ plays the role of a metric in the fiber, and $(f, g)$ is obtained by integrating the fiber scalar product over all of $\CP^{n-1}$ with the natural measure $d\mu$.}:
\begin{empheq}[box=\fbox]{align}
\hspace{1em}\vspace{5em}
\label{scalprod}
&(f,g)=\,\int\; \frac{\overline{f(z)} \; g(z)}{\left(1+\sum\limits_{\alpha=1}^{n-1} \,|z^\alpha|^2 \right)^p} \; d\mu(z,\bar{z}),\quad\quad\quad \\ \nonumber  &\textrm{where} \quad\quad (d\mu)_{\CP^{n-1}}\sim\,\left(1+\sum\limits_{\alpha=1}^{n-1} \,|z^\alpha|^2 \right)^{-n}\, \prod\limits_{\alpha=1}^{n-1}\,(i\,dz^\alpha\wedge d\overline{z^\alpha})
\end{empheq}
The measure \(d\mu\sim \Omega_{\mathrm{FS}}^{n-1}\) is in fact the volume form on \(\CP^{n-1}\), proportional to a power of the Fubini-Study form~(\ref{FSform}). In the second line of~(\ref{scalprod}) one has its expression in  the inhomogeneous coordinates.

The coherent state $|v\rangle$, when viewed as an inhomogeneous polynomial, will be denoted  $\phi_{\bar{v}}(z)$ (this notation is borrowed from \cite{Berezin}).  For example, if we take the state~(\ref{CPNcohstates}) with $n=2$, which in Fock space language is $(\bar{v}_1 a_1^\dagger+\bar{v}_2 a_2^\dagger)^4|0\rangle$, the corresponding polynomial would be $\phi_{\bar{v}}(z)\sim(1+\bar{v}z)^4$, where we have set $v:={v_2\over v_1}$ and dropped an overall factor.

\section{Integrability of the complex structure}\label{complstructapp}

Here we wish to prove two claims made in section~\ref{compstructsec0}. The first one is: 

\vspace{0.3cm}\noindent \hspace{0.3cm}
\fbox{\parbox{14cm}{
\centering 
If the restriction to $\mathfrak{m}$ of the adjoint-invariant metric $\langle\bullet, \bullet\rangle$ on $\mathfrak{u}(n)$  is Hermitian w.r.t. an almost complex structure $\mathscr{J}$,  integrability of $\mathscr{J}$ (viewed as an almost complex structure on the flag manifold) is equivalent to
\bea\label{algcond0app}
[\mathfrak{m}_+, \mathfrak{m}_+]\subset \mathfrak{m}_+,\quad\quad [\mathfrak{m}_-, \mathfrak{m}_-]\subset \mathfrak{m}_-\,.
\eea
\vspace{-0.8cm}
}
}

\vspace{0.3cm}
In general the integrability of an almost complex structure means that $[\mathfrak{m}_+, \mathfrak{m}_+]\subset \mathfrak{m}_+\oplus\, \mathfrak{h}$. To see this, note that an almost complex structure $\mathscr{J}$ is defined by the conditions $\mathscr{J}\circ J_\pm=\pm i \, J_\pm$, where $J_\pm$ are the components of a Maurer-Cartan current:
\bea\label{Jcurrdecomp}
J=-g^{-1}dg=J_\mathfrak{h}+J_++J_-,\quad\quad J_\pm\in \mathfrak{m}_\pm\,.
\eea
Since $dJ-J\wedge J=0$, we get $$dJ_-=\big[-J_0\wedge J_0+(\textrm{terms with}\,J_-) - J_+\wedge J_+\big]_{\mathfrak{m}_-}\,.$$ Therefore for the integrability of $\mathscr{J}$ one should have $[J_+\wedge J_+]_{\mathfrak{m_-}}=0$, i.e. $[\mathfrak{m}_+, \mathfrak{m}_+]\subset \mathfrak{m}_+\oplus \mathfrak{h}$. We see that the conditions  (\ref{algcond0app}) therefore define an integrable complex structure. Conversely suppose we have an integrable complex structure on $G/H$, and $\mathfrak{m}_\pm$ are its respective holomorphic/anti-holomorphic subspaces. Then $[a, b]=c+\gamma$, where $a, b, c\in \mathfrak{m}_+$ and $\gamma \in \mathfrak{h}$. Since $\boldlangle \mathfrak{m}_+, \mathfrak{h}\boldrangle=0$, computing the scalar product with a generic element $\gamma'\in \mathfrak{h}$, we obtain $\boldlangle \gamma', [a,b] \boldrangle=\boldlangle \gamma', \gamma \boldrangle$. Using the identity $\boldlangle [a, \gamma'], b \boldrangle+\boldlangle \gamma', [a, b] \boldrangle=0$, we get $\boldlangle \gamma', \gamma \boldrangle=-\boldlangle [a, \gamma'], b \boldrangle=\boldlangle a', b \boldrangle$, and $a'=[\gamma', a]\in \mathfrak{m}_+$. The subspace $\mathfrak{m}_+$ is isotropic, if the metric $\langle\bullet, \bullet \rangle$ is Hermitian, therefore $\boldlangle \gamma', \gamma \boldrangle=0$ for all $\gamma'\in \mathfrak{h}$, which implies $\gamma=0$ due to the non-degeneracy of $\langle\bullet, \bullet \rangle$. The result $[\mathfrak{m}_+, \mathfrak{m}_+]\subset \mathfrak{m}_+$ follows.

\vspace{0.3cm}
The second statement used in section~\ref{compstructsec0} is:

\vspace{0.3cm}\noindent \hspace{0.3cm}
\fbox{\parbox{14cm}{
\centering 
There are exactly $n!$ acyclic tournament diagrams.
}
}

\begin{figure}[h]
    \centering 
    \includegraphics[width=0.6\textwidth]{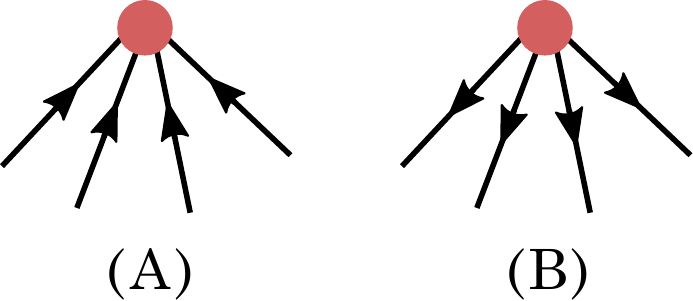}
  \caption{(A) The `sink' vertex,\quad (B) The `source' vertex\,.}
  \label{sourcesinkvertex}
\end{figure}

\vspace{0.3cm}\noindent
The statement implies that there is only one combinatorial type of diagrams, and all acyclic tournament diagrams (in such diagrams, by definition, all pairs of nodes are connected) may be obtained from any one of them by the action of the permutation group  $S_n$. Let us describe this combinatorial type. Every acyclic diagram has a `source'-vertex, in which all the lines are outgoing, and a `sink'-vertex, in which all lines are incoming  (see Fig.~\ref{sourcesinkvertex}). Indeed, if that were not so, every vertex would contain at least, say, one outgoing line. Then one can start at any vertex and follow outgoing lines, until a loop is formed. Let us consider the `source'-vertex. The diagram formed by the remaining $n-1$ vertices together with the edges joining them can be an arbitrary acyclic diagram (as the chosen vertex is a `source', there cannot be cycles containing it). Therefore we have performed the first step of the induction. The subsequent steps consist in finding the `source' vertex in the reduced diagram. It is therefore clear that there always exists a vertex with $i$ outgoing lines for all $i=0, \ldots, n-1$. This statement completely describes the combinatorial structure of the diagram. Equivalently, there is a total ordering on the set of vertices. Different diagrams differ just by a relabeling of the vertices.

\section{Proving the $\mathbb{Z}_m$-`symmetry' of integrable models}\label{integrcyclsymm}

In this Appendix we prove the $\mathbb{Z}_m$ symmetry property of the integrable flag manifold models, introduced in section~\ref{Zmsymmintmod}. The statement is that the e.o.m. of two models, in which the complex structures differ by a \emph{cyclic} permutation of the subspaces $\CC^{n_1}, \cdots,  \CC^{n_m}$, are the same. In this case the two actions differ only by a topological term.

Let us call $\mathscr{J}$ the standard complex structure, whose holomorphic subspace $\mathfrak{m}_+$ is given by upper-block-triangular matrices. Then $\mathscr{J}_1=\sigma_1(\mathscr{J})$ and $\mathscr{J}_2=\sigma_2(\mathscr{J})$ for some permutations $\sigma_{1}, \sigma_2\in S_m$. We recall the notation $J_{AB}$ from (\ref{jdecomp}). The corresponding K\"ahler forms are
\bear
\omega_1=i\,\sum\limits_{A<B}\,\mathrm{Tr}(J_{\sigma_1(A)\sigma_1(B)}\wedge J_{\sigma_1(B)\sigma_1(A)})\\
\omega_2=i\,\sum\limits_{A<B}\,\mathrm{Tr}(J_{\sigma_2(A)\sigma_2(B)}\wedge J_{\sigma_2(B)\sigma_2(A)})
\eear
Upon introducing the notation $\sigma(J_{AB}):=J_{\sigma(A)\sigma(B)}$, we may write the difference of the two forms as
\bea\label{omegadiff1}
\omega_1-\omega_2=i\,\sigma_1\left(\sum\limits_{A<B}\,\mathrm{Tr}(J_{AB}\wedge J_{BA})-\sum\limits_{A<B}\,\mathrm{Tr}(J_{\tau^{-1}(A)\tau^{-1}(B)}\wedge J_{\tau^{-1}(B)\tau^{-1}(A)})\right),\; 
\eea
where $\tau^{-1}=\sigma_1^{-1}\sigma_2\,.$ This reduces the problem to that of $\mathscr{J}_1=\mathscr{J}$ and $\mathscr{J}_2=\tau^{-1}(\mathscr{J})$. Note that the exterior derivative commutes with the permutation $\sigma$, due to the following simple fact following from the Maurer-Cartan equation:
$\sigma(dJ_{AB})=\sigma(\sum\limits_C\,J_{AC}\wedge J_{CB})=\sum\limits_C\,\sigma(J_{AC})\wedge \sigma(J_{CB})=d\sigma(J_{AB})$, where to arrive at the last equality one has to make a change of the dummy summation index $C\to \sigma(C)$.

\vspace{0.3cm}
We wish to show that $d(\omega_1-\omega_2)=0$ implies that $\tau$ is a cyclic permutation.
To this end we rewrite the above difference as follows:
\bear
&& \omega_1-\omega_2=i\,\sigma_1\left( \sum\limits_{A, B}\,\alpha_{AB}\,\mathrm{Tr}(J_{AB}\wedge J_{BA})\right),\\ && \textrm{where}\quad \alpha_{AB}={1\over 2}\big(\mathrm{sgn}(B-A)-\mathrm{sgn}(\tau(B)-\tau(A))\big)\in\{-1, 0, 1\}\,.
\eear
(We have made a change of dummy variables $A\to\tau(A)$ and $B\to\tau(B)$ in the second sum in (\ref{omegadiff1})).
Closedness of this form requires that (see~(\ref{closed1})-(\ref{closed2}))
\bea
\alpha_{AB}={1\over 2}\big(\mathrm{sgn}(B-A)-\mathrm{sgn}(\tau(B)-\tau(A))\big)=p_A-p_B\,.
\eea
Let us consider the case $B>A$. Then $p_A=p_B$ if $\tau(B)>\tau(A)$ and $p_A=p_B+1$ if $\tau(B)<\tau(A)$. This means that $\{p_A\}_{A=1\ldots m}$ form a non-increasing sequence, and moreover the difference between any two elements is either zero or $1$. This is only possible if the set has the form $(\underbracket[0.6pt][0.6ex]{p, \ldots, p}_{K}, \underbracket[0.6pt][0.6ex]{p-1, \ldots p-1}_{m-K})$. Accordingly the original sequence of $m$ consecutive numbers can be split into two consecutive sets:
\bea
1\ldots m=(I_1, I_2)\,.
\eea
Since $\tau(B)<\tau(A)$ for ($A\leq K$, $B>K$), the permutation acts as follows:
\bea
\tau(I_1, I_2)=(\tau(I_2), \tau(I_1))\,.
\eea
Moreover, since $\tau(A)<\tau(B)$ for $A<B\leq K$ and the image $\tau(I_1)$ is $(m-K+1,\ldots m)$, a moment's thought shows that $\tau(A)=m-K+A$ for $A=1\ldots K$. Analogously $\tau(B)=B-K$ for $B=K+1\ldots m$. Therefore $\tau$ is nothing but a $K$-fold cyclic permutation `to the left' (or $m-K$-fold to the right).

\vspace{0.3cm}
Since for $A<B$ the non-zero $\alpha_{AB}$ are the ones, for which $\tau(B)<\tau(A)$, this implies $B=K+1\ldots m$ and $A=1\ldots K$. These $\alpha_{AB}$ are equal to $1$, therefore
\bea
\omega_1-\omega_2=i\,\sigma_1 \left(\mathop{\sum_{ A=1\,\ldots\, K, }}_{ B= K+1\,\ldots\, n } \,\mathrm{Tr}(J_{AB}\wedge J_{BA})\right)\,,
\eea
which is easily seen to be proportional to the (generalized) Fubini-Study form on the Grassmannian $G_{L, N}$, where $L=\sum\limits_{A=1}^K \,n_A$. Conversely, one shows that for a cyclic permutation the difference between $\omega_1$ and $\omega_2$ is the closed form written above.

\section{Models with $\mathbb{Z}_m$-graded target spaces}\label{gradedsec}

A homogeneous space $G\over H$ is called $\mathbb{Z}_m$-graded (or $m$-symmetric), if the Lie algebra $\mathfrak{g}$ of its isometry group admits the following decomposition:
\bea\label{Zm}
\mathfrak{g}=\oplus_{i=0}^{m-1}\,\mathfrak{g}_i,\quad\quad [\mathfrak{g}_i, \mathfrak{g}_j]\subset \mathfrak{g}_{i+j\;\mathrm{mod}\;m},\quad\quad \mathfrak{g}_0=\mathfrak{h}\,.
\eea
In this language the ordinary symmetric spaces are $2$-symmetric spaces. Similarly to what happens for symmetric spaces, the e.o.m. of a certain class of  $\sigma$-models with $\mathbb{Z}_m$-graded target-spaces may be written as flatness conditions of a one-parametric family of connections. These models were introduced in \cite{Young} and subsequently studied in~\cite{BeisertLucker}. The action has the form
\bea\label{actionY}
\widetilde{\mathcal{S}}:=\int_\Sigma\,d^2 x\,\|\dd X\|^2_G+\int_\Sigma\,X^\ast \widetilde{\omega},
\eea
where $\widetilde{\omega}$ is a 2-form constructed using the $\mathbb{Z}_m$-decomposition of the Lie algebra (\ref{Zm}). Note that, if \;$\widetilde{\omega}$\; were the fundamental Hermitian form, one would obtain precisely the action  (\ref{action}). Now we come to the precise definition of $\widetilde{\omega}$. Decompose the current $J=-g^{-1}dg$ according to (\ref{Zm}):
\bea\label{curr}
J=-g^{-1}dg=\sum\limits_{i=0}^{m-1}\;J^{(i)},\quad\quad\textrm{where}\quad J^{(i)}\in \mathfrak{g}_i\;.
\eea
The form $\widetilde{\omega}$ is defined as follows:
\bea\label{omegaY}
\widetilde{\omega}={1\over 2}\sum\limits_{k=1}^{m-1}{(m-k)-k\over m}\,\tr(J^{(k)}\wedge J^{(m-k)})
\eea
This formula raises the following question. According to (\ref{omegaY}), the form $\widetilde{\omega}$ depends on the $\mathbb{Z}_m$-grading on the Lie algebra, but generally a given Lie algebra $\mathfrak{g}$ may have many different gradings (with different, or same, values of $m$). The question is: are the models defined by (\ref{actionY})-(\ref{omegaY}), corresponding to different gradings of $\mathfrak{g}$, different?

Before answering this question, we review the construction of cyclic gradings on semi-simple Lie algebras~\cite{Kac}. Let us consider, for simplicity, the case of $\mathfrak{g}=su(n)$. A cyclic grading may be constructed as follows\footnote{Here we restrict ourselves to the grading of type $A_{n-1}^{(1)}$.}: one picks a system of $n-1$ simple positive roots $\alpha_1, \ldots \alpha_{n-1}$, as well as the maximal negative root $\alpha_{n}=-\alpha_1-\ldots-\alpha_{n-1}$\footnote{In the paper of Kac \cite{Kac} the roots $\alpha_1, \ldots \alpha_{n}$ are seen as the positive simple roots of the corresponding affine Lie algebra $\widehat{A}_{n-1}$. Consider the case $n=3$. The simple positive roots of the loop algebra $su(3)(t, t^{-1})$ may be chosen as follows: $$\alpha_1=\left( \begin{array}{ccc}
0 & 1 & 0 \\
0 & 0 & 0 \\
0 & 0 & 0  
  \end{array} \right), \;\;\;\alpha_2=\left( \begin{array}{ccc}
0 & 0 & 0 \\
0 & 0 & 1 \\
0 & 0 & 0  
  \end{array} \right), \;\;\;\alpha_0=t\,\left( \begin{array}{ccc}
0 & 0 & 0 \\
0 & 0 & 0 \\
1 & 0 & 0  
  \end{array} \right)\,.$$ In this context the latter root $\alpha_0$ -- the analog of $\alpha_{n}$ -- is customarily called `imaginary'. In fact, the whole theory of cyclic Lie algebra gradings is formulated by Kac naturally in terms of affine Lie algebras and their Dynkin diagrams.}. Then one assigns to these $n$ roots arbitrary (non-negative integer) gradings $m_1, \ldots m_{n-1}, m_{n}$. The gradings of all other roots are determined by the Lie algebra structure, and the value of $m$ is calculated as
\bea\label{msum}
m=m_1+\ldots+m_{n}\,.
\eea
In usual matrix form, this grading looks as follows:
\bea\label{gradingM}\left( \begin{array}{C{0.8cm}C{0.8cm}C{0.8cm}C{0.8cm}C{0.8cm}}
0 & $m_1$ & & &\\
 & 0 & $m_2$ & &\\
 & & 0 & $\ddots$ & \\
  & &   & 0 & $m_{n-1}$  \\
$m_{n}$ & & & & 0 \end{array} \right)\eea
The subalgebra $\mathfrak{g}_0$, which determines the denominator $H$ of the quotient space $G/H$, is determined by those $m_i$'s, which are zero. For example, if all $m_i>0$, the resulting space is the manifold of complete flags $SU(n)\over S(U(1)^{n})$.

In general, for a choice of grading determined by the set $m_1, \ldots, m_{n}$ some of the subspaces $\mathfrak{g}_i$ will be identically zero. Therefore a natural restriction to adopt is to require that $\mathfrak{g}_i\neq 0$ for all $i \;(\mathrm{mod}\; m)$. We will call such a grading \emph{admissible}. This still leaves a wide range of possibilities. For example, in the case of $SU(3)$ the following is a complete list of admissible gradings (up to the action of the Weyl \mbox{group $S_3$):}
\bear
&&\mathbb{Z}_2:\;\;
\left( \begin{array}{ccc}
0 & \mathbf{0} & 1 \\
0 & 0 & \mathbf{1} \\
\mathbf{1} & 1 & 0  
  \end{array} \right), \quad \mathbb{Z}_3:\;\;
\left( \begin{array}{ccc}
0 & \mathbf{1} & 2 \\
2 & 0 & \mathbf{1} \\
\mathbf{1} & 2 & 0  
  \end{array} \right), \quad \left( \begin{array}{ccc}
0 & \mathbf{0} & 1 \\
0 & 0 & \mathbf{1} \\
\mathbf{2} & 2 & 0  
  \end{array} \right),\\ && 
    \mathbb{Z}_4:\;\; \left( \begin{array}{ccc}
0 & \mathbf{1} & 2 \\
3 & 0 & \mathbf{1} \\
\mathbf{2} & 3 & 0  
  \end{array} \right), \quad 
 \mathbb{Z}_5:\;\;\left( \begin{array}{ccc}
0 & \mathbf{1} & 3 \\
4 & 0 & \mathbf{2} \\
\mathbf{2} & 3 & 0  
  \end{array} \right),\quad \left( \begin{array}{ccc}
0 & \mathbf{1} & 2 \\
4 & 0 & \mathbf{1} \\
\mathbf{3} & 4 & 0  
  \end{array} \right),\\
&& \mathbb{Z}_6:\;\; \left( \begin{array}{ccc}
0 & \mathbf{1} & 3 \\
5 & 0 & \mathbf{2} \\
\mathbf{3} & 4 & 0  
  \end{array} \right),\quad   \mathbb{Z}_7:\;\; \left( \begin{array}{ccc}
0 & \mathbf{1} & 3 \\
6 & 0 & \mathbf{2} \\
\mathbf{4} & 5 & 0  
  \end{array} \right)
  \eear
The $\mathbb{Z}_2$-grading and the second $\mathbb{Z}_3$-grading correspond to the homogeneous space $\SU(3)/S(\tU(2)\times \tU(1))=\CP^2$, and all other gradings correspond to the flag manifold $\mathcal{F}_{1,1,1}$.

We will now give an answer to the question posed above: what is the relation between the $\sigma$-models with the action (\ref{actionY}), taken for different gradings on the corresponding Lie algebra? Our statement is~\cite{BykovCyclic}:

 \vspace{0.3cm}\noindent \hspace{0.3cm}
\fbox{\parbox{14cm}{
\centering 
For homogeneous spaces of the unitary group, the models~(\ref{actionY})-(\ref{omegaY}) with different $A_{n-1}^{(1)}$-type gradings on $\mathfrak{g}$ are classically equivalent
 to the model~(\ref{action}) with some choice of complex structure on the target-space
}
}

\vspace{0.3cm}
In fact, one has a precise statement about the relation of the $B$-fields in the two models. To formulate it, we `solve' the constraint (\ref{msum}) as follows\footnote{Formula (\ref{inneraut}) implies that the cyclic automorphism $\widehat{\sigma}$ of the Lie algebra, which defines the $\mathbb{Z}_m$ grading, can be represented as follows: $\widehat{\sigma}(a)=\sigma a \sigma^{-1}$, where $\sigma=\mathrm{diag}(e^{2\pi i \,\frac{q_1}{m}}, \ldots, e^{2\pi i \,\frac{q_{n}}{m}})$.}:
\bea\label{inneraut}
m_k=q_k-q_{k+1}\,,
\eea
where $q_k$ are integers and $q_{n+1}\equiv q_1-m$. We then have (see~\cite{BykovCyclic} for a proof):
\bea
\tilde{\omega}=\omega-2\, \sum\limits_{i=1}^{n} \,\frac{q_i}{m}\,dJ_{ii}\,,
\eea
where $J_{ii}$ are the diagonal components of the Maurer-Cartan current. We see that, irrespective of the choice of the grading (which is now encoded in the integers~$q_i$), the form $\tilde{\omega}$ differs from the K\"ahler form by a topological term. This topological term, clearly, depends on the chosen grading, but does not contribute to the equations of motion.

Although the flag manifolds (\ref{flagunitary}) are $\mathbb{Z}_m$-graded spaces, the two classes of target spaces -- $\mathbb{Z}_m$-graded and complex homogeneous spaces -- do not coincide. For example, one has the space $\frac{G_2}{SU(3)}\simeq S^6$. The stability subgroup $SU(3)$ acts on the tangent space $\mathfrak{m}=\mathbb{R}^6$ via $V\oplus \bar{V}$, where $V\simeq \CC^3$ is the standard representation. Therefore it has a unique almost complex structure, which is \emph{not} integrable (see the review~\cite{AgricolaS6}). On the other hand, it is a nearly K\"ahler manifold and is $\mathbb{Z}_3$-graded~\cite{Butruille}. On the other side of the story, one has the complex manifold $S^1\times S^3\simeq U(2)$ (see~\cite{BykovCyclic} for a discussion), which may be viewed as a $\mathbb{T}^2$-bundle over $\CP^1$ (the simplest flag manifold). This manifold is \emph{not} a $\mathbb{Z}_m$-graded homogeneous space of the group ${G=U(2)}$. 

We also note that the construction of Lax connections for models with $\mathbb{Z}_m$-graded spaces was explored in~\cite{BeisertLucker}. The relation to the Lax connections of section~\ref{complzerocurv} has been recently established in~\cite{DelducKameyama} (this is an extension to $\mathbb{Z}_m$ of our discussion in section~\ref{symmspace} regarding symmetric spaces). The fact that the integrals of motion of the models are in involution was proven, for instance, in~\cite{LacroixPhD}.

\end{document}